\documentclass[prl,aps,amssymb,amsmath]{article}
\usepackage{amssymb,amsmath,tikz}
\usepackage{amsthm}
\usepackage{mathrsfs}

\DeclareMathOperator*{\esssup}{ess\,sup}

\DeclareMathOperator*{\Tr}{Tr \,}

\DeclareMathOperator*{\cl}{cl \,}
\DeclareMathOperator*{\intt}{int \,}

\renewcommand{\qedsymbol}{$\blacksquare$}

\newtheorem{twr}{THEOREM}
\newtheorem{lem}{LEMMA}
\newtheorem{defin}{DEFINITION}
\newtheorem{rem}{REMARK}
\newtheorem{cor}{COROLLARY}
\newtheorem{ex}{EXAMPLE}

\begin{document}
\title{{\bf A Generalization of Mackey's Theory of Induced Representations}}
\author{Jaros{\l}aw Wawrzycki \footnote{Electronic address: jaroslaw.wawrzycki@wp.pl or jaroslaw.wawrzycki@ifj.edu.pl}
\\Institute of Nuclear Physics of PAS, ul. Radzikowskiego 152, 
\\31-342 Krak\'ow, Poland}
\maketitle

\newcommand{\ud}{\mathrm{d}}

\vspace{1cm}

\begin{abstract}
In this work we extend the Mackey's theory of induced unitary representations
on a wide class of Krein-isometric induced representations in Krein spaces. The subgroup theorem and the Kronecker product theorem are shown to be valid for the induced representations of this class. Among the class of representations which are subsumed by this extension there are all the representations acting in the single particle Krein-Hilbert space and in Fock-Krein spaces of the mass less gauge free fields underling the Standard Model in the local gauge BRST formalutaion, e. g. of the electromagnetic potential field.

\end{abstract}

\tableofcontents

\section{Introduction}

In this work we present a generalization of Mackey's theory of induced unitary representations. 
The generalization lies in extension of Mackey's theory on a wide class of Krein isometric representations of a separable topological group in Krein spaces, which are induced by Krein-unitary representations of the closed subgroup. Krein space is understood as the ordinary Hilbert 
space endowed with a unitary involutive operator $\mathfrak{J}$: $\mathfrak{J}^* = \mathfrak{J}, \mathfrak{J}^2 = \boldsymbol{1}$. Operator acting in the Krein space is called Krein-isometric if it is defined on a dense domain $\mathfrak{D}$, and is closable and invertible on this domain and preserves the Krein-inner product $(\mathfrak{J} \cdot, \cdot )$, where $(\cdot, \cdot)$ is the ordinary Hilbert space inner product (positive definite). Representation is called Krein-isometric in case there exists a common dense domain $\mathfrak{D}$ for all representors of the representation, and all representors are closable Krein-isomeric operators on the common domain. Krein isometric representation of a separable topological group is called Krein-unitary if all representors are (separately) bounded, with almost uniform bound (with resect to the Hilbert space norm and topology of the group). The point is that the  Krein-isometric representations we encounter in practice are induced by Krein-unitary representations of the subgroups. Perhaps the most important examples include all representaions acting in the single particle Hilbert spaces as well as in the Fock spaces of the fundamental free quantum fields, such as the free electromagnetic potential field (in the local Lorentz gauge) or the gauge 
mass less free fields underlying the Standard Model (in the BRST approach). We give the formulation
and proof of the subgroup theorem and of the Kronecker product theorem for the induced 
Krein-isometric representations in the Krein space. We have been trying, whenever possible, 
to preserve the original Mackey's proof of the said theorems, but because our representations
are defined through unbounded operators (say representors), the proof cannot be exactly the same
as that presented by Mackey. 

Presented extension of his theory allows us to reduce
the problem of decomposition of the tensor product of induced Krein-isometric representations to an essentially geometric problem concerned with the coset and double coset spaces and to the Fubini theorem, at least for the wide range of topological groups which allow decomposition into product spaces of the subgroup and the coset and double coset spaces. 

In case of the representations we encouner in Quantum Field Theory (QFT), where the groups are Lie groups, the common domain $\mathfrak{D}$ has the reach linear-topological structure of the so-called standard nuclear space. In this case we have much more effective tools at our disposal, and the achieved decompositions can be given the meaining of generalized spectral decompositions in the sense of Neumark-Lanze, with respect to the generalized spectral measures. The same can be said of differential operators associated with pseudo-riemannian structures with non trivial symmetry groups where the presented generalization can also be applied to the spectral analysis of these operators.

Although theoretical physicists believe that such an extension of Mackey theory should exist,
it has never been worked out explicitly. In particular besides the claims 
(which we give in the furher part of this work) which conjecture existence of such an extension, it was also stressed the lack of rigorous construction of the most important free fields 
which would be based on the representation theory, compare e. g. \cite{Haag}. 
Our work allows, among other things, to complete this lack \cite{wawrzyckiQFT}. Of course 
Quantum Field Theory is not the only domain of potential applicability of the presented extension
of Mackey's theory, but any analytic situation involving preservation of a not necessary positive definite inner product with non trivial symmetry group, will be closely related to our work.

\section{Preliminaries}\label{pre}

It should be stressed that the analysis we give here is inapplicable for general linear spaces with indefinite inner product. We are concerned with non-degenerate, decomposable and complete inner product spaces in the terminology of 
\cite{Bog}, which have been called Krein spaces in \cite{DutFred}, \cite{Stro}, \cite{Bog} and \cite{Wawr} for the reasons we explain below. They emerged naturally in solving physical problems concerned with quantum mechanics (\cite{Dirac}, \cite{Pauli}) and quantum field theory (\cite{Gupta}, \cite{Bleuler}) in quantization of electromagnetic field and turned up generally to be very important (and even seem indispensable) in construction of quantum fields with non-trivial gauge freedom. Similarly we have to emphasize that we are not dealing
with general unitary (i. e. preserving the indefinite inner product in Krein space) representations of the double cover 
$\mathfrak{G} = T_{4} \circledS SL(2, \mathbb{C})$ of the Poincar\'e group, but only with the exceptional representations
of {\L}opusza\'nski-type, which naturally emerge in construction of the free photon field, which have a rather exceptional structure of induced representations, and allow non-trivial analytic constructions of tensoring and decomposing, which 
is truly exceptional among Krein-unitary (preserving the indefinite product) representations in Krein spaces.

The non-degenerate, decomposable and complete indefinite inner product space $\mathcal{H}$, hereafter called \emph{Krein space}, may equivalently be described as an ordinary Hilbert space $\mathcal{H}$ with an ordinary strictly positive inner product $(\cdot, \cdot)$, together with a distinguished self-adjoint (in the ordinary Hilbert space sense) fundamental symmetry (Gupta-Bleuler operator) $\mathfrak{J} = P_+ - P_- $, where $P_+$ and $P_-$ are ordinary self-adjoint (with respect to the Hilbert space inner product $(\cdot, \cdot)$)) projections such that their sum is the identity operator: $P_+ + P_- = I$. The indefinite inner product is given by $(\cdot, \cdot)_{\mathfrak{J}} = (\mathfrak{J} \cdot, \cdot) = (\cdot, \mathfrak{J} \cdot)$. 
Recall that in our previous paper \cite{Wawr} the indefinite product was designated by $(\cdot, \cdot)$ and the ordinary Hilbert space inner product associated with the fundamental symmetry $\mathfrak{J}$ was designated by $(\cdot, \cdot)_{\mathfrak{J}}$. The indefinite and the associated definite inner product play symmetric roles in the sense that one may start with a fixed indefinite inner product in the Krein space and construct the Hilbert space associated with an admissible fundamental symmetry, or vice versa: one can start with a fixed Hilbert space and for every fundamental symmetry construct the indefinite inner product in it, both approaches are completely equivalent provided the fundamental symmetry being admissible (in the sense of \cite{Stro}) and fixed. We hope the slight change of notation will not cause any serious misunderstandings and is introduced because our analytical arguments will be based on the ordinary Hilbert space properties, so will frequently refer to the standard literature on the subject, so we designated the ordinary strictly definite inner product by $(\cdot, \cdot)$ which is customary.

Let an operator $A$ in $\mathcal{H}$ be given. The operator $A^{\dag}$ in $\mathcal{H}$ is called Krein-adjoint
of the operator $A$ in $\mathcal{H}$ in case it is adjoint in the sense of the indefinite inner product:   
$(Ax, y)_{\mathfrak{J}} = (\mathfrak{J} Ax, y) = (\mathfrak{J} x, A^{\dag}y) = (x, A^{\dag}y)_{\mathfrak{J}}$
for all $x, y \in \mathcal{H}$, or equivalently $A^{\dag} = \mathfrak{J} A^{*}\mathfrak{J}$, where $A^{*}$ is the ordinary adjoint operator with respect to the definite inner product $(\cdot, \cdot)$. The operator $U$ and its inverse $U^{-1}$ isometric with respect to the indefinite product $(\cdot, \cdot)_{\mathfrak{J}}$, e. g. 
$(Ux, Uy)_{\mathfrak{J}} = (x, y)_{\mathfrak{J}}$ for all $x, y \in \mathcal{H}$ (same for $U^{-1}$),
equivalently $UU^{\dag} = U^{\dag}U = I$, will also be called 
unitary (sometimes $\mathfrak{J}$-unitary or Krein-unitary) trusting to the context or explanatory remarks to make clear what is meant in each instance: unitarity for the indefinite inner product or the ordinary unitarity for the strictly definite Hilbert space inner product.

In particular we may consider $\mathfrak{J}$-\emph{symmetric} representations $x \mapsto A_x$ of involutive algebras,
i. e. such that $x^{*} \mapsto {A_x}^{\dag}$, where $(\cdot) \mapsto (\cdot)^{*}$ is the involution in the algebra in question. A fundamental role for the spectral analysis in Krein spaces is likewise played by commutative (Krein) self-adjoint, or $\mathfrak{J}$-symmetric weakly closed subalgebras. However their structure is far from being completely described, with the exception of the special case when the rank of $P_+$ or $P_-$ is finite dimensional (here the analysis is complete and was done by Neumark). Even in this particular case a unitary representation of a separable locally compact group in the Krein space, although reducible, may not in general be decomposable, compare \cite{Neumark1, Neumark2, Neumark3, Neumark4}.

In case the dimension of the rank 
$\mathcal{H}_+ = P_{+} \mathcal{H}$ or $\mathcal{H}_{-} = P_{-} \mathcal{H}$ of $P_+$ or $P_-$ is finite we get the spaces analysed by Pontrjagin, Krein and Neumark, compare e. g. \cite{Pontrjagin}, \cite{Krein} and the literature in \cite{Bog}.

The circumstance that the Krein space may be defined as an ordinary Hilbert $\mathcal{H}$ space with a distinguished non-degenerate fundamental symmetry (or Gupta-Bleuler operator) $\mathfrak{J} = \mathfrak{J}^{*}$, $\mathfrak{J}^2 = I$ in it , say a pair $(\mathcal{H}, \mathfrak{J})$, allows us to extend the fundamental
analytical constructions on a wide class of induced Krein-isometric representations of $\mathfrak{G} = T_{4} \circledS SL(2, \mathbb{C})$  in Krein spaces. In particular we may define a Krein-isometric representation of $T_{4} \circledS SL(2, \mathbb{C})$ induced by a Krein-unitary representation of a subgrup $H$ corresponding to a particular class of 
$SL2,\mathbb{C})$-orbits  on the dual group $\widehat{T_4}$ of $T_4$ (in our case we consider the class corresponding to the representation with the spectrum of the four-momenta concentrated on the ``light cone'') 
word for word as in the ordinary Hilbert space by replacing the representation of the subgroup $H$ by a Krein-unitary representation $L$ in a Krein space $(\mathcal{H}_L, \mathfrak{J}_L)$. This leads to a Krein-isometric representation 
$U^L$ in a Krein space $(\mathcal{H}^L , \mathfrak{J}^L)$ (see Sect. \ref{def_ind_krein}). Application of  
Lemma \ref{lop_ind_1}, Section \ref{lop_ind}, leads to the ordinary direct integral $\mathcal{H} = \int_{\mathfrak{G}/H} \, \mathcal{H}_q \, d\mu_{\mathfrak{G}/H}(q)$ of Hilbert spaces $\mathcal{H}_q = \mathcal{H}_L$ over the coset measure space $\mathfrak{G}/H$ with the measure induced by the Haar measure on $\mathfrak{G}$. One obtains in this manner the Krein space $(\mathcal{H}, U^{-1}\mathfrak{J}^L U)$ given by the ordinary Hilbert space $\mathcal{H}$ equal to the  above mentioned direct integral of the ordinary Hilbert spaces 
$\mathcal{H}_q$ all of them equal to $\mathcal{H}_L$ together with the fundamental symmetry $ \mathfrak{J} = U^{-1}\mathfrak{J}^L U$ equal to the ordinary direct integral $\int_{\mathfrak{G}/H} \, \mathfrak{J}_q \, d\mu_{\mathfrak{G}/H}(q)$ of fundamental symmetries $\mathfrak{J}_q = \mathfrak{J}_L$ as operators in $\mathcal{H}_q = \mathcal{H}_L$ and with the  representation $U^{-1}U^LU$ of $\mathfrak{G}$ in the Krein space $(\mathcal{H}, \mathfrak{J})$ (and  $U$ given by a completely analogous formula as that in Lemma \ref{lop_ind_1} of Section \ref{lop_ind}) of Wigner's form  
\cite{Wigner_Poincare} (imprimitivity system). 

This is the case for the indecomposable (although reducible) representation of 
$\mathfrak{G} = T_{4} \circledS SL(2, \mathbb{C})$ constructed by 
{\L}opusza\'nski with $H = T_4 \cdot G_\chi$, with the "small" subgroup $G_\chi \cong \tilde{E_2}$ of $SL(2, \mathbb{C})$ corresponding to the ``light-cone'' orbit in the spectrum of four-momenta operators. One may give to it the form of  representation $U^{-1}U^L U$ equivalent to an induced representation $U^L$, because the representors of the normal factor (that is of the translation subgroup $T_4$) of the semidirect product $T_{4} \circledS SL(2, \mathbb{C})$ as well as their generators, i. e. four-momentum operators $P_0, \ldots , P_3$, commute with the fundamental symmetry $\mathfrak{J} = \int_{\mathfrak{G}/H} \, \mathfrak{J}_q \, d\mu_{\mathfrak{G}/H}(q)$, so that all of them are not only $\mathfrak{J}$-unitary but unitary with respect to the ordinary Hilbert space inner product (and their generators $P_0, \ldots, P_3$ are not only Krein-self-adjoint but also self-adjoint in the ordinary sense with respect to the ordinary definite inner product of the Hilbert space $\mathcal{H}$), so the algebra generated by $P_0, \ldots , P_3$ leads to the ordinary direct integral decomposition with the decomposition corresponding to the ordinary spectral measure, contrary to what happens for general Krein-selfadjoint commuting operators in Krein space $(\mathcal{H}, \mathfrak{J})$ (for details see Sect. \ref{lop_ind}). This in case of 
$\mathfrak{G} = T_{4} \circledS SL(2, \mathbb{C})$, gives to the representation 
$U^{-1}U^L U$ of $\mathfrak{G}$ the form of Wigner \cite{Wigner_Poincare} (viz. a
\emph{system of imprimitivity} in mathematicians' parlance) with the only difference that $L$ is not unitary but 
Krein-unitary in $(\mathcal{H}_L , \mathfrak{J}_L)$.   

Another gain we have thanks to the above mentioned circumstance is that we can construct tensor product 
$(\mathcal{H}_1, \mathfrak{J}_1) \otimes (\mathcal{H}_2, \mathfrak{J}_2)$ of Krein spaces
$(\mathcal{H}_1, \mathfrak{J}_1)$ and $(\mathcal{H}_2, \mathfrak{J}_2)$ as $(\mathcal{H}_1 \otimes \mathcal{H}_2, \mathfrak{J}_1 \otimes \mathfrak{J}_2)$ where in the last expression we have the ordinary tensor products of Hilbert spaces and operators in Hilbert spaces (compare Sect. \ref{kronecker}).

Similarly having any two such ($\mathfrak{J}_1$- and  $\mathfrak{J}_2$-)isometric representations $U^L$ and $U^M$ induced by ($\mathfrak{J}_L$ and $\mathfrak{J}_M$-unitary) representations $L$ and $M$ of subgroups $G_1$ and $G_2$  in Krein spaces $(\mathcal{H}_L, \mathfrak{J}_L)$ and $(\mathcal{H}_M, \mathfrak{J}_M)$ respectively we may construct the tensor product $U^L \otimes U^M$ of Krein-isometric representations in the tensor product Krein space $(\mathcal{H}_1, \mathfrak{J}_1) \otimes (\mathcal{H}_2, \mathfrak{J}_2)$, which is likewise ($\mathfrak{J}_1 \otimes \mathfrak{J}_2$-)isometric. It turns out that the Kronecker product  $U^L \times U^M$ and $U^{L \times M}$ are (Krein-)unitary equivalent (see Sect. \ref{kronecker})
as representations of $\mathfrak{G} \times \mathfrak{G}$. Because the tensor product $U^L \otimes U^M$ as a representation of $\mathfrak{G}$ is the restriction of the Kronecker product $U^L \times U^M$ to the diagonal subgroup of 
$\mathfrak{G} \times \mathfrak{G}$ we may analyse the representation $U^L \otimes U^M$ by analysing the restriction of the induced representation $U^{L \times M}$ to the diagonal subgroup exactly as in the Mackey theory of induced representations in Hilbert spaces.  Although in general for $\mathfrak{J}_1 \otimes \mathfrak{J}_2$-unitary representations in Krein space $(\mathcal{H}_1, \mathfrak{J}_1) \otimes (\mathcal{H}_2, \mathfrak{J}_2)$ ordinary decomposability breaks down, we can nonetheless still decompose the representation $U^{L \times M}$ restricted to the diagonal into induced representations which, by the above mentioned  Krein-unitary equivalence, gives us a decomposition of the tensor product representation 
$U^L \otimes U^M$  of $\mathfrak{G}$.  Indeed, it turns out that the whole argument of Mackey \cite{Mackey} preserves its validity and effectiveness in the construction of  decomposition of tensor product of induced representations for the case in which the representations $L$ and $M$ of the subgrups $G_1$ and $G_2$ are replaced with (specific) unitary (or $\mathfrak{J}_L$- and $\mathfrak{J}_M$-unitary) representations in Krein spaces $(\mathcal{H}_L, \mathfrak{J}_L )$ and $(\mathcal{H}_M, \mathfrak{J}_M )$ respectively. We give details on the subject below in Section \ref{subgroup}.  Because the 
{\L}opusza\'nski representation is (Krein-unitary equivalent to) an induced representation in a Krein space (Sect. \ref{lop_ind}), we can decompose the tensor product of {\L}opusza\'nski representations. The specific property
of the group $T_{4} \circledS SL(2, \mathbb{C})$ is that  this decomposition  may be performed explicitly into indecomposable sub-representations.  

The Krein-isometric induced representations of $T_4 \circledS SL(2, \mathbb{C})$ 
which we describe here cover all representations important
for QFT. All the representations which act on single particle states of local fields (including zero mass gauge fields) have three important properties: 1) They are strongly continuous on a common dense invariant subdomain. 2) Translations commute with the fundamental symmetry
$\mathfrak{J}$, so that translations are unitary with respect to the Hilbert space inner product as well as are Krein-unitary, and thus compose ordinary (strongly continuous) unitary representation of the translation subgroup. 3) The representations are ``locally'' bounded with respect to the joint spectrum of translation generators. Let us formulate the requirement of ``local boundedness'' more precisey. Let us denote the translation representor $U_{(a,1)}$ just by $T(a)$ and the 
representor $U_{(0,\alpha)}$ of the 
$SL(2, \mathbb{C})$ subgroup just by $U(\alpha)$. 
Let $P^0 , \ldots , P^3$ be the respective generators of the translations (they do exist by the strong continuity assumption posed on the Krein-isometric representation -- physicist's everyday computations involve the generators 
and thus our assumption is justified).  Let $\mathcal{C}$ be the commutative $C^*$-algebra generated 
by the functions $f(P^0 , \ldots ,P^3)$ of translation generators $P^0 , \ldots , P^3$,
where $f$ is continuous on $\mathbb{R}^4$ and vanishes at infinity. Let 
\begin{equation}\label{dec-sp-P}
\mathcal{H} = \int \limits_{\textrm{Spec} \, (P^0 , \ldots , P^3)} \mathcal{H}_{p} \, \ud \mu (p)
\end{equation} 
be the direct integral decomposition of $\mathcal{H}$ corresponding to the algebra $\mathcal{C}$
(in the sense of \cite{von_neumann_dec} or \cite{Segal_dec_I}) with a spectral measure
$\mu$ on the joint spectrum $\textrm{Spec} \, (P^0 , \ldots , P^3)$ of the translation generators.
The requirement of ``local boundedness'' of the Krein-isometric representations of the semi-direct products $T_4 \circledS SL(2, \mathbb{C})$ which are preserved by the representations acting in the single particle and inFock spaces of free fields (of realistic QFT) is the following. 
The restriction $U(\alpha)$, $\alpha \in SL(2, \mathbb{C})$ to the second factor $SL(2, \mathbb{C})$ is locally bounded with respect to the direct integral decomposition (\ref{dec-sp-P}) of the Hilbert space $\mathcal{H}$,
determined by the restriction $T(a)$, $a \in G_1$, of the representation of $T_4 \circledS SL(2, \mathbb{C})$ to the abelian normal factor $T_4$. More precisely: let $\| \cdot \|$
be the ordinary Hilbert space $\mathcal{H}$ norm, then for every compact subset $\Delta$ of the 
dual $\widehat{G_1}$ and every $\alpha \in G_2$ there exists a positive constant 
$c_{\Delta, \alpha}$ (possibly depending on $\Delta$ and $\alpha$) such that
\[
\|U(\alpha) f \| < c_{\Delta, \alpha} \|f \|,
\]   
for all $f\in \mathcal{H}$ whose spectral support (in the spectral decomposition with respect to translation generators) is contained within the compact set $\Delta$.

Nonetheless the relevant representations, or the associated imprimitivity systems (e.g. {\L}opusza\'nski representation) are unbounded, and require a special care in the correct definition of the Kronecker product and moreover contain analytic subtleties which could have been omitted in the original Mackey theory.
The other difference in comparison to the original Mackey theory is that we exploit (and prove)
a decomposition/disintegration theorem for measures which are not finite, which makes the proof 
longer in comparison to Mackey's proof. In principle we could have confine ourselves after Mackey
to decomposition of finite measures (easier). However the representations encountered in QFT are naturally related to Poincar\'e invariant measures which are not finite. Avoiding them by utilizing finite measures would not be very economical for a physicist, because in further computations he had to recover then the ``Clebsch-Gordan'' coefficients relating obtained decompositions to the original representations naturally connected with infinite invariant measures.

\begin{rem}
Let us emphasize that here ``continuity'', ``density'', ``boundedness'', and other standard analytic notions,
as the ``closure of a densely defined operator'' or ``weak''or ``strong'' topologies in the algebra of bounded operators,  
refer to the ordinary Hilbert space norm and definite Hilbert space inner product in $\mathcal{H}$ of the Krein space 
$(\mathcal{H}, \mathfrak{J})$ in question. We are mainly concerned with the Lie group 
$\mathfrak{G} = T_{4} \circledS SL(2, \mathbb{C})$ but the general theory of induced representations in Krein spaces presented here is valid for general separable locally compact topological groups $\mathfrak{G}$. Thus separability and local compactness of $\mathfrak{G}$ is assumed to be valid throughout the whole paper whenever the identification
$\mathfrak{G} = T_{4} \circledS SL(2, \mathbb{C})$ is not explicitly stated.
\label{pre.1}
\end{rem}

\section{Definition of the induced representation $U^L$ in Krein 
space $(\mathcal{H}^L , \mathfrak{J}^L )$}\label{def_ind_krein}

Here by a Krein-unitary and strongly continuous representation $L: G \ni x \mapsto L_x$ of a separable 
locally compact group $G$ we shall mean a homomorphism of $G$ into the group of all (Krein-)unitary transformations of some separable Krein space $(\mathcal{H}_L, \mathfrak{J}_L)$ (i. e. with separable Hilbert space $\mathcal{H}_L$) onto itself which is:

\begin{enumerate}

\item[(a)]

Strongly continuous: for each $\upsilon \in \mathcal{H}_L$ the function $x \mapsto L_x \upsilon$ is continuous 
with respect to the ordinary strictly definite Hilbert space norm $\| \upsilon \| = \sqrt{(\upsilon, \upsilon)}$
in $\mathcal{H}_L$.

\item[(b)]
Almost uniformly bounded: there exist a compact neighbourhood $V$ of unity $e \in G$ such that the set
$\| L_{x} \|$, $x \in V \subset G$ is bounded or, what is the same thing, that
the set $\| L_{x} \|$ with $x$ ranging over a compact set $K$ is bounded for every compact 
subset $K$ of $G$. 

\end{enumerate}
Because the strong operator topology in $\mathcal{B}(\mathcal{H}_L)$ is stronger than the weak operator topology then for each $\upsilon , \varphi \in \mathcal{H}_L$ the function $x \mapsto (L_x \upsilon , \varphi )$ is continuous on $G$. One point has to be noted: because the range and domain of each $L_x$ equals 
$\mathcal{H}_L$, which as a Krein space $(\mathcal{H}_U, \mathfrak{J}_U)$ is closed and non-degenerate, then by Theorem 3.10 of \cite{Bog} each $L_x$ is continuous i. e. bounded with respect to the Hilbert space norm $\| \cdot \|$ in 
$\mathcal{H}_L$, and each $L_x$ indeed belongs to the algebra $\mathcal{B}(\mathcal{H}_L)$
of bounded  operators in the Hilbert space $\mathcal{H}_L$ (which is non-trivial as an $\mathfrak{J}_L$-isometric densely defined operator in the Krein space $(\mathcal{H}_L, \mathfrak{J}_L)$ may be discontinuous, as we will see in this Section, compare also \cite{Bog}). We also could immediately refer to a theorem which says that Krein-unitary operator is continuous, i. e. Hilbert-space-norm bounded (compare Theorem 4.1 in \cite{Bog}).

Besides in this paper will be considered a very specific class of Krein-isometric representations $U$ of $\mathfrak{G}$ in Krein spaces, to which the induced representations of $\mathfrak{G}$ in Krein spaces, hereby defined, belong. Namely
here by a Krein-isometric and strongly continuous representation  of a separable 
locally compact group $\mathfrak{G}$ we shall mean a homomorphism $U: \mathfrak{G} \ni x \mapsto U_x$ of $\mathfrak{G}$ into a group of Krein-isometric and closable operators of some separable Krein space $(\mathcal{H}, \mathfrak{J})$
with dense common domain $\mathfrak{D}$ equal to their common range in $\mathcal{H}$ and such that  
\begin{enumerate}

\item[]

$U$ is strongly continuous on the common domain $\mathfrak{D}$: for each $f \in \mathfrak{D} \subset \mathcal{H}$ the function $x \mapsto U_x f$ is continuous with respect to the ordinary strictly definite Hilbert space norm 
$\| f \| = \sqrt{(f, f)}$ in $\mathcal{H}$.  

\end{enumerate}

Let $H$ be a closed subgroup of a separable locally compact group $\mathfrak{G}$. In the applications
we have in view the right $H$-cosets, i. e. elements of $\mathfrak{G}/H$, which are exceptionally regular, and have a ``measure product property''. This is e. g. the case we encounter in decomposing tensor products of the representations of the double cover 
$\mathfrak{G}$ of the Poincar\'e group in Krein spaces encountered in QFT. 
Namely every element (with a possible exception of a subset of $\mathfrak{G}$ of 
Haar measure zero) $\mathfrak{g} \in \mathfrak{G}$ can be uniquely represented as a product
$\mathfrak{g} = h \cdot q$, where $h \in H$ and $q \in Q \cong \mathfrak{G}/H$ with a subset $Q$ 
of $\mathfrak{G}$ which is not only measurable but, outside a null set, is a sub-manifold of 
$\mathfrak{G}$, such that $\mathfrak{G}$ is the product 
$H \times \mathfrak{G}/H$ measure space, with the regular Baire measure space structure on $\mathfrak{G}/H$ 
associated to the canonical locally compact topology on $\mathfrak{G}/H$ induced by the natural
projection $\pi: \mathfrak{G} \mapsto \mathfrak{G}/H$ and with the ordinary right Haar measure space
structure $(H, \mathscr{R}_H, \mu_H)$ on $H$, which is known to be regular with the ring 
$\mathscr{R}_H$ of Baire sets. (We will need the complete measure spaces on $\mathfrak{G}, \mathfrak{G}/H$ 
but the Baire measures are sufficient to generate them by the Carath\'eodory method, 
because we have assumed the topology on $\mathfrak{G}$ to fulfil the second axiom
of countability.) In short   
$(\mathfrak{G}, \mathscr{R}_{\mathfrak{G}}, \mu) = 
(H \times \mathfrak{G}/H,\mathscr{R}_{H \times Q}, \mu_H \times \mu_{\mathfrak{G}/H} )$.   
In our applications we are dealing with pairs $H \subset \mathfrak{G} $ of Lie subgroups of 
the double cover $T_{4} \circledS SL(2, \mathbb{C})$ of the Poincar\'e group $\mathfrak{P}$ including the group 
$T_{4} \circledS SL(2, \mathbb{C})$ itself, with a sub-manifold structure of $H$ and $Q \cong \mathfrak{G}/H$.
This  opportunities allow us to reduce the analysis  of the induced representation $U^L$ in the Krein space
defined in this Section to an application of the Fubini theorem and to the von Neumann analysis of the direct integral
of ordinary Hilbert spaces. (The same assumption together with its analogue for the double cosets in $\mathfrak{G}$ simplifies also the  problem of decomposition of tensor products of induced representations of $\mathfrak{G}$ 
and reduces it mostly to an application of the Fubini theorem and harmonic analysis on the "small" subgroups: 
namely  at the initial stage we reduce the problem to the geometry of right cosets and 
double cosets with the observation that Mackey's theorem on Kronecker product and subgroup theorem of induced representations likewise work for the induced representations in Krein spaces defined here, and then apply the Fubini theorem and harmonic analysis on the "small" subgroups.). Driving by the physical examples we assume for a while 
that the ``measure product property'' is fulfilled by the right $H$-cosets in $\mathfrak{G}$. 
(We abandon soon this assumption so that our results, namely 
\emph{the subgroup theorem} and \emph{the Kronecker product theorem}, hold true for induced representations 
in Krein spaces, without this assumption.)    

Let $L$ be any ($\mathfrak{J}_L$-)unitary strongly continuous and almost uniformly bounded representation of $H$ 
in a Krein space $(\mathcal{H}_L , \mathfrak{J}_L )$. Let $\mu_H$ and $\mu_{\mathfrak{G}/H}$ be (quasi) invariant measures on $H$ and on the homogeneous space $\mathfrak{G}/H$ of right $H$-cosets in $\mathfrak{G}$ induced by the (right) Haar measure $\mu$ on $\mathfrak{G}$ by the ``unique factorization''. Let us denote by 
$\mathcal{H}^L$ the set of all functions $f: \mathfrak{G} \ni x \mapsto f_x$ from $\mathfrak{G}$ to $\mathcal{H}_L$ such that   

\begin{enumerate}

\item[(i)]
            $(f_x, \upsilon)$ is measurable function of $x \in \mathfrak{G}$ 
            for all $\upsilon \in \mathcal{H}_L$.    
               
\item[(ii)]
            $f_{hx} = L_h (f_{x})$ for all $h \in H$ and $x \in \mathfrak{G}$.

\item[(iii)] Into the linear space of functions $f$ fulfilling (i) and (ii) let us introduce the operator
             $\mathfrak{J}^L$ by the formula $(\mathfrak{J}^L f)_{x} = L_h \mathfrak{J}_L L_{h^{-1}} (f_{x})$, 
             where $x = h\cdot q$ is the unique decomposition 
             of $x \in  \mathfrak{G}$. Besides (i) and (ii) we require 
\[
\int \, (\, \mathfrak{J}_L ((\mathfrak{J}^L f)_x ), f_x \,) \, d\mu_{\mathfrak{G}/H} < \infty,
\]
             where the meaning of the integral is to be found in the fact that the integrand is constant on the right 
             $H$-cosets and hence defines a function on the coset space $\mathfrak{G}/H$.

\end{enumerate}
(Note that $L$ is written here as superscript! The $\mathcal{H}_L$ with the lower case of 
the index $L$ is reserved for the space of the representation $L$ of the subgroup $H \subset \mathfrak{G}$.)

Because every $x \in \mathfrak{G}$ has a unique factorization $x = h \cdot q$ with $h \in  H$ and 
$q \in Q \cong \mathfrak{G}/H$, then by ``unique factorization'' the functions $f \in \mathcal{H}^L$ as well as the functions $x \mapsto (f_x , \upsilon)$
with $\upsilon \in \mathcal{H}_L$, on $\mathfrak{G}$, may be treated as functions on the Cartesian product $H \times Q
\cong H \times \mathfrak{G}/H \cong \mathfrak{G}$. The axiom (i) means that the functions 
$(h,q) \mapsto (f_{h\cdot q}, \upsilon)$ for $\upsilon \in \mathcal{H}_L$ are measurable on the product measure space 
$(H \times \mathfrak{G}/H,\mathscr{R}_{H \times \mathfrak{G}/H}, \mu_H \times \mu_{\mathfrak{G}/H}) 
\cong (H \times Q,\mathscr{R}_{H \times Q}, \mu_H \times \mu_{\mathfrak{G}/H} )$. In particular let $W : q \mapsto W_q \in \mathcal{H}_L$ be a function on $Q$ such that $q \mapsto (W_q , \upsilon)$ is measurable with respect to the standard measure space $(Q, \mathscr{R}_Q , dq)$ for all $\upsilon \in \mathcal{H}_L$, and such that 
$\int \, (W_q, W_q ) \, d\mu_{\mathfrak{G}/H}(q) < \infty$. Then by the  analysis of \cite{von_neumann_dec} (compare also \cite{Neumark_dec}, \S 26.5) which is by now standard, the set of such functions $W$ (when functions equal almost everywhere are identified) compose the direct integral $\int \, \mathcal{H}_L \, d\mu_{\mathfrak{G}/H}(q)$ Hilbert space with the inner product $ (W, F) = \int \, (W_q, F_q ) \, d\mu_{\mathfrak{G}/H}(q)$. 
For every such $W \in \int \, \mathcal{H}_L \, d\mu_{\mathfrak{G}/H}(q)$ the function 
$(h,q) \mapsto f_{h\cdot q} = L_h W_q$ fulfils (i) and
(ii). (ii) is trivial. For each $\upsilon \in \mathcal{H}_L$ the function $(h,q) \mapsto (f_{h\cdot q}, \upsilon ) 
= (L_h W_q , \upsilon )$ is measurable on the product measure space $(H \times Q,\mathscr{R}_{H \times Q}, \mu_H \times \mu_{\mathfrak{G}/H} ) \cong (\mathfrak{G}, \mathscr{R}_{\mathfrak{G}}, \mu)$ because for any orthonormal basis $\{ e_n \}_{n \in \mathbb{N}}$ of the Hilbert space $\mathcal{H}_L$ we have:
\[
\begin{split}
(f_{h\cdot q}, \upsilon) = (f_{h\cdot q}, \mathfrak{J}_L \mathfrak{J}_L  \upsilon) 
= (L_h W_q, \mathfrak{J}_L \mathfrak{J}_L  \upsilon)
= (\mathfrak{J}_L L_h W_q,  \mathfrak{J}_L  \upsilon) \\
= (\mathfrak{J}_L W_q , L_{h^{-1}} \mathfrak{J}_L  \upsilon)
= \sum_{n \in \mathbb{N}} (\mathfrak{J}_L W_q ,e_n)(e_n , L_{h^{-1}} \mathfrak{J}_L  \upsilon)
\end{split}
\]
where each summand gives a measurable function 
$(h,q)$ ${}\mapsto{}$ $(\mathfrak{J}_L W_q ,e_n)(e_n , L_h \mathfrak{J}_L  \upsilon)$ on the product measure space 
$(H \times Q,\mathscr{R}_{H \times Q}, \mu_H \times \mu_{\mathfrak{G}/H} )$ by Scholium 3.9 of \cite{Segal_Kunze}. On the other hand for every function $(h,q) \mapsto (f_{h \cdot q} ,\upsilon)$ measurable on the product measure space the restricted functions $q \mapsto (f_{h \cdot q} ,\upsilon)$ and $h \mapsto (f_{h \cdot q} ,\upsilon)$, i. e. with one of the arguments $h$ and $q$ fixed, are measurable, which follows from the Fubini theorem (compare e. g. \cite{Segal_Kunze}, Theorem 3.4) and thus $q \mapsto (f_q , \upsilon)$ is measurable (i.e. with the argument $h$ fixed and equal $e$ in $(h,q) \mapsto (f_{h \cdot q} ,\upsilon)$). Because a simple computation shows that  
\[
\begin{split}
\int \, (\, \mathfrak{J}_L ((\mathfrak{J}^L f)_x ), f_x \,) \, d\mu_{\mathfrak{G}/H} 
= \int \, (\, \mathfrak{J}_L ((\mathfrak{J}^L f)_{h\cdot q} ), f_{h\cdot q} \,) \, d\mu_{\mathfrak{G}/H}(q) \\
=\int \, (f_q , f_q \,) \, d\mu_{\mathfrak{G}/H}(q),
\end{split}
\]
one can see that when functions equal almost everywhere are identified 
$\mathcal{H}^L$ becomes a Hilbert space with the inner product 

\begin{equation}\label{inn_ind_def}
(f, g) = \int \, (\, \mathfrak{J}_L ((\mathfrak{J}^L f)_x ), g_x \,) \, d\mu_{\mathfrak{G}/H}. 
\end{equation}
(In fact because the values of $f \in \mathcal{H}^L$ are in the fixed Hilbert space $\mathcal{H}_L$ we do not
have to tangle into the the whole machinery of direct integral Hilbert spaces of von Neumann. It suffices
to make obvious modifications in the corresponding proof that $L^2(\mathfrak{G}/H)$ is a Hilbert space, compare
\cite{Neumark_dec}, , \S 26.5.)

A simple verification shows that $\mathfrak{J}^L$ is a bounded self-adjoint operator in the Hilbert space $\mathcal{H}^L$
with respect to the definite inner product (\ref{inn_ind_def}) and that $(\mathfrak{J}^L)^2 = I$. Therefore
$(\mathcal{H}^L , \mathfrak{J}^L )$ is a Krein space with the indefinite product
\begin{equation}\label{ind_ind}
\big(f, g \big)_{\mathfrak{J}^L} = (\mathfrak{J}^L f, g) = \int \, (\, \mathfrak{J}_L (f_x ), g_x \,) \, d\mu_{\mathfrak{G}/H}
\end{equation}
which is meaningful because the integrand is constant on the right $H$-cosets, i. e. it is a function of $q \in Q \cong \mathfrak{G}/H$. 

Let the function $[x] \mapsto \lambda ([x], g)$ on $\mathfrak{G}/H$ be the Radon-Nikodym derivative 
$\lambda( \cdot ,g) = \frac{\ud (R_g \mu)}{\ud \mu}(\cdot)$, where $[x]$ stands for the right 
$H$-coset $Hx$ of $x \in \mathfrak{G}$
($\mu$ stands for the (quasi) invariant measure $\mu_{\mathfrak{G}/H}$ on $\mathfrak{G}/H$ induced by the assumed ``factorization'' property from the Haar measure $\mu$ on $\mathfrak{G}$ and $R_g \mu$
stands for the right translation of the measure $\mu$: $R_g \mu (E) = \mu(Eg)$). 

For every $g_0 \in \mathfrak{G}$ let us consider a densely defined operator $U^L_{g_0}$.
Its domain $\mathfrak{D}(U^L_{g_0})$  is equal to the set of all those  $f \in  \mathcal{H}^L$ 
for which the function
\[
x \mapsto   f'_x = \sqrt{\lambda ([x], g_0)} \, f_{xg_0}
\]
has finite Hilbert space norm (i. e. ordinary norm with respect to the ordinary definite 
inner product (\ref{inn_ind_def})) in $\mathcal{H}^L$:
\[
\begin{split}
 \big( f' , f' \big) = \int \, (\, \mathfrak{J}_L ((\mathfrak{J}^L f')_x ), f'_x \,) \, d\mu_{\mathfrak{G}/H} \\
= \int \, (\, \mathfrak{J}_L L_{h(x)} \mathfrak{J}_L L_{h(x)^{-1}} \sqrt{\lambda ([x], g_0)} \, f_{xg_0}  , 
\sqrt{\lambda ([x], g_0)} \, f_{xg_0} \,) \, 
d\mu_{\mathfrak{G}/H}(x) < \infty,
\end{split}
\]
where $h(x) \in H$ is the unique element corresponding to $x$ such that $h(x)^{-1}x \in Q$; and whenever $ f \in \mathfrak{D} (U^{L}_{g_0})$ we put 
\[
(U^L_{g_0} f)_x = \sqrt{\lambda ([x], g_0)} \, f_{xg_0}.
\]

$U^L$, after restriction to a suitable sub-domain, becomes a group homomorphism of 
$\mathfrak{G}$ into a group of densely defined $\mathfrak{J}^L$-isometries of the Krein space 
$(\mathcal{H}^L , \mathfrak{J}^L )$. Let us formulate this statement more precisely in a form of a Theorem:

\begin{twr}
The operators $U^L_{g_0}$, $g_0 \in \mathfrak{G}$, are closed and $\mathfrak{J}^L$-isometric with dense domains
$\mathfrak{D}(U^L_{g_0})$, dense ranges $\mathfrak{R}(U^L_{g_0})$ and dense intersection $\bigcap_{g_0 \in \mathfrak{G}} \mathfrak{D}(U^L_{g_0}) = \bigcap_{g_0 \in \mathfrak{G}} \mathfrak{R}(U^L_{g_0})$. $U^{L}_{g_0 k_0}$ is equal to the closure of the composition $\widetilde{U^{L}_{g_0}} \, \widetilde{U^{L}_{k_0}} = 
\widetilde{U^{L}_{g_0 k_0}}$
of the restrictions $\widetilde{U^{L}_{g_0}}$ and $\widetilde{U^{L}_{k_0}}$ of $U^{L}_{g_0}$ and $U^{L}_{k_0}$
to the domain $\bigcap_{g_0 \in \mathfrak{G}} \mathfrak{D}(U^L_{g_0})$, i. e.  the map $g_0 \mapsto \widetilde{U^L_{g_0}}$ is a Krein-isometric representation of 
$\mathfrak{G}$. There exists a dense sub-domain $\mathfrak{D} \subset \bigcap_{g_0 \in \mathfrak{G}} \mathfrak{D}(U^L_{g_0})$ such that $U^L_{g_0} \mathfrak{D} = \mathfrak{D}$, $U^L_{g_0}$ is the closure of the restriction 
$\widetilde{\widetilde{U^{L}_{g_0}}}$ of $U^L_{g_0}$ to the sub-domain $\mathfrak{D}$, and  $g_0 \mapsto \widetilde{\widetilde{U^L_{g_0}}}$ is strongly continuous Krein-isometric representation of $\mathfrak{G}$ on its domain 
$\mathfrak{D}$.
\label{def_ind_krein:twr.1}
\end{twr}

\qedsymbol \, 
Let us introduce the class $C^{L}_{00} \subset \mathcal{H}^L$ of functions 
$h\cdot q \mapsto f_{h \cdot q} = L_h W_q$ with 
$q \mapsto W_q \in \mathcal{H}_L$ continuous and compact support on $Q \cong \mathfrak{G}/H$. Of course 
each such function $W$ is an element of the direct integral Hilbert space 
$\int \, \mathcal{H}_L \, d\mu_{\mathfrak{G}/H}$. One easily verifies that all the conditions of Lemma \ref{lem:dense.4}
of (the next) Sect. \ref{dense} are true for the class $C^{L}_{00}$. Therefore $C^{L}_{00}$ is 
dense in $\mathcal{H}^L$. Let $h\cdot q \mapsto f_{h \cdot q} = L_h W_q$ be an element of $C^{L}_{00}$ and let 
$K$ be the compact support of the function $W$.  Using the ``unique factorization'' let us introduce the functions
$(q,h_0 , q_0 ) \mapsto h'_{{}_{q,h_0 , q_0}} \in H$ 
and $(q,h_0 , q_0 ) \mapsto q'_{{}_{q,h_0 , q_0}} \in Q \cong \mathfrak{G}/H$
in the following way. Let $g_0 = q_0 \cdot h_0$. We define $ h'_{{}_{q,h_0 , q_0}} \in H$ 
and $q'_{{}_{q,h_0 , q_0}} \in Q \subset \mathfrak{G}$ 
to be the elements, uniquely corresponding to $(q,h_0 , q_0 )$, such that 
\begin{equation}\label{q'_h'}
q\cdot h_0 \cdot q_0 = h'_{{}_{q,h_0 , q_0}} \cdot q'_{{}_{q,h_0 , q_0}}.
\end{equation} 
Finally let $c_{K, g_0} = \sup_{q \in K} \big\| L_{h'_{{}_{q,h_0 , q_0}}} \big\|$, which is finite
outside a null set, on account of the almost uniform boundedness of the representation $L$,
and because  $q \mapsto h'_{{}_{q,h_0 , q_0}}$ is continuous outside a $\mu_{\mathfrak{G}/H}$-null set
(easily seen by the ``measure product property'', but it holds true even if the ``measure product property is not assumed'' -- compare the comments below in this Section.). 
\begin{equation}\label{def_ind_krein:ineq}
\begin{split}
\| U^{L}_{h_0 \cdot q_0} f \|^2 =  \big(U^{L}_{h_0 \cdot q_0} f , U^{L}_{h_0 \cdot q_0} f \big) \\
= \int \, (\, \mathfrak{J}_L ((\mathfrak{J}^L U^{L}_{h_0 \cdot q_0}f)_{h\cdot q} ), (U^{L}_{h_0 \cdot q_0}f)_{h\cdot q} \,) \, 
d\mu_{\mathfrak{G}/H}(q) \\
= \int \, (\, L_{h'_{q, h_0 , q_0}}f_{q'_{q, h_0 , q_0}} , L_{h'_{q, h_0 , q_0}}f_{q'_{q, h_0 , q_0}} \,) \, 
d\mu_{\mathfrak{G}/H}(q'_{q, h_0 , q_0}) \\
\leq  c_{K,g_0}^{2} \, \int \, (\, f_{q'_{q, h_0 , q_0}} , f_{q'_{q, h_0 , q_0}} \,) \, 
d\mu_{\mathfrak{G}/H}(q'_{q, h_0 , q_0}) \\
= c_{K,g_0}^{2} \, \| f \|^2 , \,\,\,\,\,\,\, 
g_0 = h_0 \cdot q_0 \in \mathfrak{G}.  
\end{split}
\end{equation}
Thus it follows that $C^{L}_{00} \subset \mathfrak{D}(U^{L}_{g_0})$ for every $g_0 \in \mathfrak{G}$. Similarly it is easily verifiable that $C^{L}_{00} \subset \mathfrak{R}(U^{L}_{g_0})$ whenever the Radon-Nikodym derivative 
$\lambda ([x], g_0)$ is continuous in $[x]$. It follows from definition
that for $f \in \mathcal{H}^L$ being a member of $\bigcap_{g_0 \in \mathfrak{G}} \mathfrak{D}(U^L_{g_0})$ is equivalent to being a member of $\bigcap_{g_0 \in \mathfrak{G}} \mathfrak{R}(U^L_{g_0})$.

We shall show that $\big( \, U^{L}_{g_0} \, \big)^\dagger = U^{L}_{{g_0}^{-1}}$, 
where $T^\dagger$ stands for the adjoint of the operator $T$ in the sense of Krein \cite{Bog}, page 121: 
for any linear operator $T$ with dense domain $\mathfrak{D}(T)$ the vector 
$g \in \mathcal{H}^L$ belongs to $\mathfrak{D}(T^\dagger)$ if and only if there exists a $k \in \mathcal{H}^L$ 
such that 
\[
(\mathfrak{J}^L Tf , g) = (\mathfrak{J}^L f , k), \,\,\, \textrm{for all} \,\, f \in \mathfrak{D}(T),
\] 
and in this case we put $T^\dagger g = k$, with the unique $k$ as $\mathfrak{D}(T)$ is dense
(i. e. same definition as for the 
ordinary adjoint with the definite Hilbert space inner product $(\cdot, \cdot)$ given by (\ref{inn_ind_def})
replaced with the indefinite one $(\mathfrak{J}^L \cdot, \cdot)$, given by (\ref{ind_ind})). 

Now let $g$ be arbitrary in $\mathfrak{D}\big( \, \big( \,U^{L}_{g_0} \, \big)^\dagger \, \big)$, 
and let $\big(U^{L}_{g_0} \big)^\dagger g = k$. 
The inclusion $\big( \, U^{L}_{g_0} \, \big)^\dagger \subset U^{L}_{{g_0}^{-1}}$ is equivalent
to the equation $U^{L}_{{g_0}^{-1}} g = k$. By the definition of the Krein adjoint of an operator, for any
$f \in \mathfrak{D}\big(U^{L}_{g_0}\big)$ we have 
\[
\begin{split}
\big( \, \mathfrak{J}^L \, U^{L}_{g_0} f , g \,\big) = (\mathfrak{J}^L \, f, k), \\
\,\,\, \textrm{i. \, e.}
\int \, \Big( \mathfrak{J}_L \, \big( \, U^{L}_{g_0} \, f \big)_x , g_x  \, \Big) \, d\mu_{\mathfrak{G}/H}(x)
= \int \, (\mathfrak{J}_L \, f_x , k_x ) \, d\mu_{\mathfrak{G}/H}(x);
\end{split}
\]
which by the definition of $U^{L}_{g_0}$ and quasi invariance of the measure $\mu_{\mathfrak{G}/H}$
means that
\[
\begin{split}
\int \, \Big( \mathfrak{J}_L \, f_x , \sqrt{\frac{d\mu_{\mathfrak{G}/H}(x{g_0}^{-1})}{d\mu_{\mathfrak{G}/H}(x)}} 
g_{x{g_0}^{-1}}  \, \Big) \, d\mu_{\mathfrak{G}/H}(x) \\
= \int \, (\mathfrak{J}_L \, f_x , k_x ) \, d\mu_{\mathfrak{G}/H}(x) \,\,\, \textrm{for all} \,\, f \in 
\mathfrak{D}\big(U^{L}_{g_0} \big);
\end{split}
\]
i. e. the function $u$
\[
x \mapsto u_x =  \sqrt{\frac{d\mu_{\mathfrak{G}/H}(x{g_0}^{-1})}{d\mu_{\mathfrak{G}/H}(x)}} 
g_{x{g_0}^{-1}} - k_x
\]
is $\mathfrak{J}^L$-orthogonal to all elements of $\mathfrak{D}\big( \big( \,U^{L}_{g_0} \big)$: $(\mathfrak{J}^L f , u) = 0$ for all $f \in \mathfrak{D}\big(U^{L}_{g_0} \big)$.
Because $\mathfrak{D}\big(U^{L}_{g_0}\big)$ is dense in $\mathcal{H}^L$, and $\mathfrak{J}^L$ is unitary with respect to the ordinary Hilbert space inner product (\ref{inn_ind_def}) in $\mathcal{H}^L$ it follows that $\mathfrak{J}^L \mathfrak{D}\big(U^{L}_{g_0}\big)$ is dense in 
$\mathcal{H}^L$. Therefore $u$ must be zero as a vector orthogonal to 
$\mathfrak{J}^L \mathfrak{D}\big(U^{L}_{g_0} \big)$ in the sense of the Hilbert space inner product (\ref{inn_ind_def}). Thus
\[
\sqrt{\frac{d\mu_{\mathfrak{G}/H}(x{g_0}^{-1})}{d\mu_{\mathfrak{G}/H}(x)}} 
g_{x{g_0}^{-1}} = k_x
\] 
almost everywhere, and because by definition $(k, k) < \infty$, we have shown that $U^{L}_{{g_0}^{-1}} g = k$.

Next we show that $\big( \, U^{L}_{g_0} \, \big)^\dagger \supset U^{L}_{{g_0}^{-1}}$. Let $g$ be arbitrary
in $\mathfrak{D}(U^L_{{g_0}^{-1}})$ and let $U^{L}_{{g_0}^{-1}} g = k$. It must be shown that for any 
$f \in \mathfrak{D}\big(U^{L}_{g_0} \big)$,
$(\mathfrak{J}^L \, U^{L}_{g_0} f  , g) = (\mathfrak{J}^L \, f, k)$. This is the same as showing that 
\[
\int \, \Big( \mathfrak{J}_L \, \big( \, U^{L}_{g_0} \, f \big)_x , g_x  \, \Big) \, d\mu_{\mathfrak{G}/H}(x)
= \int \, (\mathfrak{J}_L \, f_x , k_x ) \, d\mu_{\mathfrak{G}/H}(x),
\] 
which again easily follows from definition of $U^{L}_{g_0}$ and quasi invariance of the measure 
$d\mu_{\mathfrak{G}/H}(x)$:
\[
\begin{split}
\int \, \Big( \mathfrak{J}_L \, \big( \, U^{L}_{g_0} \, f \big)_x , g_x  \, \Big) \, d\mu_{\mathfrak{G}/H}(x) 
= \int \, \sqrt{\frac{d\mu_{\mathfrak{G}/H}(xg_0)}{d\mu_{\mathfrak{G}/H}(x)}} \big(\mathfrak{J}_L \, f_{xg_0} , g_x \big) \,
d\mu_{\mathfrak{G}/H}(x) \\
=  \int \, \sqrt{\frac{d\mu_{\mathfrak{G}/H}(x{g_0}^{-1}g_0)}{d\mu_{\mathfrak{G}/H}(x{g_0}^{-1})}} \Big(\mathfrak{J}_L \, f_{x{g_0}^{-1}g_0} , g_{x{g_0}^{-1}} \Big) \,
\, \frac{d\mu_{\mathfrak{G}/H}(x{g_0}^{-1})}{d\mu_{\mathfrak{G}/H}(x)} \, d\mu_{\mathfrak{G}/H}(x) \\
= \int \,  \Big( \mathfrak{J}_L \, f_{x} , \sqrt{\frac{d\mu_{\mathfrak{G}/H}(x{g_0}^{-1}}{d\mu_{\mathfrak{G}/H}(x)}} g_{x{g_0}^{-1}} \Big)  \, d\mu_{\mathfrak{G}/H}(x) \\
=  \int \,  \Big( \mathfrak{J}_L \, f_{x} , \big(U^{L}_{{g_0}^{-1}} g\big)_{x} \Big)  \, 
d\mu_{\mathfrak{G}/H}(x)  
= \int \, (\mathfrak{J}_L \, f_x , k_x ) \, d\mu_{\mathfrak{G}/H}(x).
\end{split}
\] 
Thus we have shown that $\big(U^{L}_{g_0} \big)^\dagger = U^{L}_{{g_0}^{-1}}$. 

Because $C^{L}_{00} \subset \mathfrak{D}(U^L_{{g_0}^{-1}})$ then $\mathfrak{D}(U^L_{{g_0}^{-1}})$
is dense, thus  $U^{L}_{{g_0}^{-1}}$, equal to $\big(U^{L}_{g_0} \big)^\dagger$, is closed by Theorem 2.2 of \cite{Bog} (Krein adjoint $T^\dagger$ is always closed,
as it is equal $\mathfrak{J}^L  T^* \mathfrak{J}^L$ with the ordinary adjoint $T^*$ operator, 
and because the fundamental symmetry $\mathfrak{J}^L$
is unitary in the associated Hilbert space $\mathcal{H}^L$, compare Lemma 2.1 in \cite{Bog}).

In order to prove the second statement it will be sufficient to show that 
$\big( \, \widetilde{U^{L}_{g_0}} \, \big)^\dagger = U^{L}_{{g_0}^{-1}}$
because the homomorphism property of the map $g_0 \mapsto U^{L}_{g_0}$ restricted to $\bigcap_{g \in \mathfrak{G}}
\mathfrak{D}(U^L_{g})$
is a simple consequence of the definition of $U^{L}_{g_0}$. But the proof of the equality 
$\big( \, \widetilde{U^{L}_{g_0}} \, \big)^\dagger = U^{L}_{{g_0}^{-1}}$ runs exactly the same way as the 
proof of the equality $\big( \, U^{L}_{g_0} \, \big)^\dagger = U^{L}_{{g_0}^{-1}}$, with the trivial 
replacement of $\mathfrak{D}\big(U^{L}_{g_0}\big)$ by $\mathfrak{D}$, as it is valid for any dense sub-domain 
$\mathfrak{D}$ contained in $\mathfrak{D}\big(U^{L}_{g_0}\big)$ instead of $\mathfrak{D}\big(U^{L}_{g_0}\big)$.  
Then by Theorem 2.5 of \cite{Bog}
it follows that $\big( \, \widetilde{U^{L}_{g_0}} \, \big)^{\dagger \dagger} 
= \big( \, U^{L}_{{g_0}^{-1}} \, \big)^\dagger$ is equal to the closure $\overline{\widetilde{U^{L}_{g_0}}}$
of the operator $\widetilde{U^{L}_{g_0}}$. Because $\big( \, U^{L}_{{g_0}^{-1}} \, \big)^\dagger
= U^{L}_{g_0}$, we get $U^{L}_{g_0} = \overline{\widetilde{U^{L}_{g_0}}}$. 

By the above remark we also have $U^{L}_{g_0} = \overline{\widetilde{\widetilde{U^{L}_{g_0}}}}$ for any
restriction $\widetilde{\widetilde{U^{L}_{g_0}}}$ of $U^{L}_{g_0}$ to a dense sub-domain $\mathfrak{D}
\subset \mathfrak{D}(U^L_{g_0})$

In order to prove the third statement, let us introduce a dense sub-domain $C^{L}_{0} \subset C^{L}_{00}$
of continuous functions with compact support on $\mathfrak{G}/H$. Its full definition and properties are given in the 
next Section. In particular $U^{L}_{g_0} C^{L}_{0} = C^{L}_{0}$ whenever the Radon-Nikodym derivative 
$\lambda ([x], g_0)$ is continuous in $[x]$. For each element $f^0$ of $C^{L}_{0}$ we have the inequality
shown to be valid in the course of proof of Lemma \ref{lem:dense.1}, Sect. \ref{dense}: 
\[
\| f^{0}_{x_1} - f^{0}_{x_2} \|^2 \leq \sup_{h \in H} \big\|  f^{L, V}_{(h,e) \cdot (e,x_1 )} - f^{L,V}_{(h,e)\cdot (e,x_2 )} \big\|^2 \, 2 \,\sup_{x \in \mathfrak{G}} \mu_{H}(Kx^{-1} \cap H)
\]
where $f^{L, V}$ is a function depending on $f^0$, continuous on the direct product group $H \times \mathfrak{G}$
and with compact support $K_H \times V$ with $V$ being a compact neighbourhood of the two points $x_1$ and $x_2$.
Because any such function $f^{L, V}$ must be uniformly continuous, the strong continuity of $U^{L}$ on 
the sub-domain $C^{L}_{0}$ follows. Because $U^{L}_{g_0} C^{L}_{0} = C^{L}_{0}$, the third statement is proved
with $\mathfrak{D} = C^{L}_{0}$ (In case the Radon-Nikodym derivative was not continuous and ``measure product property'' not satisfied it would be sufficient to use all finite sums 
$U^{L}_{g_1} f^1 + \ldots U^{L}_{g_n} f^n , f^k \in C^{L}_{0}$ as the common sub-domain 
$\mathfrak{D}$ instead of $C^{L}_{0}$).
\qed

\begin{rem}
By definition of the Krein-adjoint operator and the properties: 
1) $U^L_{g} \mathfrak{D} = \mathfrak{D}$, $g \in \mathfrak{G}$,  2)
$\big( \, U^{L}_{g} \, \big)^\dagger = U^{L}_{g^{-1}}$, $g \in \mathfrak{G}$, it easily follows that for each $g \in \mathfrak{G}$ 
\begin{equation}\label{UD=D}
\big( \, U^{L}_{g} \, \big)^\dagger \, U^{L}_{g} = I \,\,\, \textrm{and} \,\,\,
U^{L}_{g} \, \big( \, U^{L}_{g} \, \big)^\dagger = I
\end{equation}
on the domain $\mathfrak{D}$. We may easily modify the common domain $\mathfrak{D}$ so as to achieve
the additional property: 3) $\mathfrak{J}^L \mathfrak{D} = \mathfrak{D}$ together with 1) and 2)
and thus with (\ref{UD=D}). 
Indeed, to achieve this one may define
$\mathfrak{D}$ to be the linear span of the set $\Big\{ \, \Big( (\mathfrak{J}^L)^{m_1} U^{L}_{g_1}   \ldots 
U^{L}_{g_n} (\mathfrak{J}^L)^{m_{n+1}} \Big) f \, \Big\}$: with $g_k$ ranging over $\mathfrak{G}$, $f \in  C^{L}_{0}$, 
$n \in \mathbb{N}$ and $k \mapsto m_{k}$ over the sequences with $m_k$ equal 0 or 1. 
In case the Radon-Nikodym derivative $\lambda$ is continuous and the 
``measure product property'' fulfilled, $\mathfrak{D} = C^{L}_{0}$ meets all the requirements. 

\label{rem:def_ind_krein.1}
\end{rem}

\begin{cor}
For every $U^{L}_{g_0}$ there exists a unique unitary (with respect to the definite inner product 
(\ref{inn_ind_def})) operator $U_{g_0}$ in $\mathcal{H}^L$ and unique selfadjoint (with respect to
(\ref{inn_ind_def})) positive operator $H_{g_0}$, with dense domain $\mathfrak{D}(U^{L}_{g_0})$ and dense range such that $U^{L}_{g_0} = U_{g_0} H_{g_0}$.
\label{def_ind_krein:cor.1}
\end{cor}
\qedsymbol \,
Immediate consequence of the von Neumann polar decomposition theorem and closedness of $U^{L}_{g_0}$.
\qed

\vspace*{0.2cm}

Of course the ordinary unitary operators $U_{g_0}$ of the Corollary do not compose any representation in general as the 
operators $U_{g_0}$ and $H_{g_0}$ of the polar decomposition do not commute if $U_{g_0}$ is non normal.

\begin{twr}
$L$ and $\mathfrak{J}_L$ commute  if and only if $U^L$ and $\mathfrak{J}^L$ commute.
If $U^L$ and $\mathfrak{J}^L$ commute, then $L$ is not only $\mathfrak{J}_L$-unitary but also unitary in the ordinary sense for the definite inner product in the Hilbert space $\mathcal{H}_L$.
If $U^L$ and $\mathfrak{J}^L$ commute then $U^L$ is not only $\mathfrak{J}^L$-isometric
but unitary with respect to the ordinary Hilbert space inner product (\ref{inn_ind_def}) in $\mathcal{H}^L$, 
i . e. the operators $U^{L}_{g_0}$ are bounded and unitary with respect to (\ref{inn_ind_def}).
The representation $L$ is uniformly bounded if and only if the induced representation $U^{L}$ is Krein-unitary 
(with each $U^{L}_{g_0}$ bounded) and uniformly bounded.
\label{def_ind_krein:twr.2}
\end{twr}

\qedsymbol \,
Using the functions $(q,h_0 , q_0 ) \mapsto h'_{{}_{q,h_0 , q_0}} \in H$ and 
$(q,h_0 , q_0 ) \mapsto q'_{{}_{q,h_0 , q_0}} \in Q \cong \mathfrak{G}/H$ defined by (\ref{q'_h'}),
one easily verifies that $L$ and $\mathfrak{J}_L$ commute (and thus $L$ is not only $\mathfrak{J}_L$-unitary but also unitary in the ordinary sense for the definite inner product in the Hilbert space 
$\mathcal{H}_L$) if and only if $U^L$ and $\mathfrak{J}^L$ commute (i. e. when $U^L$ is not only $\mathfrak{J}^L$-isometric
but unitary with respect to the ordinary Hilbert space norm (\ref{inn_ind_def}) in $\mathcal{H}^L$). To this end we utilize 
the fact that for each fixed $x$, $f_x$ with $f$ ranging over $C^{L}_{00}$ has $\mathcal{H}^L$ as their 
closed linear span. We leave details to the reader.  
\qed

\vspace*{0.2cm}

\begin{cor}
If $N \subset H \subset \mathfrak{G}$ is a normal subgroup of $\mathfrak{G}$ such that the restriction of $L$
to $N$ is uniformly bounded (or commutes with $\mathfrak{J}_L$) then the restriction of $U^L$ to the subgroup
$N$ is a Krein-unitary representation of the subgroup with each $U^{L}_{n}, n \in N$ bounded uniformly in $n$ 
(or $U^{L}$ restricted to $N$ commutes with $\mathfrak{J}^L$ and is an ordinary unitary representation of $N$ in the Hilbert space $\mathcal{H}^L$). 

\label{def_ind_krein:cor.2}
\end{cor}
\qed

In the proof of the strong continuity of $U^L$ on the dense domain $\mathfrak{D}$ we have used a specific dense subspace 
$C^{L}_{0}$ of $\mathcal{H}^L$. In the next Subsection we give its precise definition and provide the remaining relevant analytic underpinnings which we introduce after Mackey. In the proof of strong continuity we did as in the classical proof of strong continuity of the right regular representation of $\mathfrak{G}$ in $L^2 (\mathfrak{G})$ or in $L^2 (\mathfrak{G}/H)$ (of course with the obvious Radon-Nikodym factor in the latter case), with the necessary modifications required for the Krein space. In our  proof of strong continuity on $\mathfrak{D}$ the strong continuity of the representation $L$ plays a much more profound role in comparison to the original Mackey's theory.

The additional assumption posed on right $H$-cosets, i. e. ``measure product property'' is unnecessary. In order to give to this paper a more independent character we point out that the above construction of the induced representation in Krein space is possible without this assumption which may be of use for spectral analysis for (unnecessary elliptic) operators on manifolds uniform for more general semi-direct product Lie groups preserving indefinite pseudo-riemann structures. Namely for any closed subgroup $H \subset \mathfrak{G}$ (with the ``measure product property'' unnecessary fulfilled)
the right action of $H$ on $\mathfrak{G}$ is proper and both $\mathfrak{G}$ and $\mathfrak{G}/H$ are metrizable so that a theorem of Federer and Morse \cite{Federer_Morse} can be applied (with the regular Baire (or Borel) Haar measure space structure $(\mathfrak{G}, \mathscr{R}_{\mathfrak{G}}, \mu)$ on $\mathfrak{G}$) 
in proving that  there exists a Borel subset $B \subset \mathfrak{G}$ such that: 
(a) $B$ intersects each right $H$-coset in exactly one point and (b) for each compact subset $K$ of $\mathfrak{G}$,
$\pi^{-1}(\pi(K))\cap B$ has a compact closure (compare Lemma 1.1 of \cite{Mackey}). 
In short $B$ is a ``regular Borel section of $\mathfrak{G}$ with respect to $H$''. In particular it follows that any 
$\mathfrak{g} \in \mathfrak{G}$ has unique factorization $\mathfrak{g} = h \cdot b$, $h \in H, b \in B$.  Using the Lemma and extending a technique of A. Weil used in studying relatively invariant measures Mackey gave in \cite{Mackey} a general construction of quasi invariant measures in $\mathfrak{G}/H$ (all being equivalent).

\vspace*{0.2cm}

{\bf REMARK}.
The general construction of quasi invariant (standard) Baire (or Borel) measures on the locally compact homogeneous space $\mathfrak{G}/H$ was proposed in a somewhat shortened form in \S 1 of \cite{Mackey}, where the technique of A. Weil was adopted and developed into a $\rho$- and $\lambda$-functions construction. Today it is known as a standard construction of \emph{the quotient of a measure space by a group}, detailed exposition can be found e.g. in \cite{Bourbaki}. Only for sake of completeness let us remind the main Lemmas and Theorem of \S 1 of  \cite{Mackey} (details omitted in the exposition of \cite{Mackey} are to be found e. g. in \cite{Bourbaki} with the trivial interchanging of left and right). Let 
$L_g \mu$ and $R_g \mu$ be the left and right translations of a measure $\mu$ on $\mathfrak{G}/H$: $L_g \mu (E) = \mu (gE)$ and $R_g \mu(E) = \mu (Eg)$. Let $\mu$ be the right Haar measure on $\mathfrak{G}$. Denoting the the constant Radon-Nikodym derivative of the right Haar measure $L_g \mu$ with respect to $\mu$ by $\Delta_{\mathfrak{G}}(g)$, and similarly defined constant Radon-Nikodym derivative  for the closed subgroup $H$ by $\Delta_{H}(g)$ we have the the following Lemmas and Theorems. 
\begin{enumerate}

\item[] LEMMA. \emph{Let $\mu$ be a non-zero measure on $\mathfrak{G}/H$ and $\mu_0 = \mu_\mathfrak{G}$ be the right Haar measure on $\mathfrak{G}$. The following conditions are equivalent}:

\item[a)] \emph{$\mu$ is quasi invariant with respect to $\mathfrak{G}$;}

\item[b)] \emph{a set $E \subset \mathfrak{G}/H$ is of $\mu$-measure zero if and only if $\pi^{-1}(E)$ is
of $\mu_0$-measure zero;}

\item[c)] \emph{the ``pseudo-counter-image'' measure $\mu^{\sharp}$ is equivalent to $\mu_0$.} 

\item[] \emph{Assume one (and thus all) of the conditions to be fulfilled and thus let $\mu^{\sharp} = 
\rho \cdot \mu_0$, where $\rho$ is a Baire (or Borel) $\mu$-measurable function non zero everywhere on $\mathfrak{G}$.
Then for every $s \in \mathfrak{G}$ the Radon-Nikodym derivative $\lambda( \cdot, s)$ of the measure 
$R_s \mu$ with respect to the measure $\mu$ is equal to
\[
\lambda( \pi(x), s) = \frac{\ud (R_s \mu)}{\ud \mu}(\pi(x)) = \rho(xs)/\rho(x)
\]
almost $\mu$-everywhere on $\mathfrak{G}$.}

\end{enumerate}

\begin{enumerate}

\item[] THEOREM. a) \emph{Any two non zero quasi invariant measures on $\mathfrak{G}/H$ are equivalent.}

\item[b)] \emph{If $\mu$ and $\mu'$ are two non zero quasi invariant measures on $\mathfrak{G}/H$
and $\ud (R_s \mu)/ \ud \mu = d(R_s \mu')/ \ud \mu'$ 
almost $\mu$-everywhere (and thus almost $\mu'$-everywhere), 
then $\mu' = c \cdot \mu$,
where $c$ is a positive number.}

\end{enumerate}

\begin{enumerate}

\item[] LEMMA. \emph{Measure $\rho \cdot \mu_0$ has the form $\mu^{\sharp}$ if and only if for each $h \in H$
the equality 
\[
\rho (hx) = \frac{\Delta_{H}(h)}{\Delta_{\mathfrak{G}}(h)} \rho(x)
\]
is fulfilled almost $\mu_0$-everywhere on $\mathfrak{G}$.}

\end{enumerate}

\begin{enumerate}

\item[] THEOREM. a) \emph{There exist functions $\rho$ fulfilling the conditions of the preceding Lemma, for example}

\[
\rho(x) = \frac{\Delta_H (h(x))}{\Delta_{\mathfrak{G}}(h(x))}, 
\]
\emph{where $h(x) \in H$ is the only element of $H$ corresponding to 
$x \in \mathfrak{G}$ such that $h(x)^{-1}x \in B$.}

\item[b)] \emph{ $\rho$ can be chosen to be continuous.}

\item[c)] \emph{One may chose the regular section $B$ to be continuous outside a discrete countable set in $\mathfrak{G}/H$
whenever $\mathfrak{G}$ is a topological manifold with $H$ as closed topological sub-manifold; thus $x \mapsto h(x)$
becomes continuous outside a set of measure zero in $\mathfrak{G}$.}

\item[d)] \emph{Given such a function $\rho$ one can construct a quasi invariant measure $\mu$ on $\mathfrak{G}/H$
such that $\mu^{\sharp} = \rho \cdot \mu_0$.}

\item[e)] \emph{$\rho(xs)/\rho(x)$ with $s,x \in \mathfrak{G}$ does not depend on $x$ within the class $\pi(x)$ and determinates a function $(\pi(x), s) \mapsto \lambda(\pi(x), s)$
on $\mathfrak{G}/H \times \mathfrak{G}$ equal to the Radon-Nikodym derivative 
$\ud (R_s \mu)/ \ud \mu (\pi(x))$.}

\item[f)] \emph{Given any Baire (or Borel) function $\lambda (\cdot, \cdot)$ on $\mathfrak{G}/H \times \mathfrak{G}$
fulfilling the general properties of Radon-Nikodym derivative: (i) for all $x, s, z \in \mathfrak{G}$, 
$\lambda(\pi(z), xs) = \lambda(\pi(zx), s) \lambda(\pi(z), x)$, (ii) for all $h \in H$, $\lambda(\pi(e), h)
= \Delta_H (h)/\Delta_{\mathfrak{G}}(h)$, (iii) $\lambda(\pi(e), s)$ is bounded on compact sets as a function of $s$,
one can construct a quasi invariant measure $\mu$ on $\mathfrak{G}/H$ such that 
$\ud (R_s \mu)/ \ud \mu (\pi(x)) = \lambda(\pi(x), s)$, 
almost $\mu$-everywhere with respect to $s,x$ on $\mathfrak{G}$.}

\end{enumerate}

Thus every non zero quasi invariant measure $\mu$ on $\mathfrak{G}/H$ gives rise to a $\rho$-function
and $\lambda$-function and vice versa every ``abstract Radon-Nikodym derivative'' i.e. $\lambda$-function 
(or equivalently every $\rho$-function)
gives rise to a quasi invariant measure $\mu$ on $\mathfrak{G}/H$ determined up to a non zero constant factor. 
Every quasi invariant measure $\mu$ on $\mathfrak{G}/H$ is thus a pseudo-image of the right Haar measure $\mu$
on $\mathfrak{G}$ under the canonical projection $\pi$ in the terminology of \cite{Bourbaki}. In particular if 
the groups $\mathfrak{G}$ and $H$ are unimodular (i. e. $\Delta_{\mathfrak{G}} = 1_{\mathfrak{G}}$ and $\Delta_H = 1_H$) then among quasi invariant measures on $\mathfrak{G}/H$ there exists a strictly invariant measure. 
\qed

\vspace*{0.2cm}

The measure space structure of 
$\mathfrak{G}/H$ uniform for the group $\mathfrak{G}$ may be transferred to $B$ together with the uniform structure, such that $(\mathfrak{G}/H,\mathscr{R}_{\mathfrak{G}/H}, \mu_{\mathfrak{G}/H}) \cong (B,\mathscr{R}_{B}, \mu_B )$. The set $B$ plays the role of the sub-manifold $Q$ in the ``measure product property''. This however would be insufficient, and we have to prove  a kind of regularity of right $H$-cosets instead of ``measure product property''. Namely let us 
define $h(x) \in H$,
which corresponds uniquely to $x \in \mathfrak{G}$, such that $h(x)^{-1}x \in B$. We have to prove that the functions 
$x \mapsto h(x)$ and  $x \mapsto h(x)^{-1}x$ are Borel (thus in particular measurable), which however was carried through in the proof of Lemma 1.4 of \cite{Mackey}. Now the only point which has to be changed is the definition of the fundamental symmetry operator $\mathfrak{J}^L$ in $\mathcal{H}^L$. We put
\[
(\mathfrak{J}^L f)_x = L_{h(x)} \mathfrak{J}_{L} L_{h(x)^{-1}} \, f_x .
\] 
We define $\mathcal{H}^L$ as 
the set of functions $\mathfrak{G} \mapsto \mathcal{H}_L$ fulfilling the conditions (i), (ii)  and such that 
\[
\int \limits_{B} \, (\, \mathfrak{J}_L ((\mathfrak{J}^L f)_x ), f_x \,) \, d\mu < \infty.
\]
The proof that $\mathcal{H}^L$ is a Hilbert space with the inner product 
\[
(f, g) = \int \, (\, \mathfrak{J}_L ((\mathfrak{J}^L f)_x ), g_x \,) \, d\mu
=\int \limits_{B} \, (f_b , g_b \,) \, d\mu_{B}(b), \,\,\, \textrm{where} \,\, b \in B, 
\]
is the same in this case with the only difference that the regularity of $H$-cosets is used instead of the Fubini theorem in reducing the problem to the von Neumann's direct integral Hilbert space construction. Namely we define a unitary map $V:
f \mapsto W^{f} = f\vert_B$
from the space $\mathcal{H}^L$ to the direct integral Hilbert space $\int \, \mathcal{H}_L \, d\mu_B$  of functions $b \mapsto W_b \in \mathcal{H}_L$ by a simple restriction to $B$ which is ``onto'' in consequence of the regularity of $H$-cosets. Its isometric character is trivial. $V$ has the inverse $W \mapsto f^W$ with $\big(f^{W}\big)_{x} = 
L_{h(x)}W_{h(x)^{-1}x}$. In particular $f^W$ is measurable on $\mathfrak{G}$ as for an orthonormal basis $\{ e_n \}_{n \in \mathbb{N}}$ of the Hilbert space $\mathcal{H}_L$ and any $\upsilon \in \mathcal{H}_L$ we have:
\[
\begin{split}
(f^{W}_{x}, \upsilon) = (f^{W}_{x}, \mathfrak{J}_L \mathfrak{J}_L  \upsilon) 
= (L_{h(x)} W_{h(x)^{-1}x}, \mathfrak{J}_L \mathfrak{J}_L  \upsilon)
= (\mathfrak{J}_L L_{h(x)} W_{h(x)^{-1}x},  \mathfrak{J}_L  \upsilon) \\
= (\mathfrak{J}_L W_{h(x)^{-1}x} , L_{h(x)^{-1}} \mathfrak{J}_L  \upsilon)
= \sum_{n \in \mathbb{N}} (\mathfrak{J}_L W_{h(x)^{-1}x} ,e_n)(e_n , L_{h(x)^{-1}} \mathfrak{J}_L  \upsilon)
\end{split}
\]
which, as a point-wise convergent series of measurable (again by Scholium 3.9 of \cite{Segal_Kunze}) functions in $x$ 
is measurable in $x$. 
We have to prove in addition that the induced representations $U^L$ in Krein spaces $(\mathcal{H}^L , \mathfrak{J}^L )$ corresponding to different choices of regular Borel sections $B$ are (Krein-)unitary equivalent. Namely let $B_1$ and $B_2$ be the two Borel sections in question. The Krein-unitary operator 
$U_{12}: (U_{12}f)_x = L_{h_{12}(x)} f_x $, where $h_{12}(x) \in H$ transforms the intersection point of the right $H$-coset $Hx$ with the section $B_1$ into the intersection point of the same coset $Hx$ with the Borel section $B_2$,
gives the Krein-unitary equivalence. The proof is similar to the proof of Lemma \ref{lop_ind_1} of Sect. \ref{lop_ind}. 

Therefore from now on everything which concerns induced representations in Krein spaces, with the group $\mathfrak{G}$
not explicitly assumed to be equal $T_{4} \circledS SL(2, \mathbb{C})$, does not assume 
``measure product property''. Also Theorems \ref{def_ind_krein:twr.1} and \ref{def_ind_krein:twr.2} and Corollaries 
\ref{def_ind_krein:cor.1} and \ref{def_ind_krein:cor.2}
remain true without the ``measure product property'' for any locally compact and separable
$\mathfrak{G}$ and its closed subgroup $H$. Indeed using the regular Borel section $B$
of $\mathfrak{G}$ the functions (\ref{q'_h'}):
$(q,h_0 , q_0 ) \mapsto h'_{{}_{q,h_0 , q_0}}$ and $(q,h_0 , q_0 ) \mapsto q'_{{}_{q,h_0 , q_0}}$ 
may likewise be defined in this more general situation. Moreover, by Lemma 1.1 and the proof 
of Lemma 1.4 of \cite{Mackey}, $h'_{{}_{q,h_0 , q_0}}$ ranges within a compact subset of $H$, 
whenever $q$ ranges within in a compact subset of $\mathfrak{G}$, so that the proofs remain unchanged.

The construction of the induced representation in Krein space has also another invariance property: it does not depend
on the choice of a quasi invariant measure $\mu$ on $\mathfrak{G}/H$ in the unique equivalence class, 
provided the Radon-Nikodym derivative $ \frac{\ud \mu'}{\ud \mu}$ corresponding to measures 
$\mu'$ and $\mu$ in the class is essentially ``upper'' and ``lower'' bounded: there exist two positive numbers $\delta$ and $\Delta$ such that 
\[
\esssup \frac{\ud \mu'}{\ud \mu} < \Delta \,\,\,\textrm{and} \,\,\,
\esssup \frac{\ud \mu}{\ud \mu'}< \delta. 
\]
This boundedness condition of Radon-Nikodym derivative is unnecessary in case of the original Mackey's theory of
induced representations, which are unitary in the ordinary sense. 
Introducing the left-handed-superscript $\mu$ in ${}^{\mu}\mathcal{H}^L$ and ${}^{\mu}U^L$ 
for indicating the measure used in the construction of $\mathcal{H}^L$ and $U^L$, we may formulate a Theorem:

\begin{twr}
Let $\mu'$ and $\mu$ be quasi invariant measures in $\mathfrak{G}/H$ with Radon-Nikodym
derivative $\psi = \frac{\ud \mu'}{\ud \mu}$ essentially ``upper'' and ``lower'' bounded.
Then there exists a Krein-unitary transformation $V$ from ${}^{\mu}\mathcal{H}^L$ onto 
${}^{\mu'}\mathcal{H}^L$ such that $V \big( {}^{\mu}U^{L}_y \big) V^{-1} = {}^{\mu'}U^{L}_y$
for all $y \in \mathfrak{G}$; that is the representations ${}^{\mu}U^L$ and ${}^{\mu'}U^L$
are Krein-unitary equivalent.
\label{def_ind_krein:twr.3}
\end{twr}

\qedsymbol \,
Let $f$ be any element of ${}^{\mu}\mathcal{H}^L$ and let $\pi$ be the canonical map 
$\mathfrak{G} \mapsto \mathfrak{G}/H$. Boundedness condition of the Radon-Nikodym derivative $\psi$ ensures 
$(\sqrt{\psi \circ \pi} \,f , \sqrt{\psi \circ \pi} \,f )$ to be finite, i. e. ensures
$\sqrt{\psi \circ \pi} \,f$  to be a member of ${}^{\mu'}\mathcal{H}^L$ as $\sqrt{\psi \circ \pi}$
is measurable; and moreover the Krein-square-inner product
$(\sqrt{\psi \circ \pi} \,f , \sqrt{\psi \circ \pi} \,f)_{\mathfrak{J}^L}$ in ${}^{\mu'}\mathcal{H}^L$
is equal to that $(f , f)_{\mathfrak{J}^L}$ in ${}^{\mu}\mathcal{H}^L$. Moreover boundedness of $\psi$
guarantees that every $g$ in ${}^{\mu'}\mathcal{H}^L$ is evidently of the form $\sqrt{\psi \circ \pi} \,f$
for some $f \in {}^{\mu}\mathcal{H}^L$. Let $V$ be the operator of multiplication by $\sqrt{\psi \circ \pi}$.
Then $V$ defines a Krein-unitary map of ${}^{\mu}\mathcal{H}^L$ onto ${}^{\mu'}\mathcal{H}^L$. 
The verification that $V \big( {}^{\mu}U^{L} \big) V^{-1} = {}^{\mu'}U^{L}$ is immediate. 
\qed

Finally we mention the following easy but useful 
\begin{twr}
Let $L$ and $L'$ be Krein-unitary representations in $(\mathcal{H}_L , \mathfrak{J}_L)$,
which are Krein-unitary and unitary equivalent, then the induced representations $U^{L}$ and $U^{L'}$ are 
Krein-unitary equivalent.  
\label{def_ind_krein:twr.4}
\end{twr}

\section{Certain dense subspaces of $\mathcal{H}^L$}\label{dense}

We present here some lemmas of analytic character which we shall need later and which we have used in the proof
of Thm. \ref{def_ind_krein:twr.1} of Sect. \ref{def_ind_krein}.
Let $\mu_H$ be the right invariant Haar measure on $H$. Let $C^L$ denote the set of all functions
$f: \mathfrak{G} \ni x \mapsto f_x \in \mathcal{H}_L$, which are continuous with respect to the Hilbert space 
norm $\| \cdot \| = \sqrt{(\cdot , \cdot)}$ in the Hilbert space $\mathcal{H}_L$, and with compact support.
Let us denote the support of $f$ by $K_f$.
\begin{lem}
For each $f \in C^L$ there is a unique function $f^0$ from $\mathfrak{G}$ to $\mathcal{H}_L$ such that 
$\int \, ( \mathfrak{J}_L L_{h^{-1}} f_{hx} , \upsilon) \, d\mu_H (h) = (\mathfrak{J}_L f^{0}_{x}, \upsilon )$ for all
$x \in \mathfrak{G}$ and all $\upsilon \in \mathcal{H}_L$. This function is continuous and it is a member
of $\mathcal{H}^L$. The function $\mathfrak{G}/H \ni [x] \mapsto \big(\mathfrak{J}_L 
(\mathfrak{J}^Lf^{0})_{x}, f^{0}_{x} \big)$ as well as the function $\mathfrak{G}/H \ni [x] \mapsto f^{0}_x$ 
has a compact support. Finally $\sup_{x \in \mathfrak{G}} 
\big(\mathfrak{J}_L (\mathfrak{J}^Lf^{0})_{x}, f^{0}_{x} \big) = \sup_{b \in B} (f^{0}_{b} , f^{0}_{b}) < \infty$, 
where $B$ is a regular Borel section 
of $\mathfrak{G}$ with respect to $H$ of Sect. \ref{def_ind_krein}. 
\label{lem:dense.1}
\end{lem}

\qedsymbol \,
 Let $f \in C^L$. For each fixed $x \in \mathfrak{G}$ consider the anti-linear functional
\[ 
\upsilon \mapsto F_{x}(\upsilon) =  \int \, ( \mathfrak{J}_L L_{h^{-1}} f_{hx} , \upsilon) \, d\mu_H (h)
\] 
on $\mathcal{H}_L$.  From the Cauchy-Schwarz inequality for the Hilbert space inner product $(\cdot , \cdot )$ in the Hilbert space $\mathcal{H}_L$ and unitarity of $\mathfrak{J}_L$  with respect to the inner product 
$(\cdot , \cdot )$ in $\mathcal{H}_L$, one gets 
\[
\begin{split} 
|F_{x}(\upsilon)| \leq  \int \, 
|( \mathfrak{J}_L L_{h^{-1}} f_{hx} , \upsilon)| \, d\mu_H (h) 
\leq   \int \, \| \mathfrak{J}_L  L_{h^{-1}}f_{hx} \| \|  \upsilon \| \, d\mu_H (h) \\
=  \Big( \int \, \| \mathfrak{J}_L  L_{h^{-1}}f_{hx} \|  \, d\mu_H (h) \Big) \,\| \upsilon \|
=  \Big( \int \, \| L_{h^{-1}}f_{hx} \|  \, d\mu_H (h) \Big) \,\| \upsilon \|;
\end{split}
\] 
where the integrand in the last expression is a compactly supported continuous function of $h$ as a consequence of the strong continuity of the representation $L$ and because $f$ is compactly supported norm continuous. Therefore the integral in the last expression is finite, so that the functional $F_x$ is continuous. Thus by Riesz's theorem (in the conjugate version) there exists a unique element $g_x$ of $\mathcal{H}_L$  (depending of course on $x$) such that for all $\upsilon \in \mathcal{H}_L : F_x (\upsilon) = (g_x , \upsilon)$. We put $f^{0}_{x} = \mathfrak{J}_L g_x$, so that 
$F_x (\upsilon) = (\mathfrak{J}_L f^{0}_{x} , \upsilon)$, $\upsilon \in \mathcal{H}_L$. We have to show that 
$f^{0}: x \mapsto f^{0}_{x}$ has the desired properties. 

That $f^{0}_{h'x} = L_{h'} f^{0}_{x}$ for all $h' \in H$ and $x \in \mathfrak{G}$  follows from right invariance of 
the Haar measure $\mu_H$ on $H$:
\[
\begin{split}
(\mathfrak{J}_L L_{h'}f^{0}_{x} , \upsilon) = (\mathfrak{J}_L f^{0}_{x} , L_{h'^{-1}} \upsilon)
= \int \, ( \mathfrak{J}_L L_{h^{-1}} f_{hx} , L_{h'^{-1}} \upsilon) \, d\mu_H (h) \\
= \int \, ( \mathfrak{J}_L L_{h'} L_{h^{-1}} f_{hx} ,  \upsilon) \, d\mu_H (h) 
= \int \, ( \mathfrak{J}_L L_{(hh'^{-1})^{-1}} f_{hx} ,  \upsilon) \, d\mu_H (h) \\
= \int \, ( \mathfrak{J}_L L_{(hh'h'^{-1})^{-1}} f_{hh'x} ,  \upsilon) \, d\mu_H (hh')
= \int \, ( \mathfrak{J}_L L_{(hh'h'^{-1})^{-1}} f_{hh'x} ,  \upsilon) \, d\mu_H (h) \\
= \int \, ( \mathfrak{J}_L L_{(h)^{-1}} f_{hh'x} ,  \upsilon) \, d\mu_H (h)
= (\mathfrak{J}_L f^{0}_{h'x} , \upsilon), 
\end{split}
\]
for all $\upsilon \in \mathcal{H}_L$, $ h' \in H$, $x \in \mathfrak{G}$.

Denote the compact support of $f$ by $K$. From the strong continuity of the representation $L$ it follows immediately that the function
\[
(h, x) \mapsto f^{L}_{(h,x)} = L_{h^{-1}}f_{hx}
\]
is a norm continuous function on the direct product group $H \times \mathfrak{G}$
and compactly supported with respect to the first variable, i. e. for every $x \in \mathfrak{G}$ the 
function $h \mapsto f^{L}_{(h,x)}$
has compact support equal $K x^{-1} \cap H$. It is therefore 
uniformly norm continuous on the direct product group $H \times \mathfrak{G}$ with respect to the first variable.
For any compact subset $V$ of $\mathfrak{G}$ let $\phi_V$ be a real continuous function on $\mathfrak{G}$ with compact support equal 1 everywhere on $V$ (there exists such a function because $\mathfrak{G}$ as a topological 
space is normal). For $f \in C^L$ and any compact $V \subset \mathfrak{G}$ we introduce a 
norm continuous function on the direct product group $H\times \mathfrak{G}$ as a product $f^L \, \phi_V$:
\[
(h, x) \mapsto f^{L, V}_{(h,x)} = f^{L}_{(h,x)}  \phi_V (x),
\] 
which in addition is compactly supported and has the property that 
\[
f^{L, V}_{(h,x)} = L_{h^{-1}}f_{hx}
\]
for $(h,x) \in H \times V \subset H \times \mathfrak{G}$. In particular $f^{L,V}$ as compactly supported is not
only norm continuous but uniformly continuous on the direct product group $H \times \mathfrak{G}$ (i. e. uniformly 
in both variables jointly). 
Let $\{e_n\}_{n \in \mathbb{N}}$ be an orthonormal basis in the Hilbert space 
$\mathcal{H}_L$ and let $\mathcal{O} \subset \mathfrak{G}$ be any open set containing $x_1 , x_2 \in \mathfrak{G}$ 
with compact closure $V$. From the definition of $f^{0}$ it follows that
\[
\begin{split}
\| f^{0}_{x_1} - f^{0}_{x_2} \|^2 = \| \mathfrak{J}_L (f^{0}_{x_1} - f^{0}_{x_2}) \|^2 
= \sum_{n \in \mathbb{N}} \big| (\mathfrak{J}_L (f^{0}_{x_1} - f^{0}_{x_2}) , e_n ) \big|^2 \\
= \sum_{n \in \mathbb{N}}  \Big| \int \, ( \mathfrak{J}_L L_{h^{-1}} (f_{hx_1} - f_{hx_2}) , e_n) \, d\mu_H (h) \Big|^2 \\
\leq \sum_{n \in \mathbb{N}}   \int \, 
\big|( \mathfrak{J}_L L_{h^{-1}} (f_{hx_1} - f_{hx_2}) , e_n) \big|^2 \, d\mu_H (h) \\
=    \int \, \sum_{n \in \mathbb{N}}
\big|( \mathfrak{J}_L L_{h^{-1}} (f_{hx_1} - f_{hx_2}) , e_n) \big|^2 \, d\mu_H (h) \\
=  \int \, \big\| \mathfrak{J}_L  L_{h^{-1}}(f_{hx_1} - f_{hx_2}) \big\|^2 \, d\mu_H (h)
=  \int \, \big\|  L_{h^{-1}}(f_{hx_1} - f_{hx_2}) \big\|^2 \, d\mu_H (h) \\
\leq \sup_{h \in H} \big\|  L_{h^{-1}}f_{hx_1} - L_{h^{-1}}f_{hx_2} \big\|^2 \, 
\mu_{H}\big((Kx_{1}^{-1} \cap H) \cup (Kx_{2}^{-1} \cap H) \big) \\G_2 \cap \, ({x_0}^{-1}G_1 x_0)
\leq \sup_{h \in H} \big\|  f^{L, V}_{(h,e) \cdot (e,x_1 )} - f^{L,V}_{(h,e)\cdot (e,x_2 )} \big\|^2 \, 2 \,\sup_{x \in \mathfrak{G}} \mu_{H}(Kx^{-1} \cap H).
\end{split}
\]
Because the function $f^{L, V}$ is norm continuous on $H \times \mathfrak{G}$ and the continuity is uniform and $\sup_{x \in \mathfrak{G}} \mu_{H}(Kx^{-1} \cap H)< \infty$
(\cite{Mackey}, proof of Lemma 3.1) the norm continuity of $f^0$ is proved.
 
Similarly we get
\[
\begin{split}
 \| f^{0}_{x} \|^2 \leq  \sup_{h \in H} \big\|  L_{h^{-1}}f_{hx} \big\|^2 \,  \mu_{H}(Kx^{-1} \cap H) \\ 
= \sup_{h \in H} \big\|  f^{L, V^x }_{(h,e) \cdot (e,x )} \big\|^2 \,  \mu_{H}(Kx^{-1} \cap H) < \infty,
\end{split}
\]
because $K$ is compact and $f^{L, V^x}$ is norm continuous on $H \times \mathfrak{G}$ and compactly supported,
where $V^x$ is a compact neighbourhood of $x \in \mathfrak{G}$. Therefore $\| f^{0}_{x} \| = 0$ for all $x \notin HK$.
Thus as a function on $\mathfrak{G}/H$: $[x] \mapsto \big((\mathfrak{J}^Lf^{0})_{x}, f^{0}_{x} \big)$ and \emph{a fortiori} the function $[x] \mapsto \big(\mathfrak{J}_L (\mathfrak{J}^Lf^{0})_{x}, f^{0}_{x} \big)$ vanishes outside the compact canonical image of $HK$ in $\mathfrak{G}/H$. 

Finally let us note that if $h(x)$ is the element of $H$ defined in Sect. \ref{def_ind_krein} 
corresponding to $x \in \mathfrak{G}$, then 
\[
\big(\mathfrak{J}_L (\mathfrak{J}^Lf^{0})_{x}, f^{0}_{x} \big) =  (f^{0}_{b} , f^{0}_{b})
\] 
with $b = h(x)^{-1}x$ -- the unique intersection point of the coset $Hx$ with the Borel section $B$.
Because $f$ is continuous with compact support, then the last assertion of the Lemma follows from 
Lemma 1.1 of \cite{Mackey}.
\qedsymbol \,

\vspace*{0.5cm}

We shall denote the class of functions $f^0$ for $f \in C^L$ of Lemma \ref{lem:dense.1} by $C^{L}_{0}$.

\begin{lem}
For each fixed $x \in \mathfrak{G}$ the vectors $f^{0}_{x}$ for $f^{0} \in C^{L}_{0}$ form a dense linear subspace
of $\mathcal{H}_L$.
\label{lem:dense.2}
\end{lem}

\qedsymbol \,
Note that if $f^{0} \in C^{L}_{0}$ and $R_s f$ is defined by the equation $(R_s f)_x = f_{xs}$
for all $x$ and $s$ in $\mathfrak{G}$ then $R_s f^{0} = (R_s f)^{0}$ so that for all $f \in C^L$ and $s \in \mathfrak{G}$, 
$R_s f^0 \in C^{L}_{0}$. Therefore the set $\mathcal{H}_{L}''$ of vectors $f^{0}_x$ for $f^0 \in C^{L}_{0}$ and $x$ fixed is independent of $x$. Let $\mathcal{H}_{L}'$ be the $\mathfrak{J}_L$-orthogonal complement of $\mathcal{H}_{L}''$,
i. e. the set of all $\upsilon \in \mathcal{H}_L$ such that $(\mathfrak{J}_L g, \upsilon) = 0$ for all 
$g \in \mathcal{H}_{L}''$. Then if $\upsilon \in \mathcal{H}_{L}'$ we have $(f^{0}_{x}, \upsilon) = 0$ for all
$f^0 \in C^{L}_{0}$ and all $x \in \mathfrak{G}$. Therefore $(\mathfrak{J}_L f^{0}_{hx}, \upsilon) 
= (\mathfrak{J}_L f^{0}_{x}, L_{h^{-1}}\upsilon) = 0$ for all $f^0$ in $C^{L}_{0}$, all $x$ in $\mathfrak{G}$ and  
all $h \in H$. Hence $\mathcal{H}_{L}'$ is invariant under the representation, as $L$ is $\mathfrak{J}_L$-unitary.
Let $L'$ be the restriction of $L$ to $\mathcal{H}_{L}'$. Suppose that there exists a non zero member $f^0$ of
$C^{L'}_{0}$. Thus \emph{a fortiori} $f^0 \in C^{L}_{0}$ and we have a contradiction since the values of $f^0$
are all in $\mathcal{H}_{L}'$, so that we would have in $( \mathfrak{J}^L , \mathcal{H}^L )$: 
\[
\begin{split}
(f^0 , g)_{\mathfrak{J}^L} = ( \mathfrak{J}^L f^0, g ) \\
= \int \, (\mathfrak{J}_L f^{0}_{x} , g_x \, d\mu_{\mathfrak{G}/H} = 0
\end{split}
\]   
for all $g \in \mathcal{H}^L$, which would give us $f^0 = 0$, because the Krein space 
$( \mathfrak{J}^L , \mathcal{H}^L )$ of the induced representation $U^L$ is non degenerate (or $\mathfrak{J}^L$ invertible). Thus in order to show that 
$\mathcal{H}_{L}' = 0$ we need only show that when $\mathcal{H}_{L}' \neq 0$ there exists a non zero member 
$f^0$ of $C^{L'}_{0}$. But if none existed then 
\[
\int \, (\mathfrak{J}_L L'_{h^{-1}} f_{hx}, \upsilon) \, d\mu_H (h)
\] 
would be zero for all $x$, all $\upsilon$ in $\mathcal{H}_L$ and all $f$ in $C^{L'}$. In particular the integral would be 
zero for $f = u \upsilon'$, for all continuous complex functions  $u$ on $\mathfrak{G}$ of compact support and all 
$\upsilon' \in \mathcal{H}_{L}'$, i .e  
\[
\int \, u(hx) (\mathfrak{J}_L L'_{h^{-1}} \upsilon' , \upsilon ) \, d\mu_H (h)
\] 
would be zero for all $x$, all $\upsilon$ in $\mathcal{H}_L$, all $\upsilon'$ in $\mathcal{H}_{L}'$ and all 
complex continuous $u$ of compact support on $\mathfrak{G}$, which, because $L$ (and thus $L'$) 
is strongly continuous, would imply that 
\[
(\mathfrak{J}_L L'_{h^{-1}} \upsilon' , \upsilon ) = 0
\]
for all $\upsilon$ in $\mathcal{H}_L$, all $\upsilon'$ in $\mathcal{H}_{L}'$ and all $h \in H$.
This is impossible because the Krein space $( \mathfrak{J}_L , \mathcal{H}_L )$ of the representation $L$ is non degenerate and $L'_{h^{-1}}$ non-singular as a Krein-unitary operator. Thus we have proved that $\mathcal{H}_{L}' = 0$. This means that
$\mathfrak{J}_L \mathcal{H}_{L}''$ is dense in the Hilbert space $\mathcal{H}_L$, and because $\mathfrak{J}_L$ is
unitary in $\mathcal{H}_L$ with respect to the ordinary definite inner product $(\cdot , \cdot)$, this means that 
$\mathcal{H}_{L}''$ is dense in the Hilbert space $\mathcal{H}_L$.
\qed

\vspace*{0.5cm}

\begin{lem}
Let $C$ be any family of functions from $\mathfrak{G}$ to $\mathcal{H}_L$ such that:
\begin{enumerate}

\item[(a)]
$C \subset \mathcal{H}^L$.

\item[(b)]
For each $s \in \mathfrak{G}$ there exists a positive Borel function $\rho_s$ such that for all $f \in C$,
$\rho_s R_s f \in C$ where $(R_s f)_x = f_{xs}$.

\item[(c)]
If $f \in C$ then $gf \in C$ for all bounded continuous complex valued functions $g$ on $\mathfrak{G}$ which
are constant on the right $H$-cosets.

\item[(d)]

There exists a sequence $f^1 , f^2 , \ldots$ of members of $C$ and a subset $P$ of $\mathfrak{G}$ of positive 
Haar measure such that for each $x \in P$ the members $f^{1}_{x} , f^{2}_{x} , \ldots$ of $\mathcal{H}_L$ 
have $\mathcal{H}_L$ as their closed linear span.

\end{enumerate}

Then the members of $C$ have $\mathcal{H}^L$ as their closed linear span.
\label{lem:dense.3}
\end{lem}

\qedsymbol \,
 Choose $f^1 , f^2 , \ldots$ as in the condition (d). Let $u$ be any member of $\mathcal{H}^L$
which is $\mathfrak{J}^L$-orthogonal to all members of $C$:  
\[
\big(f, u \big)_{\mathfrak{J}^L} = (\mathfrak{J}^L f, u) = \int \, (\, \mathfrak{J}_L (f_x ), u_x \,) \, d\mu_{\mathfrak{G}/H} = 0
\]
for all $f \in C$. Then
\[
\begin{split}
(\mathfrak{J}^L (\rho_s g)(R_s f^j), u) = \int \, (\, \mathfrak{J}_L ((\rho_s g)(x)(R_s f^j)_x ), u_x \,) \, d\mu_{\mathfrak{G}/H} = 0
\end{split}
\] 
for every $j \in \mathbb{N}$, all $s$ and every bounded continuous $g$ on $\mathfrak{G}$ which is constant on the right $H$-cosets. It follows at once that for all $s$ and all $j \in \mathbb{N}$ $(\mathfrak{J}_L f^{j}_{xs}, u_x) = 0$ for almost all $x \in \mathfrak{G}$. Since $x \mapsto (\mathfrak{J}_L f^{j}_{x}, u_x)$ is a Borel function on $\mathfrak{G}$ the
function 
\[
(x, s) \mapsto (\mathfrak{J}_L f^{j}_{xs}, u_x) = \sum_{n \in \mathbb{N}}(\mathfrak{J}_L f^{j}_{xs}, e_n ) (e_n , u_x) 
\]
is Borel on the product measure space $\mathfrak{G} \times \mathfrak{G}$ on repeating the argument of Sect. \ref{def_ind_krein} 
(Scholium 3.9 of \cite{Segal_Kunze}) and joining it with the fact that composition of a measurable (Borel) function on 
$\mathfrak{G}$ with the
continuous function $\mathfrak{G} \times \mathfrak{G} \ni (x, s) \mapsto xs \in \mathfrak{G}$ is measurable (Borel)
on the product measure space $\mathfrak{G} \times \mathfrak{G}$ (compare e. g. \cite{Segal_Kunze}). Thus we may apply 
the Fubini theorem (Thm. 3.4 in \cite{Segal_Kunze}) and conclude that for almost all $x$, $(\mathfrak{J}_L f^{j}_{xs}, u_x)$ is zero for almost all $s$. Since $j$ runs over a countable class 
we may select a single null set $N \subset \mathfrak{G}$ such that for each $x \notin N$,  
$(\mathfrak{J}_L f^{j}_{xs}, u_x)$ is, for almost all $s$, zero for all $j \in \mathbb{N}$. It follows that for each 
$x \notin N$ there exists $s \in x^{-1}P$ such that $(\mathfrak{J}_L f^{j}_{xs}, u_x) = 0$ for $j \in \mathbb{N}$
and hence that $u_x = 0$ because $\mathfrak{J}_L$ is unitary with respect to the ordinary definite Hilbert space inner product in the Hilbert space $\mathcal{H}_L$. Thus $u$ is almost everywhere zero and $\mathfrak{J}^L C$ must be 
dense in $\mathcal{H}^L$. 
Because $\mathfrak{J}^L$ is unitary in the ordinary sense with respect to the definite inner product 
(eq. (\ref{inn_ind_def}) of Sect. \ref{def_ind_krein}) in $\mathcal{H}^L$, $C$ must be dense in $\mathcal{H}^L$.  
\qed

\vspace*{0.5cm}

\begin{lem}
Let $C^1$ be any family of functions from $\mathfrak{G}$ to $\mathcal{H}_L$ such that:
\begin{enumerate}

\item[(a)]
For each $f \in C^1$ there exists a positive Borel function $\rho$ on $\mathfrak{G}$ such that
\[
\Big(\mathfrak{J}_L \frac{1}{\rho (x)}f_x , \upsilon \Big) = \Big(\frac{1}{\rho (x)}\mathfrak{J}_L f_x , \upsilon \Big)
= \frac{1}{\rho (x)} \Big(\mathfrak{J}_L f_x , \upsilon \Big)
\]
is continuous as a function of $x$ for all $\upsilon \in \mathcal{H}_L$.

\item[(b)]
$C^1 \subset \mathcal{H}^L$.

\item[(c)]
For each $s \in \mathfrak{G}$ there exists a positive Borel function $\rho_s$ such that for all $f \in C^1$,
$\rho_s R_s f \in C^1$ where $(R_s f)_x = f_{xs}$.

\item[(d)]
If $f \in C^1$ then $gf \in C^1$ for all bounded continuous complex valued functions $g$ on $\mathfrak{G}$ which
are constant on the right $H$-cosets and vanish outside of $\pi^{-1}(K)$ for some compact subset $K$ of 
$\mathfrak{G}/H$.

\item[(e)]
For some (and hence all) $x \in \mathfrak{G}$ the members $f_{x}$ of $\mathcal{H}_L$  for $f \in C^1$
have $\mathcal{H}_L$ as their closed linear span.

\end{enumerate}

Then the members of $C^1$ have $\mathcal{H}^L$ as their closed linear span.

\label{lem:dense.4}
\end{lem}

\qedsymbol \,
 Choose $f^1 , f^2 , \ldots$ in $C^1$ so that $f^{1}_{e} , f^{2}_{e} , \ldots$ 
have $\mathcal{H}_L$ as their closed linear span; $e$ being the identity of $\mathfrak{G}$. 
Let $u$ be any member of $\mathcal{H}^L$
which is $\mathfrak{J}^L$-orthogonal to all members of $C^1$. Then
\[
\begin{split}
(\mathfrak{J}^L (\rho_s g)(R_s f^j), u) = \int \, (\, \mathfrak{J}_L ((\rho_s g)(x)(R_s f^j)_x ), u_x \,) \, d\mu_{\mathfrak{G}/H} = 0
\end{split}
\] 
for every $j \in \mathbb{N}$, all $s$ and every bounded continuous $g$ on $\mathfrak{G}$ which is constant on the 
right $H$-cosets. It follows at once that for all $s$ and all $j \in \mathbb{N}$ $(\mathfrak{J}_L f^{j}_{xs}, u_x) = 0$ for almost all $x \in \mathfrak{G}$. Since $(x, s) \mapsto (\mathfrak{J}_L f^{j}_{xs}, u_x)$ is a Borel function on 
the product measure space $\mathfrak{G} \times \mathfrak{G}$ (compare the proof of Lemma \ref{lem:dense.3}) 
we may apply the Fubini theorem as in the preceding Lemma and conclude that for almost all $x$, 
$(\mathfrak{J}_L f^{j}_{xs}, u_x)$ is zero for almost all $s$. Since $j$ runs over a countable class we may select a 
single null set $N$ in $\mathfrak{G}$ such that for each $x \notin N$, $(\mathfrak{J}_L f^{j}_{xs}, u_x)$ is for
almost all $s$ zero for all $j$. Suppose that $u_{x_1} \neq 0$ for some $x_1 \notin N$. Then 
$(\mathfrak{J}_L f^{j}_{e}, u_{x_1}) \neq 0$ for some $j$ as $\mathfrak{J}_L$ is unitary with respect to the ordinary
Hilbert space inner product $(\cdot, \cdot)$ in $\mathcal{H}_L$ (as in the proof of the preceding Lemma). But for some positive Borel function $\rho$, $(\mathfrak{J}_L f^{j}_{x}, u_{x_1})\big/\rho(x)$ is continuous in $x$. 
Hence $(\mathfrak{J}_L f^{j}_{x_1 s}, u_{x_1})\big/\rho(x_1 s) \neq 0$ for $s$ in some neighbourhood of $x_{1}^{-1}$.
Thus $(\mathfrak{J}_L f^{j}_{x_1 s}, u_{x_1}) \neq 0$ for $s$ in some neighbourhood of $x_{1}^{-1}$. But this contradicts 
the fact that $(\mathfrak{J}_L f^{j}_{x_1 s}, u_{x_1})$ is zero for almost all $s \in \mathfrak{G}$. Therefore
$u_x$ is zero almost everywhere. Thus only the zero element is orthogonal (in the ordinary positive inner product space
in $\mathcal{H}^L$) to all members of $\mathfrak{J}^L C^1$ and it follows that $\mathfrak{J}^L C^1$ must be dense
in $\mathcal{H}^L$. Because $\mathfrak{J}^L$ is unitary with respect to the ordinary  definite inner product 
$(\cdot, \cdot)$ in $\mathcal{H}^L$, it follows that $C^1$ is dense in $\mathcal{H}^L$.
\qed

\vspace*{0.5cm}

\begin{lem}
$C^{L}_{0}$ is dense in $\mathcal{H}^L$.
\label{lem:dense.5}
\end{lem}

\qedsymbol \,
The Lemma is an immediate consequence of Lemmas \ref{lem:dense.2} and \ref{lem:dense.4}. 
\qed

\vspace*{0.5cm}

{\bf REMARK}. For the reasons explained in Sect. \ref{decomposition} we are interesting 
in complete measure spaces on $\mathfrak{G}/H$ and on all other quotient spaces encountered later in this paper. 
But the Baire or Borel measure is pretty sufficient in the investigation of the associated Hilbert spaces 
$L^2 (\mathfrak{G}/H , \mu_{\mathfrak{G}/H})$ or $\mathcal{H}^L$ as all measurable sets differ from the Borel sets just by 
null sets, and the space of equivalence classes of Borel square summable functions in $L^2 (\mathfrak{G}/H , \mu_{\mathfrak{G}/H})$ is the same as the space of equivalence classes of square summable measurable functions. Recall that the Baire measure space may be completed to a Lebesgue-type measure space, e. g. using the 
Carath\'eodory method. In other words the Baire or Borel (the same in this case) measure space may be completed such that any subset of measurable null set will be measurable.
\qed

\vspace*{0.5cm}

\begin{lem}
There exists a sequence $f^1 , f^2 , \ldots$ of elements $C^{L}_{0} \subset \mathcal{H}^L$ such that for each fixed $x \in \mathfrak{G}$ the vectors $f^{k}_{x}$, $k = 1, 2, \ldots$ form a dense linear subspace of $\mathcal{H}_L$.
\label{lem:dense.6}
\end{lem}

\qedsymbol \,
 We have seen in the previous Sect. that as a Hilbert space $\mathcal{H}^L$  is unitary equivalent to the direct integral Hilbert space $\int \, \mathcal{H}_L \, d\mu_{\mathfrak{G}/H}$ over the $\sigma$-finite and regular Baire  (or Borel) measure space $(\mathfrak{G}/H , \mathscr{R}_{\mathfrak{G}/H}, \mu_{\mathfrak{G}/H})$ with 
separable $\mathcal{H}_L$. Because $\mathfrak{G}/H = \mathfrak{X}$ is locally compact metrizable and fulfils the second axiom of countability its minimal (one point or Alexandroff) compactification $\mathfrak{X}_+$ is likewise metrizable
(compare e. g. \cite{Engelking}, Corollary 7.5.43). Thus the Banach algebra $C(\mathfrak{X}_+)$ is separable,
compare e. g. \cite{Krein}, Thm. 2 or \cite{Gelfand_Silov}). Because $C(\mathfrak{X}_+)$ is equal to the minimal unitization
$C_0 (\mathfrak{X})^+$ of the Banach algebra $C_0 (\mathfrak{X})$ of continuous functions on $\mathfrak{X}$ 
vanishing at infinity (compare \cite{Neumark_dec}), thus by the construction of minimal unitization
it follows that $C_0 (\mathfrak{X})$ is separable (of course with respect to the supremum norm in 
$C_0 (\mathfrak{X})$) as a closed ideal
in $C_0 (\mathfrak{X})^+$ of codimension one. Because the  
measure space $(\mathfrak{G}/H , \mathscr{R}_{\mathfrak{G}/H}, \mu_{\mathfrak{G}/H})$ is the regular Baire measure space, 
induced by the integration lattice $C_\mathcal{K}(\mathfrak{X}) \subset C_0 (\mathfrak{X})$ of continuous functions with compact support (compare \cite{Segal_Kunze}), it follows
from Corollary 4.4.2 of \cite{Segal_Kunze} that the Hilbert space 
$L^2 (\mathfrak{G}/H , \mu_{\mathfrak{G}/H})$ of square summable functions over $\mathfrak{X} = \mathfrak{G}/H$
is separable. Let $\{e_n \}_{n \in  \mathbb{N}}$ be an orthonormal basis in $\mathcal{H}_L$. Using standard -- by now -- Hilbert space (\cite{Neumark_dec}) and measure space 
(e. g. Fubini theorem, compare eq. (\ref{dir_int_L^2:decompositions}) of Sect.\ref{decomposition}) techniques  and the results of \cite{von_neumann_dec} one can prove that 
\[
\begin{split}
\int \, \mathcal{H}_L \, d\mu_{\mathfrak{G}/H} 
= \bigoplus \limits_{n \in \mathbb{N}} \, \int \, \mathbb{C} e_n \, d\mu_{\mathfrak{G}/H} \\
= \bigoplus \limits_{n \in \mathbb{N}} \, \mathcal{H}_n
\,\,\,\textrm{where} \,\,\, \mathcal{H}_n \cong L^2 (\mathfrak{G}/H , \mu_{\mathfrak{G}/H}).
\end{split}
\]
Thus  $\int \, \mathcal{H}_L \, d\mu_{\mathfrak{G}/H}$ itself must be separable and therefore 
$\mathcal{H}^L$ is separable. Thus we may choose a sequence $f^1 , f^2 , \ldots$ of elements 
$C^{L}_{0} \subset \mathcal{H}^L$ such that for each $f \in C^{L}_{0}$ there exists a subsequence 
$f^{n_1} , f^{n_2} , \ldots$ which converges in norm $\| \cdot \|$ of $\mathcal{H}^L$ to $f$. Then a slight and obvious
modification of the standard proof of the Riesz-Fischer theorem (e. g. \cite{Segal_Kunze}, Thm. 4.2) gives a sub-subsequence $f^{n_{m_1}} , f^{n_{m_2}} , \ldots$ which, after restriction to the regular Borel section $B \cong \mathfrak{G}/H$ converges almost uniformly to the restriction of $f$ to $B$ (where $B \cong \mathfrak{G}/H$ is locally compact with the natural topology induced by the canonical projection $\pi$, with the Baire measure space structure 
$(\mathfrak{G}/H,\mathscr{R}_{\mathfrak{G}/H}, \mu_{\mathfrak{G}/H}) \cong (B,\mathscr{R}_{B}, \mu_B )$) obtained by Mackey's technique of quotiening the measure space $\mathfrak{G}$ by the group $H$
recapitulated shortly in Sect. \ref{def_ind_krein}. As $f^k , f$ are continuous and compactly supported as functions on 
$B \cong \mathfrak{G}/H$, the convergence is uniform on $B$. The Lemma now, for $x \in B$, is an immediate consequence of Lemma \ref{lem:dense.2}. Because for each $x \in \mathfrak{G}$ we have $f^{k}_{x} = L_{h(x)} f^{k}_{h(x)^{-1}x}, f_{x} = L_{h(x)} f_{h(x)^{-1}x}$ with $h(x)^{-1}x \in B$ and because $L_h$ is invertible (and bounded) for every $h \in H$, the Lemma is proved.   
\qed

\vspace*{0.5cm}














\section{{\L}opusza\'nski representation as an induced representation}\label{lop_ind}

Let $\mathfrak{G}$ be a separable locally compact group and $H$ its closed subgroup.
In this section we shall need Lemma \ref{lop_ind_1} (below), which we prove assuming 
the ``measure product property'',
because it is sufficient for the analysis of the {\L}opusza\'nski representation
of the double covering of the Poincar\'e group. However it can be proved without this assumption,
as the reader will easily see by recalling the respective remarks of Sect. \ref{def_ind_krein}.

Thus we assume (for simplicity)
that the right Haar measure space $\Big( \, \mathfrak{G}\, , \,\, \mathscr{R}_{{}_{\mathfrak{G}}} \, , \,\,
\mu_{{}_{\mathfrak{G}}} \, \Big)$ 
be equal to the product measure space  $\Big( \, H \times \mathfrak{G}/H, \,\, 
\mathscr{R}_{{}_{H \times \mathfrak{G}/H}}\, , \,\, \mu_{{}_{H}} \times \mu_{{}_{\mathfrak{G}/H}} \, \Big)$
with $\Big( \, H , \,\, 
\mathscr{R}_{{}_{H}}\, , \,\, \mu_{{}_{H}} \, \Big)$ equal to the right Haar measure space on $H$
and with the Mackey quotient measure space $\Big( \, \mathfrak{G}/H \, , \,\, \mathscr{R}_{{}_{\mathfrak{G}/H}} \, , \,\,
\mu_{{}_{\mathfrak{G}/H}} \, \Big)$ on $\mathfrak{G}/H$ (described briefly in Sect \ref{def_ind_krein}).
In most cases of physical applications both $\mathfrak{G}$ and $H$ are unimodular.  
Let $g = h \cdot q$ be the corresponding unique factorization of $g \in \mathfrak{G}$
with $h \in H$ and $q \in Q \subset \mathfrak{G}$ representing the class $[g] \in \mathfrak{G}/H$. Uniqueness of the factorization allows us to introduce the following functions ((already mentioned in Sect. \ref{def_ind_krein}) 
$(q,h_0 , q_0 ) \mapsto h'_{{}_{q,h_0 , q_0}} \in H$ 
and $(q,h_0 , q_0 ) \mapsto q'_{{}_{q,h_0 , q_0}} \in Q \cong \mathfrak{G}/H$, where for any
$g_0 = q_0 \cdot h_0 \in \mathfrak{G}$ we define $ h'_{{}_{q,h_0 , q_0}} \in H$ 
and $q'_{{}_{q,h_0 , q_0}} \in Q \subset \mathfrak{G}$ 
to be the elements, uniquely corresponding to $(q,h_0 , q_0 )$, such that 
\[
q\cdot h_0 \cdot q_0 = h'_{{}_{q,h_0 , q_0}} \cdot q'_{{}_{q,h_0 , q_0}}.
\] 
In particular if  $g = h q$, then $q$ represents $[g] \in \mathfrak{G}/H$, and  
$q'_{{}_{q,h_0 , q_0}}$ represents $[gg_0]$, i.e. the right action of $\mathfrak{G}$ on $\mathfrak{G}/H$.
It is easily verifiable that $(q,h_0 , q_0 ) \mapsto h'_{{}_{q,h_0 , q_0}}$ behaves like a multiplier,
i.e. denoting $h'_{{}_{q,h_0 , q_0}}$ and $q'_{{}_{q,h_0 , q_0}}$ just by $h'_{{}_{q,g_{{}_0}}}$
and $q'_{{}_{q,g_{{}_0}}}$ we have
\[
\boxed{h'_{{}_{q,\, g_{{}_0}}} \cdot h'_{{}_{q'_{{}_{q,g_{{}_0}}}, \, g_{{}_1}}} = h'_{{}_{q , \, g_{{}_0} g_{{}_1}}}.}
\] 

Let $U^L$ be the Krein isometric representation of $\mathfrak{G}$ induced by an almost uniformly bounded Krein-unitary
representation of $H$ in the Krein space $(\mathcal{H}_L , \mathfrak{J}_L)$, defined as in Sect. \ref{def_ind_krein}. 
Let us introduce the Hilbert space
\begin{equation}\label{hilbert_system_prim}
\mathcal{H} = \int \limits_{\mathfrak{G}/H} \, \mathcal{H}_L \, \ud \mu_{{}_{\mathfrak{G}/H}}
\end{equation}
and the fundamental symmetry $\mathfrak{J}$
\begin{equation}\label{fund_sym_dec}
\mathfrak{J} = \int \limits_{\mathfrak{G}/H} \, \mathfrak{J}_L \, \ud \mu_{{}_{\mathfrak{G}/H}}
\end{equation}
in $\mathcal{H}$, i.e. operator decomposable with respect to the decomposition (\ref{hilbert_system_prim}) 
whose all components in its decomposition are equal $\mathfrak{J}_L$. Because
 $\mathfrak{J}_{L}^{*} = \mathfrak{J}_L$
and $\mathfrak{J}_{L}^{*} \mathfrak{J}_L = \mathfrak{J}_L \mathfrak{J}_{L}^{*} = I$, then by \cite{von_neumann_dec}
the same holds true of the operator $\mathfrak{J}$, i.e. it is unitary and selfadjoint, i.e. $\mathfrak{J}^* = \mathfrak{J}$
and $\mathfrak{J}^* \mathfrak{J} = \mathfrak{J} \mathfrak{J}^* = I$, so that  $\mathfrak{J}^2 = I$ and 
$\mathfrak{J}$ is a fundamental symmetry. We may therefore introduce the Krein space $(\mathcal{H}, \mathfrak{J})$.

\begin{lem}
Let $\mathfrak{G}$ be a separable locally compact group and $H$ its closed subgroup. Assume (for simplicity)
that the ''measure product property'' is fulfilled by $\mathfrak{G}$ and $H$. Then the operators
\[
U : \mathcal{H} \mapsto \mathcal{H}^L, \,\,\, \textrm{and} \,\,\,
S : \mathcal{H}^L \mapsto \mathcal{H},
\]
defined as follows 
\[
\Big( U W \Big)_{h\cdot q} = L_h W_{{}_q}, \,\,\, \textrm{and} \,\,\,
\Big( S f \Big)_{q} = L_{{}_{h^{-1}}} f_{{}_{h\cdot q}},
\] 
for all $W \in \mathcal{H}$ and $f \in \mathcal{H}^L$, are well defined operators, both are isometric
and Krein-isometric between $(\mathcal{H}, \mathfrak{J})$ and $(\mathcal{H}^L , \mathfrak{J}^L)$ and 
moreover $US = I$ and $SU = I$ and moreover 
\[
U^{-1} \mathfrak{J}^L U = \mathfrak{J},
\]
so that $U$ and $S$ are unitary and Krein-uinitary. We have 
\[
\Big( V_{g_{{}_0}} W \Big)_{q} = \Big( U^{-1} U^{L}_{g_{{}_0}} U W \Big)_{q} \\
= \sqrt{\lambda(q, g_{{}_0})} L_{{}_{h'_{{}_{q,g_{{}_0}}}}} W_{{}_{q'_{{}_{q,g_{{}_0}}}}};
\] 
or equivalently
\[
\Big( V_{g_{{}_0}} W \Big)_{[g]} = \Big( U^{-1} U^{L}_{g_{{}_0}} U W \Big)_{[g]} \\
= \sqrt{\lambda([g], g_{{}_0})} L_{{}_{h'_{{}_{[g],g_{{}_0}}}}} W_{{}_{[g \cdot g_{{}_0}]}}.
\]  
In short: $U^L$ is unitary and Krein unitary equivalent to the Krein-isometric representation
$V$ of $\mathfrak{G}$ in $(\mathcal{H}, \mathfrak{J})$.  

\label{lop_ind_1}
\end{lem}

\qedsymbol \,
That the functions $UW$, $W \in \mathcal{H}$, and $U^{-1}f$, $f \in \mathcal{H}^L$ fulfil the required measurability conditions has been already shown in Sect. \ref{def_ind_krein}). Verification of the isometric and Krein-isometric
character of both $U$ and $S$ is easy, and we leave it to the reader. Checking $US = I$ and $SU = I$ as well as the
last equality is likewise simple.   
\qed

Now let us turn our attention to the construction of semi-direct product groups and their specific class of 
Krein-isometric representations to which the {\L}opusza\'nski representation belong
together with the related systems of imprimitivity in the Krein space $(\mathcal{H}, \mathfrak{J})$,
say of Lemma \ref{lop_ind_1}.
Let $G_1$ and $G_2$ be separable locally compact groups and let $G_1$ be abelian 
($G_1$ plays the role of four translations subgroup $T_4$ and $G_2$ plays the role of the
$SL(2, \mathbb{C})$ subgroup of the double covering $\mathfrak{G} = T_4 \circledS SL(2,\mathbb{C})$
of the Poincar\'e group). Let there be given a homomorphism
of $G_2$ into the group of automorphisms of $G_1$ and let $y [x] \in G_1$ be the action of the automorphism
corresponding to $y$ on $x \in G_1$. We assume that $(x,y) \mapsto y [x]$ is jointly continuous in both variables.
We define the semi-direct product $\mathfrak{G} = G_1 \circledS G_2$ as the topological product
$G_1 \times G_2$ with the multiplication rule $(x_1 , y_1)(x_2 , y_2) = (x_1 y_{{}_1} [x_2], y_1 y_2)$. 
$\mathfrak{G} = G_1 \circledS G_2$ under this operation is a separable locally compact group. Recall that the subset of elements $(x,e)$ with $x \in G_1$ and $e$ being the identity is a closed subgroup of the semi direct product $\mathfrak{G}$ naturally isomorphic to $G_1$ and similarly the set of elements
$(e,y)$, $y \in G_2$ is a closed subgroup of $\mathfrak{G} = G_1 \circledS G_2$ naturally isomorphic to
$G_2$. Let us identify those subgroups with $G_1$ and $G_2$ respectively. Since $(x,e)(e,y) = (x,y)$
it follows at once that any Krein-isometric representation $(x,y) \mapsto V_{(x,y)}$ of 
$\mathfrak{G} = G_1 \circledS G_2$ in the Krein space  $(\mathcal{H}, \mathfrak{J})$ is determined by its 
restrictions $N$ and $U$ to the subgroups $G_1$ and $G_2$ respectively:
$V_{(x,y)} = N_x U_y$. Conversely if $N$ and $U$ are Krein-isometric representations of $G_1$
and $G_2$ which act in the same Krein space $(\mathcal{H}, \mathfrak{J})$
and with the same core invariant domain $\mathfrak{D}$, and moreover if the representation $N$
commutes with the fundamental symmetry $\mathfrak{J}$ and is therefore unitary, then one easily checks that
$(x,y) \mapsto N_x U_y$ defines a Krein-isometric representation if and only if $U_y N_x U_{y^{-1}} = N_{ y[x]}$.
Indeed the ``if'' part is easy. Assume then that $ V_{(x,y)} = N_x U_y$ is a representation. Then for any 
$(x,y), (x',y') \in G_1 \circledS G_2$ one has $N_x U_y N_{x'} U_{y^{-1}} U_{yy'} = N_x N_{y[x']}U_{yy'}$
on the core dense set $\mathfrak{D}$.
Because $N_x$ is unitary it follows that $U_y N_{x'} U_{y^{-1}} U_{yy'} = N_{y[x']}U_{yy'}$ on $\mathfrak{D}$.
Because $U_y \mathfrak{D} = \mathfrak{D}$ for all $y \in y \in G_2$ and $U_y U_{y^{-1}} = I$ on $\mathfrak{D}$,
then it follows that $U_y N_{x'}U_{y^{-1}} = N_{y[x']}$ on $\mathfrak{D}$ for all $x,x' \in G_1$ and all $y \in G_2$.
Because the right hand side is unitary, then $U_y N_{x'} U_{y^{-1}}$ can be extended to a unitary operator,
although $U$ is in general unbounded.  
 Now assume (which is the case for representations of translations acting in one particle states in QFT,
for example this is the case for the restriction of the {\L}opusza\'nski representation to the translation subgroup) 
that the representation $N$ of the abelian subgroup $G_1$ commutes with the fundamental symmetry 
$\mathfrak{J}$ in $\mathcal{H}$, and thus it is not only Krein-isometric but unitary in $\mathcal{H}$ 
in the usual sense. Moreover the restrictions  $N$ of representations acting in one particle states are in 
fact of uniform (even finite) multiplicity. Because $N$ is a unitary representation 
of a separable locally compact abelian group $G_1$ in the Hilbert space the Neumak's theorem is applicable,
which says that $N$ is determined by a projection valued (spectral) measure $S \mapsto E_S$ 
(which as we will see may be associated with the direct integral decomposition 
(\ref{hilbert_system_prim}) with the appropriate subgroup $H$), defined on the Borel (or Baire)
sets $S$ of the character group $\widehat{G_1}$ of $G_1$:  
\[
N_x = \int \limits_{\widehat{G_1}} \chi(x) \, dE(\chi).
\]

It is readily verified that 
$N$ and $U$ satisfy the above identity if and only if the spectral measure $E$ and the 
representation $U$ satisfy $U_y E_S U_{y^{-1}} = E_{[S]y}$, for all $y \in G_2$ and
all Borel sets $S \subset \widehat{G_1}$; where the action $[\chi]y$ of $y \in G_2$
on $\chi \in \widehat{G_1}$ is defined
by the equation $\langle [\chi]y, x \rangle = \langle \chi, y^{-1}[x] \rangle$ (with $\langle \chi, x \rangle$
denoting the value of the character $\chi \in \widehat{G_1}$ on the element $x \in G_1$). 
Indeed:
\begin{multline}\label{ueu}
U_y N_x U_{y^{-1}} = \int \limits_{\widehat{G_1}} \chi(x) \, d(U_yE(\chi)U_{y^{-1}}) 
= N_{y[x]} = \int \limits_{\widehat{G_1}} \chi(y[x]) \, dE(\chi) \\ 
=\int \limits_{\widehat{G_1}} \big([\chi]y^{-1}\big)(x) \, dE(\chi) = \int \limits_{\widehat{G_1}} \chi(x) \, dE([\chi]y) .
\end{multline}
We call such $E$, $N$, and $U$
a \emph{system of imprimitivity in the Krein space} $(\mathcal{H}, \mathfrak{J})$, after Mackey \cite{Mackey_imprimitivity}
who defined the structure for representations $N$ and $U$ in Hilbert space $\mathcal{H}$
which are both unitary in the ordinary sense.

Consider now the action of $G_2$ on $\widehat{G_1}$. If the spectral measure $E$ is concentrated in one
of the orbits of $\widehat{G_1}$ under $G_2$ let $\chi_0$ be any member of this orbit $\mathscr{O}_{\chi_0}$ and
let $G_{\chi_0}$ be the subgroup of all $y \in G_2$ for which $[\chi_0]y = \chi_0$. 
Then $y \mapsto [\chi_0]y$ defines a one-to-one Borel set preserving map between the points
of this orbit $\mathscr{O}_{\chi_0}$ and the points of the homogeneous space $G_2 / G_{\chi_0} = \mathfrak{G}/H$,
where $H = G_1 \cdot G_{\chi_0}$. In this way 
$E$, $N$, $U$, becomes a system of imprimitivity based on the homogeneous space $\mathfrak{G} /H$.
Now when $E$ is concentrated on a single orbit the assumption of uniform multiplicity of $N$ would be unnecessary,
but instead we may require $U$ to be ``locally bounded'': $||U_y f || < c_{\Delta} ||f ||$ for all $f \in \mathcal{H}$ 
whose spectral support (in their decomposition with respect to $E$) is contained within compact subset $\Delta \subset G_2 / G_{\chi_0} = \mathfrak{G}/H$, with a positive constant $c_{\Delta}$ depending on $\Delta$. 
(In fact we have implicitly used the ``local boundedness'' in the first equality of (\ref{ueu}).) Then using ergodicity
of the action of $G_2$ (resp. $\mathfrak{G}$) on $G_2 / G_{\hat{x}_0}$ (resp. $\mathfrak{G}/H$) one can prove
uniform multiplicity of the spectral measure $E$. 
A computation similar to that performed by Mackey in \cite{Mackey_imprimitivity} (compare also
\cite{Mackey1}, \S 6 or \cite{Mackey2}, \S 3.7) shows that the representation $V_{(x,y)} = N_x U_y$ defined by the system is just equal to the Krein-isometric 
representation $V$ of $\mathfrak{G} = G_1 \circledS G_2$ in the 
Krein space $(\mathcal{H}, \mathfrak{J})$ of the Lemma (\ref{lop_ind_1}) with a representation $L$ of the
subgroup $H$, which is easily checked to be Krein-unitary in case the multiplicity of $N$ is
assumed to be finite. Thus it follows the following theorem

\begin{twr}
Let $E$, $N$, $U$ be a system of imprimitivity giving a Krein-isometric representation 
$V_{(x,y)} = N_x U_y$ of a semi direct product $\mathfrak{G} = G_1 \circledS G_2$ 
of separable locally compact groups $G_1$ and $G_2$ with $G_1$ abelian in a Krein space
$(\mathcal{H}, \mathfrak{J})$ and with the representation
$N$ commuting with $\mathfrak{J}$ and thus being unitary in $\mathcal{H}$, 
for which the following assumptions are satisfied:
\begin{enumerate}

\item[1)]
The spectral measure is concentrated on a single orbit $\mathscr{O}_{\chi_0}$ in $\widehat{G_1}$ under $G_2$.

\item[2)]
The representation $U$ (equivalently the representation $V$)  is ``locally bounded''
with respect to $E$. 

\end{enumerate}

Then the representation $N$ (and equivalently the spectral measure $E$) is of uniform multiplicity.
The fundamental symmetry $\mathfrak{J}$ is decomposable with respect to the decomposition
of $\mathcal{H}$ associated (in the sense of \cite{von_neumann_dec}) to the spectral measure $E$
of the system, and has a decomposition of the form (\ref{fund_sym_dec}).
Assume moreover that:

\begin{enumerate}

\item[3)]
The representation $N$ has finite multiplicity.

\end{enumerate}

Then $V$ is unitary and Krein-unitary equivalent to a Krein-isometric representation $U^L$ 
induced by a Krein unitary representation $L$ of the subgroup $H = G_1 \cdot G_{\chi_0}$ 
associated to the orbit. 
\label{lop_ind:twr.1}
\end{twr}
\qed

This theorem may be given a more general form by discarding 3), but the given version is sufficient 
for the representations acting in one particle states
of free fields with non trivial gauge freedom, and thus acting in Krein spaces (with the 
fundamental symmetry operator $\mathfrak{J}$ called Gupta-Bleuler operator in physicists parlance),
where the representations $L$ act in Krein spaces $(\mathcal{H}_L , \mathfrak{J}_L)$
of finite dimension.

Consider for example the double covering $\mathfrak{G} = T_4 \circledS SL(2,\mathbb{C})$
of the Poincar\'e group with the semi direct product structure defined
by the following homomorphism: $\alpha [t_x]= \alpha x \alpha^*$,
where the translation $t_x:  (a_0 , a_1 , a_2 , a_3) \mapsto (a_0 , a_1 , a_2 , a_3) + (x_0 , x_1 , x_2 , x_3)$
is written as a hermitian matrix 
\[
x = \left( \begin{array}{cc} x_0 + x_3 & x_1 - i x_2 \\ 
                            x_1 + i x_2 & x_0 - x_3  \end{array}\right)
\] 
in the formula $\alpha x \alpha^*$ giving $\alpha [t_x]$ and $\alpha^*$ is the hermitian 
adjoint of $\alpha \in SL(2, \mathbb{C})$. 

Characters $\chi_p \in \widehat{T_4}$ of the group $T_4$ have the following form
\[
\chi _p (t_x) = e^{i( - p_0a_0 + p_1 x_1 + p_2 x_2 + p_3 x_3)},
\]
for $p = (p_0 , p_1 , p_2 , p_3)$ ranging over $\mathbb{R}^4$. For each character $\chi_p \in \widehat{T_4}$
let us consider the orbit $\mathscr{O}_{\chi_p}$ passing through $\chi_p$,
under the action $\chi_p \mapsto [\chi_p]\alpha$, $\alpha \in SL(2, \mathbb{C})$,
where $[\chi_p]\alpha$ is the character given by the formula
\[
T_4 \ni t_x \xrightarrow{[\chi_p]\alpha} \big( [\chi_p]\alpha \big)(t_x)
= \chi_p (\alpha^{-1} [t_x])  = \chi_p (\alpha^{-1} x {\alpha^{*}}^{-1}) = \chi_{\alpha p \alpha^{*}}(x)
= \chi_{\alpha p\alpha^{*}}(t_x),
\]
where in the formulas $\alpha p \alpha^*$ and $\alpha^{-1}x{\alpha^{*}}^{-1}$, $x$ and $p$
are regarded as hermitian $2 \times 2$ matrices:
\[
x = \left( \begin{array}{cc} x_0 + x_3 & x_1 - i x_2 \\ 
                            x_1 + i x_2 & x_0 - x_3  \end{array}\right) \,\,\, \textrm{and} \,\,\,
p = \left( \begin{array}{cc} p_0 + p_3 & p_1 - i p_2 \\ 
                            p_1 + i p_2 & p_0 - p_3  \end{array}\right).
\] 
Let $G_{\chi_p}$ be the stationary subgroup of the point $\chi_p \in \widehat{T_4}$. 
Let $H = H_{\chi_p} = T_4 \cdot G_{\chi_p}$, and let $L'$ be a Krein-unitary representation 
of the stationary group $G_{\chi_p}$. Then $L$ given by 
\[
L_{t_x \cdot g} = \chi_p (t_x) L'_{g}, \,\,\, t_x \in T_4 , g \in G_{\chi_p}
\]
is a well defined Krein-unitary representation of $H_{\chi_p} = T_4 \cdot G_{\chi_p}$ because $G_{\chi_p}$
is the stationary subgroup for the point $\chi_p$. 
The functions $(q,h_0 , q_0 ) \mapsto h'_{{}_{q,h_0 , q_0}} \in H$ 
and $(q,h_0 , q_0 ) \mapsto q'_{{}_{q,h_0 , q_0}} \in Q \cong \mathfrak{G}/H_{\chi_p}$ 
corresponding to the respective $H= H_{\chi_p}$ or the respective orbits $\mathscr{O}_{\chi_p}$ are known
for all orbits in $\widehat{T_4}$ under $SL(2, \mathbb{C})$ and may be explicitly computed.

For example for $p = (1, 0, 0, 1)$ lying on the light cone in the joint spectrum sp$(P_0 , \ldots P_3)$
of the canonical generators of one parameter subgroups of translations, the stationary
subgroup $G_{\chi_p} = G_{\chi_{{}_{(1, 0, 0, 1)}}}$ is equal to the group of matrices 
\[
\left( \begin{array}{cc} e^{i\phi/2} & e^{i \phi/2}z \\ 
                                     0 & e^{-i\phi/2}  \end{array}\right), \,\,\,
 0 \leq \phi < 4\pi , \,\,\, z \in \mathbb{C} 
\]
isomorphic to (the double covering of) the 
symmetry group $E_2$ of the Euclidean plane and with the orbit $\mathscr{O}_{\chi_{{}_{(1, 0, 0, 1)}}}$
equal to the forward cone with the apex removed.

Consider then the Hilbert space $\mathcal{H}_L$ to be equal $\mathbb{C}^4$
with the standard inner product and with the fundamental symmetry equal
\[
\mathfrak{J}_L
 = \left( \begin{array}{cccc} -1 & 0 & 0 & 0 \\ 
                                           0 & 1 & 0 & 0 \\
                                           0 & 0 & 1 & 0 \\
                                           0 & 0 & 0 & 1  \end{array}\right).
\] 

Finally let $L'$ be the following Krein-unitary representation
\begin{equation}\label{L'}
\left(\begin{array}{cc} e^{i\phi/2} & e^{i \phi/2}z \\ 
                                     0 & e^{-i\phi/2}  \end{array}\right) \xrightarrow{L'_{(z, \phi)}}
\left( \begin{array}{cccc} 
1 + \frac{1}{2}|z|^2                     & \frac{1}{\sqrt{2}}z & \frac{1}{\sqrt{2}}\overline{z} & -\frac{1}{2}|z|^2 \\ 
\frac{1}{\sqrt{2}} e^{-i\phi}\overline{z} & e^{-i\phi} & 0                    & \frac{1}{\sqrt{2}} e^{-i\phi}\overline{z} \\
\frac{1}{\sqrt{2}} e^{i\phi}z            & 0         & e^{i\phi}             & -\frac{1}{\sqrt{2}} e^{i\phi}z  \\
\frac{1}{2}|z|^2                          & \frac{1}{\sqrt{2}}z & \frac{1}{\sqrt{2}}\overline{z} & 1 
- \frac{1}{2}|z|^2  \end{array}\right)
\end{equation}
of $G_{\chi_{{}_{(1, 0, 0, 1)}}} \cong \widetilde{E_2}$ in the Krein space $(\mathcal{H}_L , \mathfrak{J}_L)$ and 
define the Krein-unitary representation $L$: 
$H = T_4 \cdot G_{\chi_{{}_{(1, 0, 0, 1)}}} \ni t_x \cdot (z,\phi) \xrightarrow{L_{t_x \cdot (z, \phi)}}
\chi_{{}_{(1, 0, 0, 1)}}(t_x) L'_{(z, \phi)}$ corresponding to the Krein-unitary representation $L'$
of $G_{\chi_{{}_{(1, 0, 0, 1)}}}$. 
Then one obtains in this way the system of imprimitivity with the representation
$V$ of the Lemma \ref{lop_ind_1} equal to the {\L}opusza\'nski representation acting in the 
one particle states of the free photon field in the momentum representation, having exactly Wigner's form
\cite{Wigner_Poincare} with the only difference that $L$ is not unitary but Krein-unitary.

\vspace*{1cm}

Several remarks are in order.

1) In case of $\mathfrak{G} = T_4 \circledS SL(2,\mathbb{C})$,
$\widehat{T_4} = \mathbb{R}^4$ with the natural smooth action of $SL(2,\mathbb{C})$ 
giving it the Lorentz structure. The possible orbits  
$\mathscr{O}_{\chi_p} \subset \widehat{T_4} = \mathbb{R}^4$ 
are: the single point $(0,0,0,0,)$ -- the apex of ``the light-cone'', the upper/lower half
of the light cone (without the apex), the upper/lower sheet of the paraboloid, and
the one-sheet hyperboloid. Thus all of them are smooth manifolds (with the exclusion of the apex, of course). 
Joining this with the
Mackey analysis of quasi invariant measures on  homogeneous $\mathfrak{G}/H$ spaces one can see that
the spectral measures of the translation generators (for representations with the joint spectrum
sp$(P_0 , \ldots P_3)$ concentrated on single orbits) are equivalent to measures  induced by the 
Lebesgue measure on $\mathbb{R}^4 = \widehat{T_4}$ (of course with the exclusion of the representations corresponding
the the apex -- the single point orbit, with the zero $(0,0,0,0)$
as the only value of the joint spectrum sp$(P_0 , \ldots P_3)$.

2) Note that for the system of imprimitivity $E$, $N$, $U$ in the Krein space the condition:
\begin{multline*}
V_{(x,y)} E_S V_{(x,y)^{-1}} = N_x U_y E_S U_{y^{-1}} N_{x^{-1}} \\
= N_x E_{[S]y} N_{x^{-1}}
= E_{[S]y} \,\,\, \textrm{for all} \,\,\, (x,y) \in G_1 \circledS G_2 \,\,\, \textrm{and all Borel sets}
\,\,\, S \subset \widehat{G_1}
\end{multline*}
holds, and is essentially equivalent to the condition: 
\[
U_y E_S U_{y^{-1}} = E_{[S]y}, \,\,\, \textrm{for all} \,\,\, y \in G_2 , 
 \,\,\, \textrm{and all Borel sets} \,\,\, S \subset \widehat{G_1}.
\]
We may write it as 
$V_{(x,y)} E_S V_{(x,y)^{-1}} = E_{[S](x,y)}$, with the trivial action 
$[\chi](x,e) = \chi$, $x \in G_1$ and $[\chi](e,y) = [\chi]y$.
It is more convenient to relate the system of imprimitivity immediately to $V$ and inspired by Mackey put
the following more general definition.

Let $V$ be a Krein-isometric representation of a separable locally compact group $\mathfrak{G}$
in a Krein space $(\mathcal{H}, \mathfrak{J})$. By a system of imprimitivity for $V$, we 
mean the system $E$, $B$, $\varphi$ consisting of
\begin{enumerate}

\item[a)]
an analytic Borel set $B$;

\item[b)]
an anti-homomorphism $\varphi$ of $\mathfrak{G}$ into the group of all Borel automorphisms of
$B$ such that $(y,b) \mapsto (y,[b]y)$ is a Borel automorphism of $\mathfrak{G} \times B$;
here we have written $[b]y$ for the action of the automorphism $\varphi(y)$ on $b \in B$.

\item[c)]
The spectral measure $E$ consists of selfadjont and Krein selfadjoint projections commuting 
with $\mathfrak{J}$ in $(\mathcal{H}, \mathfrak{J})$, and is such that 
$V_y E_S {V_{y}}^{-1} = E_{[S]y^{-1}}$.

\item[d)]  
The representation $V$ is ``locally bounded'' with respect to $E$.

\end{enumerate} 
   
Any induced Krein-isometric representation ${}^{\mu}U^L$  possesses a canonical system of
imprimitivity in $(\mathcal{H}^L , \mathfrak{J}^L)$ related to it. Namely
let $S$ be a Borel set on $\mathfrak{G}/H$, and let $S'$ be its inverse under the 
quotient map $\mathfrak{G} \rightarrow \mathfrak{G}/H$. Let $1_{S'}$ be the characteristic
function of $S'$. Then $f \xrightarrow{E_S}  1_{S'} f$, $f \in \mathcal{H}^L$
is a self adjoint and Krein self adjoint projection, which commutes with $\mathfrak{J}^L$.
Thus $S \mapsto E_S$ is a spectral measure based on the analytic Borel space $\mathfrak{G}/H$.
By the inequality (\ref{def_ind_krein:ineq}) in the proof of Theorem \ref{def_ind_krein:twr.1}
the representation ${}^{\mu}U^L$ is ``locally bounded'', i.e. fulfils condition 3) 
of Theorem \ref{lop_ind:twr.1} or condition d).

The representation $V$ of Lemma \ref{lop_ind_1} in the Krein space $(\mathcal{H},\mathfrak{J})$
together with the spectral measure $E'$ on $B = \mathfrak{G}/H$ associated with the decomposition 
(\ref{hilbert_system_prim}) is a system of imprimitivity in Krein space which by Lemma \ref{lop_ind_1}
is Krein-unitary and unitary equivalent to the canonical system of imprimitivity $U^L$, $E$, $\varphi$ 
defined above.  That $V, E'$ of Lemma \ref{lop_ind_1} with $\varphi_{g_{{}_0}}(q) = q'_{{}_{q,g_{{}_0}}}$ 
composes a system of imprimitivity can be checked directly using the multiplier property of the function 
$(q,g_{{}_0}) \mapsto h'_{{}_{q,g_{{}_0}}}$.

3) The plan for further computations is the following. First we start with the systems of imprimitivity fulfilling the conditions 1)-3) of Theorem \ref{lop_ind:twr.1} sufficient for accounting for the representations acting in one particle states of free fields. Then we prove the ``subgroup'' and ``Kronecker product theorems'' for the induced representations
in order to achieve decompositions of tensor products of these representations into direct integrals of representations
connected with imprimitivity systems concentrated on single orbits (using Mackey double-coset-type technics).
The component representations of the decomposition will not in general have the standard form of induced 
representations (contrary to what happens for tensor products of induced representations of Mackey which are unitary in ordinary sense). But then we back to Theorem  \ref{lop_ind:twr.1} applied again to each of the component representations
in order to restore the standard form of induced representation in Krein space to each of them separately. 
In this way we may repeat
the procedure of decomposing tensor product of the component representations (now in the standard form)
and continue it potentially in infinitum.
It turns out that the condition 3) of finite multiplicity will have to be abandoned in further stages of this process, 
but we have all the grounds for the condition 2) of ``local boundedness'' to be preserved in all
cases at all levels of the decomposition.
Indeed recall that the spectral values $(p_0, \ldots p_3 )$ of the translation generators (four-momentum operators) in the tensor product of representations corresponding to imprimitivity systems concentrated on single orbits
$\mathscr{O}',\mathscr{O}'' \subset \widehat{T_4}$, are the sums
$(p'_0, \ldots p'_3 ) + (p''_0, \ldots p''_3 )$, with the spectral values $(p'_0, \ldots p'_3 )$ and 
$(p''_0, \ldots p''_3 ) $ ranging over $\mathscr{O}'$ and $\mathscr{O}''$ respectively. Now the geometry 
of the orbits in case of $\mathfrak{G}
= T_4 \circledS SL(2,\mathbb{C})$ is such that the sets of all values  $(p'_0, \ldots p'_3 )$ and 
$(p''_0, \ldots p''_3 )$ for which $(p_0, \ldots p_3 )$ ranges over a compact set, are compact 
(discarding irrelevant null sets of $(p_0, \ldots p_3 )$ not belonging to the joint spectrum of momentum operators of the tensor product representation -- the light cones -- in the only case of tensoring representation corresponding to the positive energy light cone orbit with the representation corresponding to the negative energy light cone).

4) In fact the representation of one particle states in the Fock space (with the Gupta-Bleuler or fundamental
symmetry operator) is induced by the following representation $L''$ in the
above defined Krein space $(\mathcal{H}_L , \mathfrak{J}_L)$ of the double covering
of the symmetry group of the Euclidean plane:
\begin{equation}\label{L''}
L''_{(z, \phi)} =
\left( \begin{array}{cccc} 
1 + \frac{1}{2}|z|^2                     & \frac{1}{2}(\overline{z}+z) & \frac{i}{2}(z-\overline{z}) & -\frac{1}{2}|z|^2 \\ 
\frac{1}{2} (e^{-i\phi}\overline{z} +e^{i\phi}z) & \cos \phi & \sin \phi & -\frac{1}{2}(e^{-i\phi}\overline{z} +e^{i\phi}z \\
\frac{i}{2}(e^{i\phi}z-e^{-i\phi}\overline{z}) & -\sin \phi & \cos \phi & -\frac{1}{2}(e^{i\phi}z- e^{-\phi}\overline{z}) \\
\frac{1}{2}|z|^2  & \frac{1}{2}(\overline{z} + z) & \frac{i}{2}(z-\overline{z}) & 1- \frac{1}{2}|z|^2  \end{array}\right),
\end{equation}
compare e.g. \cite{Weinberg1, Weinberg2}, or \cite{lop1,lop2}. But the operator
\[
\left( \begin{array}{cccc} 1 &    0       &          0  & 0 \\ 
                           0 & 1/\sqrt{2} & i/\sqrt{2}  & 0 \\
                           0 & 1/\sqrt{2} & -i/\sqrt{2} & 0 \\
                           0 &    0       &    0        & 1  \end{array}\right).
\] 
which is Krein-unitary and unitary in $(\mathcal{H}_L , \mathfrak{J}_L)$ sets up 
Krein-unitary and unitary equivalence between the representation $L'$ of (\ref{L'})
and the representation $L''$ of (\ref{L''}) as well as between the associated representations $L$.
By Theorem \ref{def_ind_krein:twr.4} it makes no difference which one we use, but for some technical reasons we
prefer the representation $L$ associated with (\ref{L'}).

5) The representation which we have called by the name of {\L}opusza\'nski have appeared in physics
rather very early, compare \cite{Wigner}, and then in relation to the Gupta-Bleuler quantization
of the free photon field:  \cite{Weinberg1, Weinberg2}, \cite{Halpern}, \cite{Kupersztych}. 
But it was {\L}opusza\'nski \cite{lop1,lop2} who initiated a systematic study of the relation
of the representation with the Gupta-Bleuler 
formalism. That's why we call the representation after him.

\section{Kronecker product of induced representations in Krein spaces}\label{kronecker}

In this Section we define the outer Kronecker product and inner Kronecker product of Krein
isometric (and Krein unitary) representations and give an important theorem concerning 
Krein isometric representation induced by a Kronecker product of Krein-unitary representations.

The whole construction is based on the ordinary tensor product of the associated Hilbert spaces 
and operators in the Hilbert spaces. We recapitulate shortly a specific realization of the 
tensor product of Hilbert spaces as trace class conjugate-linear operators, in short we realize it by the  Hilbert-Schmidt 
class of conjugate-linear operators with the standard 
operator $L^2$-norm, for details we refer the reader to the original
paper by Murray and von Neumann \cite{Murray_von_Neumann}. Alternatively one may consider linear Hilbert-Schmidt class operators, but
replace one of the Hilbert spaces in question by its conjugate space, compare \cite{Mackey}, \S 5.

Let $\mathcal{H}_1$ and $\mathcal{H}_2$ be two
separable Hilbert spaces over $\mathbb{C}$ (recall that by the proof of Lemma \ref{lem:dense.6}
the Hilbert space $\mathcal{H}^L$ of the Krein-isometric representation $U^L$ of a separable locally compact group 
$\mathfrak{G}$ induced by a Krein-unitary representation $L$ of a closed subgroup $G_1 \subset \mathfrak{G}$
is separable). A mapping $T$ of $\mathcal{H}_2$ to $\mathcal{H}_1$
is conjugate-linear iff $T(\alpha f + \beta g) = \overline{\alpha} \, T(f) + \overline{\beta} \, T(g)$
for all $f,g \in \mathcal{H}_2$ and all complex numbers $\alpha$ and $\beta$, with the ``over-line'' sign standing
for complex conjugation. For any such conjugate-linear operator $T$ we define the conjugate 
version of its adjoint $T^{{}^\circledast}$, namely this is the operator fulfilling 
$(Tg, f) = (T^{{}^\circledast} f, g)$ for all $f \in \mathcal{H}_1$ and all $g \in \mathcal{H}_2$.
In particular if $T$ is bounded, conjugate-linear, finite-rank operator so is its conjugate
adjoint $T^{{}^\circledast}$. If $U_1$ and $U_2$ are bounded operators in $\mathcal{H}_1$
and $\mathcal{H}_2$ respectively then $U_1 T U_2$ is a finite rank operator from $\mathcal{H}_2$
into $\mathcal{H}_1$. One easily verifies that $(A T B)^{{}^\circledast} = B^* T^{{}^\circledast} A^*$,
where $A$ and $B$ are linear operators in $\mathcal{H}^1$ and $\mathcal{H}^2$ with $A^*$ and
$B^*$ equal to their ordinary adjoint operators. If $U_1$ and $U_2$ are densely defined operators in $\mathcal{H}_1$
and $\mathcal{H}_2$ respectively on linear domains $\mathfrak{D}_1 \subset \mathcal{H}_1$
and $\mathfrak{D}_2 \subset \mathcal{H}_2$ and $T$ is finite rank operator with the rank contained
in $\mathfrak{D}_1$ and supported in $\mathfrak{D}_2$, then $U_1 T U_2$ is a well defined finite rank operator.
Let $\mathcal{H}' = \mathcal{H}_1 \otimes' \mathcal{H}_2$ be the linear space of finite rank 
conjugate-linear operators $T$ of $\mathcal{H}_2$ into $\mathcal{H}_1$. For any two such operators
$T$ and $S$ the operator $T S^{{}^\circledast}$ is linear from $\mathcal{H}_1$
into $\mathcal{H}_1$ and of finite rank (similarly $T^{{}^\circledast} S$ is linear and finite rank
from $\mathcal{H}_2$ into $\mathcal{H}_2$). We may therefore introduce the following 
inner product in $\mathcal{H}'$:  
\begin{multline*}
\langle T, S \rangle = \Tr [\, T S^{{}^\circledast} \,] = \sum \limits_{n} (T \, S^{{}^\circledast} e_n, e_n) \\
= \sum \limits_{n} (T^{{}^\circledast} e_n , S^{{}^\circledast} e_n) 
= \sum \limits_{m} (T \varepsilon_m , S \varepsilon_m)    \\
\sum \limits_{m} (T^{{}^\circledast} S \varepsilon_m , \varepsilon_m)
= \Tr [\, T^{{}^\circledast} S \,], 
\end{multline*}
where $\{ e_n \}_{n \in \mathbb{N}}$ and $\{ \varepsilon_m \}_{m \in \mathbb{N}}$ are orthonormal bases
in $\mathcal{H}_1$ and $\mathcal{H}_2$ respectively. The completion of $\mathcal{H}'$ with respect to
this inner product composes the tensor product $\mathcal{H}  = \mathcal{H}_1 \otimes \mathcal{H}_2$.

Let $A$ and $B$ be bounded operators in $\mathcal{H}_1$
and $\mathcal{H}_2$. Their tensor product $A \otimes B$ acting in $\mathcal{H}_1 \otimes \mathcal{H}_2$
is defined as the operator $T \mapsto ATB^*$, for $T \in \mathcal{H}_1 \otimes \mathcal{H}_2$. 
In particular if for any $f \in \mathcal{H}_1$ and $g \in \mathcal{H}_2$ we define the finite rank 
conjugate-linear operator $T_{{}_{f, g}}: w \mapsto f \, \cdot \, (g, w)$ supported on the 
linear subspace generated by $g$ with the range generated by $f$, then 
$T_{{}_{f, g}} \in \mathcal{H}_1 \otimes \mathcal{H}_2$ is written as $f \otimes g$ and we have
$(f_1 \otimes g_1 , f_2 \otimes g_2) = \Tr \big[ T_{{}_{f_1, g_1}} \, 
\big(T_{{}_{f_2, g_2}}\big)^{{}^\circledast} \big]
= \Tr \big[ T_{{}_{f_1, g_1}} \, T_{{}_{g_2, f_2}} \big] = (f_1 , f_2)\cdot (g_1 , g_2)$ 
because $\big(T_{{}_{f_2, g_2}}\big)^{{}^\circledast} = T_{{}_{g_2, f_2}}$.

If $(\mathcal{H}_1 , \mathfrak{J}_1)$ and $(\mathcal{H}_2 , \mathfrak{J}_2)$ are two Krein spaces,
then we define their tensor product as the Krein space 
$(\mathcal{H}_1 \otimes \mathcal{H}_2 , \mathfrak{J}_1 \otimes \mathfrak{J}_2)$; verification
of the self-adjointness of $\mathfrak{J}_1 \otimes \mathfrak{J}_2$ and the property 
$\big( \mathfrak{J}_1 \otimes \mathfrak{J}_2 \big)^2 = I$ is immediate.

We say an operator $T$ from $\mathcal{H}_2$ into $\mathcal{H}_1$ is supported by finite dimensional (or 
more generally: closed) 
linear subspace $\mathfrak{M} \subset \mathcal{H}_2$ or by the projection $P_{{}_\mathfrak{M}}$, in case 
$T = TP_{{}_\mathfrak{M}}$, where $P_{{}_\mathfrak{M}}$ is the self adjoint projection with range $\mathfrak{M}$.
Similarly we say an operator $T$ from $\mathcal{H}_2$ into $\mathcal{H}_1$ has range in a finite dimensional
(or more generally: closed) linear subspace $\mathfrak{N} \subset \mathfrak{H}_1$, in case 
$T = P_{{}_\mathfrak{N}} T$, where $P_{{}_\mathfrak{N}}$ is the self adjoint projection with range $\mathfrak{N}$. 
One easily verifies the following tracial property. Let $B$ be any finite rank and \emph{linear} operator
from $\mathcal{H}_1$ into $\mathcal{H}_1$ supported on a finite dimensional linear subspace of the domain $\mathfrak{D}_1$ and with the range also finite dimensional and lying in $\mathfrak{D}_1$. Then for any linear
operator defined on the dense domain $\mathfrak{D}_1 \subset \mathcal{H}_1$ and preserving 
it, i. e.  with $\mathfrak{D}_1$ contained in the common domain of $A$ and its 
adjoint $A^*$, we have the tracial property
\[
\Tr [B A] = \Tr [A B]. 
\]  
Indeed any such linear $B$ is a finite linear combination of the operators $\mathbb{T}_{{}_{f, f'}}$
defined as follows: $\mathbb{T}_{{}_{f, f'}} (w) = (w, f) \cdot f'$. 
By linearity it will be sufficient to establish the tracial property for the linear operator
$B$ of the form $B = \mathbb{T}_{{}_{f_{{}_1}, f_{{}_2}}} + \mathbb{T}_{{}_{f_{{}_3}, f_{{}_4}}}$
with $f_i \in \mathfrak{D}_1$, $i = 1,2,3,4$. Using the Gram-Schmidt orthogonalization 
we construct an orthonormal basis $\{ e_n \}_{n \in \mathbb{N}}$ of $\mathcal{H}_1$ with
$e_n \in \mathfrak{D}_1$. We have in this case
\begin{multline}\label{tr_BA}
\Tr [B A] = \Tr \Big[ \big( \mathbb{T}_{{}_{f_{{}_1}, f_{{}_2}}} 
+ \mathbb{T}_{{}_{f_{{}_3}, f_{{}_4}}} \big) A \Big] 
= \Tr \Big[ \mathbb{T}_{{}_{f_{{}_1}, f_{{}_2}}} A \big]
+ \Tr \Big[ \mathbb{T}_{{}_{f_{{}_3}, f_{{}_4}}}  A \Big]  \\
= \sum \limits_n \big( \mathbb{T}_{{}_{f_{{}_1}, f_{{}_2}}} A e_n , e_n \big) 
+ \sum \limits_n \big( \mathbb{T}_{{}_{f_{{}_3}, f_{{}_4}}}  A e_n , e_n \big)  \\
= \sum \limits_n (A e_n , f_1) \cdot (f_2 , e_n) + \sum \limits_n (A e_n , f_3) \cdot (f_4 , e_n) \\
= \sum \limits_n ( e_n , A^* f_1) \cdot (f_2 , e_n) + \sum \limits_n ( e_n , A^* f_3) \cdot (f_4 , e_n)
= (f_2 , A^* f_1) + (f_4 , A^* f_3)   \\
= (A f_2 , f_1) + (A f_4 , f_3) < \infty,
\end{multline}
because by the assumed properties of the operator $A$ the vectors $f_1 , f_3 \in \mathfrak{D}_1$ are
contained in the domain of $A^*$ and likewise the vectors $f_2 , f_4 \in \mathfrak{D}_1$ lie in the domain of $A$. 
Similarly we have:
\begin{multline}\label{tr_AB}
\Tr [A B] = \Tr \Big[ \big( A \mathbb{T}_{{}_{f_{{}_1}, f_{{}_2}}} 
+  \mathbb{T}_{{}_{f_{{}_3}, f_{{}_4}}} \big) \Big] 
= \Tr \Big[ A \mathbb{T}_{{}_{f_{{}_1}, f_{{}_2}}} \big]
+ \Tr \Big[ A \mathbb{T}_{{}_{f_{{}_3}, f_{{}_4}}}  \Big]  \\
= \sum \limits_n \big(  A \mathbb{T}_{{}_{f_{{}_1}, f_{{}_2}}} e_n , e_n \big) 
+ \sum \limits_n \big( A \mathbb{T}_{{}_{f_{{}_3}, f_{{}_4}}}  e_n , e_n \big)  \\
= \sum \limits_n ( e_n , f_1) \cdot ( A f_2 , e_n) + \sum \limits_n ( e_n , f_3) \cdot ( A f_4 , e_n)
= (A f_2 , f_1) + (A f_4 , f_3) < \infty.
\end{multline}
Comparing (\ref{tr_BA}) and (\ref{tr_AB}) we obtain the tracial property.

Now let  $U_1 = U^{L}_{x}$ and $U_2 = U^{M}_{y}$ be densely defined and closable Krein isometric 
operators of the respective Krein isometric induced representations of the groups $\mathfrak{G}_1$ and $\mathfrak{G}_2$ 
in $\mathcal{H}_1 = \mathcal{H}^L$
and $\mathcal{H}_2 = \mathcal{H}^M$ respectively with linear domains $\mathfrak{D}_i \subset \mathcal{H}_i$,
$i = 1,2$, equal to the corresponding
domains $\mathfrak{D}$ of Theorem \ref{def_ind_krein:twr.1} and Remark \ref{rem:def_ind_krein.1}
and with the respective fundamental symmetries $\mathfrak{J}_1 = \mathfrak{J}^L$, $\mathfrak{J}_2 = \mathfrak{J}^M$. 
Therefore by Theorem \ref{def_ind_krein:twr.1} and Remark \ref{rem:def_ind_krein.1} 
$U^i (\mathfrak{D}_i) = \mathfrak{D}_i$ and 
$\mathfrak{J}_i (\mathfrak{D}_i) = \mathfrak{D}_i$, $i = 1, 2$, so that $\mathfrak{D}_i$ is contained in the domain
of ${U_i}^*$ and ${U_i}^* (\mathfrak{D}_1 ) = \mathfrak{D}_i$.
Finally let $T,S$ be any finite rank operators in the linear subspace $\mathfrak{D}_{12}
= \textrm{linear\,span}\{ T_{{}_{f, g}}, f \in \mathfrak{D}_1 , g \in \mathfrak{D}_2 \}$ of finite rank 
operators supported in $\mathfrak{D}_2$ and with ranges in $\mathfrak{D}_1$. In particular for each 
$S \in \mathfrak{D}_{12}$, $S^{{}^\circledast}$ is supported in $\mathfrak{D}_1$ and has rank in $\mathfrak{D}_2$.
By the known property of Hilbert Schmidt operators $\mathfrak{D}_1 \otimes \mathfrak{D}_2 = \mathfrak{D}_{12}$
is dense in $\mathcal{H}_1 \otimes \mathcal{H}_2$.
We claim that $U_1 \otimes U_2$ is well defined on $\mathfrak{D}_1 \otimes \mathfrak{D}_2 = \mathfrak{D}_{12}$.
Indeed, by the Gram-Schmidt orthonormalization we may construct an orthonormal base $\{e_n\}_{n \in \mathbb{N}}$
of $\mathcal{H}_1$ with each $e_n$ being an element of the linear dense domain $\mathfrak{D}_1$. 
For any $f_1 , f_2 \in \mathfrak{D}_1$ and $g_1 , g_2 \in \mathfrak{D}_2$ we have 
\begin{multline*}
\Big\| \big( U_1 \otimes U_2 \big) \big( f_1 \otimes g_1 + f_2 \otimes g_2 \big) \Big\|^2  \\
= \Big\langle \,\, U_1 \big( T_{{}_{f_{{}_1}, g_{{}_1}}} + T_{{}_{f_{{}_2}, g_{{}_2}}} \big){U_2}^* \,\, , \,\,\,\, 
U_1 \big( T_{{}_{f_{{}_1}, g_{{}_1}}} + T_{{}_{f_{{}_2}, g_{{}_2}}} \big)  {U_2}^* \,\, \Big\rangle  \\
= \Tr \Big[ \, U_1 \big( T_{{}_{f_{{}_1}, g_{{}_1}}} \big) {U_2}^* 
\big(  U_1 \big( T_{{}_{f_{{}_1}, g_{{}_1}}}  \big) {U_2}^* \big)^{{}^\circledast} \, \Big]
+ \Tr \Big[ \, U_1 \big( T_{{}_{f_{{}_1}, g_{{}_1}}} \big) {U_2}^* 
\big(  U_1 \big( T_{{}_{f_{{}_2}, g_{{}_2}}}  \big) {U_2}^* \big)^{{}^\circledast} \, \Big] \\
+ \Tr \Big[ \, U_1 \big( T_{{}_{f_{{}_2}, g_{{}_2}}} \big) {U_2}^* 
\big(  U_1 \big( T_{{}_{f_{{}_1}, g_{{}_1}}}  \big) {U_2}^* \big)^{{}^\circledast} \, \Big]
+ \Tr \Big[ \, U_1 \big( T_{{}_{f_{{}_2}, g_{{}_2}}} \big) {U_2}^* 
\big(  U_1 \big( T_{{}_{f_{{}_2}, g_{{}_2}}} \big) {U_2}^* \big)^{{}^\circledast} \, \Big] \\
= \sum \limits_n \big( U_1 f_1, e_n  \big) \cdot \big( g_1, {U_2}^* U_2 g_1  \big) 
\cdot \big( {U_1}^* e_n, f_1  \big) \\
+ \sum \limits_n \big( U_1 f_1, e_n  \big) \cdot \big( g_1, {U_2}^* U_2 g_2  \big) 
\cdot \big( {U_1}^* e_n, f_2  \big) \\
+ \sum \limits_n \big( U_1 f_2, e_n  \big) \cdot \big( g_2, {U_2}^* U_2 g_1  \big) 
\cdot \big( {U_1}^* e_n, f_1  \big) \\
+ \sum \limits_n \big( U_1 f_2, e_n  \big) \cdot \big( g_2, {U_2}^* U_2 g_2  \big) 
\cdot \big( {U_1}^* e_n, f_2  \big).
\end{multline*}  
Because $\mathfrak{D}_1$ is in the domain of ${U_1}^*$ and ${U_1}^* (\mathfrak{D}_1) = U_1 (\mathfrak{D}_1)
= \mathfrak{D}_1$ and similarly for $U_2$, the last expression is equal to 
\begin{multline*}
\sum \limits_n \big( U_1 f_1, e_n  \big) \cdot \big( U_2 g_1,  U_2 g_1  \big) 
\cdot \big(  e_n, U_1 f_1  \big) \\
+ \sum \limits_n \big( U_1 f_1, e_n  \big) \cdot \big( U_2 g_1,  U_2 g_2  \big) 
\cdot \big(  e_n, U_1 f_2  \big) \\
+ \sum \limits_n \big( U_1 f_2, e_n  \big) \cdot \big( U_2 g_2,  U_2 g_1  \big) 
\cdot \big(  e_n, U_1 f_1  \big) \\
+ \sum \limits_n \big( U_1 f_2, e_n  \big) \cdot \big( U_2 g_2,  U_2 g_2  \big) 
\cdot \big(  e_n, U_1 f_2  \big) \\
= \big( U_1 f_1,   U_1 f_1 \big) \cdot \big( U_2 g_1,  U_2 g_1  \big) 
+ \big( U_1 f_1,  U_1 f_2 \big) \cdot \big( {U_2} g_1,  U_2 g_2  \big) \\ 
+ \big( U_1 f_2, U_1 f_1  \big) \cdot \big( U_2 g_2,  U_2 g_1  \big) 
+ \big( U_1 f_2, U_1 f_2  \big) \cdot \big( U_2 g_2,  U_2 g_2  \big) < \infty,
\end{multline*}
so that 
\begin{multline*}
\Big\| \big( U_1 \otimes U_2 \big) \big( f_1 \otimes g_1 + f_2 \otimes g_2 \big) \Big\|^2  \\
= \Big\langle \,\, U_1 \big( T_{{}_{f_{{}_1}, g_{{}_1}}} + T_{{}_{f_{{}_2}, g_{{}_2}}} \big){U_2}^* \,\, , \,\,\,\, 
U_1 \big( T_{{}_{f_{{}_1}, g_{{}_1}}} + T_{{}_{f_{{}_2}, g_{{}_2}}} \big)  {U_2}^* \,\, \Big\rangle < \infty 
\end{multline*}  
and $\big( U_1 \otimes U_2 \big) \big( f_1 \otimes g_1 + f_2 \otimes g_2 \big)$ is well defined. By induction
for each $T \in \mathfrak{D}_{12}$,  
$\big( U_1 \otimes U_2 \big) \big( T \big)= U_1 T { U_2}^*$ is well defined conjugate-linear operator of 
Hilbert-Schimdt class, so that $U_1 \otimes U_2$ is well defined on the linear domain $\mathfrak{D}_{12}$
dense in $\mathcal{H}_1 \otimes \mathcal{H}_2$.  By the Proposition of Chap. VIII.10, page 298 of 
\cite{Reed_Simon} it follows that $U_1 \otimes U_2$ is closable.  
Next, let $T, S \in \mathfrak{D}_{12}$, then by Theorem \ref{def_ind_krein:twr.1} 
and Remark \ref{rem:def_ind_krein.1}
\begin{multline*}
\mathfrak{J}_i (\mathfrak{D}_i) = \mathfrak{D}_i \,\,\, \textrm{and} \,\,\, 
U_i (\mathfrak{D}_i) = \mathfrak{D}_i  \,\,\, \textrm{and} \,\,\, \textrm{and} \,\,\,
{U_i}^* (\mathfrak{D}_i) = \mathfrak{D}_i  \\
\big( \, U_{i} \, \big)^\dagger \, U_{i} = \mathfrak{J}_i {U_i}^* \mathfrak{J}_i U_{i} 
= I  \,\,\, \textrm{and} \,\,\,
U_{i} \, \mathfrak{J}_i {U_i}^* \mathfrak{J}_i = I \,\,\, \textrm{on} \,\,\, \mathfrak{D}_i .
\end{multline*}
Thus for each $T, S \in \mathfrak{D}_{12}$ the following expressions are well defined and 
(e. g. for $T = T_{{}_{f_1, g_1}}$ and $S = T_{{}_{f_2, g_2}}$) 
\begin{multline*}
\Big( \,\, (\mathfrak{J}_1 \otimes \mathfrak{J}_2) \, (U_1 \otimes U_2)  \, (f_1 \otimes g_1) \,\, , 
\,\,\,\, (U_1 \otimes U_2) \,  (f_2 \otimes g_2) \,\,  \Big) \\
= \Big\langle \,\, \mathfrak{J}_1 U_1 T {U_2}^*\mathfrak{J}_2 \,\, , \,\,\,\,  U_1 S {U_2}^*  \,\, \Big\rangle  
= \Tr \Big[ \, \mathfrak{J}_1 U_1 T {U_2}^* \mathfrak{J}_2 \big(  U_1 S {U_2}^*  \big)^{{}^\circledast} \, \Big]  \\
= \Tr \Big[ \, \mathfrak{J}_1 U_1 T {U_2}^* \mathfrak{J}_2  U_2 S^{{}^\circledast} {U_1}^*   \, \Big] 
= \Tr \Big[ \, \mathfrak{J}_1 U_1 T \mathfrak{J}_2 \mathfrak{J}_2 {U_2}^* \mathfrak{J}_2  U_2 S^{{}^\circledast} {U_1}^*   \, \Big] \\ 
= \Tr \Big[ \,  \mathfrak{J}_1 U_1 T \mathfrak{J}_2 \{ \mathfrak{J}_2 {U_2}^* \mathfrak{J}_2  U_2 \} 
S^{{}^\circledast} {U_1}^*  \, \Big] 
= \Tr \Big[ \, \{ \mathfrak{J}_1 U_1 \} T \mathfrak{J}_2 S^{{}^\circledast} {U_1}^* \, \Big] \\
= \Tr \Big[ \,  T \mathfrak{J}_2 S^{{}^\circledast} {U_1}^* \{ \mathfrak{J}_1 U_1 \} \, \Big] 
= \Tr \Big[ \,  T \mathfrak{J}_2 S^{{}^\circledast} \mathfrak{J}_1 \{ \mathfrak{J}_1  {U_1}^* \mathfrak{J}_1 U_1 \} \, \Big]  \\ 
= \Tr \Big[ \,  T \mathfrak{J}_2 S^{{}^\circledast} \mathfrak{J}_1 \, \Big] 
= \Tr \Big[ \, \mathfrak{J}_1 T \mathfrak{J}_2 S^{{}^\circledast}  \, \Big] \\
= \Big( \,\, (\mathfrak{J}_1 \otimes \mathfrak{J}_2)  \, (f_1 \otimes g_1) \,\, , \,\,\,\,   f_2 \otimes g_2  \Big), 
\end{multline*}  
because the tracial property is applicable to the pair of operators 
\[
B =  T \mathfrak{J}_2 S^{{}^\circledast} {U_1}^*
\,\,\, \textrm{and} \,\,\,
A = \mathfrak{J}_1 U_1 
\]
as well as to the pair of operators
\[
B =  T \mathfrak{J}_2 S^{{}^\circledast}
\,\,\, \textrm{and} \,\,\,
A = \mathfrak{J}_1, 
\]
as both the operators $B$ are linear finite rank operators supported on finite dimensional subspaces contained 
in $\mathfrak{D}_1$ and with finite dimensional ranges contained in $\mathfrak{D}_1$
and for the operators $A$ indicated to above the linear domain $\mathfrak{D}_1$ is contained in the common domain of $A$
and $A^*$;  and moreover $\mathfrak{J}_1 (U_1)^* \mathfrak{J}_1 U_1$ and 
$ \mathfrak{J}_2 {U_2}^* \mathfrak{J}_2  U_2$ are well defined unit operators on the domains $\mathfrak{D}_1$
and $\mathfrak{D}_2$ respectively. Therefore $U_1 \otimes U_2 = U^L \otimes U^M$ is Krein-isometric on its
domain $\mathfrak{D}_{12}$ which holds by continuity for its closure.

We may therefore define the outer Kronecker product Krein-isometric representation 
$U^L \times U^M : \mathfrak{G}_1 \times \mathfrak{G}_2 \ni (x,y) \mapsto U^{L}_{x} \otimes U^{M}_{y}$ 
of the product group $\mathfrak{G}_1 \times \mathfrak{G}_2$, which is Krein isometric in the
Krein space $\big(\mathcal{H}^L \otimes \mathcal{H}^M  , \,\, \mathfrak{J}^L \otimes \mathfrak{J}^M \big)$.
All the more, if $U_1$ and $U_2$ are Krein-unitary representations of $G_1$ and $G_2$, 
respectively in $(\mathcal{H}_1 , \mathfrak{J}_1)$ and 
$(\mathcal{H}_2 , \mathfrak{J}_2)$, so is $U_1 \times U_2$ in the Krein space
$\big(\mathcal{H}_1 \otimes \mathcal{H}_2  , \,\, \mathfrak{J}_1 \otimes \mathfrak{J}_2 \big)$. Similarly
one easily verifies that $U_1 \times U_2$ is almost uniformly bounded whenever  $U_1$ and $U_2$ are.
In particular if $G_1$ and $G_2$ are two closed subgroups of the separable locally compact groups 
$\mathfrak{G}_1$ and $\mathfrak{G}_2$ respectively and $L$ and $M$ their Krein unitary and uniformly
bounded representations, then we may define the outer Kronecker
product representation $L \times M$ of the product group $G_1 \times G_2$ by the ordinary formula 
$\mathfrak{G}_1 \times \mathfrak{G}_2 \ni (\xi , \eta) \mapsto  L_\xi \otimes M_\eta$, which is 
Krein unitary and almost uniformly bounded in the Krein space
$\big(\mathcal{H}_L \otimes \mathcal{H}_M  , \,\, \mathfrak{J}_L \otimes \mathfrak{J}_M \big)$
whenever $L$ and $M$ are in the respective Krein spaces $(\mathcal{H}_L, \mathfrak{J}_L)$
and $(\mathcal{H}_M, \mathfrak{J}_M)$. We may therefore define the Krein-isometric 
representation ${}^{\mu_1 \times \mu_2}U^{L \times M}$ of the group $\mathfrak{G}_1 \times \mathfrak{G}_2$
in the Krein space $\mathcal{H}^{L \times M}$ induced by the representation $L \times M$ of the 
closed subgroup $G_1 \times G_2$, where $\mu_i$ are the respective quasi invariant measures in 
$\mathfrak{G}_i /G_i$.

Let us make an observation used in the proof of the Theorem of this Section.
Let $B_1$ be a Borel section of $\mathfrak{G}_1$ with respect to 
$G_1$ and respectively $B_2$ a Borel section of $\mathfrak{G}_2$ with respect to $G_2$ 
defined as in Section \ref{def_ind_krein} with the associated Borel functions
$h_1 : \mathfrak{G}_1 \ni x \mapsto h_1 (x) \in G_1$ such that ${h_1 (x)}^{-1}x \in B_1$ and
$h_2: \mathfrak{G}_2 \ni y \mapsto h_2 (y) \in G_2$ such that ${h_2 (y)}^{-1}y \in B_2$.
Then $B_1 \times B_2$ is a Borel section of $\mathfrak{G}_1 \times \mathfrak{G}_2$ with respect
to the closed subgroup $G_1 \times G_2$ with the associated Borel function
$h: (x, y) \mapsto h(x,y) \in G_1 \times G_2$ such that ${h(x, y)}^{-1} (x,y) \in B_1 \times B_2$, equal to
$h(x, y) = \big( \, h_{{}_1} (x) \, , \,\, h_{{}_2} (y) \, \big) \, = \, h_{{}_1} (x) \times h_{{}_2} (y)$. 
Let $w \in \mathcal{H}^{L \times M}$. Thus the corresponding operator $\mathfrak{J}^{L \times M}$
acts as follows 
\begin{multline*}
\big(\mathfrak{J}^{L \times M} w \big)_{{}_{(x, y)}} = 
(L \times M)_{{}_{h_1 (x) \times h_2 (y)}} \circ \big( \mathfrak{J}_{L \times M} \big) \circ
(L \times M)_{{}_{{h_{{}_1} (x)}^{-1} \times {h_{{}_2} (y)}^{-1}}} w_{{}_{(x,y)}} \\
= \big( L_{{}_{h_{{}_1} (x)}} \otimes M_{{}_{h_{{}_2} (y)}} \big) \circ
\big( \mathfrak{J}_L \otimes \mathfrak{J}_M  \big) \circ
\big( L_{{}_{{h_{{}_1} (x)}^{-1}}} \otimes M_{{}_{{h_{{}_2} (y)}^{-1}}} \big) w_{{}_{(x, y)}} \\
= \big( L_{{}_{h_{{}_1} (x)}} \mathfrak{J}_L   L_{{}_{{h_{{}_1} (x)}^{-1}}} \big) \otimes 
\big( M_{{}_{h_{{}_2} (y)}} \mathfrak{J}_M  M_{{}_{{h_{{}_2} (y)}^{-1}}} \big) w_{{}_{(x, y)}}.      
\end{multline*}
Thus the vector $\mathfrak{J}_{L \times M} \big( \mathfrak{J}^{L \times M}  w \big)_{{}_{(x, y)}}$ in the integrand in the formula for the inner product in $\mathcal{H}^{L \times M}$
\[
(w, u)
= \int \limits_{\mathfrak{G}_1 \times \mathfrak{G}_2} 
\Big( \mathfrak{J}_{L \times M} \big( \mathfrak{J}^{L \times M}  w \big)_{{}_{(x, y)}} , 
\big(u \big)_{{}_{(x, y)}}  \Big)  \,\, \ud (\mu_1 \times \mu_2 ) ([(x,y)])
\]
may be written as follows
\begin{multline*}
\mathfrak{J}_{L \times M} \big( \mathfrak{J}^{L \times M}  w \big)_{{}_{(x,y)}}
= \big( \mathfrak{J}_L \otimes \mathfrak{J}_M \big) \circ \big( \mathfrak{J}^{L \times M}  w \big)_{{}_{(x, y)}} \\
=  \big( \mathfrak{J}_L L_{{}_{h_{{}_1} (x)}} \mathfrak{J}_L   L_{{}_{{h_{{}_1} (x)}^{-1}}} \big) \otimes 
\big( \mathfrak{J}_M  M_{{}_{h_{{}_2} (y)}} \mathfrak{J}_M  M_{{}_{{h_{{}_2} (y)}^{-1}}} \big) w_{{}_{(x, y)}}
= \big( {}_{{}_x}\mathfrak{J}^L \otimes {}_{{}_y}\mathfrak{J}^M  \big) w_{{}_{(x, y)}},
\end{multline*}
where we have introduced the following self-adjoint operators 
\[
{}_{{}_x}\mathfrak{J}^L = \mathfrak{J}_L L_{{}_{h_{{}_1} (x)}} \mathfrak{J}_L   L_{{}_{{h_{{}_1} (x)}^{-1}}}
\,\,\, \textrm{and} \,\,\,
{}_{{}_y}\mathfrak{J}^M = \mathfrak{J}_M  M_{{}_{h_{{}_2} (y)}} \mathfrak{J}_M  M_{{}_{{h_{{}_2} (y)}^{-1}}}
\]
acting in $\mathcal{H}_L$ and $\mathcal{H}_M$, respectively, with the ordinary tensor product 
operator ${}_{{}_x}\mathfrak{J}^L \otimes {}_{{}_y}\mathfrak{J}^M $ acting in the tensor product
$\mathcal{H}_L \otimes \mathcal{H}_M$ Hilbert space.

Checking their self-adjointness is immediate. Indeed, because $L$ is Krein unitary in $(\mathcal{H}_L , \mathfrak{J}_L)$
we have (and similarly for the rep. $M$):
\[
L_{{}_{{h_{{}_1} (x)}^{-1}}} = \big( L_{{}_{h_{{}_1} (x)}} \big)^\dagger 
= \mathfrak{J}_L \big( L_{{}_{h_{{}_1} (x)}} \big)^* \mathfrak{J}_L .
\]
Therefore 
\[
{}_{{}_x}\mathfrak{J}^L = \mathfrak{J}_L L_{{}_{h_{{}_1} (x)}} \big( L_{{}_{h_{{}_1} (x)}} \big)^* \mathfrak{J}_L ,  
\]
because $\big( \mathfrak{J}_L \big)^2 = I$. Because $\mathfrak{J}_L$ is self-adjoint, self-adjointness
of ${}_{{}_x}\mathfrak{J}^L$ is now immediate (self-adjointness of ${}_{{}_y}\mathfrak{J}^M$
follows similarly).   

We are ready now to formulate the main goal of this Section:

\begin{twr}
Let $L$ and $M$ be Krein-unitary strongly continuous and almost uniformly bounded representations of the
closed subgroups $G_1$ and $G_2$ of the separable locally compact groups $\mathfrak{G}_1$ and $\mathfrak{G}_2$,
respectively, in the Krein spaces $(\mathcal{H}_L , \mathfrak{J}_L)$ and $(\mathcal{H}_M , \mathfrak{J}_M)$.
Then the Krein isometric representation ${}^{\mu_1 \times \mu_2}U^{L \times M}$ of the group
$\mathfrak{G}_1 \times \mathfrak{G}_2$ with the representation space equal to the Krein space 
$(\mathcal{H}^{L \times M} , \mathfrak{J}^{L \times M})$ is unitary and Krein-unitary equivalent to the 
Krein-isometric representation ${}^{\mu_1}U^L \, \times \, {}^{\mu_2}U^M$ of the group 
$\mathfrak{G}_1 \times \mathfrak{G}_2$
with the representation space equal to the Krein space 
$(\mathcal{H}^L \otimes \mathcal{H}^M , \mathfrak{J}^L \otimes \mathfrak{J}^M)$. More precisely:
there exists a map $V: \mathcal{H}^L \otimes \mathcal{H}^M \mapsto \mathcal{H}^{L \times M}$
which is unitary between the indicated Hilbert spaces and Krein-unitary between the Krein spaces 
$(\mathcal{H}_L \otimes \mathcal{H}_M , \mathfrak{J}_L \otimes \mathfrak{J}_M)$ and
$(\mathcal{H}^{L \times M} , \mathfrak{J}^{L \times M})$ and such that 
\begin{equation}\label{ulxm=ulxum}
\boxed{V^{-1} \, \Big( \, {}^{\mu_1 \times \mu_2}U^{L \times M} \, \Big) \, V \,\, 
= \,\, {}^{\mu_1}U^L \, \times \, {}^{\mu_2}U^M .}  
\end{equation}  

\label{twr.1:kronecker}
\end{twr}
\qedsymbol \,
Let $T$ be any member of $\mathcal{H}^L \otimes \mathcal{H}^M$, regarded as a \emph{conjugate-linear} operator
from ${}^{\mu_2}\mathcal{H}^M$ into ${}^{\mu_1}\mathcal{H}^L$, with the corresponding \emph{linear} operator 
$T \, T^{{}^\circledast}$ on  ${}^{\mu_1}\mathcal{H}^L$ having finite trace. Let moreover $T$ be a finite rank
operator. Then there exist $f_{{}_1}, f_{{}_2}, \ldots , f_{{}_n} \in {}^{\mu_1}\mathcal{H}^L$ and 
$g_{{}_1}, g_{{}_2}, \ldots , g_{{}_n} \in {}^{\mu_2}\mathcal{H}^M$ such that
$T(g) = T_{{}_{f_{{}_1}, g_{{}_1}}} (g) + \ldots + T_{{}_{f_{{}_n}, g_{{}_n}}} (g)
= f_{{}_1} \cdot \big( g_{{}_1} , w \big) + \ldots + f_{{}_n} \cdot \big( g_{{}_n} , g \big)$. For each
$(x, y) \in \mathfrak{G}_1 \times \mathfrak{G}_2$ we may define a conjugate-linear finite rank operator 
$\big(V(T)\big)_{{}_{(x, y)}}$ from $\mathcal{H}_M$ into $\mathcal{H}_L$ as follows. Let $\upsilon \in \mathcal{H}_M$,
then we put $\big(V(T)\big)_{{}_{(x, y)}}(\upsilon) = f_{{}_1} \cdot \big( g_{{}_1} , \upsilon \big) + \ldots + 
f_{{}_n} \cdot \big( g_{{}_n} , \upsilon \big)$. Note, please, that $\big(V(T)\big)_{{}_{(\xi x, \eta y)}} \, = \,
L_\xi \big(V(T)\big)_{{}_{(x, y)}} (M_\eta)^*$ for all $(x, y) \in \mathfrak{G}_1 \times \mathfrak{G}_2$

and all $(\xi, \eta) \in G_1 \times G_2$, so that the function $V(T): 
\mathfrak{G}_1 \times \mathfrak{G}_2 \ni (x, y) \mapsto \big(V(T)\big)_{{}_{(x, y)}} \in \mathcal{H}_L 
\otimes \mathcal{H}_M$ fulfils $\big(V(T)\big)_{{}_{(\xi x, \eta y)}} = \big( L_\xi \otimes M_\eta \big) 
\big(V(T)\big)_{{}_{(x, y)}}$ for all $(x, y) \in \mathfrak{G}_1 \times \mathfrak{G}_2$
and all $(\xi, \eta) \in G_1 \times G_2$.  

We shall show that the function $V(T)$ is a member of $\mathcal{H}^{L\times M}$ and moreover, that $V$ is unitary. To this
end we observe first, that $V$ is isometric (for the ordinary definite inner products), i. e. 
$\| V(T) \| = \| T \|$. Indeed, let $\{e_k\}_{k \in \mathbb{N}}$ be an orthonormal basis in $\mathcal{H}_L$. 
Using the observation we have made just before the formulation of the Theorem, self-adjointness
of the operators ${}_{{}_x}\mathfrak{J}^L$ and ${}_{{}_y}\mathfrak{J}^M$ and Scholium 3.9 and 5.3 of
\cite{Segal_Kunze}, we obtain:
\begin{multline*}
\| T \|^2 =  \big( \, f_{{}_1}\otimes g_{{}_1} + \ldots +  f_{{}_n} \otimes g_{{}_n} \, , \,\, 
f_{{}_1}\otimes g_{{}_1} + \ldots +  f_{{}_n} \otimes g_{{}_n} \, \big)   \\
= \Tr \Big[ \big( T_{{}_{f_{{}_1}, g_{{}_1}}} + \ldots + T_{{}_{f_{{}_n}, g_{{}_n}}} \big) 
\big( T_{{}_{g_{{}_1}, f_{{}_1}}} + \ldots + T_{{}_{g_{{}_n}, f_{{}_n}}} \big) \Big] 
= \sum \limits_{i,j = 1}^{n} (f_i , f_j) \cdot (g_i , g_j)    \\
= \sum \limits_{i,j = 1}^{n} 
\bigg( 
\int \limits_{\mathfrak{G}_1} 
\Big( \mathfrak{J}_L  \big( \mathfrak{J}^L  f_i \big)_{{}_{x}} , 
\big(f_j \big)_{{}_{x}}  \Big)  \,\, \ud \mu_1 ([x]) \,\, \bigg) \cdot
\bigg( 
\int \limits_{\mathfrak{G}_2} 
\Big( \mathfrak{J}_M  \big( \mathfrak{J}^M  g_i \big)_{{}_{y}} , 
\big(g_j \big)_{{}_{y}}  \Big)  \,\, \ud \mu_2 ([y]) \,\, 
\bigg)      \\
= \int \limits_{\mathfrak{G}_1 \times \mathfrak{G}_2} \, \bigg( \sum \limits_{i,j = 1}^{n}  
\Big( \mathfrak{J}_L  \big( \mathfrak{J}^L  f_i \big)_{{}_{x}} , 
\big(f_j \big)_{{}_{x}}  \Big) \cdot
\Big( \mathfrak{J}_M  \big( \mathfrak{J}^M  g_i \big)_{{}_{y}} , 
\big(g_j \big)_{{}_{y}}  \Big) 
\bigg)  
\,\,\,\, \ud (\mu_1 \times \mu_2 ) ([(x,y)]) \\
= \int \limits_{\mathfrak{G}_1 \times \mathfrak{G}_2} 
\bigg( \sum \limits_{i,j = 1}^{n}  
\Big( {}_{{}_x}\mathfrak{J}^L \big( f_i \big)_{{}_{x}} , 
\big(f_j \big)_{{}_{x}}  \Big) \cdot
\Big( {}_{{}_y}\mathfrak{J}^M \big( g_i \big)_{{}_{y}} , 
\big(g_j \big)_{{}_{y}}  \Big) 
\bigg)  
\,\,\,\, \ud (\mu_1 \times \mu_2 ) ([(x,y)])   \\
= \int \, \bigg( \sum \limits_{i,j = 1}^{n} \sum \limits_{k \in \mathbb{N}}  
\Big( {}_{{}_x}\mathfrak{J}^L \big( f_i \big)_{{}_{x}} , \, e_k \, \Big) \cdot \Big( \, e_k  , \,
\big(f_j \big)_{{}_{x}}  \Big) \cdot
\Big( {}_{{}_y}\mathfrak{J}^M \big( g_i \big)_{{}_{y}} , 
\big(g_j \big)_{{}_{y}}  \Big) 
\bigg)  
\,\,\,\, \ud (\mu_1 \times \mu_2) ([(x,y)])  \\
= \int \, \bigg( \sum \limits_{i,j = 1}^{n} \sum \limits_{k \in \mathbb{N}}  
\Big( \, e_k  , \, \big(f_j \big)_{{}_{x}}  \Big) \cdot  
\Big( \big( g_i \big)_{{}_{y}} , {}_{{}_y}\mathfrak{J}^M  \big(g_j \big)_{{}_{y}}  \Big) \cdot 
\Big( {}_{{}_x}\mathfrak{J}^L \big( f_i \big)_{{}_{x}} , \, e_k \, \Big)
 \bigg)  
\,\,\,\, \ud (\mu_1 \times \mu_2) ([(x,y)])  \\
= \int \, \bigg( \sum \limits_{i,j = 1}^{n} \sum \limits_{k \in \mathbb{N}}  
\Big( \, e_k  , \, \big(f_j \big)_{{}_{x}}  \Big) \cdot   
\Big( {}_{{}_x}\mathfrak{J}^L \Big( 
\big( f_i \big)_{{}_{x}} \cdot \Big( \big( g_i \big)_{{}_{y}} , {}_{{}_y}\mathfrak{J}^M  \big(g_j \big)_{{}_{y}}  \Big)
\Big) 
\, , \, e_k  \, \Big)
 \bigg)  
\,\,\,\, \ud (\mu_1 \times \mu_2 ) ([(x,y)])  \\
= \int \, \bigg( \sum \limits_{i,j = 1}^{n} \sum \limits_{k \in \mathbb{N}}  
\Big( \, e_k  , \, \big(f_j \big)_{{}_{x}}  \Big) \cdot   
\Big( \, {}_{{}_x}\mathfrak{J}^L \, \circ \, T_{{}_{(f_{{}_i})_{{}_x}, (g_{{}_i})_{{}_y} }} \, \circ \,
{}_{{}_y}\mathfrak{J}^M \big( \big(g_j \big)_{{}_{y}} \big) \, ,  \, e_k \, \Big) 
 \bigg)  
\,\,\,\, \ud (\mu_1 \times \mu_2 ) ([(x,y)])  \\
\end{multline*}
\begin{multline*}
= \int \, \bigg( \sum \limits_{i,j = 1}^{n} \sum \limits_{k \in \mathbb{N}}   
\Big( \, {}_{{}_x}\mathfrak{J}^L \, \circ \, T_{{}_{(f_{{}_i})_{{}_x}, (g_{{}_i})_{{}_y} }} \, \circ \,
{}_{{}_y}\mathfrak{J}^M \, \circ \, T_{{}_{(g_{{}_j})_{{}_x}, (f_{{}_j})_{{}_y} }}
\big( e_k \big) \, ,  \, e_k \, \Big)
\bigg)  
\,\,\,\, \ud (\mu_1 \times \mu_2 ) ([(x,y)])  \\
= \int \, \bigg( \sum \limits_{i,j = 1}^{n}  
\Tr \Big[ \, {}_{{}_x}\mathfrak{J}^L \, \circ \, T_{{}_{(f_{{}_i})_{{}_x}, (g_{{}_i})_{{}_y} }} \, \circ \,
{}_{{}_y}\mathfrak{J}^M \, \circ \, T_{{}_{(g_{{}_j})_{{}_x}, (f_{{}_j})_{{}_y} }} \, \Big]
\bigg)  
\,\,\,\, \ud (\mu_1 \times \mu_2 ) ([(x,y)])  \\
= \int \, \bigg( \sum \limits_{i,j = 1}^{n}  
\Tr \Big[ \, {}_{{}_x}\mathfrak{J}^L \, \circ \, T_{{}_{(f_{{}_i})_{{}_x}, (g_{{}_i})_{{}_y} }} \, \circ \,
{}_{{}_y}\mathfrak{J}^M \, \circ \, \big( T_{{}_{(f_{{}_j})_{{}_x}, (g_{{}_j})_{{}_y} }} \big)^{{}^\circledast} \, \Big]
\bigg)  
\,\,\,\, \ud (\mu_1 \times \mu_2 ) ([(x,y)])  \\
=\int \limits_{\mathfrak{G}_1 \times \mathfrak{G}_2} 
\Bigg( \, \Big( {}_{{}_x}\mathfrak{J}^L \otimes {}_{{}_y}\mathfrak{J}^M \Big) \Big( V(T) \Big)_{{}_{(x, y)}}  \, , \,
\Big( V(T) \Big)_{{}_{(x, y)}} \,  \Bigg)  \,\, \ud (\mu_1 \times \mu_2 ) ([(x,y)]) \\
=\int \limits_{\mathfrak{G}_1 \times \mathfrak{G}_2} 
\Bigg( \, \mathfrak{J}_{L \times M} \Big( \mathfrak{J}^{L \times M}  V(T) \Big)_{{}_{(x, y)}} \, , \,
\Big( V(T) \Big)_{{}_{(x, y)}} \,  \Bigg)  \,\, \ud (\mu_1 \times \mu_2 ) ([(x,y)]) = \| V(T) \|^2.
\end{multline*}
(The unspecified domain of integration in the above formulas is of course equal $\mathfrak{G}_1 \times \mathfrak{G}_2$.)
Therefore $V$ is isometric and $V(T) \in \mathcal{H}^{L\times M}$ for the indicated $T$, as the 
required measurability conditions
again easily follow from Scholium 3.9 of \cite{Segal_Kunze}. 
Now by the properties of Hilbert-Schmidt operators, the finite rank conjugate-linear operators 
$T: {}^{\mu_2}\mathcal{H}^M \mapsto {}^{\mu_1}\mathcal{H}^L$ are dense 
in ${}^{\mu_1}\mathcal{H}^L \otimes {}^{\mu_2}\mathcal{H}^M$  (compare e. g. \cite{Murray_von_Neumann}, Chap. II 
or \cite{Segal_Kunze}, Chap. 14.2 or \cite{Segal}). Thus the domain of the operator $V$ is dense. 

In order to show that the range of $V$ is likewise dense, consider the closure $C^1$ under the norm in 
$\mathcal{H}^{L\times M}$ of the linear set  of all functions $V(T)$,
where $T = T_{{}_{f_{{}_1}, g_{{}_1}}} + \ldots + T_{{}_{f_{{}_n}, g_{{}_n}}}$ with $f_i$ ranging over  $C^{L}_{0}
\subset \mathcal{H}^L$ and  $g_j$ over the corresponding set $C^{M}_{0} \subset \mathcal{H}^M$. Because $V$
is isometric it can be uniquely extended so that $C^1$ lies in the range of this unique extension. Let us denote the 
extension likewise by $V$. (For a densely defined Krein-isometric map this
would in general be impossible because $V$ could be discontinuous, this is the reason why we need to know if $V$ is continuous, i. e.  bounded for the ordinary positive definite inner products.)

Now by the property of Hilbert-Schmidt operators (mentioned above) the linear span of operators 
$T_{\upsilon, \textrm{v}} : \mathcal{H}_M \mapsto \mathcal{H}_L$
with $\upsilon$ and $\textrm{v}$ ranging over dense subsets of $\mathcal{H}_L$ and $\mathcal{H}_M$, respectively,
is dense in $\mathcal{H}_L \otimes \mathcal{H}_M$. This property of Hilbert-Schmidt operators together with
a repeated application of Lemma \ref{lem:dense.2} and \ref{lem:dense.5} of Sect. \ref{dense} and Scholium 3.9 and 5.3 of  \cite{Segal_Kunze} will show that all the conditions, (a)-(e), of Lemma \ref{lem:dense.4} are satisfied for 
$C^1 \subset \mathcal{H}^{L \times M}$. 

Let $\pi$ be the canonical quotient map $\mathfrak{G}_1 \times \mathfrak{G}_2 \mapsto 
(\mathfrak{G}_1 \times \mathfrak{G}_2 )/(G_1 \times G_2)$.
In particular if $\psi$ is a complex valued continuous function on $\mathfrak{G}_1 \times \mathfrak{G}_2$
which is constant on the right $G_1 \times G_2$ cosets and vanish outside of $\pi^{-1}(K)$
for some compact subset $K$ of $(\mathfrak{G}_1 \times \mathfrak{G}_2 )/(G_1 \times G_2)$, then it is measurable
and $\psi \in L^2 ((\mathfrak{G}_1 \times \mathfrak{G}_2 )/(G_1 \times G_2), \mu_1 \times \mu_2)$ 
and by Scholim 3.9 and 5.3 of  \cite{Segal_Kunze} it is an $L^2$-limit  of continuous such functions of ``product
form'' $\phi \cdot \varphi : \mathfrak{G}_1 \times \mathfrak{G}_2  \ni (x,y) \mapsto
\phi(x) \cdot \varphi(y)$. Thus the condition (d) of Lemma \ref{lem:dense.4} follows. The above mentioned
property of Hilbert-Schmidt operators and Lemma \ref{lem:dense.2} applied to $C^{L}_{0}
\subset \mathcal{H}^L$ and to $C^{M}_{0} \subset \mathcal{H}^M$, proves condition (e) of 
Lemma \ref{lem:dense.4}. Condition (b) follows from the the fact that $V(T) \in \mathcal{H}^{L\times M}$
for finite rank operators $T$, proved in the first part of the proof. An application of
the Lusin Theorem (Corollary 5.2.2 of \cite{Segal_Kunze}, together with an obvious adaptation
of the the standard proof of the Riesz-Fischer theorem already used in the proof of Lemma \ref{lem:dense.5}) 
proves condition (a) of Lemma \ref{lem:dense.4}. 
By the remark opening the proof of Lemma \ref{lem:dense.2} the linear sets $ C^{L}_{0}$
and $C^{M}_{0}$ of functions are closed with respect to right $\mathfrak{G}_1$ and $\mathfrak{G}_2$-translations,
respectively. Thus it easily follows that the linear 
set of functions $V\big( T_{{}_{f_{{}_1}, g_{{}_1}}} + \ldots + T_{{}_{f_{{}_n}, g_{{}_n}}} \big)$
with $f_{{}_i} \in C^{L}_{0}$, $g_{{}_j} \in C^{M}_{0}$ is closed under the right  
$\mathfrak{G}_1 \times \mathfrak{G}_2$-translations. Then, a simple continuity argument shows that 
$C^1$ is closed under right $\mathfrak{G}_1 \times \mathfrak{G}_2$-translations. Thus condition
(c) of Lemma \ref{lem:dense.4} is satisfied with trivial functions $\rho_s$ all equal identically to
the constant unit function. 
 
Thus  Lemma \ref{lem:dense.4} may be applied to $C^1$ lying in the range of $V$, so that the range
is dense in $\mathcal{H}^{L \times M}$. Therefore $C^1 = \mathcal{H}^{L \times M}$ and $V$ is unitary.  

We shall show that $V$ is Krein-unitary.  By the unitarity of $V$, it will be sufficient by continuity
to show that $V$ is Krein-isometric on finite rank operators $T \in \mathcal{H}^L \otimes \mathcal{H}^M$.
By self-adjointness of $\mathfrak{J}^L$ and $\mathfrak{J}^M$ we have  the following equalities for $T$ of the form indicated to above: 
\begin{multline*}
\big( \, \| T \|_{{}_{\mathfrak{J}^L \otimes \mathfrak{J}^M}} \,\big)^2 
=  \Big( \big(\mathfrak{J}^L \otimes \mathfrak{J}^M \big) \, \big( f_{{}_1}\otimes g_{{}_1} + \ldots +  
f_{{}_n} \otimes g_{{}_n} \big) \, , 
\,\, f_{{}_1}\otimes g_{{}_1} + \ldots +  f_{{}_n} \otimes g_{{}_n} \, \Big)   \\
= \Tr \Big[ \mathfrak{J}^L \big( T_{{}_{f_{{}_1}, g_{{}_1}}} + \ldots + T_{{}_{f_{{}_n}, g_{{}_n}}} \big) \mathfrak{J}^M
\big( T_{{}_{f_{{}_1}, g_{{}_1}}} + \ldots + T_{{}_{f_{{}_n}, g_{{}_n}}}  \big)^{{}^\circledast}  \Big]   \\
= \sum \limits_{i,j = 1}^{n} (\mathfrak{J}^L f_i , f_j) \cdot (\mathfrak{J}^M g_i , g_j)    \\
= \sum \limits_{i,j = 1}^{n} 
\bigg( 
\int \limits_{\mathfrak{G}_1} 
\Big( \mathfrak{J}_L  \big(f_i \big)_{{}_{x}} , 
\big(f_j \big)_{{}_{x}}  \Big)  \,\, \ud \mu_1 ([x]) \,\, \bigg) \cdot
\bigg( 
\int \limits_{\mathfrak{G}_2} 
\Big( \mathfrak{J}_M  \big( g_i \big)_{{}_{y}} , 
\big(g_j \big)_{{}_{y}}  \Big)  \,\, \ud \mu_2 ([y]) \,\, 
\bigg)      \\
= \int \limits_{\mathfrak{G}_1 \times \mathfrak{G}_2} \, \bigg( \sum \limits_{i,j = 1}^{n}  
\Big( \mathfrak{J}_L  \big( f_i \big)_{{}_{x}} , 
\big(f_j \big)_{{}_{x}}  \Big) \cdot
\Big( \mathfrak{J}_M  \big( g_i \big)_{{}_{y}} , 
\big(g_j \big)_{{}_{y}}  \Big) 
\bigg)  
\,\,\,\, \ud (\mu_1 \times \mu_2 ) ([(x,y)]) \\
=\int \limits_{\mathfrak{G}_1 \times \mathfrak{G}_2} 
\Tr \Big[ \mathfrak{J}_L \big( T_{{}_{(f_{{}_1})_{{}_x}, (g_{{}_1})_{{}_y} }}
+ \ldots + 
T_{{}_{(f_{{}_n})_{{}_x}, (g_{{}_n})_{{}_y} }}
\big) \mathfrak{J}_M
\big( T_{{}_{(f_{{}_1})_{{}_x}, (g_{{}_1})_{{}_y} }}
+ \ldots       \\
\ldots + T_{{}_{(f_{{}_n})_{{}_x}, (g_{{}_n})_{{}_y} }}
\big)^{{}^\circledast} \Big]
  \,\, \ud (\mu_1 \times \mu_2 ) ([(x,y)])     \\
=\int \limits_{\mathfrak{G}_1 \times \mathfrak{G}_2} 
\Bigg( \, \Big( \mathfrak{J}_L \otimes \mathfrak{J}_M \Big) \Big( V(T) \Big)_{{}_{(x, y)}} \, , \,
\Big( V(T) \Big)_{{}_{(x, y)}} \,  \Bigg)  \,\, \ud (\mu_1 \times \mu_2 ) ([(x,y)])     \\
=\int \limits_{\mathfrak{G}_1 \times \mathfrak{G}_2} 
\Bigg( \, \mathfrak{J}_{L \times M} \Big( V(T) \Big)_{{}_{(x, y)}} \, , \,
\Big( V(T) \Big)_{{}_{(x, y)}} \,  \Bigg)  \,\, \ud (\mu_1 \times \mu_2 ) ([(x,y)])   \\
= \big( \, \| V(T) \|_{{}_{\mathfrak{J}^{L \times M}}} \,\big)^2 .
\end{multline*}

Recall that  the domain $\mathfrak{D}_{12}$ (common for all $(x,y) \in \mathfrak{G}_1 \times \mathfrak{G}_2$) 
of the operators $U_1 \otimes U_2 = {}^{\mu_1}U^{L}_{x} \, \otimes \, {}^{\mu_2}U^{M}_{y}
= \big( {}^{\mu_1}U^{L} \, \times \, {}^{\mu_2}U^{M} \big)_{(x,y)}$  representing $(x,y) \in \mathfrak{G}_1 \times \mathfrak{G}_2$,
is invariant for the operators $U_1 \otimes U_2 = {}^{\mu_1}U^{L}_{x} \, \otimes \, {}^{\mu_2}U^{M}_{y}
= \big( {}^{\mu_1}U^L \, \times \, {}^{\mu_2}U^M \big)_{(x, y)}$. For each
$(x,y)$ let us denote the closure of ${}^{\mu_1}U^L \, \times \, {}^{\mu_2}U^M
=  {}^{\mu_1}U^{L}_{x} \, \otimes \, {}^{\mu_2}U^{M}_{y}$ likewise by ${}^{\mu_1}U^L \, \times \, {}^{\mu_2}U^M$.
Note that $V(T)$, $T \in \mathfrak{D}_{12}$ compose an invariant domain of the representation
${}^{\mu_1 \times \mu_2}U^{L \times M}$. Denote the closures of the operators 
${}^{\mu_1 \times \mu_2}U^{L \times M}_{(x, y)}$ with the common invariant domain $V(\mathfrak{D}_{12})$ likewise by 
${}^{\mu_1 \times \mu_2}U^{L \times M}_{(x, y)}$.  

The equality (\ref{ulxm=ulxum}) is regarded as equality for the closures of the operators
${}^{\mu_1 \times \mu_2}U^{L \times M}$ and ${}^{\mu_1}U^L \, \times \, {}^{\mu_2}U^M$. 

By Theorem \ref{def_ind_krein:twr.1} and its proof the closures of ${}^{\mu_1 \times \mu_2}U^{L \times M}$ do not depend
on the choice of the dense common invariant domain. Therefore in order to show the equality 
(\ref{ulxm=ulxum}) it is sufficient that the respective closed operators in (\ref{ulxm=ulxum}) coincide
on the domain of all finite rank operators $T \in \mathfrak{D}_{12}$. This however is immediate.
Indeed, let $T = T_{{}_{f_{{}_1}, g_{{}_1}}} + \ldots + T_{{}_{f_{{}_n}, g_{{}_n}}}$
with $f_i \in \mathfrak{D}_1$ and $g_j \in \mathfrak{D}_2$. Then
\begin{multline}\label{ultum*}
\big( {}^{\mu_1}U^{L} \, \times \, {}^{\mu_2}U^{M} \big)_{(x_0 ,y_0 )} (T) 
= \big( {}^{\mu_1}U^{L}_{x_0} \, \otimes \, {}^{\mu_2}U^{M}_{y_0} \big) (T) \,\,
= \,\, {}^{\mu_1}U^{L}_{x_0} \,\,\, T \,\,\, \big( {}^{\mu_2}U^{M}_{y_0} \big)^* \\
= \sqrt{\lambda_1(\cdot, x_0)} \sqrt{\lambda_2 (\cdot, y_0)} \,
\big(T_{{}_{R_{x_0}f_{{}_1}, R_{y_0}g_{{}_1}}} + \ldots 
+  T_{{}_{R_{x_0}f_{{}_n}, R_{y_0}g_{{}_n}}} \big).
\end{multline}
On the other hand we have:
\begin{multline*}
 \Big( {}^{\mu_1 \times \mu_2}U^{L \times M}_{(x_0 ,y_0 )} V(T) \Big)_{{}_{(x, y)}} 
=\sqrt{\lambda_1([x], x_0)} \sqrt{\lambda_2 ([y], y_0)} \,
\Big( V(T) \Big)_{{}_{(x \cdot x_0, y \cdot y_0)}} \\
= \sqrt{\lambda_1([x], x_0)} \sqrt{\lambda_2 ([y], y_0)} \,
\Big(
T_{{}_{(R_{x_0}f_{{}_1})_{{}_x}, (R_{y_0}g_{{}_1})_{{}_y} }} + \ldots
\ldots + T_{{}_{(R_{x_0}f_{{}_n})_{{}_x}, (R_{y_0}g_{{}_n})_{{}_y} }}
\Big),
\end{multline*}
so that 
\begin{multline*}
\Big( V^{-1} \, \big( \, {}^{\mu_1 \times \mu_2}U^{L \times M} \, \big) \, V \Big) \big( T \big) \\
= \sqrt{\lambda_1(\cdot, x_0)} \sqrt{\lambda_2 (\cdot, y_0)} \,
\big(T_{{}_{R_{x_0}f_{{}_1}, R_{y_0}g_{{}_1}}} + \ldots 
+  T_{{}_{R_{x_0}f_{{}_n}, R_{y_0}g_{{}_n}}} \big).
\end{multline*}
Comparing it with (\ref{ultum*}) one can see that (\ref{ulxm=ulxum}) holds
on $\mathfrak{D}_{12}$. Thus the proof of (\ref{ulxm=ulxum}) is complete now. The Theorem
is hereby proved completely.
\qed

Presented proof of Theorem \ref{twr.1:kronecker} is an extended and modified version of the
Mackey's proof of Theorem 5.2 in \cite{Mackey}.  

Note, please, that the equality (\ref{ulxm=ulxum}) for the closures of the operators 
${}^{\mu_1 \times \mu_2}U^{L \times M}$ and ${}^{\mu_1}U^L \, \times \, {}^{\mu_2}U^M$ 
is non trivial. Indeed, recall that in general almost all kinds of pathology not excluded 
by general theorems can be shown to exist for unbounded operators. In particular two \emph{distinct} 
and closed operators may still coincide on a dense domain. This is why we need to be careful 
in proving (\ref{ulxm=ulxum}).  This in particular shows that the fundamental theorems of the original
Mackey theory by no means are automatic for the induced Krein-isometric representations, where the
representors are in general densely defined and unbounded. Here we saw it for the Theorem 
\ref{twr.1:kronecker}. But differences in the proofs arise likewise in the latter part of the theory. 
In particular if we 
want to prove the \emph{subgroup theorem} and the so called \emph{Kronecker product theorem}
for the induced Krein-isometric representations with precisely the same assumptions
posed on the group as in Mackey's theory, then some additional analysis will have to be made 
in treating decompositions of non finite quasi invariant measures. Compare Sect. \ref{decomposition}.

\section{Subgroup theorem in Krein spaces. Preliminaries}\label{subgroup.preliminaries}

This Section is a word for word repetition of the argument of \S 6 of \cite{Mackey}. That 
the general Mackey's argument may be applied to induced representations in Krein spaces is the whole point.  
Although  it is rather clear that his general argumentation is applicable in the Krein space, we  
restate it here because it lies at the very heart of the presented method of decomposition
of tensor product of induced representations, and will make the paper self contained. It should be noted however
that it requires some additional analysis in decomposing non finite quasi invariant measures,
which makes a difference in proving the existence of the corresponding direct integral decompositions.

The circumstance that the {\L}opusza\'nski representation of $\mathfrak{G}$ is equivalent to an induced 
representation in a Krein space greatly simplifies the problem of decomposing tensor product of {\L}opusza\'nski 
representations and reduces it largely to the geometry of right cosets and double cosets in the group $\mathfrak{G}$ and to 
a ``Fubini-like'' theorem, just like for the ordinary induced representations of Mackey. Similar decomposition
method of quotiening by a subgroup in construction of complete sets of unitary representations of semi simple 
Lie groups was applied by Gelfand and Neumark, and by several authors in constructing harmonic analysis on classical Lie groups. The main gain is that the subtle analytic properties of the {\L}opusza\'nski representation (unboundedness)  
does not intervene dramatically after this reduction to geometry of cosets and double cosets.

Our main theorem asserts the existence of a certain useful direct integral decomposition of
the tensor product $U^L \otimes U^M$ of two induced representations of a group $\mathfrak{G}$
in a Krein space, whose construction is completely analogous to that of Mackey for ordinary
unitary representations, compare \cite{Mackey}. By definition $U^L \otimes U^M$ is obtained 
from the outer Kronecker product representation $U^L \times U^M$ of $\mathfrak{G} \times \mathfrak{G}$
by restricting $U^L \times U^M$ to the diagonal subgroup $\overline{\mathfrak{G}} \cong \mathfrak{G}$
of all $(x,y) \in \mathfrak{G}\times \mathfrak{G}$ with $x = y$. By the Theorem of Sect. \ref{kronecker},
$U^L \times U^M$ is Krein-unitary equivalent to $U^{L \times M}$. Thus $U^L \otimes U^M$ can be analysed by analysing
the restriction of $U^{L \times M}$ to the diagonal subgroup $\overline{\mathfrak{G}} \cong \mathfrak{G}$. 
Our theorem on tensor product decomposition follows (just as in \cite{Mackey}) from these remarks and
a theorem on restriction to a subgroup of an induced representation in a Krein space, say a \emph{subgroup theorem}.
\emph{Subgroup theorem} gives a decomposition of the restriction of an induced representation (in a Krein space)
to a closed subgroup, with the component representations in the decomposition themselves Krein-unitary equivalent to induced representations. Namely, let $L$ be strongly continuous almost uniformly bounded Krein-unitary representation of the
closed subgroup $G_1$ of $\mathfrak{G}$ and consider the restriction ${}_{G_2}U^L$ of $U^L$ to a second 
closed subgroup $G_2$. While $\mathfrak{G}$ acts transitively on the homogeneous space $\mathfrak{G}/G_1$ of
right $G_1$-cosets this will not be true in general of $G_2$. Moreover, and this is the main advantage of
induced representations, any division of $\mathfrak{G}/G_1$ into two parts $S_1$ and $S_2$, each a Baire (or Borel)
set which is not a null set (with respect to any, and hence every quasi invariant measure on $\mathfrak{G}/G_1$),
and each invariant under $G_2$ leads to a corresponding direct sum decomposition of ${}_{G_2}U^L$. Indeed
the closed subspaces $\mathcal{H}^{L}_{S_1}$ and $\mathcal{H}^{L}_{S_2}$ of all $f \in \mathcal{H}^L$ which vanish
respectively outside of $\pi^{-1}(S_1)$ and $\pi^{-1}(S_2)$ are invariant and are orthogonal complements of each other
with respect to the ordinary (as well as the Krein) inner product on $\mathcal{H}^L$.

Assume for a while, just for illustrative purposes, that there is a null set $N$ in $\mathfrak{G}/G_1$ whose complement
is the union of countably many non null orbits $C_1 , C_2 , \ldots$ of $\mathfrak{G}/G_1$ under $G_2$.  
Then by the above remarks we obtain a direct sum decomposition of ${}_{G_2}U^L$ into as many parts as there are non null orbits. Our analysis reaches its goal after analysing the nature of these parts. Analysis of these parts is our goal of the rest of this Section.

In our paper we shall consider a more general case in which all of the orbits can be null sets and the sum becomes an integral and we have to use the von Neumann theory of direct integral Hilbert spaces \cite{von_neumann_dec}.
Of course according to the definition given above (with $S_1$ or $S_2$ equal to 
a $G_2$ orbit $C$ in $\mathfrak{G}/G_1$), $\mathcal{H}^{L}_{C}$ will be zero dimensional whenever the orbit 
$C$ is a null set. However it is possible to reword the definition so that it always gives a non zero Hilbert
space (with the respective Krein structure) and so that when $C$ is not a null set this definition is essentially the same as that already given, compare \cite{Mackey}, \S 6. Indeed note that when $C$ is a non null set then $\mathcal{H}^{L}_{C}$
may be equivalently defined as follows. Let $x_c$ be any member of $\mathfrak{G}$ such that $\pi(x_c) \in C$
and consider the set ${\mathcal{H}^{L}_{C}}'$ of all functions $f$ from the double coset $G_1 x_c G_2$ to $\mathcal{H}_L$
such that: (i) $x \mapsto (f_x ,\upsilon)$ is a Borel function for all $\upsilon  \in \mathcal{H}_L$, 
(ii) $f_{\xi x} = L_\xi (f_x )$ for all $\xi \in G_1$ and all $x \in G_1 x_c G_2$ and 
(iii): 
\[
\| f \|_C = \int \limits_{C} \, (\, \mathfrak{J}_L ((\mathfrak{J}^L f)_x ), f_x \,) \, d\mu_{\mathfrak{G}/G_1}
= \int \limits_{(G_1 x_c G_2 ) \, \cap B} \, (f_b , f_b \,) \, d\mu_{B}(b) < \infty,
\]  
where $B$ is the regular Borel section of $\mathfrak{G}$ with respect to $G_1$ of Sect. \ref{def_ind_krein}
(we could use as well the sub-manifold $Q$ of Sect. \ref{def_ind_krein} but we prefer to proceed generally and independently of the ``factorization'' assumption). The operator $\mathfrak{J}^L$
in $\mathcal{H}^{L}_{C}$ is given by simple restriction, and its definition on 
${\mathcal{H}^{L}_{C}}'$ is obvious:
\[
(\mathfrak{J}^{L,C} f)_x = L_{h(x)} \mathfrak{J}_{L} L_{h(x)^{-1}} \, f_x ;
\] 
with the obvious definition of the Krein inner product in ${\mathcal{H}^{L}_{C}}'$
\[
\big(f, g \big)_{\mathfrak{J}^{L,C}} = (\mathfrak{J}^{L,C} f, g) 
= \int \limits_{C} \, (\, \mathfrak{J}_L (f_x ), g_x \,) \, d\mu_{\mathfrak{G}/G_1}, \,\,\, f,g \in {\mathcal{H}^{L}_{C}}'.
\]
Similarly we define the operator $U^{L, C}_{\xi}$ in $\mathcal{H}^{L}_{C}$ for $\xi \in G_2$ as the restriction of 
$U^{L}_{\xi}$ to $\mathcal{H}^{L}_{C}$, i. e. to the functions supported by the orbit $C$, and its definition giving an equivalent representation on ${\mathcal{H}^{L}_{C}}'$ is likewise obvious:
\[
(U^{L, C}_{\xi} f)_x = \sqrt{\lambda ([x], \xi)} \, f_{x \xi},
\]
with the $\lambda$-function of the quasi invariant measure $\mu$ restricted to $C \times G_2$. 

Moreover, and this is the whole point, the measure in $C$
need not be defined by restricting $\mu = \mu_{\mathfrak{G}/G_1}$ to $C$. There exists a non zero measure $\mu_C$ on $C$ quasi invariant with respect to $G_2$ determined up to a constant factor, whose
Radon-Nikodym function $\ud(R_\eta \mu_C )/\ud \mu_C$, $\eta \in G_2$ 
(i. e. the associated $\lambda_C$-function) is equal to the restriction to the subspace $C \times G_2$ of the 
$\lambda$-function, i. e. Radon-Nikodym derivative $\ud (R_\eta \mu)/\ud \mu$ , associated with $\mu = \mu_{\mathfrak{G}/G_1}$. Indeed, although $C$ does not have the form of a quotient of a group by its closed subgroup, it follows from Theorem 3, page 253 of \cite{Kuratowski} that the map $x \mapsto \pi(x_c x)$ induces a Borel isomorphism
$\psi$ of the quotient space $G_2 /G_{x_c}$ onto $C$, where $G_{x_c} = G_2 \cap ({x_c}^{-1}G_1 x_c)$ is the closed subgroup of all $x \in G_2$ such that $\pi(x_c x) = \pi(x_c)$. Thus $C \times G_2 \cong G_2 /G_{x_c} \times G_2$
as Borel spaces under the indicated isomorphism and moreover if $[x] \in G_2 /G_{x_c}$ and $[z] = \pi(x_c x)$ correspond
under this isomorphism and $\eta \in G_2$ then $[x]\eta$ and $[z]\eta$ do also, where $[x]\eta = [x\eta]$ and 
$[z]\eta = [z\eta]$ denote the action of $\eta \in G_2$ on $[x] \in G_2 / G_{x_c}$ and $[z] \in C$ respectively. 
Thus the existence of the quasi invariant measure  
$\mu_C$ on $C$ follows from the general Mackey classification of quasi invariant 
measures on the quotient of a locally compact group by a closed subgroup, compare the respective Theorem of 
Sect. \ref{def_ind_krein}. Using the quasi invariant measure $\mu_C$ on $C$ gives a non trivial space 
${\mathcal{H}^{L}_{C}}'$ for every orbit $C$, which in case of a non null
orbit $C$ is trivially equivalent to $\mathcal{H}^{L}_{C}$.  

\vspace*{0.5cm}

{\bf REMARK}. Recall that Borel $\psi$ is a Borel isomorphism
with respect to the Borel structure on $C$ induced from the surrounding space $\mathfrak{G}/G_1$:
we define $E \subset C$ to be Borel iff $E = E' \cap C$ for a Borel set $E'$ in $\mathfrak{G}/G_1$.
However our assumptions concerning the group $\mathfrak{G}$ and the subgroups $G_1$ and $G_2$ are exactly the same as those of Mackey, and they do not even guarantee the local compactness of the orbits $C$, compare Sect. \ref{decomposition}.
\qed

\vspace*{0.5cm}

We are now in a position to formulate the main goal 
of this Section: 

\begin{lem}
Let $C$ be any orbit in $\mathfrak{G}/G_1$ under $G_2$ and let $x_c$ be such that $\pi(x_c) \in C$. Let 
${\mathcal{H}^{L}_{C}}'$ be defined as above. Let ${^{\mu^{x_c}}}U^{L^{x_c}}$ be the representation of $G_2$ induced by the
strongly continuous almost uniformly bounded Krein-unitary representation $L^{x_c}: \eta \mapsto 
L_{x_c \eta {x_c}^{-1}}$ of $G_2 \cap ({x_c}^{-1}G_1 x_c )$ with 
the representation space of $L^{x_c}$ equal to $\mathcal{H}_{L^{x_c}} = \mathcal{H}_L$ and the fundamental symmetry 
$\mathfrak{J}_{L^{x_c}} = \mathfrak{J}_L$; and with the quasi invariant measure $\mu^{x_c}$ in the homogeneous space
$G_2 / (G_2 \cap \, ({x_c}^{-1}G_1 x_c))$ equal to the transfer of the measure $\mu_C$ in $C$ over to the homogeneous space by the map $\psi$.
Let ${}^{\mu^{x_c}}\mathcal{H}^{L^{x_c}}$ be the Krein space of the induced 
representation ${}^{\mu^{x_c}}U^{L^{x_c}}$. We assume the fundamental symmetry $\mathfrak{J}_{x_c}$ in 
${}^{\mu^{x_c}}\mathcal{H}^{L^{x_c}}$ to be defined by the equation $(\mathfrak{J}_{x_c} g)_t = 
L_{h(x_c t)} \mathfrak{J}_L L_{h(x_c t)^{-1}} g_t$ and the Krein inner product given by the ordinary formula
\[
\int \limits_{G_2 \big{/} \big(G_2 \cap \, ({x_c}^{-1}G_1 x_c)\big)} \, 
(\mathfrak{J}_{L} \tilde{f}_{t} , \tilde{f}_{t} ) \,\, \ud \mu^{x_c}([t]), \,\,\,t \in G_2 .
\]
Then there is a Krein-unitary map $V_{x_c}$ of ${\mathcal{H}^{L}_{C}}'$ onto ${}^{\mu^{x_c}}\mathcal{H}^{L^{x_c}}$
such that if $g \in {}^{\mu^{x_c}}\mathcal{H}^{L^{x_c}}$ corresponds to 
$f \in {\mathcal{H}^{L}_{C}}'$ then ${}^{\mu^{x_c}}U^{L^{x_c}}_s g$ corresponds to $U^{L, C}_{s} f$ where 
$(U^{L, C}_{s} f)_x = f_{xs} \sqrt{\lambda_C ([x],s)}$
for all $x \in C$ and all $s \in G_2$. 

\label{lem:subgroup.preliminaries.1}
\end{lem}

\qedsymbol \,
 For each function $f$ on $G_1 x_c G_2$ satisfying the conditions (i) and (ii) of 
the definition of ${\mathcal{H}^{L}_{C}}'$ let $\tilde{f}$ be defined by $\tilde{f}_t = f_{x_c t}$
for all $t \in G_2$. Then $(\tilde{f}_t , \upsilon )$ is a Borel function of $t$ on $G_2$ for all 
$\upsilon \in \mathcal{H}_L$. If $\eta \in G_{x_c} = G_2 \cap \, ({x_c}^{-1}G_1 x_c)$ then if $\xi = x_c \eta {x_c}^{-1}$
we have $\tilde{f}_{\eta t} = \tilde{f}_{{x_c}^{-1}\xi x_c t} = f_{\xi x_c t} = L_\xi \tilde{f}_t
= L_{x_c \eta {x_c}^{-1}}(\tilde{f}_t)$; that is 
\begin{equation}\label{g_2_induced}
\tilde{f}_{\eta t} = L_{x_c \eta {x_c}^{-1}}(\tilde{f}_t)
\end{equation}
for all $t \in G_2$ and all $\eta \in G_2 \cap \, ({x_c}^{-1} G_1 x_c)$. Conversely let $g$ be any function from $G_2$ to 
$\mathcal{H}_L$ which is Borel in the sense that $x \mapsto (g_x ,\upsilon)$ is a Borel function on $G_2$ for all 
$\upsilon  \in \mathcal{H}_L$ and which satisfies (\ref{g_2_induced}). We define the corresponding function $f$ by the equation $f_{\xi x_c t} = L_\xi (g_t)$ for all $\xi \in G_1$ and $t \in G_2$. 
If $\xi_1 x_c t_1 = \xi_2 x_c t_2$ then ${\xi_2}^{-1}\xi_1 = x_c t_2 {t_1}^{-1}{x_c}^{-1}$ so that 
$g_{t_2{t_1}^{-1}t} = L_{{\xi_2}^{-1}\xi_1}(g_t)$. Therefore $L_{\xi_2}(g_{t_2}) = L_{\xi_1}(g_{t_1})$
and $f$ is well defined. Next we show that $(f_x , \upsilon)$ is Borel function of $x$ on $G_1 x_c G_2$ 
for all $\upsilon \in \mathcal{H}_L$. Let $f'$ be the function on $G_1
 \times G_2$ defined by 
$f'(\xi , \eta) = L_\xi (g_\eta)$ for all $(\xi , \eta) \in G_1 \times G_2$. Choose now an orthonormal
basis $\{ \varphi_i \}_{i \in \mathbb{N}}$ in $\mathcal{H}_L$. Then we have 
$(f'(\xi, \eta), \upsilon) = (f'(\xi, \eta), \mathfrak{J}_L \mathfrak{J}_L \upsilon) = 
(\mathfrak{J}_L f'(\xi, \eta),  \mathfrak{J}_L \upsilon) 
= (\mathfrak{J}_L L_\xi (g_\eta) ,  \mathfrak{J}_L \upsilon) =  
(\mathfrak{J}_L  g_\eta , L_{\xi^{-1}} \mathfrak{J}_L \upsilon)
= \sum_{i = 1}^{\infty} = ( \mathfrak{J}_L  g_\eta , \varphi_i ) 
( \varphi_i , L_{\xi^{-1}} \mathfrak{J}_L \upsilon )$. By Scholium 3.9 of \cite{Segal_Kunze} 
$(f'(\xi, \eta), \upsilon)$ is a Borel function of $(\xi , \eta)$ on $G_1 \times G_2$ regarded as
the product measure space, for all $\upsilon \in \mathcal{H}_L$. Let us introduce after Mackey a new
group operation in $G_1 \times G_2$ putting $(\xi_1 , \eta_1) (\xi_2 , \eta_2) = (\xi _1 \xi_2 , \eta_2 \eta_1)$
and call the resulting group $G_3$. Then $\xi_1 x_c \eta_1 = \xi_2 x_c \eta_2$ if and only if 
$(\xi_2 , \eta_2)^{-1} (\xi_1 , \eta_1) = ({\xi_2}^{-1}\xi_1 , \eta_1 {\eta_2}^{-1})$ has the form 
$(\xi , {x_c}^{-1}\xi^{-1}x_c)$. The set of all  $(\xi , {x_c}^{-1}\xi^{-1}x_c)$, $\xi \in G_1$ is a subgroup
$G_4$ of $G_3$. Thus the map $(\xi ,\eta ) \mapsto \xi x_c \eta$ sets up a one-to-one correspondence 
between the points of the homogeneous space $G_3 /G_4$ of left $G_4$-cosets and the points of the double
coset $G_1 x_c G_2$. The map is continuous and on account of the assumed separability it follows again from
Theorem 3, page 253 of \cite{Kuratowski} that the map sets up a Borel isomorphism. Moreover the function
$(\xi , \eta) \mapsto (f'(\xi , \eta) , \upsilon)$ is constant on left $G_4$-cosets in $G_3$, as an easy
computation shows that $(f'((\xi , \eta)\omega_0) , \upsilon)
= (f'(\xi , \eta) , \upsilon)$ for all $\omega_0 = (\xi_0 ,{x_c}^{-1}{\xi_0}^{-1}x_c ) \in G_4$. 
Therefore $(\xi , \eta) \mapsto (f'(\xi , \eta) , \upsilon)$ defines a function on $G_3 / G_4$
which by Lemma 1.2 of \cite{Mackey} must be Borel because $(\xi , \eta) \mapsto (f'(\xi , \eta) , \upsilon)$ itself is Borel on $G_3$. That $(f_x , \upsilon)$ is a Borel function of $x \in G_1 x_c G_2$ now follows from the fact 
that the mapping of $G_3 / G_4$ onto $G_1 x_c G_2$ is a Borel isomorphism and preserves Borel sets. Finally observe that
$\tilde{f} = g$. Therefore $f \mapsto \tilde{f}$ is a one-to-one map of functions satisfying 
(i) and (ii) of the definition of ${\mathcal{H}^{L}_{C}}'$ onto Borel functions satisfying (\ref{g_2_induced}).
Consider the mapping $t \mapsto \pi(x_c t)$ of $G_2$ onto $C$. It defines one-to-one and Borel set preserving map $\psi$ from $G_2 / (G_2 \cap \, ({x_c}^{-1}G_1 x_c))$ onto $C$ and such that if $[t] = \pi'(t)$ and $[z] = \pi(z)$ 
correspond under the map $\psi$ and 
$\eta \in G_2$ then $[x]\eta$ and $[z]\eta$ do also ($\pi'$ stands for the canonical projection 
$G_2 / (G_2 \cap \, ({x_c}^{-1}G_1 x_c)) \mapsto G_2$). Finally $z \mapsto (\mathfrak{J}_L f_z , f_z)$ and 
$t \mapsto (\mathfrak{J}_{L} \tilde{f}_t , \tilde{f}_t)$ define functions
$\pi(z) \mapsto (\mathfrak{J}_L f_{\pi(z)} , f_{\pi(z)})$ and 
$\pi'(t) \mapsto (\mathfrak{J}_{L} \tilde{f}_{\pi'(t)} , \tilde{f}_{\pi'(t)})$ on $C$ and 
$G_2 / (G_2 \cap \, ({x_c}^{-1}G_1 x_c))$ respectively which correspond under the same map $\psi$:
$(\mathfrak{J}_L f_{\psi(\pi'(t))} , f_{\psi(\pi'(t))}) = (\mathfrak{J}_L f_{\pi(x_c t)} , f_{\pi(x_c t)})
= (\mathfrak{J}_L f_{x_c t} , f_{x_c t}) = (\mathfrak{J}_{L} \tilde{f}_{\pi'(t)} , \tilde{f}_{\pi'(t)})$.  
If we use this same map $\psi$ to transfer the measure $\mu_C$ on $C$ over to the homogeneous space
$G_2 / (G_2 \cap \, ({x_c}^{-1}G_1 x_c))$ we will get a quasi invariant measure $\mu^{x_c}$ there such that 
\[
\begin{split}
\int \limits_{C} \, (\mathfrak{J}_L f_z , f_z) \,\, \ud \mu_C ([z]) = 
\int \limits_{C} \, (\mathfrak{J}_L f_{[z]} , f_{[z]}) \,\, \ud \mu_C ([z]) \\
\int \limits_{C} \, (\mathfrak{J}_L f_{\psi([t])} , f_{\psi([t])}) \,\, \ud \mu_C (\psi([t])) 
= \int \limits_{G_2 / (G_2 \cap \, ({x_c}^{-1}G_1 x_c))} \, 
(\mathfrak{J}_{L} \tilde{f}_{[t]} , \tilde{f}_{[t]}) \,\, \ud \mu^{x_c}([t]) \\
= \int \limits_{G_2 / (G_2 \cap \, ({x_c}^{-1}G_1 x_c))} \, 
(\mathfrak{J}_{L} \tilde{f}_{t} , \tilde{f}_{t}) \,\, \ud \mu^{x_c}([t]).
\end{split} 
\]
Thus by the polarization identity (compare e. g. \cite{Segal_Kunze}, \S 8.3, page 222 or \cite{Bog}, page 4) the map $f \mapsto \tilde{f}$ sets up the Krein-unitary transformation $V_	{x_c}$ demanded by the Lemma as the verification 
of $V_{x_c} U^{L, C}_{s} V_{x_c}^{-1} = {}^{\mu^{x_c}}U^{L^{x_c}}_s$, $s \in G_2$, and 
$V_{x_c}\mathfrak{J}^{L,C} V_{x_c}^{-1}  = \mathfrak{J}_{x_c}$ is almost immediate
as $V_{x_c}$ is bounded, which we show below in Lemma \ref{lem:subgroup.preliminaries.2}. 
Similarly verification that $\mathfrak{J}_{x_c} \mathfrak{J}_{x_c} = I$ and that $\mathfrak{J}_{x_c}$ is self adjoint
with respect to the definite inner product    
\begin{equation}\label{g_2'_def_inn}
(\tilde{f}, \tilde{g})_{x_0} =  \int \limits_{G_2 / (G_2 \cap \, ({x_c}^{-1}G_1 x_c))} \, 
\Big(\mathfrak{J}_L (\mathfrak{J}_{x_0} \tilde{f}_t), \tilde{g}_t \Big) \,\, \ud \mu^{x_c}([t])
\end{equation}
in the Hilbert space ${}^{\mu^{x_c}}\mathcal{H}^{L^{x_c}}$, is likewise immediate. 
\qed

\vspace*{0.5cm}

Note that in general the norm and topology induced by the inner product (\ref{g_2'_def_inn}) defined by 
$\mathfrak{J}_{x_c}$ is not equivalent to the norm 
\[
\| \tilde{f} \|^2 = (\tilde{f}, \tilde{f}) =  \int \limits_{G_2 \big{/} \big(G_2 \cap \, ({x_c}^{-1}G_1 x_c) \big)} \, 
\Big(\mathfrak{J}_L (\mathfrak{J}^{L^{x_c}} \tilde{f}_t), \tilde{f}_t \Big) \,\, \ud \mu^{x_c}([t])
\]
and topology defined by the ordinary fundamental symmetry $\mathfrak{J}^{L^{x_c}}$
of Sect. \ref{def_ind_krein} (of course with $\mathfrak{G}$ and $H$ replaced with $G_2$ and 
$G_2 \cap \, ({x_c}^{-1}G_1 x_c)$): 
\[
\mathfrak{J}^{L^{x_c}} \tilde{f}_t = L^{x_c}_{h_{x_c}(t)} \mathfrak{J}_{L} L^{x_c}_{h_{x_c}(t)^{-1}} \, \tilde{f}_t ,
\] 
where $h_{x_c}(t) \in G_2 \cap \, ({x_c}^{-1}G_1 x_c)$ is defined as in Remark \ref{rem:def_ind_krein.1}
by a regular Borel section $B_{x_c}$ of $G_2$ with respect to the subgroup $G_2 \cap \, ({x_c}^{-1}G_1 x_c)$.
However if for each $t \in G_2$, $h(x_c t) \in G_{x_c}$, then the two topologies coincide. Similarly whenever the homogeneous space $G_2 \big{/} \big(G_2 \cap \, ({x_c}^{-1}G_1 x_c) \big)$
is compact then the two topologies coincide (but this case is not interesting). 

\vspace*{0.5cm}

\begin{lem}
The operators $V_{x_c}$ of the preceding Lemma are also isometric  with respect to the norms 
$\| \cdot \|_C$ in ${\mathcal{H}^{L}_{C}}'$ and $\| \cdot \|_{x_c} = \sqrt{(\cdot , \cdot)_{x_c}}$ in 
${}^{\mu^{x_c}}\mathcal{H}^{L^{x_c}}$, where $(\cdot, \cdot)_{x_c}$ is defined as by (\ref{g_2'_def_inn}), 
giving the norm in ${}^{\mu_{x_c}}\mathcal{H}^{L^{x_c}}$ 
induced by $\mathfrak{J}_{x_c}$. In particular we have $\| V_{x_c} \| = 1$ for all $x_c$.
\label{lem:subgroup.preliminaries.2}
\end{lem}

\qedsymbol \,
Denote the subgroup $G_2 \cap \, ({x_c}^{-1}G_1 x_c)$ by $G_{x_c}$. 
The Lemma is an immediate consequence of definitions of $\| \cdot \|_C$, $V_{x_c}$ 
and (\ref{g_2'_def_inn}) giving the norm $\| \cdot \|$ in ${}^{\mu^{x_c}}\mathcal{H}^{L^{x_c}}$:
\begin{equation}\label{norm_tilde}
\begin{split}
{\| V_{x_c}f \|_{{x_c}}}^2 = (\tilde{f}, \tilde{f})_{x_c} =  \int \limits_{G_2 / G_{x_c}} \, 
\big( \mathfrak{J}_L (\mathfrak{J}_{x_c} \tilde{f})_t , \tilde{f}_t \big) \,\, \ud \mu^{x_c}([t]) \\ 
= \int \limits_{G_2 / G_{x_c}} \, \big( \mathfrak{J}_L (V_{x_c}^{-1} \mathfrak{J}^{L,C} 
V_{x_c} V_{x_c}^{-1} f)_t , (V_{x_c}^{-1}f)_t \big) \,\, \ud \mu^{x_c}([t])  \\
=  \int \limits_{G_2 / G_{x_c}} \, \big( \mathfrak{J}_L (V_{x_c}^{-1} \mathfrak{J}^{L,C} 
 f)_t , (V_{x_c}^{-1}f)_t \big) \,\, \ud \mu^{x_c}([t])  
\end{split}
\end{equation}
and because $V_{x_c}$ is Krein-unitary, i. e. isometric for the Krein inner products
\[
\int \limits_{C} \, \big( \mathfrak{J}_L (\cdot)_z , (\cdot)_z \big) \,\, \ud \mu_C ([z])
\,\,\,\textrm{and} \,\,\,
\int \limits_{G_2 / G_{x_c}} \, \big( \mathfrak{J}_L (\cdot)_t , (\cdot)_t \big) \,\, \ud \mu^{x_c}([t]), 
\]
the last integral in (\ref{norm_tilde}) is equal to
\[
\int \limits_{C} \, \big( \mathfrak{J}_L (\mathfrak{J}^{L,C} f)_z , f_z \big) \,\, \ud \mu_C ([z]) 
= \| f \|_{C}^{2}.
\]
\qed

\vspace*{0.5cm}

Note, please, that the Lemmas of Sect. \ref{dense}, i. e. Lemmas \ref{lem:dense.1} -- \ref{lem:dense.6},
are equally applicable to the Krein space $(\mathcal{H}^{L^{x_c}}, \mathfrak{J}_{x_c})$,
with $\mathfrak{J}^{L^{x_c}}$ replaced by $\mathfrak{J}_{x_c}$, and with the section $B_{x_c}$
replaced with the image of $G_2 \big{/} \big(G_2 \cap \, ({x_c}^{-1}G_1 x_c)\big)$ under the inverse of the map
$t \mapsto x_c t$. We formulate this remark as a separate 

\begin{lem}
The Lemmas \ref{lem:dense.1} -- \ref{lem:dense.6} are true for the Hilbert space $\mathcal{H}^{L^{x_c}}$
of the Krein space  $(\mathcal{H}^{L^{x_c}}, \mathfrak{J}_{x_c})$, i.  e. with $L$ replaced by  $L^{x_c}$,
$\mathfrak{J}_L$ replaced by $\mathfrak{J}_{L^{x_c}}= \mathfrak{J}_L$,
$\mathcal{H}_L$replaced with $\mathcal{H}_{L^{x_c}} = \mathcal{H}_L$, $\mathfrak{J}^{L} = \mathfrak{J}^{L^{x_c}}$ replaced by $\mathfrak{J}_{x_c}$ and finally with the section $B_{x_c}$ replaced with the image of 
$G_2 \big{/} \big(G_2 \cap \, ({x_c}^{-1}G_1 x_c)\big)$ under the inverse of the map
$t \mapsto x_c t$.   
\label{lem:subgroup.preliminaries.3}
\end{lem} 
\qedsymbol \, The proofs remain unchanged.
\qed

\vspace*{0.5cm}

In Subsection \ref{subgroup} we explain why we are using $\mathfrak{J}_{x_c}$
in ${^{\mu^{x_c}}}\mathcal{H}^{L^{x_c}}$  instead of $\mathfrak{J}^{L^{x_c}}$.

\section{Decomposition (disintegration) of measures}\label{decomposition}

Let $\mathfrak{G}$, $G_1$ and $G_2$ be such as in Sect. \ref{subgroup.preliminaries}. 
Because the base of the system of neighbourhoods of unity in $\mathfrak{G}$ is countable, the uniform space 
$\mathfrak{X} = \mathfrak{G}/G_1$ is metrizable (compare e. g. \cite{Weil}, \S 2) for any closed subgroup 
$G_1 \subset \mathfrak{G}$. The right action of $G_1$ on $\mathfrak{G}$ is proper and the quotient map
$\pi: \mathfrak{G} \mapsto \mathfrak{G}/G_1$ is open, so that the space $\mathfrak{X} = \mathfrak{G}/G_1$ of 
right $G_1$ orbits ($G_1$ cosets) automatically has the required regularity: measurability of the equivalence
relation defined by the $G_1$ orbits. In particular the quotient 
space $\mathfrak{X}$ is Hausdorff, separable and locally compact and the measure $\rho \cdot \mu_0$ 
(with the $\rho$-function of Sect. \ref{def_ind_krein} and right Haar measure $\mu_0$ on $\mathfrak{G}$)
is decomposable into a direct integral of measures $\rho \cdot \mu_0 = \int \limits_{\mathfrak{G}/G_1} \, \beta_{[x]} 
\, \ud \mu ([x])$ with the component measures 
$\beta_{[x]}$ of the decomposition concentrated in the $G_1$ orbit (right coset) $[x]$
and with Radon-Nikodym derivative associated with the action of the subgroup $G_1$ (i. e. $\lambda_{[x]}$-function) corresponding to $\beta_{[x]}$ equal to the
restriction to the orbit $[x]$ and to the subgroup $G_1$ of the Radon-Nikodym (i. e. $\lambda$-function) corresponding to the measure  $\rho \cdot \mu_0$. This in particular gives us the quasi invariant regular Baire (or Borel) measure 
$\mu = \mu_{\mathfrak{G}/G_1}$ on the uniform space $\mathfrak{X}$ corresponding to $\rho$,
i. e. the factor measure of $\rho \cdot \mu_0$ (Mackey's method of constructing 
general regular quasi invariant measure on the quotient space $\mathfrak{X} = \mathfrak{G}/G_1$).  

This is not the case if we replace $\mathfrak{G}$ with $\mathfrak{X} = \mathfrak{G}/G_1$ acted on by a second closed subgroup $G_2 \subset \mathfrak{G}$. The quotient space $\mathfrak{X}/G_2$ 
is in general a badly behaved non Hausdorff space with non measurable equivalence relation defined in $\mathfrak{X}$
with the $G_2$ orbits as equivalence classes. We require a regularity condition in order to achieve an effective
tool for constructing effectively a dual of the group
$\mathfrak{G}$ in question with the help of decomposition of tensor product of induced representations.  

Let $\mathfrak{X}$, for example $\mathfrak{X} = \mathfrak{G}/G_1$, be any separable locally compact metrizable space
with an equivalence relation $R$ in $\mathfrak{X}$, for example with the equivalence classes given by right $G_2$-orbits in
$\mathfrak{X} = \mathfrak{G}/G_1$ under the right action of a second closed subgroup $G_2 \subset \mathfrak{G}$.  
Let the equivalence classes form a set $\mathfrak{C}$ and for each
$\mathfrak{x} \in \mathfrak{X}$ let $\pi_{\mathfrak{X}}(\mathfrak{x}) \in \mathfrak{C}$ denote the equivalence class of 
$\mathfrak{x}$. Let $\mathfrak{X}$ be endowed with a regular measure $\mu$ (quasi invariant in case 
$\mathfrak{X} = \mathfrak{G}/G_1$).
We define following \cite{Rohlin} the relation $R$ to be measurable 
if there exists a countable family 
$E_0 , E_1 , E_2 , \ldots$ of subsets of $\mathfrak{C}$ such that $\pi_{\mathfrak{X}}^{-1}(E_i)$ is a Baire (or Borel) set for each $i$ and such that $\mu (\pi_{\mathfrak{X}}^{-1}(E_0 )) = 0$, and such that each point $C$ of $\mathfrak{C}$ not belonging to $E_0$ is the intersection of the $E_i$ which contain it. Strictly speaking in Rohlin's definition
of measurability of $R$, accepted by Mackey in \cite{Mackey}, the set $E_0$ is empty and $\pi_{\mathfrak{X}}^{-1}(E_k )$, $k\geq 1$, are just $\mu$-measurable
and not necessary Borel. But the difference is unessential as we explain below in this Sect..
Under this assumption of measurability $\mu$ may be decomposed (disintegrated) as an integral $\mu = \int \limits_{\mathfrak{C}} \, \mu_{C} 
\, d\nu(C)$ over $\mathfrak{C}$ of measures $\mu_C$, with each $\mu_C$ concentrated on the corresponding equivalence class $C$, i .e $G_2$ orbit in case $\mathfrak{X}= \mathfrak{G}/G_1$, with a regular measure $\nu = \mu_{\mathfrak{X}/G_2}$ on 
$\mathfrak{C} = \mathfrak{X}/G_2$ i. e. the factor measure of $\mu$, which we may call the ``double factor measure'' $\mu_{(\mathfrak{G}/G_1)/G_2}$ of $\mu_0 = \mu_\mathfrak{G}$ in case  $\mathfrak{X}= \mathfrak{G}/G_1$; 
and moreover in this case  when $\mathfrak{X}= \mathfrak{G}/G_1$ the Radon-Nikodym derivative (i. e. $\lambda_C$-function) corresponding to $\mu_C$ and associated with action of the subgroup $G_2$ is equal to the restriction to the orbit $C$
and to the subgroup $G_2$ of the Radon Nikodym derivative ($\lambda$-function) corresponding to $\mu$.
In this case we say after Mackey that the subgroups $G_1$ and $G_2$ are \emph{regularly related}. In short:
the orbits in $\mathfrak{G}/G_1$ under the right action of $G_2$  form the equivalence classes of a measurable equivalence relation. (Using literally Rohlin's definition of measurability: almost all of the orbits in $\mathfrak{G}/G_1$ under the right action of $G_2$  form the equivalence classes of a Rohlin-measurable equivalence relation.)    

Let us explain the meaning of the regularity condition. Even if $G_1$ and $G_2$ were not regularly related we could
of course find a countable set $E_1 , E_2 , \ldots$ of Borel unions of orbits which generate the $\sigma$-ring
of all measurable unions of orbits. The unique equivalence relation $R$ such that 
$\mathfrak{x} \in \mathfrak{G}/G_1$ and $\mathfrak{y} \in \mathfrak{G}/G_1$ are in the relation 
whenever $\mathfrak{x}$ and $\mathfrak{y}$ are in the same sets $E_j$ will be measurable. This equivalence relation
gives us a decomposition of the quasi invariant measure $\mu$ into quasi invariant component measures 
$\mu_P$ concentrated on subsets $P \subset \mathfrak{C}$, but in this general non regular situation
the subsets $P$ are unions of many orbits $C \in \mathfrak{C}$. This would give us decomposition of $U^L$ 
restricted to $G_2$, but in this decomposition the component representations will not be associated with single 
orbits, i. e. with single double cosets $G_1 x_0 G_2$ and will not be identifiable as 
``induced representations''
$U^{L^{c_c}}$ of 
$G_2$ of Lemma \ref{lem:subgroup.preliminaries.1} of Sect. \ref{subgroup.preliminaries}. Little or nothing is 
known of such component representations related to non transitive systems of imprimitivity. In fact the regularity
of the $G_2$-orbits in $\mathfrak{G}/G_1$ is essentially equivalent for the group $\mathfrak{G}$
to be of type I. Because of the bi-unique correspondence between $G_2$ orbits in $\mathfrak{G}/G_1$ 
and double cosets $G_1 x G_2$ in $\mathfrak{G}$, and because of the relation between Borel structures
on $\mathfrak{X} = \mathfrak{G}/G_1$ and on $\mathfrak{X}/G_2$, we may reformulate the regularity condition as follows.
We assume that there exists a sequence $E_0 , E_1 , E_2 , \ldots$ of measurable subsets of $\mathfrak{G}$
each of which is a union of double cosets such that $E_0$ has Haar measure zero and each double coset not 
in $E_0$ is the intersection of the $E_j$ which contain it (compare Lemma \ref{lem:decomposition.10}).

\vspace*{0.5cm}

{\bf REMARK}. In fact the representations $U^{L^{x_c}}$ of Lemma 
\ref{lem:subgroup.preliminaries.1} of Sect. \ref{subgroup.preliminaries} do not have the standard form of induced representations defined in Sect. \ref{def_ind_krein} as $\mathfrak{J}_{x_c} \neq \mathfrak{J}^{L^{x_c}}$, but in relevant cases of representations encountered in QFT they may be shown to be Krein-unitary equivalent to standard induced representations (in the sense of Sect. \ref{def_ind_krein}).
Anyway they are concentrated in single orbits. 

Note also that One may characterise the space of orbits by considering the respective group algebra or the associated universal enveloping $C^*$algebra. Connes developed
a general theory of cross-product $C^*$-algebras and von Neumann algebras associated with foliations,
strongly motivated by the Mackey theory of induced representations, compare \cite{Connes} and references there in.
\qed

\vspace*{0.5cm}

\begin{ex}

The equivalence relation on the two-torus $\mathfrak{X} = \mathbb{R}^2 / \mathbb{Z}^2$ given by the 
leaves of the Kronecker foliation associated to an irrational number $\theta$, i. e. given by the differential
equation
\[
dy = \theta dx,
\] 
is not measurable. The leaves, i. e. equivalence classes, can be viewed as orbits of the additive group $\mathbb{R}$ on the two-torus
$\mathfrak{X} = \mathbb{R}^2 / \mathbb{Z}^2$.  
\label{ex:decomposition.1}
\end{ex}

\vspace*{0.5cm}

In the original Mackey's theory the induced representations ${}^{\mu}U^L$ and ${}^{\mu'}U^L$
are unitary equivalent whenever the quasi invariant measures $\mu$ and $\mu'$ on 
$\mathfrak{G}/G_1$ are equivalent, which is always the case, as all such measures are equivalent. In this case 
we may assume all measures $\mu$ in the induced representations ${}^{\mu}U^L$ to be finite without
any lost of generality. In particular, and this simplifies matter, we may restrict ourself to finite 
measures $\mu$ on $\mathfrak{G}/G_1$, as Mackey did in \cite{Mackey},  in constructing decomposition (disintegration) $\mu = \int \limits_{\mathfrak{C}} \, \mu_{C} \, \ud \nu(C)$ with each of the measures 
$\mu_{C}$  concentrated on the corresponding orbit $C$ and the corresponding Radon-Nikodym derivative associated with $\mu_{C}$ under the action of $G_2$ equal to the restriction to the orbit $C$ and to the subgroup $G_2$ 
of the Radon Nikodym derivative associated with $\mu$ (this is proved in \S 11 of \cite{Mackey}). 
This is not the case for the induced representations
${}^{\mu}U^L$ and ${}^{\mu'}U^L$ in Krein spaces defined here, as already indicated by Theorem
\ref{def_ind_krein:twr.3} and its proof: they are (Krein-unitary) inequivalent whenever the quotient space 
$\mathfrak{G}/G_1$ is not compact and the Radon-Nikodym derivative $\ud \mu' / \ud \mu$ is not 
``lower'' or ``upper'' bounded. Therefore we cannot restrict ourself to finite measures  $\mu$
in construction of the decomposition $\mu = \int \limits_{\mathfrak{C}} \, \mu_{C} \, \ud \nu(C)$
with the above mentioned properties. Because Mackey's construction of decomposition of finite measure 
$\mu$ is sufficient for the theory of unitary group representations (as well as 
for the extension of the construction of induced representation to representations of $C^*$-algebras 
along the lines proposed by Rieffel) decomposition having the above mentioned properties of a quasi invariant measure 
$\mu$ which is not finite has not been constructed explicitly in the classical mathematical literature, 
at least the author was not able to find it
(in the Bourbaki's course on integration \cite{Bourbaki}, Chap. 7.2.1-7.2.3  
decomposition of this type is constructed but under stronger assumption than measurability of the equivalence relation given by right $G_2$ action on $\mathfrak{X} = \mathfrak{G}/G_1$
where it is assumed instead that the action is proper and moreover where it is assumed that the measure 
$\mu$ is relatively invariant and not merely quasi invariant -- assumptions too strong for us).
Because the required decomposition of not necessary finite quasi invariant measure $\mu$ on 
$\mathfrak{G}/G_1$ is important for the decomposition of the restriction of the induced representation 
${}^{\mu}U^L$ in a Krein space to a closed subgroup (and \emph{a fortiori} to a decomposition
of tensor product of induced representations
${}^{\mu}U^L$ and ${}^{\mu}U^M$ in Krein spaces) we present here its construction explicitly
only for the sake of completeness. 
The construction presented here uses a localization procedure in reducing the problem 
of decomposition to the Mackey-Godement decomposition (\cite{Mackey}, \S 11) of a finite quasi invariant measure.

\vspace*{0.2cm}

{\bf REMARK}.
Whenever the action of $G_2$ on $\mathfrak{X} = 
\mathfrak{G}/G_1$ is proper one can just replace the continuous homomorphism $\chi: G_2 \mapsto \mathbb{R}_+$ in \cite{Bourbaki}, Chap. 7.2.1-7.2.3, 
by the Radon-Nikodym derivative associated with the measure $\mu$ on $\mathfrak{X} = \mathfrak{G}/G_1$ in this case. Using the Federer and Morse theorem \cite{Federer_Morse} one constructs a regular Borel section of $\mathfrak{X}$ with respect to $G_2$ which enables the construction of the factor measure $\nu$ on the quotient $\mathfrak{C} = \mathfrak{X}/G_2$ of the space $\mathfrak{X}$ by the group $G_2$ with the method of \cite{Bourbaki} changed in  minor points only.

\vspace*{0.2cm}

Let $\mathfrak{X}$ be the separable locally compact metrizable (in fact complete metric) space 
$\mathfrak{G}/G_1$ equipped with a regular 
quasi invariant measure $\mu$. Let $R$ be the equivalence relation 
in $\mathfrak{X}$ given by the right action of a second closed subgroup $G_2$ with the associated quotient map 
$\pi_\mathfrak{X} : \mathfrak{X} \mapsto \mathfrak{X}/R = \mathfrak{X}/G_2$, 
and let $K$ be a compact subset of $\mathfrak{X}$. There is canonically defined equivalence 
relation $R_K$ on $K$ induced by $R$ on $K$ with the associated quotient map 
$\pi_K : K \mapsto K/R_K$ equal to the restriction of $\pi_\mathfrak{X}$ to the subset $K$.
Note please that for an equivalence relation $R$ in the separable locally compact and metrizable space 
$\mathfrak{X} = \mathfrak{G}/G_1$ the above mentioned (Rohlin's \cite{Rohlin}) condition of measurability of $R$ is equivalent to the following condition: the family $\mathfrak{K}$ of those compact sets $K \subset \mathfrak{X}$ 
for which the quotient space $K/R_K$ is Hausdorff is $\mu$-dense, i. e. one of the following and equivalent conditions is fulfilled:  
\begin{enumerate}

\item[(I)]  For a subset $A \subset \mathfrak{X}$ to be locally $\mu$-negligible it is necessary and 
            sufficient that $\mu(A \cap K) = 0$ for all $K \in \mathfrak{K}$.

\item[(II)]  For any compact subset $K_0$ of $\mathfrak{X}$ and for any $\epsilon > 0$ there exists a subset
           $K \in \mathfrak{K}$ contained in $K_0$ and such that $\mu(K_0 - K) \leq \epsilon$.

\item[(III)] For each compact subset $A$ of $\mathfrak{X}$ there exists a partition of $A$ into 
             a $\mu$-negligible subset $N$ and a sequence $\{ K_n \}_{n \in \mathbb{N}}$ of compact subsets 
             belonging to $\mathfrak{K}$.

\item[(IV)]  For each compact subset $K$ of $\mathfrak{X}$ there exists an increasing 
             $H_1 \subseteq H_2 \subseteq \ldots$ sequence $\{ H_n \}_{n \in \mathbb{N}}$ of compact 
             sets belonging to $\mathfrak{K}$ contained in $K$ and such that the set 
             $Z = K - \bigcup \limits_{n \in \mathbb{N}} H_n$ is $\mu$-negligible.

\end{enumerate}

Indeed, because the the system of neighbourhoods of unity in $\mathfrak{G}$ is countable, the uniform space 
$\mathfrak{X} = \mathfrak{G}/G_1$ is completely metrizable and locally compact (compare e. g. \cite{Weil}, \S 2) for any closed subgroup $G_1 \subset \mathfrak{G}$. Therefore Proposition 3 of \cite{Bourbaki_i}, Chap. VI, \S 3.4, is
applicable. By this Proposition we need only show that using the family $\mathfrak{K}$ one can
construct the sets $E_0, E_1 , \ldots$ of the Rohlin's measurability condition of $R$, for which 
$\pi_{\mathfrak{X}}^{-1}(E_k )$, $k\geq 1$, are not only $\mu$-measurable but moreover Borel. This however follows
from the fact that $\mathfrak{X}$ is countable at infinity: there exists a sequence of compact subsets 
$K_1 \subset K_2 \subset \ldots$ of $\mathfrak{X}$ such that $\mathfrak{X} = \cup_i K_i$ and moreover we may assume that 
they are regular closed sets: $\cl \intt K_m = K_m$.   

Indeed, let $\{ \mathcal{O}_k \}_{k \in \mathbb{N}}$ be a countable base of the topology in $\mathfrak{X}$,
such that the closure $\overline{\mathcal{O}_k }$ of each $\mathcal{O}_k$ is compact (there exists such a base because
$\mathfrak{X}$ is second countable and locally compact). For each $\overline{\mathcal{O}_k }$ choose
a sequence $\{K_{kl}\}_{l \in \mathbb{N}}$ of compact sets belonging to $\mathfrak{K}$ and a $\mu$-negligible
subset $M_k$ giving the partition $\overline{\mathcal{O}_k } = M_k \dot{\cup} K_{k1} \dot{\cup} K_{k2} \dot{\cup} \ldots$
of $\overline{\mathcal{O}_k }$, existence of which is assured by the condition (III). 
Define the $\mu$-negligible set $M = \cup_k M_k$ and a maximal subset $M_0$ of $M$ invariant under the action of $G_2$
on $\mathfrak{X}$. 

By the condition (IV) we can construct for each $K_{m}$ 
a sequence $H_{m1} \subset H_{m2} \subset H_{m3} \subset \ldots $ of compact subsets of $K_m$ belonging to
$\mathfrak{K}$ and a $\mu$-negligible subset $Z_m$ such that $K_{m} = Z_m 
\dot{\cup} \big( \cup_n K_{mn}\big)$. Define the $\mu$-negligible set $Z = \bigcup \limits_{n \in \mathbb{N}} Z_m$
and the maximal subset $Z_0$ of $Z$ invariant under the action of $G_2$.

 Let us define a countable family of sets $E_0 = \pi_\mathfrak{X}(Z_0 \cup M_0)$, $E_{mn} = 
\pi_\mathfrak{X}(K_{mn}) = \pi_{K_{mn}}(K_{mn})
= K_{mn} / R_{K_{mn}}$, $m,n \in \mathbb{N}$ in $\mathfrak{X}/G_2 = \mathfrak{X}/R$, where 
$K_{mn} / R_{K_{mn}}$ is Hausdorff by assumption. 

Now let $\mathfrak{x}_1$ and $\mathfrak{x}_2$
be two elements of $\mathfrak{X}$ not in $N_0 = Z_0 \cup M_0$ such that  $\pi_\mathfrak{X}(\mathfrak{x}_1 ) \neq 
\pi_\mathfrak{X}(\mathfrak{x}_2 )$. Then by construction there exists $H_{mn} \in \mathfrak{K}$ containing  
$\mathfrak{x'}_1$ and $\mathfrak{x'}_2$ with $\pi_\mathfrak{X}(\mathfrak{x'}_1 ) = \pi_\mathfrak{X}(\mathfrak{x}_1 )$ and 
$\pi_\mathfrak{X}(\mathfrak{x'}_2 ) = \pi_\mathfrak{X}(\mathfrak{x}_2 )$. 

$H_{mn} / R_{H_{mn}}$ containing $\pi_\mathfrak{X}(\mathfrak{x}_1 ) = \pi_\mathfrak{X}(\mathfrak{x'}_1 )$ and $\pi_\mathfrak{X}(\mathfrak{x}_2 ) = \pi_\mathfrak{X}(\mathfrak{x'}_2 )$ is Hausdorff by construction. Thus there exist two compact non intersecting neighbourhoods $\overline{\mathcal{O}}_{\mathfrak{x'}_1}$ 
and $\overline{\mathcal{O}}_{\mathfrak{x'}_2}$ of $\mathfrak{x'}_1$ and $\mathfrak{x'}_2$ respectively
such that for $K_{\mathfrak{x'}_1} = \overline{\mathcal{O}}_{\mathfrak{x'}_1} \cap H_{mn}$
and $K_{\mathfrak{x'}_2} = \overline{\mathcal{O}}_{\mathfrak{x'}_2} \cap H_{mn}$
we have $\pi_\mathfrak{X}^{-1} (K_{\mathfrak{x'}_1}) \cap \pi_\mathfrak{X}^{-1} (K_{\mathfrak{x'}_2}) = \emptyset$.
By construction we may choose  $K_{{m_1}{n_1}} \subset K_{\mathfrak{x}_1}$ and $K_{{m_2}{n_2}} \subset K_{\mathfrak{x}_2}$ in $\mathfrak{K}$ such that 
$\mathfrak{x'}_1 \in K_{{m_1}{n_1}}$  and $\mathfrak{x'}_2 \in K_{{m_2}{n_2}}$. Of course we have 
$E_{{m_1}{n_1}} \cap E_{{m_2}{n_2}} = \pi_\mathfrak{X}^{-1} (K_{{m_1}{n_1}}) \cap 
\pi_\mathfrak{X}^{-1} (K_{{m_2}{n_2}}) = \emptyset$. Thus the intersection of all $E_{mn} \in \mathfrak{K}$ containing
$\pi_\mathfrak{X}(\mathfrak{x}_1 ) \in \mathfrak{X}/G_2$ is equal $\{ \pi_\mathfrak{X}(\mathfrak{x}_1 ) \}$.
We have to show that $\pi_\mathfrak{X}^{-1}(E_{mn}) = \pi_\mathfrak{X}^{-1}(\pi_\mathfrak{X}(K_{mn}))$
are Baire (or Borel) sets. To this end observe please that $\pi_\mathfrak{X}^{-1}(\pi_\mathfrak{X}(K_{mn}))$
is equal to the saturation of $K_{mn}$, i. e. $\pi_\mathfrak{X}^{-1}(\pi_\mathfrak{X}(K_{mn}))
= K_{mn} \cdot G_2$. Choose a compact neighbourhood $V$ of the unit in $G_2$ such that $V = V^{-1}$. Then
if $G_2$ is connected then $G_2 = \bigcup \limits_{n \in \mathbb{N}} V^n$; if $G_2$ is not connected
then it is still a countable sum of connected components of the form $\bigcup \limits_{n \in \mathbb{N}} V^n \eta_m$, 
with $\eta_m \in G_2$  chosen from $m$-th connected component $G_{2m}$ of $G_2$. Thus in each case $G_2$ is a countable sum
$\bigcup \limits_{k,l \in \mathbb{N}} V_{kl}$ of compact sets $V_{kl}$. Therefore
$\pi_\mathfrak{X}^{-1}(E_{mn}) = K_{mn} \cdot G_2 = \bigcup \limits_{k,l \in \mathbb{N}} K_{mn} \cdot V_{kl}$      
being a countable sum of compact sets is contained in the $\sigma$-ring generated by the compact sets and all the more 
it is a Borel set contained in the $\sigma$-ring generated by the closed sets. Thus both definitions of measurability
of the equivalence relation $R$ on $\mathfrak{X}$ are equivalent.

\vspace*{0.5cm}

\begin{lem}
There exists a Borel set $B_0$ in $\mathfrak{X} = \mathfrak{G}/G_1$ 
and a $\mu$-negligible subset $N_0 \subset \mathfrak{X}$ consisting of $G_2$ orbits in 
$\mathfrak{X} = \mathfrak{G}/G_1$
such that $B_0$ intersects each $G_2$ orbit not contained in $N_0$ in exactly one point.
\label{lem:decomposition.1}
\end{lem}

\qedsymbol \,
For the proof compare e. g. \cite{Bourbaki_i}, Chap. VI, \S 3.4, Thm. 3.
\qed

\vspace*{0.5cm}

Adding to $B_0$ any section of the $\mu$-negligible set $N_0$ we obtain a measurable section $B_{00}$
for the whole space $\mathfrak{X}$. For equivalence relations $R$ on smooth manifold $\mathfrak{X}$
defined by foliations on $\mathfrak{X}$ (i. e. smooth and integrable sub-bundles of $T\mathfrak{X}$)
existence of a measurable section is equivalent for the foliation to be of type I: i. e. the von Neumann algebra associated
to the foliation is of type I iff the foliation admits a Lebesgue measurable section, compare \cite{Connes}, 
Chap. I.4.$\gamma$, Proposition 5.    

Because the Borel space $\mathfrak{X} = \mathfrak{G}/G_1$ is standard it follows by the second Theorem on page 74 of
\cite{Mackey2} that the quotient Borel structure on $\mathfrak{X}/G_2 - N_0 /G_2$ is likewise standard;
i. e. there exits a Borel isomorphism $\psi_0$: $(\mathfrak{X} - N_0)/G_2 \rightarrow S_0 \subset$ onto a 
Borel subset $S_0$  of a complete separable metric space $S$.

The space $(\mathfrak{X} - N_0 )/G_2$ however  need not be locally compact and it is not if the action of
$G_2$ on $\mathfrak{X} = \mathfrak{G}/G_1$ is not proper but only measurable, i.e. with measurable equivalence relation determined by the action of $G_2$. Similarly $G_2$-orbit $C$ in $\mathfrak{X}$ as a subset of a locally compact space
$\mathfrak{X}$ need not be closed if the action of $G_2$ is not proper and thus need not be locally compact with the 
topology induced from the surrounding space $\mathfrak{X}$.

\vspace*{0.5cm}

\begin{lem}

Let $N_0$ be as in Lemma \ref{lem:decomposition.1}. A necessary and sufficient condition that a 
subset $E$ of $\mathfrak{X}/G_2 - N_0 /G_2$ be a Borel set is that $\pi_{\mathfrak{X}}^{-1}(E)$
be a Borel set in $\mathfrak{X} - N_0$. A necessary and sufficient condition that a function 
$f$ on $\mathfrak{X}/G_2 - N_0 /G_2$ be a Borel function is that $f \circ \pi_{\mathfrak{X}}$
be a Borel function on $\mathfrak{X} - N_0$.  

\label{lem:decomposition.3}
\end{lem}

\qedsymbol \, Let $p_0$ be the Borel function $\psi_0 \circ  \pi_{\mathfrak{X}} : \mathfrak{X} - N_0
\rightarrow S_0$. Let $E'$ be any subset of $S_0$ such that $p_{0}^{-1}(E')$ is a Borel set. Let 
$B_0$ be the Borel section of $\mathfrak{X} - N_0$ with respect to $G_2$, existence of which has been
proved in Lemma \ref{lem:decomposition.1}. Then $p_0 (p_{0}^{-1}(E') \cap B_0) = E'$, and thus $E'$ is a Borel set
by Theorem 3, page 253 of \cite{Kuratowski}, compare likewise the Theorems on pages 72-73 of \cite{Mackey2}, 
because $p_0$ is one-to-one Borel function on $B_0$. 
Conversely: if $E'$ is Borel in $S_0$ then because $p_0$ is a Borel function, so is the set $p_{0}^{-1}(E')$. The first part of the Lemma  follows now from this and from definition 
of the Borel structure induced on $\psi_0 ((\mathfrak{X} - N_0 )/G_2)$ and \emph{a fortiori} on $\mathfrak{X}/G_2 - N_0 /G_2$. The remaining part of the Lemma is an immediate consequence of the first part.   
\qed

\vspace*{0.5cm}

We have the following disintegration theorem for the (not necessarily finite) measure $\mu$
and any of its pseudo image measures $\nu$ on $\mathfrak{X}/G_2$ (for definition of pseudo image measure $\nu$
compare e. g. \cite{Bourbaki_i}, Chap. VI.3.2):

\vspace*{0.5cm}

\begin{lem}
For each orbit $C = \pi_{\mathfrak{X}}^{-1}(d_0) \subset \mathfrak{X}$ with $d_0 \in \mathfrak{X}/G_2$ there exists a 
Borel measure $\mu_C$ in $\mathfrak{X}$ concentrated on the orbit $C$, i. e. $\mu_C (\mathfrak{X} - C)
= \mu_C (\mathfrak{X} - \pi_{\mathfrak{X}}^{-1}(d_0)) = 0$.
For any $g \in L^1 (\mathfrak{X} , \mu)$ 
the set of all those $G_2$ orbits $C$ for which $g$ is not 
$\mu_C$-integrable is $\nu$-negligible and the function 
\[
C \mapsto \int \, g(x) d\mu_C(x)
\]
is $\nu$-summable and $\nu$-measurable, and
\begin{equation}\label{dec_m}
\int \, d\nu(C) \, \int \, g(x) \, d\mu_C (x)  = \int \, g(x) \, d \mu(x).
\end{equation}
In short
\[
\mu = \int \, \mu_C (x) \, d\nu(C).
\]

\label{lem:decomposition.4}
\end{lem}

\begin{rem}
For each orbit $C$ the measure $\mu_C$ may also be naturally viewed as a measure on the 
 $\sigma$-ring $\mathscr{R}_C$ of measurable subsets of $C$ induced from the surrounding space $\mathfrak{X}$:
$E \in \mathscr{R}_C$ iff $E = E' \cap C$ for some $E' \in \mathscr{R}_{\mathfrak{X}}$, i. e. with the 
subspace Borel structure. 
\label{rem:decomposition.1}
\end{rem}

\qedsymbol \,
 For the proof we refer the reader e. g.  to \cite{Bourbaki_i}, Chap. VI, \S 3.5. 
\qed

\vspace*{0.5cm}

We shall show that for each $C$ the measure $\mu_C$ is quasi invariant and that for all $\eta \in G_2$ 
the Radon-Nikodym derivative $\lambda_C (\cdot , \eta) = 
\frac{\ud (R_\eta \mu_C)}{\ud \mu_C}(\cdot)$ is equal to the restriction of the Radon-Nikodym 
derivative  $\lambda (\cdot , \eta) = \frac{\ud (R_\eta \mu)}{\ud \mu}(\cdot)$
to the orbit $C$. In doing so we prefer reducing the problem to the
Mackey-Godement decomposition  of a finite measure (\cite{Mackey}, \S 11) 
using a localization of the measure space 
$(\mathfrak{X}, \mathscr{R}_{\mathfrak{X}}, \mu)$ and its disintegration.  
Toward this end we need some further Lemmas.   

\vspace*{0.5cm}

\begin{lem}
Let $\mu$, $\mu_C$ and $\nu$ be as in the preceding Lemma. Let $K$ be a compact subset of $\mathfrak{X}$.
Then $\pi_{\mathfrak{X}} (K)$ is measurable on $\mathfrak{X}/G_2$.

\label{lem:decomposition.5'}
\end{lem}

\qedsymbol \,
Let $K$ be any compact subset of $\mathfrak{X}$ and let $Z , K_{n}$ be the subsets of condition
(IV), i. e. $K_n \in \mathfrak{K}$ is an increasing sequence of compact subsets of $K$, and $Z$ is $\mu$-negligible subset of $K$ such that $K = Z \dot{\cup} \big( K_1 \cup K_2 \cup \ldots  \big)$. Let us define the subset (if any) $Z_0 \subset Z$
consisting of intersections of full  $G_2$-orbits with $K$, i. e.  the maximal subset of $Z$ invariant under the action
of $G_2$ on $\mathfrak{X}$. 
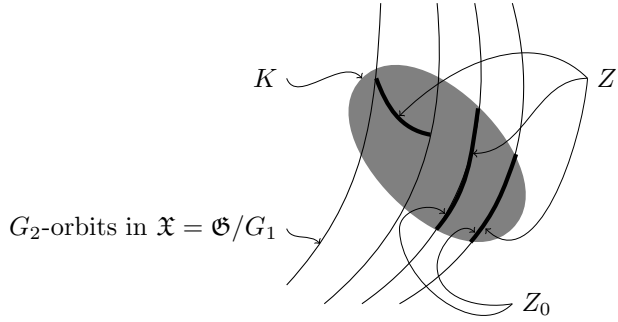
\begin{center}
\begin{tikzpicture}

\path[fill=gray] (2,4) to [out=45,in=45] (4,2)
to [out=-135,in=-135] (2,4);

\draw[thin] (1,1.25) to [out=47.5,in=-92.5] (2.25,5);



\draw[thin] (1.5,1) to [out=40,in=-85] (3,5);


\draw[thin] (2,1) to [out=35,in=-80] (3.5,5);


\draw[thin] (2.5,1) to [out=30,in=-75] (4,5);


\draw[ultra thick] (2.19,4) to [out=-70,in=170] (2.91,3.25);

\draw[ultra thick] (3.445,1.82) to [out=50,in=-109] (4.05,2.99);

\draw[ultra thick] (2.99,1.99) to [out=50,in=-98] (3.545,3.6);

\draw [->,very thin] (1,4) to [out=-70,in=130] (2, 4);

\draw [->,very thin] (1,2) to [out=-70,in=130] (1.44, 1.8);

\draw [<-,very thin] (3.48,3) to [out=0,in=180] (5,4);

\draw [<-,very thin] (2.5,3.5) to [out=45,in=135] (5,4);

\draw [<-,very thin] (3.645,2) to [out=-45,in=-95] (5,4);

\draw [<-,very thin] (3.1,2.2) to [out=135,in=90] (2.5,2) to [out=-90,in=-135] (4,1);

\draw [<-,very thin] (3.5,1.95) to [out=135,in=90] (3,1.5) to [out=-90,in=190] (4,1);

\node [right] at (4,1) {$Z_{0}$};

\node [right] at (5,4) {$Z$};

\node [left] at (1,2) {$G_2$-orbits in $\mathfrak{X} = \mathfrak{G}/G_1$};

\node [left] at (1,4) {$K$};

\end{tikzpicture} 
\end{center}
Then $\pi_\mathfrak{X} (K - Z) = \pi_\mathfrak{X} (K - Z_{0})$. We shall show that
$\mu(Z_{0} \cdot G_2) = \mu( \pi_{\mathfrak{X}}^{-1} (\pi_\mathfrak{X} ( Z_{0}) ) = 0$. Toward this end observe
that because $\mathfrak{X}$ is metrizable and separable we may assume the elements $\mathcal{O}_m$, $m \in \mathbb{N}$,
of basis of topology to be the balls with compact closure $\overline{\mathcal{O}_m}$; and the $\sigma$-ring of Borel sets on $\mathfrak{X}$
generated by the open $\mathcal{O}_m$ or closed $\overline{\mathcal{O}_m}$ balls.
\begin{center}
\begin{tikzpicture}

\path [fill=gray] (5,2) circle (0.5);

\draw[thick] (5,2) ellipse (0.7 and 0.7);

\draw[thick] (4.293,4) ellipse (0.6 and 0.5);

\draw[thick] (3.368,5) ellipse (0.575 and 0.4);

\draw[thin] (5,0) to [out=60.75,in=-90] (5.7,2)
to [out=90,in=-45] (3,6);

\draw[thin] (4.7,0) to [out=60.75,in=-90] (5.4,2)
to [out=90,in=-45] (2.7,6);

\draw[thin] (4.6,0) to [out=60.75,in=-90] (5.3,2)
to [out=90,in=-45] (2.6,6);

\draw[thin] (3.6,0) to [out=60.75,in=-90] (4.3,2)
to [out=90,in=-45] (1.6,6);

\draw[ultra thick] (5.4,1.7) to [out=89,in=-89] (5.4,2.3);

\draw[ultra thick] (5.3,1.6) to [out=88,in=-88] (5.3,2.4);


\draw [->,very thin] (3,2) to [out=-70,in=130] (4.505, 2.495);

\draw [->,very thin] (3,4) to [out=-20,in=135] (3.8, 4.3);

\draw [->,very thin] (3,1) to [out=45,in=-135] (4.6465, 1.6465);

\draw [<-,very thin] (4.7,0) to [out=-45,in=135] (6, 0);


\node [right] at (6,0) {$Z_{0} \cdot G_2$};

\node [left] at (3,4) {$\mathcal{O}_\epsilon \cdot \eta$,};

\node [left] at (3.5,3.6) {$\eta \in G_2$};

\node [left] at (3,2) {$\mathcal{O}_\epsilon$};

\node [left] at (3,1) {$K$};

\node [right] at (-2.75,4) {For each $\epsilon > 0$ };

\node [right] at (-2.75,3.5) {there exists open $\mathcal{O}_\epsilon 
\supset K$ };

\node [left] at (0.75,3) {with: $\mu(\mathcal{O}_\epsilon 
- K) < \epsilon$};

\node [right] at (-2.75,2.5) {by regularity of $\mu$};

\end{tikzpicture} 
\end{center}
By the regularity and quasi invariance of the measure $\mu$ it easily follows that the $\mu$-measure of the intersection of 
$Z_{0} \cdot G_2$ with any open set in $\mathfrak{X}$ is equal zero, and thus again by the regularity
of $\mu$ and second countability of $\mathfrak{X}$ it easily follows that 
$\mu(Z_{0} \cdot G_2) = \mu ({\pi_\mathfrak{X}}^{-1} (\pi_\mathfrak{X} ( Z_{0}) )) = 0$. Thus 
$\pi_\mathfrak{X} ( Z_{0})$ is a subset of a measurable null set, and so must be a measurable set with
$\nu(\pi_\mathfrak{X} ( Z_{0})) = 0$, because $\nu$ is a pseudo-image measure of $\mu$ under $\pi_\mathfrak{X}$. Moreover, we have: 
\[
\pi_\mathfrak{X}(K-Z) 
= \pi_\mathfrak{X}(K - Z_0) 
=  \pi_\mathfrak{X}(K) - \pi_\mathfrak{X}(Z_0), 
\]
because $Z_0$ consists of intersections of $G_2$-orbits with 
$K$. 

On the other hand
\[
\psi_0 \circ \pi_\mathfrak{X} (K -Z) 
\]
is a Borel set in $S$, and thus $\pi_\mathfrak{X} (K -Z)$ is a Borel set in $\mathfrak{X}/G_2$ as $\psi_0$
is a Borel isomorphism. Indeed, because images preserve the set theoretic sum operation we have
\[
\psi_0 \circ \pi_\mathfrak{X} (K -Z) =  \bigcup \limits_{n \in \mathbb{N}} \psi_0 \circ \pi_\mathfrak{X} (K_i). 
\]
Because $K_j \in \mathfrak{K}$ then $K_{j} / R_{K_{j}}$ is Hausdorff and the quotient map $\pi_{K_j}$ is closed and thus the quotient space $K_{j} / R_{K_{j}}$ is homeomorphic to the compact space $\pi_{K_j}(K_j)$, and moreover because  $K_j$ is compact and metrizable (as a subspace of the metrizable space $\mathfrak{X}$) the quotient space  $K_{j} / R_{K_{j}}$ is likewise metrizable (\cite{Engelking}, Thm. 7.5.22). We can therefore apply the Federer and Morse Theorem 5.1 
of \cite{Federer_Morse} in order to prove the existence for each $j$ of a Borel subset $B_{j} \subset K_j$ such that $\pi_{K_j}(B_j) = \pi_{K_j}(K_j) (= \pi_{\mathfrak{X}}(K_j))$ 
and such that $\pi_{K_j}$ is one-to-one on $B_j$. Therefore $\psi_0 \circ \pi_{\mathfrak{X}}$ is one-to-one Borel
function on a Borel subset $B_j$ of complete separable metric space  $\mathfrak{X}$ to a complete separable metric
space $S$. Therefore again by the Theorem on page 253 of \cite{Kuratowski} (compare likewise the Theorem on page 72 
of \cite{Mackey2}), it follows that $\psi_0 \circ \pi_{\mathfrak{X}}(B_j) = \psi_0 \circ \pi_{\mathfrak{X}}(K_j)$ 
is a Borel set. Because $\psi_0$ is a Borel isomorphism it follows that $\pi_{\mathfrak{X}}(K_j)$ is a Borel set
in $\mathfrak{X}/G_2$. 

Thus $\pi_\mathfrak{X}(K)$ differs from a Borel set
$\pi_\mathfrak{X}(K- Z)$ by a measurable $\nu$-negligible subset $\pi_\mathfrak{X}(Z_{0})
\subset \pi_\mathfrak{X}(K)$; so we have shown that  $\pi_\mathfrak{X}(K)$ is measurable. 
\qed

\vspace*{0.5cm}

Note that the Lemma \ref{lem:decomposition.5'} is non trivial. 
By the well known theorem of Suslin -- continuous image of a Borel set is not always Borel,
but it is always measurable, compare e. g. \cite{Jech}, Lemm. 11.6, page 142 and Thm. 11.18, page 150,
where the references to the original literature are provided. However this argument would be insufficient for 
$\pi_{\mathfrak{X}}(K)$ to be measurable in $\mathfrak{X}/G_2$
for any compact set $K \subset \mathfrak{X}$. Indeed it would in addition require to be shown that the quotient Borel structure on $\mathfrak{X}/G_2$ is equal to the $\sigma$-ring of Borel sets generated by the closed (open)
sets of the quotient topology on $\mathfrak{X}/G_2$.

\vspace*{0.5cm}

\begin{lem}
Let $\mu, \mu_C , \nu$ be as in Lemma \ref{lem:decomposition.4} and let $K$ be a compact subset of $\mathfrak{X}$.
Let $\eta \in G_2$ and let $\mathscr{R}_K$ be the $\sigma$-ring  of Borel subsets of $K$ induced form the surrounding measure space $\mathfrak{X}$. Let $(\mu)'_K$ and $(\mu_C)'_K$ denote the restrictions of 
$\mu$ and $\mu_C$ to $K$ defined on the $\sigma$-ring $\mathscr{R}_K$ respectively, and let $R_\eta \mu , R_\eta \mu_C$ denote their right translations; and similarly let $(\nu)'_{\pi_\mathfrak{X}(K)}$ be the restriction of the measure $\nu$ to the subset $\pi_{\mathfrak{X}}(K)$. 
Then

\begin{enumerate}

\item[(a)] 
\[
(\mu)'_K = \int \, (\mu_C )'_K  \, d (\nu)'_{\pi_\mathfrak{X}(K)}(C)
\]
with each $(\mu_C)'_K$ concentrated on $C \cap K$.

\item[(b)]
\[
R_\eta \mu = \int \, R_\eta \mu_C  \, d\nu(C).
\]

\end{enumerate}

\label{lem:decomposition.5}
\end{lem}
(Note that the $\sigma$-ring of Borel sets with a regular measure on this ring is sufficient to recover all measurable subsets and their measures obtained by the standard completion of the Borel measure space.)
\qedsymbol \,
  Part (a) of the Lemma is an immediate consequence 
of Lemmas  \ref{lem:decomposition.4} and  \ref{lem:decomposition.5'} with $1_K \cdot g$ 
inserted for $g$ in the formula (\ref{dec_m}), where $1_K$ is the characteristic
function of the compact set $K$. The only non-trivial part of the proof lies in showing that 
$\pi_\mathfrak{X}(K)$ is measurable, which was proved in Lemma \ref{lem:decomposition.5'}. 

For (b) observe that if $R_{\eta^{-1}}g \in L^1 (\mathfrak{X}, \mu)
\Leftrightarrow g \in L^1 (\mathfrak{X}, R_\eta \mu)$, then by Lemma \ref{lem:decomposition.4}:
\[
\begin{split}
\int g(x) \, d ( R_\eta \mu) = \int g(x \cdot \eta^{-1}) \, d \mu 
= \int d \nu (C) \, \int g(x \cdot \eta^{-1}) \, d \mu_C (x) \\
= \int d \nu (C) \, \int g(x) \, d (R_\eta \mu_C ) (x), 
\end{split}
\] 
thus 
\[
R_\eta \mu = \int R_\eta \mu_C \, d \nu (C).
\]
\qed

\vspace*{0.5cm}

Note that the operations of restriction $(\cdot)'_{K}$
to $K$ and right translation $R_\eta (\cdot)$ do not commute. Indeed if we write $R_\eta \circ (\cdot)'_{K}$
for $ R_\eta((\cdot)'_K )$, then $R_\eta \circ (\cdot)'_{K} 
= (\cdot)'_{K \cdot \eta^{-1}} \circ R_\eta = (R_\eta (\cdot))'_{K \cdot \eta^{-1}}$
i. e. first restrict to $K$ and then translate $R_\eta$ is the same as first translate $R_\eta$ and then
restrict to $K \cdot \eta^{-1}$ (and not to $K$).   

\vspace*{0.5cm}

\begin{rem}
Let $Op(\mu)$ denote a repeated application of several restrictions to compact sets and translations:
$(\cdot)'_{K_1}, R_{\eta_1}(\cdot), \ldots$ performed on the measure $\mu$. Then the repeated application of 
Lemma \ref{lem:decomposition.5} (a) and (b) gives

\[
Op( \mu) = \int  Op( \mu_C ) \,\, \ud \widetilde{Op}(\nu)(C),
\]
where $\widetilde{Op}(\nu)$ denotes the restriction $()'_{\pi_\mathfrak{X}(K)}$ with the compact set
$K \subset \mathfrak{X}$ which arises in the following way: $(\cdot)'_K$ is the restriction which arises from 
$Op$ by commuting all translations to the right (so as to be performed first) and all restrictions to the left
(so as to be performed after all translations): $Op = (\cdot)'_K \circ R_\eta (\cdot)$ or 
$Op (\cdot) = (R_\eta (\cdot))'_K$.
\label{rem:decomposition.2}
\end{rem}

\vspace*{0.5cm}

\begin{lem}
Let $K, (\mu)'_K, (\mu_C)'_K, (\nu)'_{\pi_\mathfrak{X}(K)}$ be as in the preceding Lemma. 
For any bounded and $(\mu)'_K$-measurable function  $g$ 
and for any $f \in L^1 (\pi_{\mathfrak{X}}^{-1}(K) , (\nu)'_{\pi_\mathfrak{X}(K)})$
the set of all those $G_2$ orbits $C$ having non empty intersection $C \cap K$ for which $g$ is not 
$\mu_C$-integrable is $\nu$-negligible and the the function 
\[
C \mapsto \int \, g(x) d (\mu_C)'_K (x)
\]
on this set of orbits $C$ is $(\nu)'_{\pi_\mathfrak{X}(K)}$-summable and $(\nu)'_{\pi_\mathfrak{X}(K)}$-measurable, and
\begin{equation}\label{dec_m_1}
\int f(C) \int \, g(x) \, d (\mu_C)'_K (x) \, d (\nu)'_{\pi_\mathfrak{X}(K)} (C)   
= \int \, f(\pi_{\mathfrak{X}}(x) )g(x) \, d (\mu)'_K (x).
\end{equation}

\label{lem:decomposition.6}

\end{lem}

\qedsymbol \,
 The Lemma is an immediate consequence of the preceding Lemma. The only non-trivial part of the proof is
is to show that $f$ is measurable on $\mathfrak{X}/G_2$ if and only if $f\circ \pi_{\mathfrak{X}}$ 
is measurable on $\mathfrak{X}$. But this is an immediate consequence of Lemma \ref{lem:decomposition.3}.
\qed

\vspace*{0.5cm}

In order to simplify notation  let us denote the operation of 
restriction $(\cdot)'_K$ to $K$ just by $(\cdot)'$ in the next Lemma and its proof.
In all other restrictions $(\cdot)'_D$ the sets $D$ will be specified explicitly. 

\begin{lem}
Let $\mu, \mu_C$ be as in Lemma \ref{lem:decomposition.4} and let $K$ be a compact subset
of $\mathfrak{X}$. Let $\eta \in G_2$
and let $C$ be any $G_2$-orbit having non empty intersection $C \cap K \cdot \eta^{-1} \cap K$.
Then for the respective measures obtained by right translations and restrictions  
performed on $\mu$ and $\mu_C$ respectively we have:
\begin{enumerate}

\item[(a)] The measures $((\mu_C)')'_{K \cdot \eta^{-1}}$ and $(R_\eta (\mu_C)')'$, 
defined on measurable subsets of $C \cap K \cap  K \cdot \eta^{-1}$, are equivalent.

\item[(b)]
\[
\begin{split}
\lambda_C (\cdot , \eta) = 
\frac{d(R_\eta \mu_C)}{d\mu_C}(\cdot) \\
= \frac{\ud \,(R_\eta (\mu_C)')'}{\ud \, ((\mu_C)')'_{K \cdot \eta^{-1}}}(\cdot) 
= \frac{\ud \, (R_\eta \mu')'}{d (\mu')'_{K \cdot \eta^{-1}}}(\cdot) \\
= \frac{\ud \, R_\eta \mu}{\ud \mu}(\cdot)
= \lambda (\cdot , \eta)
\end{split}
\]
on $C \cap K \cap  K \cdot \eta^{-1}$.
\end{enumerate}

\label{lem:decomposition.7}
\end{lem}

\qedsymbol \,
 In addition to the operations of translation and restriction let us introduce after Mackey, \cite{Mackey}, \S 11, one more operation $\widetilde{\cdot}$ defined on finite measures $\mu$ on $\mathfrak{X}$, giving measures
$\widetilde{\mu}$ on $\mathfrak{X}/G_2$. Namely we put $\widetilde{\mu}(E) = \mu(\pi_\mathfrak{X} ^{-1} (E))$. 
$\widetilde{\mu'}$ is well defined for any quasi invariant measure $\mu$ on $\mathfrak{X}/G_2$ because $\mu'$ is finite.  More precisely $\widetilde{\mu'}$ is defined on the 
$\sigma$-ring of measurable subsets $E$ of $\pi_\mathfrak{X}(K)$ by the formula: 
$\widetilde{\mu'}(E) = \mu'(\pi_{\mathfrak{X}}^{-1}(E)) = \mu(K \cap  \pi_{\mathfrak{X}}^{-1}(E))$. 
A simple verification of definitions shows that $\widetilde{\mu'}$ is a pseudo image measure
of the measure $\mu'$ under $\pi_\mathfrak{X}$, so that 
\[
\mu' = \int \,  \mu'_C \, d\widetilde{\mu'}(C), 
\]
on measurable subsets of $K$ and 
where the integral is over the orbits $C$ having non void intersection with $K$
and with $\mu'_C$ concentrated on $C \cap K$. 
Similarly we have for the pairs of measures 
\begin{equation}\label{pairs}
\Big( \,\, (\mu')'_{K  \eta^{-1}} \,\, , \,\,\,
\widetilde{(\mu')'_{K  \eta^{-1}} } \,\, \Big)
 \,\, \textrm{and} \,\, 
\Big( \,\, (R_\eta \mu')' \,\, , \,\,\, \widetilde{(R_\eta \mu')'} \,\, \Big):
\end{equation}
\[
(\mu')'_{K  \eta^{-1}} 
= \int \,  \Big( (\mu')'_{K \eta^{-1}} \Big)_{{}_C} 
\,\,\,\, \ud \, \widetilde{(\mu')'_{K \eta^{-1}}} \,\, (C)
\]
and
\[
(R_\eta \mu')' = \int \Big( R_\eta \mu')' \Big)_{{}_C} \,\,\,\, \ud \, \widetilde{(R_\eta \mu')'} \,\, (C),
\]
both $(\mu')'_{K  \eta^{-1}}$ and $(R_\eta \mu')'$ defined on measurable subsets of $K \cdot \eta^{-1} \cap K$ 
(instead of $K$): with the measure $(R_\eta \mu')'$ equal to the measure $R_\eta \mu$
restricted to $K \cdot \eta^{-1} \cap K$, and $(\mu')'_{K  \eta^{-1}}
= (\mu)'_{K  \eta^{-1} \cap K}$ equal to the measure $\mu$ restricted to 
the same compact subset $K \cdot \eta^{-1} \cap K$; and with the corresponding tilde measures 
both defined on measurable subsets of the measurable (Lemma \ref{lem:decomposition.5'}) set 
$\pi_\mathfrak{X}(K \cdot \eta^{-1} \cap K)$; namely 
\[
\begin{split}
\widetilde{(R_\eta \mu')'} \, (E) = (R_\eta \mu')'(\pi_{\mathfrak{X}}^{-1}(E))
= R_\eta \mu'(K \cap \pi_{\mathfrak{X}}^{-1}(E)) \\ 
=  (R_\eta \mu )'_{K \eta^{-1}}(K \cap \pi_{\mathfrak{X}}^{-1}(E)) 
= R_\eta \mu (K \eta^{-1} \cap K \cap \pi_{\mathfrak{X}}^{-1}(E))
\end{split}
\]
and 
\[
\widetilde{(\mu')'_{K  \eta^{-1}} } \, (E) = \widetilde{(\mu)'_{K  \eta^{-1} \cap K} } \, (E)
= \mu (K \eta^{-1} \cap K \cap \pi_{\mathfrak{X}}^{-1}(E)).  
\] 

Note please that our Lemma \ref{lem:decomposition.6}  holds true for any pseudo-image
measure $\nu$ of $\mu$. By Lemma \ref{lem:decomposition.5'}, any pseudo-image measure
of the restriction $\mu'$ is a restriction $(\nu)'_{\pi_\mathfrak{X}(K)}$ of a pseudo-image measure of $\mu$.
It follows that the Lemma  \ref{lem:decomposition.6} is applicable to the pairs of measures (\ref{pairs}).
Indeed it is sufficient to insert $K \cdot \eta^{-1} \cap K$ instead of $K$ in the Lemma \ref{lem:decomposition.6}
and apply it to $(\mu')'_{K  \eta^{-1}} = (\mu)'_{K  \eta^{-1} \cap K}$ (or to $(R_\eta \mu')'
= (R_\eta \mu )'_{K \eta^{-1} \cap K}$)
instead of $\mu'$, because for an appropriate $\nu$,  $(\nu)'_{\pi_\mathfrak{X}(K \cdot \eta^{-1} \cap K)}$ gives
the pseudo-image measure $\widetilde{(\mu')'_{K  \eta^{-1}} }$ (or respectively $\widetilde{(R_\eta \mu')'}$) of 
$(\mu')'_{K  \eta^{-1}}$ (or respectively of $(R_\eta \mu' )'$). 
We may thus apply Lemma 11.4 of \cite{Mackey}, \S 11, to the pairs of measures (\ref{pairs}). 
Because $\mu$ is quasi invariant, the measures $(\mu')'_{K  \eta^{-1}} = (\mu)'_{K  \eta^{-1} \cap K}$ and 
$(R_\eta \mu')' = (R_\eta \mu )'_{K \eta^{-1} \cap K}$ are equivalent as measures 
on $K \cdot \eta^{-1} \cap K$, and thus by Lemma 11.4 of \cite{Mackey} 
it follows that $\widetilde{(\mu')'_{K  \eta^{-1}}}$ and $\widetilde{(R_\eta \mu')'}$ are equivalent as measures
on $\pi_\mathfrak{X}(K \cdot \eta^{-1} \cap K)$. Introducing the corresponding measurable weight function 
$f_1$ on $\mathfrak{X}/G_2$ which is non zero on $\pi_\mathfrak{X}(K \cdot \eta^{-1} \cap K)$, 
we have
\[
f_1 \cdot \ud \, \widetilde{(\mu')'_{K  \eta^{-1}}} = \ud \, \widetilde{(R_\eta \mu')'}
\]
and 
\begin{equation}\label{pair2}
(R_\eta \mu')' = \int f_1 (C) \Big(  R_\eta \mu')' \Big)_{{}_C} \,\,\,\, 
\ud \, \widetilde{(\mu')'_{K \eta^{-1}}} \,\, (C),
\end{equation}
\begin{equation}\label{pair1}
(\mu')'_{K  \eta^{-1}} 
= \int \,  \Big( (\mu')'_{K \eta^{-1}} \Big)_{{}_C} 
\,\,\,\, \ud \, \widetilde{(\mu')'_{K \eta^{-1}}} \,\, (C).
\end{equation}
Now applying again the Lemma 11.4 of \cite{Mackey} to the pairs of measures:
\[ 
\Big( \,\, (\mu')'_{K  \eta^{-1}} \,\, , \,\,\,
\widetilde{(\mu')'_{K  \eta^{-1}} } \,\, \Big)
 \,\, \textrm{and} \,\, 
\Big( \,\, (R_\eta \mu')' \,\, , \,\,\, \widetilde{(\mu')'_{K  \eta^{-1}} }  \,\, \Big)
\]
with the respective decompositions (\ref{pair1}) and (\ref{pair2}) we prove that the measures 
$\Big( (\mu')'_{K \eta^{-1}} \Big)_{{}_C}$ and $\Big( R_\eta \mu')' \Big)_{{}_C}$ are equivalent and
\[
\begin{split}
f_1 (C) \cdot \frac{\ud \, \Big( R_\eta \mu')' \Big)_{{}_C} } { \ud \, \Big( (\mu')'_{K \eta^{-1}} \Big)_{{}_C}  }(\cdot) 
=  \frac{\ud \, (R_\eta \mu')'}{\ud \, (\mu')'_{K \cdot \eta^{-1}}}(\cdot) \\
= \frac{\ud \, (R_\eta \mu)}{\ud \, \mu}(\cdot) = \lambda( \cdot , \eta),
\end{split}
\]
on $C \cap K \cdot \eta^{-1} \cap K$, where the last two equalities follow from definitions and where
\[
f_1 = \frac{\ud \, \widetilde{(R_\eta \mu')'}}{\ud \, \widetilde{(\mu')'_{K  \eta^{-1}}}}.
\]

On the other hand it follows from Lemma \ref{lem:decomposition.5} and Remark \ref{rem:decomposition.2}
that 
\[
(R_\eta \mu')' = \int  (R_\eta (\mu_{{}_C})')'  \,\,\,\, \ud \, (\nu)'_{\pi_\mathfrak{X}(K \eta^{-1} \cap K)} (C)
\]
and 
\[
(\mu')'_{K \eta^{-1}} = \int  ((\mu_{{}_C})')'_{K \eta^{-1}}  \,\,\,\, 
\ud \, (\nu)'_{\pi_\mathfrak{X}(K \eta^{-1} \cap K)} (C).
\]
Thus both $\widetilde{(R_\eta \mu')'}$ and $(\nu)'_{\pi_\mathfrak{X}(K \eta^{-1} \cap K)}$ being 
pseudo-image measures of the measure $(R_\eta \mu')'$ under $\pi_\mathfrak{X}$ (of course restricted to
$K \eta^{-1} \cap K)$) are equivalent. Introducing the respective measurable, non zero on 
$\pi_\mathfrak{X}(K \eta^{-1} \cap K)$, weight function $f_2$ we have 
\[
f_2 \cdot \ud \, (\nu)'_{\pi_\mathfrak{X}(K \eta^{-1} \cap K)}  = \ud \, \widetilde{(R_\eta \mu')'},
\]
so that 
\[
\ud \, (R_\eta (\mu_{{}_C})')' = f_2 (C) \cdot \ud \,  \Big( R_\eta \mu')' \Big)_{{}_C}.  
\]
Similarly because $\mu$ is quasi invariant, the measures  $(\mu')'_{K  \eta^{-1}}$ and $(R_\eta \mu')'$ are 
equivalent, and thus again by Lemma 11.4 of \cite{Mackey} the measures $\widetilde{(\mu')'_{K \eta^{-1}}}$
and $(\nu)'_{\pi_\mathfrak{X}(K \eta^{-1} \cap K)}$ are likewise equivalent. Introducing the respective non zero
on $\pi_\mathfrak{X}(K \eta^{-1} \cap K)$ and measurable weight function $f_3$ we have 
\[
f_3 \cdot \ud \, (\nu)'_{\pi_\mathfrak{X}(K \eta^{-1} \cap K)}  = \ud \, \widetilde{(\mu')'_{K  \eta^{-1}} },
\]
so that 
\[
\ud \, ((\mu_{{}_C})')'_{K \eta^{-1}} = f_3 (C) \cdot \ud \,  \Big( (\mu')'_{K \eta^{-1}} \Big)_{{}_C}.  
\]
Joining the above equalities we obtain (the last two equalities follows from definition of $\lambda_C$ and 
from definition of Radon-Nikodym derivative, i. e. its local character)
\[
\begin{split}
\lambda( \cdot , \eta) 
= f_1 (C) \cdot \frac{\ud \, \Big( R_\eta \mu')' \Big)_{{}_C} } { \ud \, \Big( (\mu')'_{K \eta^{-1}} \Big)_{{}_C}  }(\cdot) 
= f_1 (C) \cdot \frac{1}{f_2 (C)} \cdot f_3 (C) \cdot \frac{ \ud \, (R_\eta (\mu_C)')' }{ \ud \, ((\mu_C)')'_{K \eta^{-1}} } \\
= \frac{ \ud \, (R_\eta (\mu_C)')' }{ \ud \, ((\mu_C)')'_{K \eta^{-1}} } 
= \frac{\ud \, (R_\eta \mu_C)}{\ud \, \mu_C}(\cdot) = \lambda_C (\cdot , \eta)
\end{split}
\]
on $C \cap K \cap  K \cdot \eta^{-1}$, because by the known property of Radon-Nikodym derivatives (compare e. g. Scholium 4.5 of \cite{Segal_Kunze})
\[
f_1  \cdot \frac{1}{f_2} \cdot f_3  
= \frac{\ud \, \widetilde{(R_\eta \mu')'} }{ \ud \, \widetilde{(\mu')'_{K  \eta^{-1}} } } 
\cdot \frac{ \ud \, (\nu)'_{\pi_\mathfrak{X}(K \eta^{-1} \cap K)} }{ \ud \, \widetilde{(R_\eta \mu')'} } 
\cdot \frac{ \ud \, \widetilde{(\mu')'_{K  \eta^{-1}} } }{ \ud \, (\nu)'_{\pi_\mathfrak{X}(K \eta^{-1} \cap K)} } = 1,
\]
on all orbits $C$ with non void intersection $C \cap K \cap  K \cdot \eta^{-1}$. 
\qed

\vspace*{0.5cm}

We are are now in a position to formulate the main goal of this Section.

\vspace*{0.5cm}

\begin{lem}
Let $\mu$ be any quasi invariant measure on $\mathfrak{X}$ and
let $\nu$ be any pseudo image measure of $\mu$. Then the measures $\mu_C$
in the decomposition 
\[
\mu = \int \, \mu_C (x) \, d\nu(C) 
\]
of Lemma \ref{lem:decomposition.4} are also quasi invariant and for each $\eta \in G_2$ the Radon-Nikodym 
derivative $\lambda_C (\cdot , \eta) = 
\frac{\ud (R_\eta \mu_C)}{\ud \mu_C}(\cdot)$ is equal to the restriction of the Radon-Nikodym 
derivative  $\lambda (\cdot , \eta) = \frac{\ud (R_\eta \mu)}{\ud \mu}(\cdot)$ to the orbit $C$.

\label{lem:decomposition.8}
\end{lem}

\qedsymbol \,
 Indeed, let $x$ be any point in $\mathfrak{X}$ and $\eta$ any element of $G_2$. We show that on a neighbourhood of $x$ the statement of the Theorem holds true. To this end let $\mathcal{O}_m$ be a neighbourhood of $x$ chosen from the basis of topology constructed above. Then $\mathcal{O}_m \cdot \eta$ is a neighbourhood of $x \cdot \eta$. Therefore the compact set
$K = \overline{\mathcal{O}_m} \cup (\overline{\mathcal{O}_m} \cdot \eta )$ has the property that 
$K \cap ( K \cdot \eta^{-1} )$ contains an open neighbourhood of $x$. 
Now it is sufficient to apply Lemma \ref{lem:decomposition.7} with this $K$ in order to show
that the equality of the Theorem holds true on some open neighbourhood of $x$.    
\qedsymbol \,

\begin{rem}
It has been proved in Sect. \ref{subgroup.preliminaries} that for each orbit $C$  there exists a measure $\mu_C$, 
concentrated on $C$, with the associated Radon-Nikodym derivative equal to the restriction to the orbit 
$C$ of the Radon-Nikodym derivative associated with $\mu$. This however would be insufficient because we need to know that
the measures $\mu_C$ conspire together so as to compose a decomposition of the measure $\mu$. This is why we need Lemma 
\ref{lem:decomposition.8}. Although the Lemma was not explicitly formulated in \cite{Mackey}, it easily follows
for the case of finite $\mu$ from the Lemmas of \cite{Mackey}, \S 11.  

\label{rem:decompositions.3}
\end{rem}

\vspace*{0.5cm}

Using Lemma \ref{lem:decomposition.4} and the general properties of the integral and the algebra of measurable functions one can prove a slightly strengthened version of 
Lemma \ref{lem:decomposition.4} which 
may be called a skew version of the Fubini theorem, because it extends the Fubini theorem to the case 
where we have a skew product measure $\mu$ with only one projection, i.e. the quotient map $\pi_\mathfrak{X}$:

\begin{lem}[Skew Version of the Fubini Theorem]
Let $\mu$, $\mu_C$ and $\nu$ be such as in Lemma \ref{lem:decomposition.4}. Let $g$ 
be a positive complex valued and measurable function on $\mathfrak{X}$. Then
 \begin{equation}\label{skew_fubini.1:decompositions}
C \mapsto \int \, g(x) d\mu_C(x)
\end{equation}
is measurable, and if any one of the following two integrals:
\[
\int \, d\nu(C) \, \int \, g(x) \, d\mu_C (x)  \,\,\, \textrm{and} \,\,\,  \int \, g(x) \, d \mu(x),
\]
does exist, then there exists the other and both are equal in this case. 

In particular it follows that if $g$ is integrable on $(\mathfrak{X} , \mathscr{R}_\mathfrak{X}, \mu)$
then  
\begin{equation}\label{skew_fubini.2:decompositions}
\int \, d\nu(C) \, \int \, g(x) \, d\mu_C (x)  = \int \, g(x) \, d \mu(x).
\end{equation}

\label{lem:decomposition.9}
\end{lem}

\qedsymbol \,
 For the proof compare \cite{Bourbaki_i}, Chap. VI, Remark of \S 3.4. 
Here we give only few comments:
The Lemma holds for positive and continuous $g$ with compact support as a consequence of
Lemma \ref{lem:decomposition.4}. Next we note that the class of functions which
satisfy (\ref{skew_fubini.1:decompositions}) and (\ref{skew_fubini.2:decompositions}) is closed 
under sequential convergence of increasing sequences.

The Lemma follows by repeated application of the sequential continuity of the integral for increasing sequences;
compare, please, the proof of Thm. 3.4 and Corollary 3.6.2 of \cite{Segal_Kunze}. 
\qed

\vspace*{0.5cm}

Note that the integral 
\[
\int \, g(x) d\mu_C(x)
\]
in (\ref{skew_fubini.1:decompositions}) and (\ref{skew_fubini.2:decompositions})
may be replaced with
\[
\int \limits_{C} \, g^C (x) d\mu_C(x),
\]
where $g^C$ is the restriction of $g$ to the orbit $C$, because $\mu_C$ is concentrated
on $C$. However just like in the ordinary Fubini theorem the whole difficulty in application
of the skew version of the Fubini Theorem lies in proving the measurability
of $g$ on the ``skew product''$\mathfrak{X} \xrightarrow{\pi_\mathfrak{X}}  \mathfrak{X}/G_2$ measure 
space $(\mathfrak{X}, \mathscr{R}_\mathfrak{X} , \mu)$. Indeed even if  the orbits $C$ were nice closed 
subsets and $g^C$ measurable on $C$ (with respect to the  measure structure induced from the 
surrounding space $\mathfrak{X}$)   
the function $g$ still could be non measurable on $(\mathfrak{X}, \mathscr{R}_\mathfrak{X} , \mu)$; 
for simple examples we refer e. g. to \cite{Segal_Kunze} or to any other book on measure theory.
More restrictive constrains are to be put on the separate $g^C$ as functions on the orbits $C$ 
in order to guarantee the measurability of $g$ on the measure space $\mathfrak{X}$. 
We face the same problem with the ordinary Fubini theorem. If in addition
$g^C \in L^2 (C, \mu_C)$ for each $C$ (or $\nu$-almost all orbits $C$), the required additional requirement
is just the von Neumann direct integral structure put on $C \mapsto g^C$ which
is the necessary and sufficient condition for 
the existence of a function $f \in  L^2 (\mathfrak{X}, \mu)$ such that
$f^C = g^C$ for $\nu$-almost all orbits $C$. Namely, consider the space of functions
$C \mapsto g^C \in L^2 (C, \mu_C)$, which composes 
\begin{equation}\label{dir_int_skew:decompositions}
\int \limits_{\mathfrak{X}/G_2} L^2 (C, \mu_C) \,\, \ud \nu (C),
\end{equation}  
then for every element $C \mapsto g^C$ of direct integral (\ref{dir_int_skew:decompositions}) 
there exists a function $f \in  L^2 (\mathfrak{X}, \mu)$ such that
$f^C = g^C$ for $\nu$-almost all orbits $C$. In short 
\begin{equation}\label{dir_int_skew_L^2:decompositions}
\boxed{\int \limits_{\mathfrak{X}/G_2} L^2 (C, \mu_C) \,\, \ud \nu (C) = L^2(\mathfrak{X} , \mu).}
\end{equation}
We skip proving the equality (\ref{dir_int_skew_L^2:decompositions}) because in the next Section we 
prove a more general version of (\ref{dir_int_skew_L^2:decompositions}) for vector valued functions 
$g \in \mathcal{H}^L$ on $\mathfrak{X} = \mathfrak{G}/G_1$, compare Lemma \ref{lem:subgroup.1} (a).
This strengthened version (\ref{dir_int_skew_L^2:decompositions}) of the skew Fubini theorem 
lies behind harmonic analysis on classical Lie groups and provides also an effective tool for 
tensor product decompositions of induced representations in Krein spaces. In practice 
the classical groups with the harmonic analysis relatively complete on them, have the structure of cosets and double 
cosets (corresponding to the orbits $C$) much more nice in comparison to what we have actually assumed, 
so that a vector valued version of the strengthened version of the ordinary Fubini theorem: 
\begin{equation}\label{dir_int_L^2:decompositions}
\boxed{\int \limits_{X} L^2 (Y, \mu_Y) \,\, \ud \mu_X  = L^2(X \times Y , \mu_X \times \mu_Y)}
\end{equation}
would be sufficient for our applications. Namely the  
``measure product property'' holds also in our practical applications for the double coset
space:
\begin{align*} 
\big( \mathfrak{G}, & \mathscr{R}_\mathfrak{G}, \mu_\mathfrak{G} \big) \\
= & \Big(\, G_1 \times \mathfrak{G}/G_1 \times ( \mathfrak{G}/G_1 )/G_2 \, , \,\, 
\mathscr{R}_{{}_{{}_{G_1 \times \mathfrak{G}/G_1 \times ( \mathfrak{G}/G_1 )/G_2} }} \, , \,\, 
\mu_{{}_{{}_{G_1}}} \times \mu_{{}_{{}_{\mathfrak{G}/G_1}}} \times \mu_{{}_{{}_{(\mathfrak{G}/G_1 )/G_2}}} \, \Big)
\end{align*} 
with the analogous functions (\ref{q'_h'}), measure $\mu = \mu_{{}_{{}_{\mathfrak{G}/G_1}}} $ and the pseudo image measure 
$\nu = \mu_{{}_{{}_{(\mathfrak{G}/G_1 )/G_2}}}$ effectively computable. 

Note that (\ref{dir_int_skew_L^2:decompositions}) and (\ref{dir_int_L^2:decompositions}) 
may be proved for more general measure spaces. In particular our proof of (\ref{dir_int_skew_L^2:decompositions}) may be adopted to general non-separable case, 
provided that the assertion of Lemma \ref{lem:decomposition.9} holds
true for the measures $\mu$ and $\nu$. 
Here the measure spaces are not ``too big'', so that
the associated Hilbert spaces of square summable functions are separable.

\vspace*{0.5cm}

At the end of this Section we transfer the measure structure on $\mathfrak{X}/G_2$ over
to the the set  $G_1 : G_2$ of all double cosets $G_1 x G_2$, using the natural 
bi-unique correspondence $C \mapsto D_{{}_C} = \pi^{-1}(C)$ between the orbits $C$
and double cosets $D$. Next we transfer it again to a measurable section $\mathfrak{B}$
of $\mathfrak{G}$ cutting every double coset at exactly one point and give measurability
criterion for a function on $\mathfrak{B}$ with this measure structure inherited from 
$\mathfrak{X}/G_2$. We shall use it in Sections \ref{subgroup} and \ref{Kronecker_product}.

\begin{defin}
We put $\ud \nu _0 (D) = \ud \nu(C_{{}_D})$ 
for the measure $\nu_0$ transferred over to measurable subsets of the set of all double cosets, where $C_{{}_D}$ 
is the orbit corresponding to the double coset, i. e. $D = \pi^{-1}(C)$. 
Let $B_0$ be a measurable section of $\mathfrak{X}$ with respect to $G_2$, existence of which has been proved
in Lemma \ref{lem:decomposition.1}. Let $B$ be a measurable (even Borel) 
section of $\mathfrak{G}$ with respect to $G_1$ (which exists by Lemma 1.1 of \cite{Mackey}). 
Next we define the set $\mathfrak{B} = \pi^{-1}(B_0) \cap B$. We call $\mathfrak{B}$ the section
of $\mathfrak{G}$ with respect to double cosets.

\label{def:decomposition.1}
\end{defin}

$\mathfrak{B}$ is measurable by Lemma 1.1 of \cite{Mackey} and by Lemmas 
\ref{lem:decomposition.1}, \ref{lem:decomposition.3}  of this Section. 
It has the property that every double coset intersects $\mathfrak{B}$ at exactly one point.
We may transfer the measure space structure $(\mathfrak{X}/G_2, \mathscr{R}_{\mathfrak{X}/G_2},
\nu)$ over to get $(\mathfrak{B}, \mathscr{R}_\mathfrak{B}, \nu_{{}_\mathfrak{B}})$. 

\begin{defin}

For each double coset $D$ 
there exists exactly one element $x_{{}_{D}} \in \mathfrak{B} \cap D$.
We define $\ud \nu_{{}_\mathfrak{B}}(x_{{}_D}) = \ud \nu_0(D)$. 
The same holds for orbits $C$: to each orbit $C$ there exists exactly one element 
$x_c \in \mathfrak{B} \cap \pi^{-1}(C)$. We put respectively
$\ud \nu_{{}_\mathfrak{B}} (x_c) = \ud \nu(C)$. Note that $x_c = x_{{}_D}$ 
iff $C$ and $D$ correspond.

\label{def:decomposition.2}
\end{defin}

\begin{lem}
A set $E$ of orbits $C$ is measurable iff the sum of
the corresponding double cosets, regarded as subsets of $\mathfrak{G}$, is measurable in $\mathfrak{G}$.
Thus in particular a function $g$ on $\mathfrak{B}$ is measurable iff there exists a function $f$ measurable on 
$\mathfrak{G}$ and constant along each double coset, such that the restriction of $f$ to $\mathfrak{B}$ 
is equal to $g$.  

\label{lem:decomposition.10}
\end{lem} 

\qedsymbol \, 
By Lemma 1.2 of \cite{Mackey} a set $F \subset \mathfrak{X} = \mathfrak{G}/G_1$
is measurable iff $A = \pi^{-1}(F)$ is measurable in $\mathfrak{G}$ and by Lemma \ref{lem:decomposition.3}
a subset $E \subset \mathfrak{X}/G_2$ is measurable iff $F = \pi_\mathfrak{X}^{-1}(E)$
is measurable on $\mathfrak{X}$. Thus a set $E$ of orbits $C$ is measurable iff the sum of
the corresponding double cosets, regarded as subsets of $\mathfrak{G}$, is measurable in $\mathfrak{G}$,
(as already claimed at the beginning of this Section). This proves the Lemma.
\qed

In particular if we define $s(x)$ to be the double coset containing
$x$, then we transfer the measure $\nu$ over to the subsets of double cosets correctly
if we define the set $E$ of double orbits to be measurable if and only if $s^{-1}(E)$ 
is measurable on $\mathfrak{G}$. 

Writing $x$ for the variable with values in $\mathfrak{G}$, and writing $[x]$
for $\pi(x)$ varying over $\mathfrak{X} = \mathfrak{G}/G_1$ we have

\begin{lem}
Let $\mu$, $\mu_C$ and $\nu$ be such as in Lemma \ref{lem:decomposition.4}. Let $g$ 
be a positive complex valued and measurable function on $\mathfrak{X}$.
Let $\mu_{D} = \mu_{x_{{}_D}} = \mu_{C_{{}_D}}$ be the measure concentrated on the orbit
$C_{{}_D}$ corresponding to the double coset $D$.Then: 

\begin{equation}\label{skew_fubini.11:decompositions}
D \mapsto \int \, g([x]) \,\, \ud \mu_{D} ([x]) \,\,\, \textrm{and} \,\,\,
\mathfrak{B} \ni x_{{}_D} \mapsto \int \, g([x]) \,\, \ud \mu_{x_{{}_D}} ([x])
\end{equation}
are measurable, and

\begin{enumerate}

\item[1)]
 if any one of the following two integrals:
\[
\int \, d\nu_0 (D) \, \int \, g([x]) \, \ud \mu_{D} ([x])  \,\,\, \textrm{and} \,\,\,  \int \, g([x]) \, \ud \mu([x]),
\]
does exist, then there exists the other and both are equal in this case. 

In particular it follows that if $g$ is integrable on $(\mathfrak{X} , \mathscr{R}_\mathfrak{X}, \mu)$
then  
\begin{equation}\label{skew_fubini.22:decompositions}
\int \, \ud \nu_0 (D) \, \int \, g([x]) \, \ud \mu_{D} ([x])  = \int \, g([x]) \, \ud \mu([x]).
\end{equation}

\item[2)]
Similarly if any one of the following two integrals:
\[
\int \, \ud \nu_{{}_\mathfrak{B}} (x_{{}_D}) \, \int \, g([x]) \, \ud \mu_{x_{{}_D}}  \,\,\, \textrm{and} \,\,\,  
\int \, g([x]) \, \ud \mu([x]),
\]
does exist, then there exists the other and both are equal in this case. 

In particular it follows that if $g$ is integrable on $(\mathfrak{X} , \mathscr{R}_\mathfrak{X}, \mu)$
then  
\begin{equation}\label{skew_fubini.222:decompositions}
\int \, \ud \nu_{{}_\mathfrak{B}} (x_{{}_D}) \, \int \, g([x]) \, \ud \mu_{x_{{}_D}} ([x])  = \int \, g([x]) \, 
\ud \mu([x]).
\end{equation}

\end{enumerate}

\label{lem:decomposition.11}
\end{lem} 

\qedsymbol \,
Because by definition (with $\mathfrak{x} \in \mathfrak{X} = \mathfrak{G}/G_1$ and $x \in \mathfrak{G}$)
\[
\int \limits_{C_{{}_D}} \, g(\mathfrak{x}) \,\, \ud \mu_{C} (\mathfrak{x}) 
=  \int \limits_{D} \, g([x]) \,\, \ud \mu_{D} ([x]),
\]
the Lemma is an immediate consequence of definitions Def \ref{def:decomposition.1} 
and \ref{def:decomposition.2} and Lemma \ref{lem:decomposition.9}. 
\qed

\vspace*{0.5cm}

\section{Subgroup theorem in Krein spaces}\label{subgroup}

Let $G_1$ and $G_2$ be regularly related closed subgroups of $\mathfrak{G}$ (for definition
compare Sect. \ref{decomposition}).
Consider the restriction ${}_{{}_{G_2}}U^L$ to the subgroup $G_2 \subset \mathfrak{G}$ of the representation 
${}^{\mu}U^L$ of $\mathfrak{G}$ in the Krein 
space ${}^{\mu}\mathcal{H}^L$, induced from a representation $L$ of the subgroup $H = G_1$, defined as in 
Sect \ref{def_ind_krein}.  
For each $G_2$-orbit $C$ in $\mathfrak{X} = \mathfrak{G}/G_1$ let us introduce the Krein-isometric 
representation $U^{L, C}$, 
defined in Sect. \ref{subgroup.preliminaries},
and acting in the Krein space $({\mathcal{H}^{L}_{C}}' , \mathfrak{J}^{L,C})$. 
Let $\nu$ be any pseudo image measure of 
$\mu$ on $\mathfrak{X}/G_2$, for its definition compare \cite{Bourbaki_i}, Chap. VI.3.2. For simplicity 
we drop the $\mu$ superscript in  ${}^{\mu}U^L$
and ${}^{\mu}\mathcal{H}^L$ and just write $U^L$ and $\mathcal{H}^L$. 

Let us remind the definition of the direct integral of Hilbert spaces after \cite{Segal_dec_I}, 
but compare also \cite{von_neumann_dec}:
 
\begin{defin}[Direct integral of Hilbert spaces]
Let $(\mathfrak{X}/G_2, \mathscr{R}_{\mathfrak{X}/G_2}, \nu)$ be a measure space $M$, and
suppose that for each point $C$ of $\mathfrak{X}/G_2$ there is a Hilbert space ${\mathcal{H}^{L}_{C}}'$.
A Hilbert space $\mathcal{H}^L$ is called a \emph{direct integral} of the ${\mathcal{H}^{L}_{C}}'$
over $M$, symbolically
\begin{equation}\label{dec_H^L}
\mathcal{H}^L = \int {\mathcal{H}^{L}_{C}}' \,\, \ud \nu (C), 
\end{equation}
if for each $g \in \mathcal{H}^L$ there is a function $C \mapsto g^C$ on $\mathfrak{X}/G_2$ to the disjoint union 
$\coprod \limits_{C \in \mathfrak{X}/G_2} {\mathcal{H}^{L}_{C}}'$, such that 
$g^C \in {\mathcal{H}^{L}_{C}}'$ for all $C$, and with the following properties 1) and 2):
\begin{enumerate}

\item[1)]

If $g$ and $k$ are in $\mathcal{H}^L$ and if $u = \alpha g + \beta k$, and if $\big( \cdot , \cdot \big)_C$
is the inner product in ${\mathcal{H}^{L}_{C}}'$ then 
$C \mapsto \Big( g^C , k^C \Big)_C$ is integrable on $M$, and the inner product $(g,k)$
on $\mathcal{H}^L$ is equal to 
\[
( g, k) = \int \limits_{\mathfrak{X}/G_2} \big( g^C , k^C \big)_C \,\, \ud \nu (C),  
\]
and $u^C = \alpha g^C + \beta k^C$ for almost all $C \in \mathfrak{X}/G_2$, and all $\alpha, \beta \in \mathbb{C}$.

\item[2)]

If $C \mapsto u^C$ is a function with $u^C \in {\mathcal{H}^{L}_{C}}'$ for all $C$, if $C \mapsto \big( g^C , u^C \big)_C$
is measurable for all $g \in \mathcal{H}^L$, and if $C \mapsto \big( u^C , u^C \big)_C$ is integrable
on $M$, then there exists an element $u'$ of $\mathcal{H}^L$ such that
\[
u'^C = u^C \,\,\,\textrm{almost everywhere on} \,\, M.
\]
\end{enumerate} 
The function $C \mapsto g^C$ is called the decomposition of $g$ and is symbolized by 
\[
g = \int \limits_{\mathfrak{X}/G_2} g^C \,\, \ud \nu (C).
\]

A linear operator $U$ on $\mathcal{H}^L$ is said to be decomposable with respect to the direct integral
Hilbert space decomposition (\ref{dec_H^L}) if there is a function $C \mapsto U^C$ on $\mathfrak{X}/G_2$
with $U^C$ being a linear operator in ${\mathcal{H}^{L}_{C}}'$ for each $C$, and 
\begin{enumerate}

\item[3)]
the property that for each $g$ in its domain and all $k$ in $\mathcal{H}^L$,
$(Ug)^C = U^C g^C$ almost everywhere on $M$ and the function 
$C \mapsto \big( U^C g^C , k^C \big)_C$ is integrable on $M$.

\end{enumerate}

If $U$ is densely defined the property 3) is equivalent to the following: 

\begin{enumerate}

\item[3')]

for all $g, k$ in $\mathcal{H}^L$ in the domain of $U$, $C \mapsto \big( U^C g^C , k^C \big)_C$
is integrable on $M$ and 
\[
\int \limits_{\mathfrak{X}/G_2} \big( U^C g^C , k^C \big)_C \,\, \ud \nu (C) = (Ug,k).
\]
\end{enumerate}

The function $C \mapsto U^C$ is then called the decomposition of $U$ with respect to 
(\ref{dec_H^L}) and symbolized by
\[
U = \int \limits_{\mathfrak{X}/G_2} U^C \,\, \ud \nu (C).
\]
If $C \mapsto U^C$ is almost everywhere a scalar operator, $U$ is called diagonalizable
with respect to (\ref{dec_H^L}). The totality of all bounded operators diagonalizable with respect
to (\ref{dec_H^L}) composes the commutative von Neumann algebra $\mathfrak{A}_{\mathfrak{G}/G_2}$ associated with the decomposition (\ref{dec_H^L}), compare \cite{von_neumann_dec}. A bounded operator $U$
in $\mathcal{H}^L$ is decomposable with respect to (\ref{dec_H^L}) if and only if 
it commutes with all elements of $\mathfrak{A}_{\mathfrak{G}/G_2}$ $\Leftrightarrow$
$U \in \big( \mathfrak{A}_{\mathfrak{G}/G_2} \big)'$. This condition may easily be extended on unbounded operators:
e. g. closable $U$ is decomposable with respect to (\ref{dec_H^L}) if the spectral projectors of  both the
factors in its polar decomposition commute with all elements of $\mathfrak{A}_{\mathfrak{G}/G_2}$;
or still more generally: $U$ is decomposable with respect to (\ref{dec_H^L}) $\Leftrightarrow$  $U$ is affiliated with 
the commutor $\big( \mathfrak{A}_{\mathfrak{G}/G_2} \big)'$ of $\mathfrak{A}_{\mathfrak{G}/G_2}$, 
i.e. iff it commutes with every unitary operator in the commutor
$\big( \mathfrak{A}_{\mathfrak{G}/G_2} \big)'' = \mathfrak{A}_{\mathfrak{G}/G_2}$ 
of $\big( \mathfrak{A}_{\mathfrak{G}/G_2} \big)'$.

\label{direct_int:subgroup}
\end{defin}

Note that the map $T$ which transforms $g$ into its decomposition $C \mapsto g^C$ may be viewed as a unitary
operator decomposing $U$:  
\[
T U T^{-1} = \int \limits_{\mathfrak{X}/G_2} U^C \,\, \ud \nu (C). 
\]

There are many possible realizations $T: f \mapsto T(f)$ of the Hilbert space 
$\mathcal{H}^L$ as the direct integral (\ref{dec_H^L}) all corresponding to the same commutative 
decomposition algebra $\mathfrak{A}_{\mathfrak{G}/G_2}$. However the difference between 
any two $T: f \mapsto T(f) = \Big( C \mapsto f^C \Big) $ and $T': f \mapsto T'(f) = \Big( C \mapsto \big(f^C \big)' \Big)$ 
of them is irrelevant: there exists for them a map $C \mapsto U^C$ with each $U^C$ unitary in ${\mathcal{H}^{L}_{C}}'$
and such that:
\begin{enumerate}

\item[1)]

$U^C f^C = \big(f^C \big)'$ for almost all $C$.  

\item[2)]
$C \mapsto \big(f^C , g^C \big)_C$ is measurable in realization $T$ $\Leftrightarrow$ 
$C \mapsto \Big( \, U^C f^C \, , \,\, U^C f^C \, \Big)_C$ is measurable in realization $T'$.

\end{enumerate}
(Compare \cite{von_neumann_dec}).

For the reasons explained in the footnote to Lemma \ref{lem:dense.6} it is sufficient to consider
the $\sigma$-rings $\mathscr{R}_{\mathfrak{X}/G_2}$ and $\mathscr{R}_{\mathfrak{X}}$ of Borel sets, with the 
Borel structure on $\mathfrak{X}/G_2$ defined as in Sect. \ref{decomposition}, in the investigation of the respective
Hilbert and Krein spaces.
    
We shall need a 

\begin{lem}  

\begin{enumerate}

\item[(a)] 

\[
\mathcal{H}^L \cong \int \limits_{\mathfrak{X}/G_2} {\mathcal{H}^{L}_{C}}' \,\, \ud \nu (C).
\]

\item[(b)]

\[
{}_{{}_{G_2}}U^L \cong \int \limits_{\mathfrak{X}/G_2} U^{L, C} \,\, \ud \nu (C).
\]

\item[(c)]

\[
\mathfrak{J}^L \cong \int \limits_{\mathfrak{X}/G_2} \mathfrak{J}^{L,C} \,\, \ud \nu (C).
\]

The equivalences $\cong$ are all under the same map (or realization) 
$T: \mathcal{H}^L \mapsto \int \limits_{\mathfrak{X}/G_2} {\mathcal{H}^{L}_{C}}' \,\, \ud \nu (C)$
giving the corresponding decomposition $T(f): C \mapsto f^C$
for each $f \in \mathcal{H}^L$, in which $f^C$ is the restriction of $f$ to the double 
coset $D_{{}_C} = G_1 x_c G_2 = \pi^{-1} (C)$ corresponding to $C$; \emph{i. e.} we chose $x_c \in \mathfrak{B} \subset \mathfrak{G}$ for which $\pi(x_c) \in C$, 
compare Def. \ref{def:decomposition.1} and \ref{def:decomposition.2}.

In particular $T$ is unitary and Krein-unitary map between the Krein spaces 
\[
(\mathcal{H}^L , \mathfrak{J}^L) \,\,\, \textrm{and} \,\,\, 
\Big( \int \limits_{\mathfrak{X}/G_2} {\mathcal{H}^{L}_{C}}' \,\,
 \ud \nu (C) , \int \limits_{\mathfrak{X}/G_2} \mathfrak{J}^{L,C} \,\, \ud \nu (C) \Big).
\]
\end{enumerate}
\label{lem:subgroup.1}
\end{lem}

\begin{rem}
The equivalences $\cong$ may be read in fact as ordinary equalities.
\label{rem:subgroup.1}
\end{rem}

\qedsymbol \,
Let 
\[
( \cdot , \cdot)_C = {\| \cdot \|_C}^2
= \int \limits_{C}
\big( \mathfrak{J}_L (\mathfrak{J}^{L,C} \, \cdot \,)_x , (\, \cdot \,)_x \big) \,\, \ud \mu_C (x)
\] 
be defined on ${\mathcal{H}^{L}_{C}}'$ as in Sect. \ref{subgroup.preliminaries}. 
Recall that for any element $g$ of $\int \limits_{\mathfrak{X}/G_2} {\mathcal{H}^{L}_{C}}' \,\, \ud \nu (C)$ i. e. a function $C \mapsto g^C$ from the set of $G_2$-orbits 
$\mathfrak{X}/G_2$ to the disjoint union 
$\coprod \limits_{C \in \mathfrak{X}/G_2} {\mathcal{H}^{L}_{C}}'$ such that 
$g^C \in {\mathcal{H}^{L}_{C}}'$ for all $C$, the function $C \mapsto {\| g^C \|_C}^2 =
(g^C , g^C)_C$ is $\nu$-summable and $\nu$-measurable and defines inner product 
for any $g, k \in \int \limits_{\mathfrak{X}/G_2} {\mathcal{H}^{L}_{C}}' \,\, \ud \nu (C)$ by the formula 
\begin{equation}\label{inn_dir_int_1:subgroup}
( g, k) = \int \limits_{\mathfrak{X}/G_2} \,\, \ud \nu (C) \, \int \limits_{C}
\big( \mathfrak{J}_L (\mathfrak{J}^{L,C} g^C)_x , k^{C}_{x} \big) \,\, \ud \mu_C (x)
= \int \limits_{\mathfrak{X}/G_2} ( g^C , k^C)_C \,\, \ud \nu (C).  
\end{equation}

We shall exhibit a natural unitary map $T$ from $\mathcal{H}^L$ onto 
$\int \limits_{\mathfrak{X}/G_2} {\mathcal{H}^{L}_{C}}' \,\, \ud \nu (C)$
or, what is equivalent, we shall show that the decomposition $T(f) = \big( C \mapsto f^C \big)$  
corresponding to each $f \in \mathcal{H}^L$, with $f^C$ equal to the restriction of $f$
to the double coset $D_{{}_C} = G_1 x_c G_2 = \pi^{-1}(C)$ corresponding to $C$, has all 
the properties required in Definition \ref{direct_int:subgroup}. 

Let $f$ and $k$ be any functions in $\mathcal{H}^L$. Then by Lemma \ref{lem:decomposition.4} we have 
\[
\int \limits_{\mathfrak{X}/G_2} \,\, \ud \nu (C) \, \int \limits_{C}
\big( \mathfrak{J}_L (\mathfrak{J}^L f)_x , k_x \big) \,\, \ud \mu_C (x)
= \int \limits_{\mathfrak{X}} \big( \mathfrak{J}_L (\mathfrak{J}^L f)_x , k_x \big) \,\, \ud \mu(x)
= \| f \|^2 < \infty,
\] 
with the set of all $G_2$ orbits $C$ for which $x \mapsto \big( \mathfrak{J}_L (\mathfrak{J}^L f)_x , k_x \big)$ 
is not $\mu_C$-integrable being $\nu$-negligible and the function 
\[
C \mapsto \int \limits_{C}
\big( \mathfrak{J}_L (\mathfrak{J}^L f)_x , k_x \big) \,\, \ud \mu_C (x)
\]
being $\nu$-summable and $\nu$-measurable. Moreover, because for each orbit  $C$ the measure $\mu_C$
is concentrated on $C$ (Lemma \ref{lem:decomposition.4}), the integral 
\[
\int \limits_{C}
\big( \mathfrak{J}_L (\mathfrak{J}^L f)_x , k_x \big) \,\, \ud \mu_C (x)
\] 
is equal 
\[
\int \limits_{C}
\big( \mathfrak{J}_L ((\mathfrak{J}^L f)^C)_x , (k^C)_x \big) \,\, \ud \mu_C (x)
= \int \limits_{C}
\big( \mathfrak{J}_L (\mathfrak{J}^L f^C)_x , (k^C)_x \big) \,\, \ud \mu_C (x)
\]
where $f^C$ (and similarly for $k^C$) is the restriction of $f$ to the double coset 
$D_{{}_C} = G_1 x G_2 = \pi^{-1}(C)$ corresponding to $C$. i.e. with 
any $x$ for which
$\pi(x) \in C$, say $x = x_c$, with $C \mapsto x_c \in \mathfrak{B}$ of Sect. \ref{decomposition}.
(We have chosen $x = x_c$ to belong to the measurable section $\mathfrak{B}$ of double cosets 
in $\mathfrak{G}$ constructed in Sect. \ref{decomposition}, but this is unnecessary here.) 
Because $f^C \in {\mathcal{H}^{L}_{C}}'$ and likewise $\mathfrak{J}^{L,C}$ are defined 
as the ordinary restrictions, $(\mathfrak{J}^L f)^C = \mathfrak{J}^L f^C = \mathfrak{J}^{L,C} f$ is the restriction
of $\mathfrak{J}^L f$ to the double coset 
$D_{{}_C} = G_1 x_c G_2$ corresponding to $C$. We thus obtain
\[
\int \limits_{C}
\big( \mathfrak{J}_L (\mathfrak{J}^L f)_x , k_x \big) \,\, \ud \mu_C (x) 
= \int \limits_{C}
\big( \mathfrak{J}_L (\mathfrak{J}^{L,C} f^C)_x , (k^C)_x \big) \,\, \ud \mu_C (x).
\]
Therefore it follows that the map $T: f \mapsto \big( C \mapsto f^C \big)$, where $f^C$
is the restriction of $f$ to the double coset corresponding to the orbit $C$,
fulfils the requirements of Part 1) of Definition \ref{direct_int:subgroup};
in particular $\| T(f) \| = \| f \|$ and the range $T(\mathcal{H}^L)$ is a Hilbert space
with the inner product (\ref{inn_dir_int_1:subgroup}).

We shall verify Part 2) of the Definition
\ref{direct_int:subgroup}: i. e. that the decomposition map $T(f) = \big( C \mapsto f^C \big)$
defined as above has the properties indicated in 2) of Definition
\ref{direct_int:subgroup} on its whole range 
$T(\mathcal{H}^L)$.  Toward this end let $C \mapsto u^C$ fulfil the conditions required
in 2) of Def. \ref{direct_int:subgroup}: 
\begin{equation}\label{condition1}
C \mapsto \int \limits_{C}
\Big( \mathfrak{J}_L (\mathfrak{J}^{L,C} u^C)_x , (k^{C})_{x} \Big) \,\, \ud \mu_C (x)
= \big( u^C , k^C \big)_C 
\end{equation}
is measurable for each $k \in \mathcal{H}^L$ and 
\begin{equation}\label{condition2}
C \mapsto \int \limits_{C}
\Big( \mathfrak{J}_L (\mathfrak{J}^{L,C} u^C)_x , (u^{C})_{x} \Big) \,\, \ud \mu_C (x)
=  \big( u^C , u^C \big)_C 
\end{equation}
is measurable and integrable. Consider the space $\mathfrak{F}$ of all functions $C \mapsto k^C \in {\mathcal{H}^{L}_{C}}'$ fulfilling the following conditions:
\[
C \mapsto \int \limits_{C}
\Big( \mathfrak{J}_L (\mathfrak{J}^{L,C} g^C)_x , g^{C}_{x} \Big) \,\, \ud \mu_C (x)
=  \big( k^C , k^C \big)_C 
\]
is measurable and integrable. Let $X$ be the maximal \emph{linear}  
subspace of $\mathfrak{F}$, where a subspace of $\mathfrak{F}$ we have called linear, 
whenever it is closed under formation of finite linear combinations over $\mathbb{C}$.
$X$ is not empty as it contains the subspace $T(\mathcal{H}^L)$ itself, which is a Hilbert space. Moreover
if $C \mapsto k^C, C \mapsto r^C$ are any two functions belonging to $X$ the formula
\[
\begin{split}
h\Big(\, C \mapsto k^C \, , \,\, C \mapsto r^C \, \Big) 
=  \int \limits_{\mathfrak{X}/G_2}  ( k^C , r^C)_C  \,\, \ud \nu (C)   \\
= \int \limits_{\mathfrak{X}/G_2} \Big( \int \limits_{C}
\big( \mathfrak{J}_L (\mathfrak{J}^{L,C} k^C)_x , ( r^C )_x \big) \,\, \ud \mu_C (x) \Big) \, \ud \nu(C)
\end{split}
\] 
defines a hermitian form on $X$. Thus by the Cauchy-Schwarz inequality we have:
\begin{equation}\label{Cauchy-Schwartz:subgroup}
\Big{|} \int \limits_{\mathfrak{X}/G_2}  ( k^C , r^C)_C  \,\, \ud \nu (C) \Big{|}^2
\leq \Big( \int \limits_{\mathfrak{X}/G_2}  ( k^C , k^C)_C  \,\, \ud \nu (C)  \Big)
\cdot \Big( \int \limits_{\mathfrak{X}/G_2}  ( r^C , r^C)_C  \,\, \ud \nu (C)  \Big).
\end{equation}
Now by the first part of the proof, $T(\mathcal{H}^L)$ is a Hilbert space with the inner product
(\ref{inn_dir_int_1:subgroup}) and in particular a linear subspace of $\mathfrak{F}$.
We may thus insert for $C \mapsto k^C$ in (\ref{Cauchy-Schwartz:subgroup}) any
decomposition $C \mapsto f^C$ of $f \in \mathcal{H}^{L}$, with $f^C$ equal to the restriction 
of $f$ to the double coset $D_{{}_C} = \pi^{-1} (C)$ corresponding to 
$C$. Similarly we may insert the function $C \mapsto u^C$ for the function $C \mapsto r^C$ in (\ref{Cauchy-Schwartz:subgroup}). Indeed, because of the conditions (\ref{condition1}) and (\ref{condition2}), fulfilled by 
the function $C \mapsto u^C$, the function
\[
C \mapsto \big( f^C + u^C , f^C + u^C \big)_C  =
\big( f^C , f^C \big)_C 
+ \big( f^C ,  u^C \big)_C 
+ \big( u^C , f^C  \big)_C 
+ \big(  u^C , u^C \big)_C 
\]
is measurable and by the Cauchy-Schwarz inequality integrable, for all $f \in \mathcal{H}^L$.
Therefore $C \mapsto u^C$ and $T(\mathcal{H}^L)$ are both contained in one linear subspace of 
$\mathfrak{F}$, and thus by the maximality of $X$ they are contained in $X$, so that we can insert
$C \mapsto u^C$ for $C \mapsto r^C$ in (\ref{Cauchy-Schwartz:subgroup}).
Thus the indicated insertions in the inequality (\ref{Cauchy-Schwartz:subgroup}) lead us to the inequality
\[
\Big{|} \int \limits_{\mathfrak{X}/G_2} ( f^C , u^C)_C  \,\, \ud \nu (C) \Big{|}^2
\leq \Big( \int \limits_{\mathfrak{X}/G_2}  ( f^C , f^C)_C  \,\, \ud \nu (C)  \Big)
\cdot \Big( \int \limits_{\mathfrak{X}/G_2}  ( u^C , u^C)_C  \,\, \ud \nu (C)  \Big)
\]
for all $C \mapsto f^C$ in $T(\mathcal{H}^L)$. 
Therefore the linear functional 
\[
T(f) \mapsto L \big( \, T(f)  \, \big) = L \Big( \,  C \mapsto f^C  \, \Big)
= h\Big(\, C \mapsto f^C  \, , \,\,  C \mapsto u^C  \, \Big), 
\]
on $T(\mathcal{H}^L)$ is bounded by the last inequality. Because the range $T(\mathcal{H}^L)$ of $T$
is a Hilbert space it follows by the Riesz theorem ((e. g. Corollary 8.3.2. of \cite{Segal_Kunze}) applied 
to the linear functional $L$ that there exists exactly one element $T(f')$ in the range of 
$T$ such that 
\[
\begin{split}
(f,f') = \big( \, T(f) \, , \,\, T(f') \, \big) \,\,\,\,\,\,\,\,\,\,\,\,\,\,\,\,\,\,\,\,\,\,\,\,\,\,\,
\,\,\,\,\,\,\,\,\,\,\,\,\,\,\,\,\,\,\,\,\,\,\,\,\,\,\,\,\,\,\,\,\,\,\,\,\,\,\,\,\,\,\,\,\,\,\,\,\,
\,\,\,\,\,\,\,\,\,\,\,\,\,\,\,\,\,\,\,\,\,\,\,\,\,\,\,\,\,\,\,\,\,\,\,\,  \\ 
= \int \limits_{\mathfrak{X}/G_2} ( f^C , f'^C)_C  \,\, \ud \nu (C) 
= h\Big(\, C \mapsto f^C \, , \,\, C \mapsto u^C \, \Big)
\end{split}
\] 
for all $f \in \mathcal{H}^L$. Therefore
\[
\int \limits_{\mathfrak{X}/G_2}  \big( f^C , f'^C \big)_C  \,\, \ud \nu (C) 
= \int \limits_{\mathfrak{X}/G_2}  \big( f^C , u^C \big)_C  \,\, \ud \nu (C) 
\] 
for all $f \in \mathcal{H}^L$ and for a fixed $f' \in \mathcal{H}^L$, or equivalently 
\[
\int \limits_{\mathfrak{X}/G_2}  \big( f^C , f'^C - u^C \big)_C  \,\, \ud \nu (C) = 0,
\] 
for all $f \in \mathcal{H}^L$. Inserting the definition of $\big( f^C , f'^C - u^C \big)_C $
we get:
\begin{equation}\label{f_j_dense:subgroup}
\begin{split}
\int \limits_{\mathfrak{X}/G_2} \,\, \int \limits_{C}
\Big( \mathfrak{J}_L (\mathfrak{J}^{L,C} f^C)_x , (f'^C - u^C)_{x} \Big) \,\, \ud \mu_C (x)  \,\, \ud \nu (C) 
\,\,\,\,\,\,\,\,\,\,\,\,\,\,\,\,\,\,\,\,\,\,\,\,\,\,\,\,\,\,\,\,\,\,\,\,\,\,\,\,\,\,\,\,\,\,\,\,
\,\,\,\,\,\,\,\,\,\,\,\,\,\,  \\
=\int \limits_{\mathfrak{X}/G_2} \,\, \int \limits_{C}
\Big( \mathfrak{J}_L (\mathfrak{J}^L f)_x , (f'^C - u^C)_{x} \Big) \,\, \ud \mu_C (x)  \,\, \ud \nu (C) = 0,
\end{split}
\end{equation}
for all $f \in \mathcal{H}^L$. By Lemma \ref{lem:dense.6} there exists a sequence $f^1 , f^2 , \ldots$ of elements $C^{L}_{0} \subset \mathcal{H}^L$ such that for each fixed $x \in \mathfrak{G}$ the vectors $f^{k}_{x}$, $k = 1, 2, \ldots$ form a dense linear subspace of $\mathcal{H}_L$. By the proof of the same Lemma \ref{lem:dense.6} there exists 
a sequence $g_1 , g_2 , \ldots$ of continuous complex valued functions on $\mathfrak{X} = \mathfrak{G}/G_1$ with compact 
supports, dense in $L^2 (\mathfrak{X},\mu)$ with respect to the $L^2$ norm 
$\| \cdot \|_{L^2}$. For each $g_j$ define the
corresponding function $g'_j$ on $\mathfrak{G}$ by the formula $g'_j (x) = g_j (\pi (x))$, where
$\pi$ is the canonical quotient map $\mathfrak{G} \mapsto \mathfrak{G}/G_1 = \mathfrak{X}$. Note, please, that
$\Big( \mathfrak{J}_L (\mathfrak{J}^L g'_j \cdot f)_{{}_x} , (f'^C - u^C)_{{}_x} \Big)
= (g'_j)_{{}_x} \cdot \Big( \mathfrak{J}_L (\mathfrak{J}^L \cdot f)_{{}_x} , (f'^C - u^C)_{{}_x} \Big)$ for all 
$j \in \mathbb{N}$ and all $f \in \mathcal{H}^L$. Inserting now $g'_j \cdot f^i$ for $f$ in 
(\ref{f_j_dense:subgroup}) we get
\[
\int \limits_{\mathfrak{X}/G_2} \,\, g_j (C) \cdot \int \limits_{C} 
\Big( \mathfrak{J}_L (\mathfrak{J}^L f^i)_{{}_x} , (f'^C - u^C)_{{}_x} \Big) \,\, \ud \mu_C (x)  \,\, \ud \nu (C) = 0,
\]  
for all $i,j \in \mathbb{N}$. Because $\{g_j\}_{j \in \mathbb{N}}$ is dense in $L^2 (\mathfrak{X},\mu)$
and the function
\[
C \mapsto \int \limits_{C} 
\Big( \mathfrak{J}_L (\mathfrak{J}^L f^i)_{{}_x} , (f'^C - u^C)_{{}_x} \Big) \,\, \ud \mu_C (x)
\]
by construction belongs to $L^2 (\mathfrak{X},\mu)$, it follows that outside a $\nu$-negligible
subset $N$ of orbits $C$
\[
\int \limits_{C} 
\Big( \mathfrak{J}_L (\mathfrak{J}^L f^i)_{{}_x} , (f'^C - u^C)_{{}_x} \Big) \,\, \ud \mu_C (x)  = 0,
\]  
for all $i \in \mathbb{N}$. Thus if $C \notin N$, then
\begin{equation}\label{*f_j_dense:subgroup}
\int \limits_{C} 
\Big( \mathfrak{J}_L (\mathfrak{J}^L f^i)_{{}_x} , (f'^C - u^C)_{{}_x} \Big) \,\, \ud \mu_C (x)  = 0,
\end{equation} 
for all $i \in \mathbb{N}$. Applying Lemma \ref{lem:subgroup.preliminaries.1} to this orbit $C$ 
and the associated ${\mathcal{H}^{L}_{C}}'$ we get an isomorphism of it with a Krein space 
$\mathcal{H}^{L^{x_c}}$  of an induced representation (recall that $x_c \in \mathfrak{B} \subset \mathfrak{G}$
with $\pi(x_c) \in C$, compare Def. \ref{def:decomposition.2}). 
Then (\ref{*f_j_dense:subgroup}) together with Lemma \ref{lem:subgroup.preliminaries.3} and 
Lemma \ref{lem:dense.3} or \ref{lem:dense.4} applied to $\mathcal{H}^{L^{x_c}}$
gives $f'^C - u^C = 0$. This shows that the decomposition $T: f \mapsto \big( C \mapsto f^C \big)$
fulfils Part 2) of Definition \ref{direct_int:subgroup}. We have thus proved 
Part (a) of the Lemma.

Then we have to prove that the operators $T \, {}_{{}_{G_2}}U^L \, T^{-1}$ and $T \, \mathfrak{J}^L \, T^{-1}$
are decomposable with respect to (\ref{dec_H^L}) and $C \mapsto U^{L, C}$ and
$C \mapsto \mathfrak{J}^{L,C}$ are their respective decompositions. Let  $\eta \in G_2$.
Writing $\lambda(\eta)$ for the $\lambda$-function $[x] \mapsto \lambda([x], \eta)$ 
corresponding to the measure $\mu$ and analogously writing $\lambda_C (\eta)$  for the $\lambda_C$ function 
$[x] \mapsto \lambda_C ([x], \eta)$ corresponding to $\mu_C$ we have: 
\[
\begin{split}
\Big( T \, {}_{{}_{G_2}}U^{L}_{\eta} \, T^{-1} \Big) \Big( C \mapsto f^C \Big)
 = \big( T \, {}_{{}_{G_2}}U^{L}_{\eta} \big) \big(f\big)  \\
 =  T \big( \sqrt{\lambda(\eta)}  R_\eta f \big) = \Big( \, C \mapsto \sqrt{\lambda(\eta)|_{{}_C}}  R_\eta f^C \, \Big),
\end{split}
\]
where $\lambda(\eta)|_{{}_C}$ denotes the restriction of $\lambda(\eta)$ to the orbit $C$. By 
Lemma \ref{lem:decomposition.8} the restriction $\lambda(\eta)|_{{}_C}$ of $\lambda(\eta)$ to the orbit $C$
is equal to $\lambda_C (\eta)$, so that
\[
\Big( T \, {}_{{}_{G_2}}U^{L}_{\eta} \, T^{-1} \Big) \Big( C \mapsto f^C \Big)
= \Big( C \mapsto \sqrt{\lambda_C (\eta)}  R_\eta f^C \Big)
= \Big( C \mapsto U^{L, C}_{\eta} f^C \Big),
\]
which means that
\[
{}_{{}_{G_2}}U^L \cong \int \limits_{\mathfrak{X}/G_2} U^{L, C} \,\, \ud \nu (C),
\]
and proves (b). Similarly for the operator $\mathfrak{J}^L$:
\[
\begin{split}
\Big( T \, \mathfrak{J}^L \, T^{-1} \Big) \Big( C \mapsto f^C \Big)
 = \big(T \, \mathfrak{J}^L \big) \big(f\big) \\
 =  T \big( \mathfrak{J}^L f \big) = \Big( \, C \mapsto \big(\mathfrak{J}^L f\big)^C \, \Big).
\end{split}
\]
By definition of the operator $\mathfrak{J}^{L,C}$ we have $\big(\mathfrak{J}^L f\big)^C
= \mathfrak{J}^L f^C = \mathfrak{J}^{L,C} f^C$. Therefore
\[
\Big( T \, \mathfrak{J}^L \, T^{-1} \Big) \Big( C \mapsto f^C \Big)
= \Big( \, C \mapsto \mathfrak{J}^{L,C} f^C \, \Big),
\]
which means that
\[
\mathfrak{J}^L \cong \int \limits_{\mathfrak{X}/G_2} \mathfrak{J}^{L,C} \,\, \ud \nu (C),
\]
and proves (c).

Because for each $C$, $\mathfrak{J}^{L,C}$ is unitary and self adjoint in ${\mathcal{H}^{L}_{C}}'$
and $\big( \mathfrak{J}^{L,C} \big)^2 = I$,
then by \cite{von_neumann_dec}, \S 14, the same holds true for the operator
\[
\int \limits_{\mathfrak{X}/G_2} \mathfrak{J}^{L,C} \,\, \ud \nu (C) \,\,\, \textrm{in} \,\,\, 
\int \limits_{\mathfrak{X}/G_2} {\mathcal{H}^{L}_{C}}' \,\, \ud \nu (C), 
\]
so that 
\[
\Big( \int \limits_{\mathfrak{X}/G_2} {\mathcal{H}^{L}_{C}}' \,\,
 \ud \nu (C) , \int \limits_{\mathfrak{X}/G_2} \mathfrak{J}^{L,C} \,\, \ud \nu (C) \Big),
\]
is a Krein space.

Finally we have to show that $T$ is Krein unitary. To this end observe that for each $f,g \in \mathcal{H}^L$
\[
\begin{split}
\Big( T(f), T(g) \Big)_{\int \mathfrak{J}^{L,C} \, \ud \nu (C)}
= \int \limits_{\mathfrak{X}/G_1}  \Big( \mathfrak{J}^{L,C} f^C , g^C \Big)_C \,\, \ud \nu (C) \\
= \int \limits_{\mathfrak{X}/G_1} \int \limits_{C} \Big( \mathfrak{J}_L \big( (\mathfrak{J}^{L,C})^2 f^C\big)_{{}_x} , 
\big( g^C \big)_{{}_x} \Big)_C \,\, \ud \mu_C (x) \,\, \ud \nu (C) \\ 
= \int \limits_{\mathfrak{X}/G_1} \int \limits_{C} \Big( \mathfrak{J}_L \big( f^C\big)_{{}_x} , 
\big( g^C \big)_{{}_x} \Big) \,\, \ud \mu_C (x) \,\, \ud \nu (C).
\end{split}
\] 
Because $f^C$ and $g^C$ are the ordinary restrictions of $f$ and $g$ to $G_1 x_c G_2$ and the measure
$\mu_C$ is concentrated on $C$ (Lemma \ref{lem:decomposition.4}), the integrand in the last integral may be 
replaced with $\Big( \mathfrak{J}_L \big( f \big)_{{}_x} , 
\big( g \big)_{{}_x} \Big)$. Because $f,g \in \mathcal{H}^L$, the function 
$x \mapsto \Big( \mathfrak{J}_L \big( f \big)_{{}_x} \big( g \big)_{{}_x} \Big)$ is constant 
on the right $G_1$-cosets and measurable and integrable on $\mathfrak{X} = \mathfrak{G}/G_1$
as a function of right $G_1$-cosets. Thus by Lemma \ref{lem:decomposition.9} the last integral is equal to
\[
\begin{split}
\int \limits_{\mathfrak{X}/G_1} \int \limits_{C} \Big( \mathfrak{J}_L \big( f \big)_{{}_x} , 
\big( g \big)_{{}_x} \Big) \,\, \ud \mu_C (x) \,\, \ud \nu (C) 
= \int \limits_{\mathfrak{X}} \Big( \mathfrak{J}_L \big( f \big)_{{}_x} , 
\big( g \big)_{{}_x} \Big) \,\, \ud \mu (x)
= \big( f , g \big)_{\mathfrak{J}^L},
\end{split}
\] 
so that 
\[
\Big( T(f), T(g) \Big)_{\int \mathfrak{J}^{L,C} \, \ud \nu(C)} 
= \big( f , g \big)_{\mathfrak{J}^L}.
\]
\qed

Actually we could merely use all  $g' \cdot f$, with $g \in C_{\mathcal{K}}(\mathfrak{X})$ and 
$f \in C^{L}_{0}$ instead of its denumerable subset $g'_j \cdot f^i$ , $i,j \in \mathbb{N}$
in the proof of Lemma \ref{lem:subgroup.1}. Its denumerability
shows that $\mathcal{H}^L$ is separable as the direct integral (a). This however is superfluous because 
separability of $\mathcal{H}^L$ has been already shown within the proof of Lemma \ref{lem:dense.6}.

\vspace*{0.5cm}

\begin{lem}  

Let $\mathfrak{B}$ be the section of $\mathfrak{G}$ with respect to double cosets
of Def. \ref{def:decomposition.1} and let $C \mapsto x_c \in \mathfrak{B}$, $D \mapsto x_{{}_D} \in \mathfrak{B}$
be the bi-unique maps of Def. \ref{def:decomposition.2}.
Let $\nu_0$ be the measure on the subsets of the set $G_1 : G_2$ of all double cosets $D$
equal to the transfer of the measure $\nu$ on $\mathfrak{X}/G_2$ over
to the set of double cosets by the natural bi-unique map $C \mapsto D_{{}_C} = \pi^{-1}(C)$. 
Let $\nu_{\mathfrak{B}}$ be the measure on the section $\mathfrak{B}$ equal to the transfer
of $\nu$ over to the section $\mathfrak{B}$ by the map $C \mapsto x_c$ (or equivalently
equal to the transfer of $\nu_0$ by the map $D \mapsto x_{{}_D}$). 
Let $\mu_D = \mu_{C_{{}_D}}$, where $C_{{}_D}$ is the orbit corresponding to the double coset $D$,
be the measure concentrated on $C_{{}_D}$, where $\mu_C$ is the measure of Lemma \ref{lem:decomposition.4}.
Let us denote the space of functions
${\mathcal{H}^{L}_{C}}'$ of Sect. \ref{subgroup.preliminaries}, defined on the double coset $D$
corresponding to $C$ just by $\mathcal{H}^{L}_{D}$ and similarly if $U^{L, C}$ and 
$\mathfrak{J}^{L,C}$ is the representation and the operator
of Sect. \ref{subgroup.preliminaries}, then we put $U^{L, D} = U^{L, C_{{}_D}}$
and $\mathfrak{J}^{L,D} = \mathfrak{J}^{L,C_{{}_D}}$;
analogously we define  $U^{L, x_{{}_D}} = U^{L, C_{{}_D}}$
and $\mathfrak{J}^{L, x_{{}_D}} = \mathfrak{J}^{L, C_{{}_D}}$. Then we have

\begin{enumerate}

\item[(a)] 

\[
\mathcal{H}^L \cong \int \limits_{\mathfrak{X}/G_2} {\mathcal{H}^{L}_{C}}' \,\, \ud \nu (C)
= \int \limits_{G_1 : G_2} \mathcal{H}^{L}_{D} \,\, \ud \nu_0 (D)
= \int \limits_{\mathfrak{B}} \mathcal{H}^{L}_{x_{{}_D}} \,\, \ud \nu_{{}_\mathfrak{B}} (x_{{}_D}).
\]

\item[(b)]

\[
{}_{{}_{G_2}}U^L \cong \int \limits_{\mathfrak{X}/G_2} U^{L, C} \,\, \ud \nu (C)
= \int \limits_{G_1 : G_2} U^{L, D} \,\, \ud \nu_0 (D)
= \int \limits_{\mathfrak{B}} U^{L, x_{{}_D}} \,\, 
\ud \nu_{{}_\mathfrak{B}} (x_{{}_D}).
\]

\item[(c)]

\[
\mathfrak{J}^L \cong \int \limits_{\mathfrak{X}/G_2} \mathfrak{J}^{L,C} \,\, \ud \nu (C)
= \int \limits_{G_1 : G_2} \mathfrak{J}^{L,D} \,\, \ud \nu_0 (D)
= \int \limits_{\mathfrak{B}} \mathfrak{J}^{L,x_{{}_D}} \,\, 
\ud \nu_{{}_\mathfrak{B}} (x_{{}_D}).
\]

The equivalences $\cong$ are all under the same map 
$T: \mathcal{H}^L \mapsto \int \limits_{G_1 : G_2} \mathcal{H}^{L}_{D} \,\, \ud \nu (C)$
giving the corresponding decomposition $T(f): D \mapsto f^{C_{{}_D}}$ (or respectively
$T(f): x_{{}_D}  \mapsto f^{C_{{}_D}}$)
for each $f \in \mathcal{H}^L$, in which $f^{C_{{}_D}}$ is the restriction of $f$ to the double 
coset $D = D_{{}_{C_{{}_D}}} = G_1 x_{{}_D} G_2 = \pi^{-1} (C_{{}_D})$ corresponding to $C_{{}_D}$.
In particular $T$ is unitary and Krein-unitary map between the Krein spaces 
\[
(\mathcal{H}^L , \mathfrak{J}^L)
\]
and
\[
\Big( \int \limits_{G_1 : G_2} \mathcal{H}^{L}_{D} \,\,
 \ud \nu (C) , \int \limits_{G_1 : G_2} \mathfrak{J}^{L,D} \,\, \ud \nu_0 (D) \Big)
\]
or respectively
\[
\Big( \int \limits_\mathfrak{B} \mathcal{H}^{L}_{x_{{}_D}} \,\,
 \ud \nu_{{}_\mathfrak{B}} (x_{{}_D}) , \int \limits_{\mathfrak{B}} \mathfrak{J}^{L,x_{{}_D}} \,\, 
\ud \nu_{{}_\mathfrak{B}} (x_{{}_D}) \Big).
\]

\end{enumerate}

\label{lem:subgroup.2}
\end{lem}

\qedsymbol \,
The Lemma follows from Lemma \ref{lem:subgroup.1} by a mere renaming of the 
points of the measure space $\mathfrak{X}/G_2$ of $G_2$-orbits $C$ in $\mathfrak{X}$, with the preservation
of the measure structure under the indicated renaming, which is guaranteed by Def. \ref{def:decomposition.1}
and \ref{def:decomposition.2}. 
\qed

\vspace*{0.5cm}

\begin{lem}
Let $\big( \, {}^{\mu^{x_c}}\mathcal{H}^{L^{x_c}} \, , \,\, \mathfrak{J}_{x_c} \big)$
be the Krein space of the representation ${}^{\mu^{x_c}}U^{L^{x_c}}$ of the subgroup $G_2$ defined
in Lemma \ref{lem:subgroup.preliminaries.1} with the inner product $(\cdot,\cdot)_{x_c}$ in 
${}^{\mu^{x_c}}\mathcal{H}^{L^{x_c}}$ defined by eq. (\ref{g_2'_def_inn}) in the proof of 
Lemma \ref{lem:subgroup.preliminaries.1}. For each $x_{{}_D} \in \mathfrak{B}$ we put 
${}^{\mu^{x_{{}_D}}}\mathcal{H}^{L^{x_{{}_D}}} = {}^{\mu^{x_c}}\mathcal{H}^{L^{x_c}}$,
$\mathfrak{J}_{x_{{}_D}} = \mathfrak{J}_{x_c}$, $G_{{}_{x_{{}_D}}} = G_{x_c}$
and $\big(\cdot , \cdot \big)_{x_{{}_D}} = (\cdot,\cdot)_{x_c}$
with the orbit $C$ corresponding to $D$. For each fixed element $f \in \mathcal{H}^L$
consider the following function 
\[
\mathfrak{B} \ni x_{{}_D} \mapsto {\widetilde{f}}^{{}^{{}^{x_{{}_D}}}} \in \,\,\, {}^{\mu^{x_{{}_D}}}\mathcal{H}^{L^{x_{{}_D}}}
\]
where for each $x_{{}_D}$, ${\widetilde{f}}^{{}^{{}^{x_{{}_D}}}}$ is defined as the function
\[
G_2  \ni t \mapsto 
\big( {\widetilde{f}}^{{}^{{}^{x_{{}_D}}}} \big)_{{}_t} = 
\big( f^D \big)_{{}_{x_{{}_D} \cdot t }},
\]   
with $f^D$ equal to the restriction of $f$ to $D$. The linear set $\mathcal{H}$ of all such functions 
$x_{{}_D} \mapsto {\widetilde{f}}^{{}^{{}^{x_{{}_D}}}}$ with $f$ ranging over the whole space $\mathcal{H}^L$ and with the inner product 
\begin{equation}\label{inn_prod_D:subgroup}
(\widetilde{f}, \widetilde{g}) = \int \limits_\mathfrak{B} 
\big( {\widetilde{f}}^{{}^{{}^{x_{{}_D}}}} , {\widetilde{g}}^{{}^{{}^{x_{{}_D}}}}  \big)_{x_{{}_D}} \,\,
 \ud \nu_{{}_\mathfrak{B}}(x_{{}_D}),
\end{equation}  
is equal to
\[
\int \limits_\mathfrak{B}  \,\,\,\, {}^{\mu^{x_{{}_D}}}\mathcal{H}^{L^{x_{{}_D}}} \,\,\,\,\,\, \ud \nu_{{}_\mathfrak{B}}(x_{{}_D}).
\]

\label{lem:subgroup.3}
\end{lem}
\qedsymbol \, Note, please, that by definition of the measures $\mu^{x_{{}_D}}$ and 
the operators $\mathfrak{J}_{x_{{}_D}}$ 
\begin{multline*}
\big( {\widetilde{f}}^{{}^{{}^{x_{{}_D}}}} , {\widetilde{g}}^{{}^{{}^{x_{{}_D}}}}  \big)_{x_{{}_D}} 
= \int \limits_{G_2 / G_{{}_{x_{{}_D}}}} 
\Big( \mathfrak{J}_L \big( \mathfrak{J}_{x_{{}_D}} {\widetilde{f}}^{{}^{{}^{x_{{}_D}}}} \big)_{{}_t} , 
\big( {\widetilde{g}}^{{}^{{}^{x_{{}_D}}}} \big)_{{}_t}  \Big)  \,\, \ud \mu^{x_{{}_D}} ([t]) \\
= \int \limits_{G_2 / G_{{}_{x_{{}_D}}}} 
\Big( \mathfrak{J}_L L_{h(x_{{}_D} \cdot t)}  \mathfrak{J}_L  L_{h(x_{{}_D} \cdot t)^{-1}} 
\big( f^D \big)_{{}_{x_{{}_D} \cdot t}} , 
\big( g^D \big)_{{}_{x_{{}_D} \cdot t}}  \Big) \,\, \ud \mu^{x_{{}_D}} ([t])     \\
=  \int \limits_{D}
\Big( \mathfrak{J}_L \big( \mathfrak{J}^L  f^D \big)_{{}_x} , 
\big(g^D \big)_{{}_x}  \Big)  \,\, \ud \mu_D ([x])
=\int \limits_{D}
\Big( \mathfrak{J}_L \big( \mathfrak{J}^L f\big)_{{}_x} , g_{{}_x}  \Big)  \,\, \ud \mu_D ([x])
\end{multline*}
and because 
\[
\mathfrak{G}/G_1 \ni [x] \mapsto  \Big( \mathfrak{J}_L  \big( \mathfrak{J}^L  f \big)_{{}_{[x]}} , 
 g_{{}_{[x]}}  \Big)
=  \Big( \mathfrak{J}_L \big( \mathfrak{J}^L  f \big)_{{}_x} , g_{{}_x}  \Big)
\]
is measurable it follows from (\ref{skew_fubini.1:decompositions}) of  
Lemma \ref{lem:decomposition.11} that the function
\[
x_{{}_D} \mapsto  \big( {\widetilde{f}}^{{}^{{}^{x_{{}_D}}}} , {\widetilde{g}}^{{}^{{}^{x_{{}_D}}}}  \big)_{x_{{}_D}} 
\]
is measurable for all $f,g \in \mathcal{H}^L$. Similarly by (\ref{skew_fubini.2:decompositions}) 
of part 2) of  Lemma  \ref{lem:decomposition.11}

\begin{multline*}
(\widetilde{f}, \widetilde{g}) = \int \limits_\mathfrak{B}  
\big( {\widetilde{f}}^{{}^{{}^{x_{{}_D}}}} , {\widetilde{g}}^{{}^{{}^{x_{{}_D}}}}  \big)_{x_{{}_D}} \,\,
 \ud \nu_{{}_\mathfrak{B}}(x_{{}_D}) \\
 = \int \limits_\mathfrak{B} \,\,  \int \limits_{G_2 / G_{{}_{x_{{}_D}}}} 
\Big( \mathfrak{J}_L \big( \mathfrak{J}_{x_{{}_D}} {\widetilde{f}}^{{}^{{}^{x_{{}_D}}}} \big)_{{}_t} , 
\big( {\widetilde{g}}^{{}^{{}^{x_{{}_D}}}} \big)_{{}_t}  \Big)_{x_{{}_D}}  \,\, \ud \mu^{x_{{}_D}} ([t])
 \,\,\, \ud \nu_{{}_\mathfrak{B}}(x_{{}_D})   \\
\int \limits_\mathfrak{B} \,\,  \int \limits_{D}
\Big( \mathfrak{J}_L \big( \mathfrak{J}^L f\big)_{{}_x} , g_{{}_x}  \Big)  \,\, \ud \mu_D ([x])
\,\,\, \ud \nu_{{}_\mathfrak{B}}(x_{{}_D}) 
= \int \limits_{\mathfrak{G}/G_1}
\Big( \mathfrak{J}_L \big( \mathfrak{J}^L f\big)_{{}_x} , g_{{}_x}  \Big)  \,\, \ud \mu ([x]) \\
= (f,g).
\end{multline*}
Therefore $\mathcal{H}$ is a Hilbert space with the inner product (\ref{inn_prod_D:subgroup}) as the 
isometric image of the Hilbert space 
$\mathcal{H}^L$. We need only show Part 2) of Def. \ref{direct_int:subgroup} to be fulfilled.
Toward this end let $x_{{}_D} \mapsto {u}^{{}^{{}^{x_{{}_D}}}} \in {}^{\mu^{x_{{}_D}}}\mathcal{H}^{L^{x_{{}_D}}}$ be a function fulfilling the conditions of Part 2) of Def. \ref{direct_int:subgroup} (of course with the obvious replacements 
of $C$ with $D$ and ${\mathcal{H}^{L}_{C}}'$ with ${}^{\mu^{x_{{}_D}}}\mathcal{H}^{L^{x_{{}_D}}}$). We have to show
existence of a function $f' \in \mathcal{H}^L$ such that the function 
$x_{{}_D} \mapsto {\widetilde{f'}}^{{}^{{}^{x_{{}_D}}}}$ is equal almost everywhere
to the function $x_{{}_D} \mapsto {u}^{{}^{{}^{x_{{}_D}}}}$. We proceed exactly as in the proof
of Part (a) of Lemma \ref{lem:subgroup.1} by formation of the analogous maximal linear subspace $X$ in the
space $\mathfrak{F}$ of all functions $x_{{}_D} \mapsto {k}^{{}^{{}^{x_{{}_D}}}}$ for which
\[
x_{{}_D} \mapsto \big( {k}^{{}^{{}^{x_{{}_D}}}} , 
{k}^{{}^{{}^{x_{{}_D}}}}  \big)_{x_{{}_D}} 
\] 
is measurable and integrable and then using Riesz theorem and Lemma \ref{lem:dense.3} or 
\ref{lem:dense.4} in proving the existence of $f'$ (in this case the proof is even simpler because 
the Lemma \ref{lem:subgroup.preliminaries.1} is not necessary in proving 
${\widetilde{f'}}^{{}^{{}^{x_{{}_D}}}} - {u}^{{}^{{}^{x_{{}_D}}}} = 0$ from the analogue of (\ref{*f_j_dense:subgroup});
indeed it is sufficient to apply Lemma \ref{lem:subgroup.preliminaries.3} and Lemma \ref{lem:dense.3} or 
\ref{lem:dense.4}).  
\qed

$\widehat{\mathcal{A}}$
From now on we identify the Hilbert space $\mathcal{H}^L$ with the direct integral:
\[
\mathcal{H}^L = \int \limits_{\mathfrak{B}} \mathcal{H}^{L}_{x_{{}_D}} \,\, \ud \nu_{{}_\mathfrak{B}} (x_{{}_D}).
\]
with the realization $T \mapsto T(f)$ of the direct integral equal to $T(f): x_{{}_D} \mapsto f^D$, where $f^D$
is the ordinary restriction of $f \in \mathcal{H}^L$ to the double coset $D$. Similarly by
\[
\int \limits_\mathfrak{B}  \,\,\,\, {}^{\mu^{x_{{}_D}}}\mathcal{H}^{L^{x_{{}_D}}} \,\,\,\,\,\, \ud \nu_{{}_\mathfrak{B}}(x_{{}_D}),
\]
we understand the direct integral with the realization of Lemma \ref{lem:subgroup.3}.

\begin{lem}
For each orbit $C$ let $V_{x_c}$ be the Krein-unitary map defined in Lemma \ref{lem:subgroup.preliminaries.1}.
For each $x_{{}_D} \in \mathfrak{B}$ (equivalently: each double coset $D$) let us put $V_{x_{{}_D}} = V_{x_c}$ 
with $C$ corresponding to $D$. Then $x_{{}_D} \mapsto V_{x_{{}_D}}$ is a decomposition of a well defined operator
\begin{multline*}
\mathcal{H}^L = \int \limits_{\mathfrak{B}} \mathcal{H}^{L}_{x_{{}_D}} \,\, \ud \nu_{{}_\mathfrak{B}} (x_{{}_D}) 
\xrightarrow{V}
\int \limits_\mathfrak{B}  \,\,\,\, {}^{\mu^{x_{{}_D}}}\mathcal{H}^{L^{x_{{}_D}}} \,\,\,\,\,\, 
\ud \nu_{{}_\mathfrak{B}}(x_{{}_D}) :   \\
\Big( x_{{}_D} \mapsto {f}^{{}^{{}^{x_{{}_D}}}} \Big) \mapsto 
\Big( x_{{}_D} \mapsto  V_{x_{{}_D}}  {f}^{{}^{{}^{x_{{}_D}}}}  \Big).
\end{multline*}
In short
\[
V = \int \limits_{\mathfrak{B}} V_{x_{{}_D}} \,\, 
\ud \nu_{{}_\mathfrak{B}} (x_{{}_D}).
\]
The operator $V$ is unitary and Krein-unitary between the Krein spaces
\begin{multline*}
\Big( \int \limits_\mathfrak{B}  \,\,\,\, {}^{\mu^{x_{{}_D}}}\mathcal{H}^{L^{x_{{}_D}}} \,\,\,\,\,\, 
\ud \nu_{{}_\mathfrak{B}}(x_{{}_D}) \,\,
 , \,\,\,\, \int \limits_\mathfrak{B}  \,\,\,\, \mathfrak{J}_{x_{{}_D}} \,\,\,\,\,\, 
\ud \nu_{{}_\mathfrak{B}}(x_{{}_D}) \,\, \Big)   \\       
\,\,\, \textrm{and} \,\,\, 
\Big( \int \limits_\mathfrak{B} \mathcal{H}^{L}_{x_{{}_D}} \,\,
 \ud \nu_{{}_\mathfrak{B}} (x_{{}_D}) \,\, , \,\,\,\, \int \limits_{\mathfrak{B}} \mathfrak{J}^{L,x_{{}_D}} \,\, 
\ud \nu_{{}_\mathfrak{B}} (x_{{}_D}) \,\, \Big) =(\mathcal{H}^L , \mathfrak{J}^L);
\end{multline*}
and moreover:
\[
V \Big( \, {}_{{}_{G_2}}U^L \Big) V^{-1}
= V \Big( \, \int \limits_{\mathfrak{B}} U^{L, x_{{}_D}} \,\, 
\ud \nu_{{}_\mathfrak{B}} (x_{{}_D}) \, \Big) V^{-1}
=  \int \limits_{\mathfrak{B}} \,\,\, {}^{\mu^{x_{{}_D}}}U^{L^{x_{{}_D}}} \,\,\,\,\,\, 
\ud \nu_{{}_\mathfrak{B}} (x_{{}_D})  
\]
and 
\[
V \Big( \, \mathfrak{J}^L  \Big) V^{-1}
= V \Big( \, \int \limits_{\mathfrak{B}} \mathfrak{J}^{L,x_{{}_D}} \,\, 
\ud \nu_{{}_\mathfrak{B}} (x_{{}_D}) \, \Big) V^{-1}
=  \int \limits_{\mathfrak{B}} \mathfrak{J}_{x_{{}_D}} \,\, 
\ud \nu_{{}_\mathfrak{B}} (x_{{}_D}). 
\]

\label{lem:subgroup.4}
\end{lem}
\qedsymbol \, 
Let $f$ be any element of $\mathcal{H}^L$ and $t \in G_2$. By definition we have
\[
\Big( V_{x_{{}_D}}  {f}^{{}^{{}^{x_{{}_D}}}}  \Big)_{{}_t} 
= \big(f^D \big)_{{}_{x_{{}_D} \cdot t}}  
=  \Big( {\widetilde{f}}^{{}^{{}^{x_{{}_D}}}} \Big)_{{}_t}, 
\]
with ${\widetilde{f}}^{{}^{{}^{x_{{}_D}}}}$ defined in Lemma \ref{lem:subgroup.3}. Thus by the realization of 
\[
\int \limits_\mathfrak{B}  \,\,\,\, {}^{\mu^{x_{{}_D}}}\mathcal{H}^{L^{x_{{}_D}}} \,\,\,\,\,\, \ud \nu_{{}_\mathfrak{B}}(x_{{}_D})
\]
given in Lemma \ref{lem:subgroup.3}, $V$ is onto. Moreover, by the proof of Lemma \ref{lem:subgroup.3}
\[
x_{{}_D} \mapsto \big( V_{x_{{}_D}}  {f}^{{}^{{}^{x_{{}_D}}}} , {\widetilde{g}}^{{}^{{}^{x_{{}_D}}}}  \big)_{x_{{}_D}}
= \big( {\widetilde{f}}^{{}^{{}^{x_{{}_D}}}} , {\widetilde{g}}^{{}^{{}^{x_{{}_D}}}}  \big)_{x_{{}_D}} \,\,
\]
is measurable for all $g \in \mathcal{H}^L$, and thus for all 
\[
\big( x_{{}_D} \mapsto  {\widetilde{g}}^{{}^{{}^{x_{{}_D}}}} \big)  \in
\int \limits_\mathfrak{B}  \,\,\,\, {}^{\mu^{x_{{}_D}}}\mathcal{H}^{L^{x_{{}_D}}} \,\,\,\,\,\, 
\ud \nu_{{}_\mathfrak{B}}(x_{{}_D});
\]
therefore $V$ is a well defined operator. Moreover, by the proof of Lemma \ref{lem:subgroup.3} 
\begin{multline*}
(Vf,Vg)
= \int \limits_\mathfrak{B} 
\big( V_{x_{{}_D}}  {f}^{{}^{{}^{x_{{}_D}}}} , \, V_{x_{{}_D}} {g}^{{}^{{}^{x_{{}_D}}}}  \big)_{x_{{}_D}} \,\,
 \ud \nu_{{}_\mathfrak{B}}(x_{{}_D})  \\
= \int \limits_\mathfrak{B} 
\big( {\widetilde{f}}^{{}^{{}^{x_{{}_D}}}} , {\widetilde{g}}^{{}^{{}^{x_{{}_D}}}}  \big)_{x_{{}_D}} \,\,
 \ud \nu_{{}_\mathfrak{B}}(x_{{}_D}) = (f,g),
\end{multline*}
so that $V$ is unitary (it likewise follows from Lemma \ref{lem:subgroup.preliminaries.2}). 

Again by Lemma \ref{lem:decomposition.11} we have:
\begin{multline*}
(Vf, Vg)_{{}_{\int  \mathfrak{J}_{x_{{}_D}} \,\, 
\ud \nu_{{}_\mathfrak{B}} (x_{{}_D})}} 
= \int \limits_\mathfrak{B} 
\big( \mathfrak{J}_{x_{{}_D}} {\widetilde{f}}^{{}^{{}^{x_{{}_D}}}} , 
{\widetilde{g}}^{{}^{{}^{x_{{}_D}}}}  \big)_{x_{{}_D}} \,\,
 \ud \nu_{{}_\mathfrak{B}}(x_{{}_D})   \\
= \int \limits_\mathfrak{B} 
\int \limits_{D}
\Big( \mathfrak{J}_L  \big( f^D \big)_{{}_x} , 
\big(g^D \big)_{{}_x}  \Big)  \,\, \ud \mu_D ([x]) \,\,
 \ud \nu_{{}_\mathfrak{B}}(x_{{}_D})                      \\
=  \int \limits_{\mathfrak{G}/G_1}
\Big( \mathfrak{J}_L  f_{{}_x} , g_{{}_x}  \Big)  \,\, \ud \mu_D ([x]) \,\,
 \ud \nu_{{}_\mathfrak{B}}(x_{{}_D}) 
= (f,g)_{{}_{\mathfrak{J}^L}}
\end{multline*}
which shows that $V$ is Krein unitary.

Because by Lemma \ref{lem:subgroup.preliminaries.1}
\[
V_{x_{{}_D}} \,\, U^{L, x_{{}_D}} \,\, {V_{x_{{}_D}}}^{-1}
=  \,\, {}^{\mu^{x_{{}_D}}}U^{L^{x_{{}_D}}} \,\,\, \textrm{and}  \,\,\, 
V_{x_{{}_D}} \,\, \mathfrak{J}^{L,x_{{}_D}} \,\, {V_{x_{{}_D}}}^{-1}
= \mathfrak{J}_{x_{{}_D}} ,
\] 
the rest of the Lemma is thereby proved.  
\qed

\begin{rem}
By a mere renaming of points associated to the isomorphisms
 $\mathfrak{B} \cong G_1 : G_2 \cong \mathfrak{X}/G_2$ of measure spaces, e.g introducing
$V_{{}_{D}} = V_{x_{{}_D}}$, $\mu^{D} = \mu^{x_{{}_D}}$ and the measure $\nu_0$ as in 
Def. \ref{def:decomposition.2} we may rephrase Lemma \ref{lem:subgroup.4} as follows.
$D \mapsto V_{{}_{D}}$ is a decomposition of a well defined operator
\[
V = \int \limits_{G_1 : G_2} V_{{}_{D}} \,\, 
\ud \nu_0 (D) :
\]
\begin{multline*}
\mathcal{H}^L = \int \limits_{G_1 : G_2} \mathcal{H}^{L}_{D} \,\, \ud \nu_0 (D) 
\xrightarrow{V}
\int \limits_{G_1 : G_2}  \,\,\,\, {}^{\mu^{{}^D}}\mathcal{H}^{L^{{}^D}} \,\,\,\,\,\, 
\ud \nu_0 (D) :   \\
\Big(D \mapsto f^D \Big) \mapsto 
\Big( D \mapsto  V_{{}_{D}}  \, f^D  \Big).
\end{multline*}
The operator $V$ is unitary and Krein-unitary between the Krein spaces
\begin{multline*}
\Big( \int \limits_{G_1 : G_2}  \,\,\,\, {}^{\mu^{{}^D}}\mathcal{H}^{L^{{}^D}} \,\,\,\,\,\, 
\ud \nu_0(D) \,\,
 , \,\,\,\, \int \limits_{G_1 : G_2}  \,\,\,\, \mathfrak{J}_{{}_D} \,\,\,\,\,\, 
\ud \nu_0(D) \,\, \Big)   \\       
\,\,\, \textrm{and} \,\,\, 
\Big( \int \limits_{G_1 : G_2} \mathcal{H}^{L}_{D} \,\,
 \ud \nu_0 (D) \,\, , \,\,\,\, \int \limits_{G_1 : G_2} \mathfrak{J}^{L,D} \,\, 
\ud \nu_0 (D) \,\, \Big) = (\mathcal{H}^L , \mathfrak{J}^L);
\end{multline*}
and moreover: 
\[
V \Big( \, {}_{{}_{G_2}}U^L \Big) V^{-1}
= V \Big( \, \int \limits_{G_1 : G_2} U^{L, D} \,\, 
\ud \nu_0 (D) \, \Big) V^{-1}
=  \int \limits_{G_1 : G_2} \,\,\, {}^{\mu^{{}^D}}U^{L^{{}^D}} \,\,\,\,\,\, 
\ud \nu_0 (D)  
\]
and 
\[
V \Big( \, \mathfrak{J}^L  \Big) V^{-1}
= V \Big( \, \int \limits_{G_1 : G_2} \mathfrak{J}^{L,D} \,\, 
\ud \nu_{{}_\mathfrak{B}} (x_{{}_D}) \, \Big) V^{-1}
=  \int \limits_{G_1 : G_2} \,\,\, \mathfrak{J}_{{}_D} \,\,\,\,\,\, 
\ud \nu_0 (D).  
\]

\label{rem:subgroup.2}
\end{rem}

\begin{defin}
Let $G_1$ and $G_2$ be two closed subgroups of a separable locally compact 
group $\mathfrak{G}$. Let $B$ be any Borel section of $\mathfrak{G}$ with respect
to $G_1$ and for each $x \in \mathfrak{G}$  let $h(x)$ be the unique element of $G_1$
such that $h(x)^{-1} \cdot x \in B$. 
Let $\mu$ be any quasi invariant measure $\mu$
on $\mathfrak{G}/G_1$ and let $\nu$ be any pseudo-image measure on $(\mathfrak{G}/G_1 )/G_2$ of the measure
$\mu$ under the quotient map  $\pi_{\mathfrak{G}/G_1}: \mathfrak{G}/G_1 \mapsto (\mathfrak{G}/G_1 )/G_2$;
so that:
\[ 
\mu =  \int \limits_{(\mathfrak{G}/G_1 )/G_2} \mu_C \, \ud \nu(C).
\]
Let us call any measure $\nu_0$ on measurable subsets of the set $G_1 : G_2$   of all double cosets 
\emph{admissible} iff it is equal to the transfer of $\nu$ over to $G_1 : G_2$ by the natural map 
$(\mathfrak{G}/G_1 )/G_2 \ni C \mapsto \pi^{-1}(C) \in G_1 : G_2$. Finally let $x$ be any
element of $\mathfrak{G}$ with $\pi(x) \in C$. We put $\mu^x$ for the measure on 
$G_2 / G_x$ equal to the transfer of the measure $\mu_C$ over to
$G_2 / G_x$ by the map $G_2 / G_x \ni [y] \mapsto [xy] \in C \subset \mathfrak{G}/G_1$, 
where $G_x = G_2 \cap (x^{-1} G_1 x)$
and where $[\cdot]$ denotes the respective equivalence classes. 
  
\label{defin:subgroup.2}
\end{defin}
    
Summing up we have just proved the following

\begin{twr}[Subgroup Theorem]
Let $U^L$ be the isometric representation of the separable locally compact group $\mathfrak{G}$
in the Krein space $(\mathcal{H}^L , \mathfrak{J}^L)$, induced by the Krein-unitary 
representation $L$ of the closed subgroup
$G_1$ of $\mathfrak{G}$ and the quasi invariant measure $\mu$ on 
$\mathfrak{G}/G_1$ and the Borel section $B$ of $\mathfrak{G}$
with respect to $G_1$. Then $U^L$ is independent to within Krein-unitary 
equivalence of the choice of $B$. Let $G_2$ be a second closed subgroup of $\mathfrak{G}$ and suppose
that $G_1$ and $G_2$ are regularly related. For each $x \in \mathfrak{G}$ consider the closed subgroup
$G_x = G_2 \cap (x^{-1} G_1 x)$ and let $U^{L^{x}}$ denote the representation of $G_2$
in the Krein space $(\mathcal{H}^{L^{x}}, \mathfrak{J}_{{}_x})$ induced by the Krein-unitary representation $L^{x}:
\eta \mapsto L_{x\eta x^{-1}}$ of the subgroup $G_x$ in the Krein space 
$(\mathcal{H}_L , \mathfrak{J}_L )$, where $\big( \mathfrak{J}_{{}_x} g \big)_{{}_t}
= L_{h(x\cdot t)} \mathfrak{J}_L L_{{h(x \cdot t)}^{-1}} \big( g \big)_{{}_t}$
and with the inner product in $\mathcal{H}^{L^{x}}$ and Krein-inner product 
in $(\mathcal{H}^{L^{x}}, \mathfrak{J}_{{}_x})$ defined respectively by the formulas

\[
(f,g)_{{}_x} = \int \limits_{G_2 / G_x} 
\Big( \mathfrak{J}_L \big( \mathfrak{J}_{{}_x} f \big)_{{}_t} , 
\big( g \big)_{{}_t}  \Big)  \,\, \ud \mu^x ([t])
\]

and
\[
(f,g)_{{}_{\mathfrak{J}_{{}_x}}} = (\mathfrak{J}_{{}_x} f , g)_{{}_x} = 
 \int \limits_{G_2 / G_x} 
\Big( \mathfrak{J}_L \big(f \big)_{{}_t} , 
\big( g \big)_{{}_t}  \Big)  \,\, \ud \mu^x ([t]);
\]
and with the quasi invariant measure $\mu^x$ on $G_2 /G_x$ given by Def. \ref{defin:subgroup.2}. 
Then $U^{L^{x}}$ is determined to 
within Krein-unitary and unitary equivalence by the double coset $G_1 x G_2 = s(x)$ to which 
$x$ belongs and we may write $U^{L^{{}^D}} = U^{L^{x}}$, where $D = s(x)$. Finally 
$U^L$ restricted to $G_2$ is a direct integral over $G_1 : G_2$ with respect to any admissible 
(Def. \ref{defin:subgroup.2}) measure in $G_1 : G_2$, of the representations $U^{L^{{}^D}}$.

\label{twr:subgroup.1}
\end{twr}

It may happen that all the component representations ${}^{\mu^{x_{{}_D}}}U^{L^{x_{{}_D}}}$
are bounded and thus Krein-unitary, although $U^L$ is unbounded. In this case the norms
$\big{\|} {}^{\mu^{x_{{}_D}}}U^{L^{x_{{}_D}}} \big{\|}_{x_{{}_D}}$ are unbounded functions
of $x_{{}_D}$ (resp. $D$). Unfortunately instead of $\mathfrak{J}_{x_{{}_D}}$ we cannot use any standard 
fundamental symmetry in 
${^{\mu^{x_{{}_D}}}}\mathcal{H}^{L^{x_{{}_D}}}$:
\[
\Big( \mathfrak{J}^{L^{x_{{}_D}}} 
{\widetilde{f}}^{{}^{{}^{x_{{}_D}}}} \Big)_{{}_t} = L^{x_{{}_D}}_{h_{x_{{}_D}}(t)} 
\mathfrak{J}_{L} L^{x_{{}_D}}_{h_{x_{{}_D}}(t)^{-1}} \, 
\Big({\widetilde{f}}^{{}^{{}^{x_{{}_D}}}}\Big)_{{}_t} ,
\] 
where $h_{x_{{}_D}}(t) \in G_{x_{{}_D}}$ is defined as in Sect. \ref{def_ind_krein}
by a regular Borel section $B_{x_{{}_D}}$ of $G_2$ with respect to the subgroup 
$G_{x_{{}_D}} = G_2 \cap \, ({x_{{}_D}}^{-1}G_1 x_{{}_D})$. A difficulty will arise with this 
$\mathfrak{J}^{L^{x_{{}_D}}}$.
Namely in general the norms $\big{\|} {}^{\mu^{x_{{}_D}}}U^{L^{x_{{}_D}}} \big{\|}_{x_{{}_D}}$  
are such that the operator $V$ would be unbounded with the standard fundamental symmetries
in ${^{\mu^{x_{{}_D}}}}\mathcal{H}^{L^{x_{{}_D}}}$.

It is important that in practical computations, e.g. with $\mathfrak{G}$ equal to the double covering 
of the Pouncar\'e group, much stronger regularity is preserved, e. g. the ``measure product property''
(see the end of Sect. \ref{decomposition}), with the measurable sections $B$ and $\mathfrak{B}$ as differential
sub-manifolds (if we discard unimportant null subset), so that the function $x \mapsto h(x)$ and
all the remaining functions -- analogue of (\ref{q'_h'}) -- associated 
to the measure product structure are effectively computable together with the measures
$\mu$ and $\nu_0$. This is important because together with the theorem of the next Section
give an effective tool for decomposing tensor product of induced representations of the double cover 
of the Poincar\'e group in Krein spaces.
Moreover the operator $V$ of Lemma \ref{lem:subgroup.4} and 
Remark \ref{rem:subgroup.2} is likewise effectively computable in this case.

\section{Kronecker product theorem in Krein spaces}\label{Kronecker_product}

Let ${}^{\mu_1}U^L$ and ${}^{\mu_2}U^M$ be Krein-isometric representations
of the separable locally compact group $\mathfrak{G}$ induced from Krein-unitary
representations of the closed subgroups $G_1 \subset \mathfrak{G}$ and $G_2 \subset \mathfrak{G}$
respectively. 
The Krein-isometric representation ${}^{\mu_1}U^L \, \otimes \, {}^{\mu_2}U^M$ of $\mathfrak{G}$ is obtained from the 
Krein-isometric representation ${}^{\mu_1}U^L \, \times \, {}^{\mu_2}U^M$ of $\mathfrak{G} \times \mathfrak{G}$ by restriction to the 
diagonal subgroup $\overline{\mathfrak{G}}$ of all those 
$(x,y) \in \mathfrak{G} \times \mathfrak{G}$ for which $x = y$, which is naturally isomorphic
to $\mathfrak{G}$ itself: $\overline{\mathfrak{G}} \cong \mathfrak{G}$, with
the natural isomorphism $(x,x) \mapsto x$. Thus by the natural isomorphism 
the representation ${}^{\mu_1}U^L \, \otimes \, {}^{\mu_2}U^M$ of $\mathfrak{G}$ may be identified
with the restriction of the representation ${}^{\mu_1}U^L \, \times \, {}^{\mu_2}U^M$ of 
 the group $\mathfrak{G} \times \mathfrak{G}$ to the diagonal subgroup $\overline{\mathfrak{G}}$. By 
Theorem \ref{twr.1:kronecker}, ${}^{\mu_1}U^L \, \times \, {}^{\mu_2}U^M$ is Krein-unitary and unitary equivalent to the 
Krein-isometric representation ${}^{\mu_1 \times \mu_2}U^{L \times M}$ of $\mathfrak{G} \times \mathfrak{G}$
induced by the Krein-unitary representation $L \times M$ of the closed subgroup $G_1 \times G_2$. Thus the
Krein-isometric representation $U^L \otimes U^M$ of $\mathfrak{G}$ is naturally 
equivalent to the restriction of the Krein-isometric representation ${}^{\mu_1 \times \mu_2}U^{L \times M}$ of 
$\mathfrak{G} \times \mathfrak{G}$ to the closed diagonal subgroup $\overline{\mathfrak{G}}$. Thus we are trying
to apply the \emph{Subgroup Theorem} \ref{twr:subgroup.1} inserting 
$\mathfrak{G} \times \mathfrak{G}$ for $\mathfrak{G}$, $\overline{\mathfrak{G}}$ for  $G_2$,
and the subgroup $G_1 \times G_2 \subset \mathfrak{G} \times \mathfrak{G}$ for
$G_1$ in the Subgroup Theorem. But the Subgroup Theorem 
is applicable in that way if the subgroups $G_1 \times G_2$ and 
$\overline{\mathfrak{G}}$ are regularly related.
Mackey recognized that they are indeed regularly related in 
$\mathfrak{G} \times \mathfrak{G}$ if and only if $G_1$ and $G_2$ are in $\mathfrak{G}$, 
pointing out a natural measure isomorphism between the
measure spaces $(G_1 \times G_2 ) : \overline{\mathfrak{G}}$ and $G_1 : G_2$ of double cosets
respectively in $\mathfrak{G} \times \mathfrak{G}$ and $\mathfrak{G}$.   
The isomorphism is induced by the map 
$\mathfrak{G} \times \mathfrak{G} \ni (x,y) \mapsto xy^{-1} \in \mathfrak{G}$. However his 
argumentation strongly depends on the finiteness of the quasi invariant measures in the homogeneous spaces
$(\mathfrak{G} \times \mathfrak{G}) /(G_1 \times G_2)$ and $\mathfrak{G}/G_1$ which slightly 
simplifies the construction of the $\sigma$-rings of measurable subsets in the corresponding
spaces of double cosets. Our proof that the map $(x,y) \mapsto xy^{-1}$ induces isomorphism
of the respective spaces of double cosets must have been slightly changed at this point
by addition of Lemma \ref{lem:decomposition.10}. The rest of the proof of 
Theorem \ref{twr:Kronecker_product} of this 
Section follows from the Subgroup Theorem \ref{twr:subgroup.1} in the same way as Theorem
7.2 from Theorem 7.1 in \cite{Mackey}.

By the above remarks we shall show that the measure spaces
$(G_1 \times G_2 ) : \overline{\mathfrak{G}}$ and $G_1 : G_2$ of double cosets constructed as in
Sect. \ref{decomposition} are isomorphic, with the isomorphism induced by the map 
$\mathfrak{G} \times \mathfrak{G} \ni (x,y) \mapsto xy^{-1} \in \mathfrak{G}$. Note first of all
that the indicated map sets up a one-to-one correspondence between the double cosets
in  $(G_1 \times G_2 ) : \overline{\mathfrak{G}}$ and double cosets in $G_1 : G_2$, in which the double coset
$(G_1 \times G_2 ) (x,y) \overline{\mathfrak{G}}$ corresponds to the double coset 
$G_1 xy^{-1} G_2$. Moreover in this mapping a set is measurable if and only if its 
image is measurable and \emph{vice versa}, a set is measurable if and only if its inverse image is measurable.
Thus it is an isomorphism of measure spaces. Indeed $(x_1 , x_2)$ and $(x_2 , y_2)$ 
go into the same point of $\mathfrak{G}$ under the indicated map
if and only if they belong to the same left $\overline{\mathfrak{G}}$ coset in 
$\mathfrak{G} \times \mathfrak{G}$. Now by Lemma \ref{lem:decomposition.10} of sect. \ref{decomposition} and 
by Lemma 1.2 of \cite{Mackey} (equally applicable to left coset spaces) the indicated one-to-one map
of double coset spaces is an isomorphism of measure spaces. Thus the Subgrop Theorem \ref{twr:subgroup.1} is 
applicable to ${}^{\mu_1 \times \mu_2}U^{L \times M}$ with $L$ replaced by $L \times M$, $\mathfrak{G}$
replaced by $\mathfrak{G} \times \mathfrak{G}$, $G_1$ replaced by $G_1 \times G_2$ 
and $G_2$ replaced by $\overline{\mathfrak{G}}$
and the function $\mathfrak{G} \ni x \mapsto h(x) \in G_1$
replaced by the function $\mathfrak{G} \times \mathfrak{G}
\ni (x,y) \mapsto h(x,y) = \big( h_{{}_1}(x) , h_{{}_2}(y) \big) \in G_1 \times G_2$,
where the functions $\mathfrak{G} \ni x \mapsto  h_{{}_1}(x) \in G_1$
and $\mathfrak{G} \ni y \mapsto  h_{{}_2}(y) \in G_2$
correspond to the respective Borel sections of $\mathfrak{G}$
with respect to $G_1$ and $G_2$ respectively used in the construction
of the representations ${}^{\mu_1}U^L$ and ${}^{\mu_2}U^M$ 
(compare Sect. \ref{def_ind_krein} and \ref{kronecker}).

In order to simplify formulation of the upcoming theorem let us give the following

\begin{defin}
Let $\nu_{0}^{12}$ be the admissible measure on the set of double cosets 
$\big(G_1 \times G_2 ):\overline{\mathfrak{G}}$ in $\mathfrak{G} \times \mathfrak{G}$
given by Def. \ref{defin:subgroup.2}, where we have used the product quasi invariant measure 
$\mu = \mu_1 \times \mu_2$ on the homogeneous space $\big( \mathfrak{G} \times \mathfrak{G} \big)
\big{/} \big( G_1 \times G_2 \big)$. Let us define the measure 
$\nu_{12}$ on the space $G_1 : G_2$ of double cosets in $\mathfrak{G}$ to be equal to the transfer
of $\nu_{0}^{12}$ by the map induced by 
$\mathfrak{G} \times \mathfrak{G} \ni (x,y) \mapsto xy^{-1} \in \mathfrak{G}$.
If $(\mu_1 \times \mu_2)^{(x,y)}$ is the quasi invariant measure  on
$\overline{\mathfrak{G}} / G_{(x,y)}$ given by Def. \ref{defin:subgroup.2}
with $G_{(x,y)} = \overline{\mathfrak{G}} \cap \big( (x,y)^{-1} (G_1 \times G_2) (x,y) \big)$
then we define $\mu^{x,y}$ to be the transfer of the measure $(\mu_1 \times \mu_2)^{(x,y)}$
over to the homogeneous space $\mathfrak{G} \big{/} \big( x^{-1}G_1 x \, \, \cap \, y^{-1} G_2 y \big)$
by the map $(x,x) \mapsto x$. 

\label{def:Kronecker_product.1}
\end{defin}   

Now we are ready to formulate the main goal of this paper:

\begin{twr}[Kronecker Product Theorem]
Let $G_1$ and $G_2$ be regularly related closed subgroups of the separable locally compact group 
$\mathfrak{G}$. Let $L$ and $M$ be Krein-unitary representations of $G_1$ and $G_2$ respectively
in the Krein spaces $(\mathcal{H}_L , \mathfrak{J}_L)$ and $(\mathcal{H}_M , \mathfrak{J}_M)$.
For each $(x, y) \in \mathfrak{G} \times \mathfrak{G}$ consider the Krein-unitary 
representations $L^x : s \mapsto L_{xsx^{-1}}$ and $M^y : s \mapsto M_{ysy^{-1}}$
of the subgroup $(x^{-1}G_1 x) \cap (y^{-1}G_2 y)$
in the Krein spaces $(\mathcal{H}_L , \mathfrak{J}_L)$ and $(\mathcal{H}_M , \mathfrak{J}_M)$
respectively.
Let us denote the tensor product $L^x \otimes M^y$ Krein-unitary representation
acting in the Krein space $(\mathcal{H}_L \otimes \mathfrak{J}_M , \mathfrak{J}_L \otimes \mathfrak{J}_M)$,
by $N^{x,y}$. 
Let $U^{N^{x,y}}$ be the Krein-isometric representation of $\mathfrak{G}$ induced
by $N^{x,y}$ acting in the Krein space $\big(\mathcal{H}^{N^{x,y}}, \mathfrak{J}{{}_{x,y}}\big)$, where
for each $w \in \mathcal{H}^{N^{x,y}}$
\[
\big( \mathfrak{J}_{{}_{x,y}} w \big)_{{}_{s}}
= \big( L_{h_{{}_1}(xs)} \,  \mathfrak{J}_L \,  L_{h_{{}_1}(xs)^{-1}} \big) \otimes 
\big( M_{h_{{}_2}(ys)} \, \mathfrak{J}_M \,  M_{h_{{}_2}(ys)^{-1}} \big) \big( w \big)_{{}_{s}};
\]
and with the inner product in Hilbert space $\mathcal{H}^{N^{x,y}}$ and the Krein-inner product
in the Krein space $\big(\mathcal{H}^{N^{x,y}}, \mathfrak{J}{{}_{x,y}}\big)$
given by the formulas
\[
(w,g)_{{}_{x,y}} 
= \int \limits_{\mathfrak{G} \big{/} \big( x^{-1}G_1 x \, \, \cap \, y^{-1} G_2 y \big)} 
\Big( \mathfrak{J}_L \otimes \mathfrak{J}_M \big( \mathfrak{J}_{{}_{x,y}}  w \big)_{{}_s} , 
\big( g \big)_{{}_s}  \Big)  \,\, \ud \mu^{x,y} ([s])
\]

and 
\begin{multline*}
(w,g)_{{}_{\mathfrak{J}_{{}_{x,y}}}} = (\mathfrak{J}_{{}_{x,y}} w , g)_{{}_{x,y}}   \\
= \int \limits_{\mathfrak{G} \big{/} \big( x^{-1}G_1 x \, \, \cap \, y^{-1} G_2 y \big)} 
\Big( \mathfrak{J}_L \otimes \mathfrak{J}_M \big( w \big)_{{}_s} , 
\big( g \big)_{{}_s}  \Big)  \,\, \ud \mu^{x,y} ([s]),
\end{multline*}
with the quasi invariant measure $\mu^{x,y}$ given by Def. \ref{def:Kronecker_product.1}.
Then $U^{N^{x,y}}$ is determined to within Krein-unitary equivalence by the double coset
$D =  G_1 xy^{-1}G_2$ to which $xy^{-1}$ belongs and we may write $U^{N^{x,y}} = U^D$. Finally
$U^L \otimes U^M$ is Krein-unitary equivalent to the direct integral of $U^D$ with respect to the
measure $\nu_{12}$ (Def. \ref{def:Kronecker_product.1}) on $G_1 : G_2$.

\label{twr:Kronecker_product}
\end{twr}
\qedsymbol \, 
By the above remarks the Subgroup Theorem \ref{twr:subgroup.1} is applicable to the restriction
of the representation ${}^{\mu_1 \times \mu_2}U^{L \times M}$ of $\mathfrak{G} \times \mathfrak{G}$
to the subgroup $\overline{\mathfrak{G}}$. 
By this theorem, ${}^{\mu_1 \times \mu_2}U^{L \times M}$ restricted to 
$\overline{\mathfrak{G}}$ is a direct integral over the space of  double cosets  $(G_1 \times G_2 ) (x,y) \overline{\mathfrak{G}}$
with exactly one representant $(x,y)$ for each
double coset, of the representations $U^{(L \times M)^{(x,y)}}$ of 
the subgroup $\overline{\mathfrak{G}}$ (\emph{i. e.} with $(x,y)$ ranging 
over $\mathfrak{B}_1 \times \mathfrak{B}_2$ -- the corresponding section 
of $\mathfrak{G} \times \mathfrak{G}$ with respect to double cosets 
$(G_1 \times G_2 ) : \overline{\mathfrak{G}}$). 
Each of the representations
$U^{(L \times M)^{(x,y)}}$ of $\overline{\mathfrak{G}}$ is induced by the Krein-unitary representation 
$(L \times M)^{(x,y)}: (s,s) \mapsto (L \times M)_{{}_{(x,y)(s,s)(x,y)^{-1}}}
= L_{xsy^{-1}} \otimes M_{xsy^{-1}}$
of the subgroup $G_{(x,y)} = \overline{\mathfrak{G}} \cap \big( (x,y)^{-1} (G_1 \times G_2) (x,y) \big) 
\subset \overline{\mathfrak{G}}$
in the Krein space $(\mathcal{H}_L \otimes \mathcal{H}_M , \mathfrak{J}_L \otimes \mathfrak{J}_M)$. 
Moreover $U^{(L \times M)^{(x,y)}}$ acts in the Krein space 
$\big( \mathcal{H}^{(L \times M)^{(x,y)}} , \mathfrak{J}_{{}_{(x,y)}} \big)$ where for each function $w \in \mathcal{H}^{(L \times M)^{(x,y)}}$ we have
\begin{multline*}
\big( \mathfrak{J}_{{}_{(x,y)}} w \big)_{{}_{(s,s)}}
= (L \times M)_{h((x,y)\cdot (s,s))} \mathfrak{J}_{L\times M} (L \times M)_{{h((x,y) \cdot (s,s))}^{-1}} \big( w \big)_{{}_{(s,s)}}   \\
= L \times M)_{(h_{{}_1}(xs) , h_{{}_2}(ys))} \mathfrak{J}_{L\times M} 
(L \times M)_{{(h_{{}_1}(xs)^{-1} , h_{{}_2}(ys)^{-1})}} \big( w \big)_{{}_{(s,s)}} \\
= \big( L_{h_{{}_1}(xs)} \,  \mathfrak{J}_L \,  L_{h_{{}_1}(xs)^{-1}} \big) \otimes 
\big( M_{h_{{}_2}(ys)} \, \mathfrak{J}_M \,  M_{h_{{}_2}(xs)^{-1}} \big) \big( w \big)_{{}_{(s,s)}}.
\end{multline*}
The inner product in $\mathcal{H}^{(L \times M)^{(x,y)}}$ and Krein-inner product in
$\big( \mathcal{H}^{(L \times M)^{(x,y)}} , \mathfrak{J}_{{}_{(x,y)}} \big)$ are defined by 
\begin{multline*}
(w,g)_{{}_{(x,y)}} = \int \limits_{\overline{\mathfrak{G}} / G_{(x,y)}} 
\Big( \mathfrak{J}_{L \times M} \big( \mathfrak{J}_{{}_{(x,y)}}  w \big)_{{}_{(s,s)}} , 
\big( g \big)_{{}_{(s,s)}}  \Big)  \,\, \ud (\mu_1 \times \mu_2)^{(x,y)} ([(s,s)])  \\
= \int \limits_{\overline{\mathfrak{G}} / G_{(x,y)}} 
\Big( \mathfrak{J}_L \otimes \mathfrak{J}_M \big( \mathfrak{J}_{{}_{(x,y)}}  w \big)_{{}_{(s,s)}} , 
\big( g \big)_{{}_{(s,s)}}  \Big)  \,\, \ud (\mu_1 \times \mu_2)^{(x,y)} ([(s,s)])
\end{multline*}
and
\begin{multline*}
(w,g)_{{}_{\mathfrak{J}_{{}_{(x,y)}}}} = (\mathfrak{J}_{{}_{(x,y)}} w , g)_{{}_{(x,y)}}   \\
=\int \limits_{\overline{\mathfrak{G}} / G_{(x,y)}} 
\Big( \mathfrak{J}_{L \times M} \big(w \big)_{{}_{(s,s)}} , 
\big( g \big)_{{}_{(s,s)}}  \Big)  \,\, \ud (\mu_1 \times \mu_2)^{(x,y)} ([(s,s)])  \\
= \int \limits_{\overline{\mathfrak{G}} / G_{(x,y)}} 
\Big( \mathfrak{J}_L \otimes \mathfrak{J}_M \big(w \big)_{{}_{(s,s)}} , 
\big( g \big)_{{}_{(s,s)}}  \Big)  \,\, \ud (\mu_1 \times \mu_2)^{(x,y)} ([(s,s)]);
\end{multline*}
with the quasi invariant measure $(\mu_1 \times \mu_2)^{(x,y)} $ on
$\overline{\mathfrak{G}} / G_{(x,y)}$ given by Def. \ref{defin:subgroup.2}.

Now under the natural isomorphism $(x,x) \mapsto x$ transferring $\overline{\mathfrak{G}}$ onto $\mathfrak{G}$ 
the group $G_{(x,y)} = \overline{\mathfrak{G}} \cap \big( (x,y)^{-1} (G_1 \times G_2) (x,y) \big)$
is transferred onto the subgroup $x^{-1}G_1 x \, \, \cap \, y^{-1} G_2 y$ of $\mathfrak{G}$ 
and the homogeneous space $\overline{\mathfrak{G}} / G_{(x,y)}$ with the quasi invariant
measure $(\mu_1 \times \mu_2)^{(x,y)}$ is transferred over to the homogeneous space 
$\mathfrak{G} \big{/} \big( x^{-1}G_1 x \, \, \cap \, y^{-1} G_2 y \big)$ with the quasi
invariant measure, which we denote by $\mu^{x,y}$. 
\qed

\section{Krein-isometric representations induced by decomposable Krein-unitary
representations}\label{decomposable_L}

We say a family $\mathfrak{S}$ of operators in a Hilbert space $\mathcal{H}$ is \emph{reducible}
by an idempotent $P$ (i. e. a bounded operator $P$ which satisfies the identity $P^2 = P$),
or equivalently by a closed subspace equal to the range $P\mathcal{H}$ of $P$, in case 
$PUP = UP$ for all $U \in \mathfrak{S}$. We say the family $\mathfrak{S}$ is \emph{decomposable} 
in case $PU = UP$ for all $U \in \mathfrak{S}$. In this case the Hilbert space $\mathcal{H}$ 
is the direct sum of closed subspaces $\mathcal{H}_1 = P \mathcal{H}$ and 
$\mathcal{H}_2 = (I - P)\mathcal{H}$ and every operator in $\mathfrak{S}$
is a direct sum of operators $U_1$ and $U_2$ with $U_i$ acting in $\mathcal{H}_i$, $i = 1,2$.
The closed subspaces $\mathcal{H}_i$, $i = 1,2$, are orthogonal iff $P$ is self adjoint.
Moreover if $(\mathcal{H}, \mathfrak{J})$ is a Krein space, the closed subspaces
$\mathcal{H}_i$, $i = 1,2$, are Krein-orthogonal iff the idempotent $P$ is Krein-self-adjoint:
$P^\dagger = P$.

Now the Krein-isometric representations $U^L$  inherit
decomposability from decomposability of $L$. Namely for each idempotent $P_L$ acting in the Krein space of the representation $L$ we may define a natural idempotent $P^L$
by the formula $(P^L f)_{{}_x} = P_L f_{{}_x}$ for $f \in \mathcal{H}^L$ provided $P_L$ commutes with the
representation $L$. Checking that $P^L$ is well defined (with measurable $x \mapsto \Big(\, \big(P^L f\big)_{{}_x} \, , \,\, \upsilon \, \Big)$ for each $\upsilon \in \mathcal{H}_L$ and $\big(P^L f\big)_{{}_{hx}} = L_h \big(P^L f\big)_{{}_{x}}$) 
and that $P^L$ is a bounded idempotent is immediate.
Moreover $P^L$ likewise commutes with $U^L$ and is self-adjoint whenever $P_L$ is.

Thus in particular for the standard Krein-isometric representation we have the following

\begin{twr}
Let $H$ be a closed subgroup of the separable locally compact group $\mathfrak{G}$. Let 
$U^L$ be the Krein-isometric representation of $\mathfrak{G}$ acting in the
Krein space $(\mathcal{H}^L , \mathfrak{J}^L)$, induced by the Krein-unitary
representation $L$ of the subgroup $H$, acting in the Krein space $(\mathcal{H}_L , \mathfrak{J}_L)$. 
Let for a measure space $(\mathbb{R}, \mathscr{R}_{\mathbb{R}}, m)$ 
the operators of the representation $L$ and the fundamental symmetry 
$\mathfrak{J}_L$ be decomposable:
\[
L = \int \limits_\mathbb{R} \,\, {}_{{}_\lambda}L \,\,\,\, \ud m (\lambda)
\,\,\,\,\,\,\,\, \textrm{and} \,\,\,\,\,\,\,\,
\mathfrak{J}_L = \int \limits_\mathbb{R} \,\, \mathfrak{J}_{{}_{{}_{{}_\lambda}L}} \,\,\,\, \ud m (\lambda)
\]
with respect to a direct integral decomposition
\begin{equation}\label{hl_dec_h_lambda}
\mathcal{H}_L = \int \limits_\mathbb{R} \,\, \mathcal{H}_{{}_{{}_\lambda}L} \,\,\,\, \ud m (\lambda),
\end{equation}
of the Hilbert space $\mathcal{H}_L$.  Then 
\begin{equation}\label{decomosable_L_dec_H^L}
\mathcal{H}^L = \int \limits_\mathbb{R} \,\, \mathcal{H}^{{}^{{}_{{}_\lambda}L}}  \,\,\,\, \ud m (\lambda)
\end{equation}
and all operators of the representation $U^L$ and the fundamental symmetry $\mathfrak{J}^L$ are
decomposable with respect to (\ref{decomosable_L_dec_H^L}), i. e.
\[
U^L \, = \, \int \limits_\mathbb{R} \,\, U^{{}^{{}_{{}_\lambda}L}}  \,\,\,\, \ud m (\lambda)
\,\,\,\,\,\,\,\, \textrm{and} \,\,\,\,\,\,\,\,
\mathfrak{J}^L \, = \, \int \limits_\mathbb{R} \,\, \mathfrak{J}^{{}^{{}_{{}_\lambda}L}}  \,\,\,\, \ud m (\lambda);
\] 
where $U^{{}^{{}_{{}_\lambda}L}}$ is the Krein-isometric representation in the Krein space 
$\big( \mathcal{H}^{{}^{{}_{{}_\lambda}L}} , \mathfrak{J}^{{}^{{}_{{}_\lambda}L}} \big)$ induced by
the Krein-unitary representation ${}_{{}_\lambda}L$ of the subgroup $H$, acting in the Krein space  
$\big( \mathcal{H}_{{}_{{}_\lambda}L} , \mathfrak{J}_{{}_{{}_{{}_\lambda}L}} \big)$. 
\label{twr:decomposable_L.1}
\end{twr}   
\qedsymbol \, [Outline of the proof.]
Let $\lambda \mapsto E (\lambda)_L$ be the spectral measure associated with the decomposition
(\ref{hl_dec_h_lambda}). Consider the direct integral decompositions 
\begin{multline*}
\mathcal{H}^L = \int \limits_\mathbb{R} \,\, \mathcal{H}^{L}(\lambda)  \,\,\,\, \ud m (\lambda), \\
\mathfrak{J}^L \, = \, \int \limits_\mathbb{R} \,\, \mathfrak{J}^{L}(\lambda)  \,\,\,\, \ud m (\lambda)
\,\,\,\,\,\,\,\, \textrm{and} \,\,\,\,\,\,\,\,
U^L \, = \, \int \limits_\mathbb{R} \,\, U^{L}(\lambda)  \,\,\,\, \ud m (\lambda),
\end{multline*}
of $\mathcal{H}^L$, $\mathfrak{J}^L$ and $U^L$,
associated with the corresponding spectral measure $\lambda \mapsto E (\lambda)^L$ 
and the same measure $m$. Using the vector-valued version of (\ref{dir_int_L^2:decompositions})
and the Fubini theorem one shows that 
$\mathcal{H}^{L}(\lambda) = \mathcal{H}_{{}_{{}_\lambda}L}$ and the equalities of the Radon-Nikodym
derivatives
\begin{multline*}
\frac{\ud \, \Big( \, E(\lambda)^L \mathfrak{J}^L f \, , \,\, E(\lambda)^L g \, \Big)}{\ud m (\lambda)} 
= \frac{\ud \, \Big( \, E(\lambda)^L \mathfrak{J}^{{}^{{}_{{}_\lambda}L}} f \, , \,\, E(\lambda)^L g \, \Big)}
{\ud m (\lambda)},  \\
\frac{\ud \, \Big( \, E(\lambda)^L U^L f \, , \,\, E(\lambda)^L g \, \Big)}{\ud m (\lambda)} 
= \frac{\ud \, \Big( \, E(\lambda)^L U^{{}^{{}_{{}_\lambda}L}} f \, , \,\, E(\lambda)^L g \, \Big)}
{\ud m (\lambda)}, 
\end{multline*}
for all $f, g \in \mathcal{H}^L$ in the domain of $U^L$, which means that 
$\mathfrak{J}^L (\lambda) = \mathfrak{J}^{{}^{{}_{{}_\lambda}L}}$ and $U^L (\lambda) = U^{{}^{{}_{{}_\lambda}L}}$.
\qed

Using the Dunford-Gelfand-Mackey \cite{Dunford, Gelfand_spectral} (or more general \cite{Foias,Lanze}) spectral measures and corresponding decompositions, we could generalize the last theorem
keeping Krein self adjointness of the idempotents of the decomposition of $L$ just using Dunford 
or more general spectral measures), 
but abandoning their commutativity with $\mathfrak{J}_L$, and thus discarding their self-adjointness.  

\vspace*{0.5cm}

But in decomposition of the Krein-isometric induced representation restricted to a closed subgroup as
in the Subgroup Theorem (or respectively in decomposition of the tensor product of Krein-isometric 
induced representations as in the Kronecker Product Theorem) we have encountered Krein-isometric induced representations 
$U^{L^{x}}$ in the Krein space  $(\mathcal{H}^{L^{x}}, \mathfrak{J}_{{}_x})$ 
with the non-standard fundamental symmetry $\mathfrak{J}_{{}_x}$ instead of the standard one
$\mathfrak{J}^{L^x}$ 
(respectively their tensor product $U^{N^{x,y}}$ acting in the tensor product Krein space 
$\big(\mathcal{H}^{N^{x,y}}, \mathfrak{J}{{}_{x,y}}\big) = 
\big(\mathcal{H}^{L^{x}} \otimes \mathcal{H}^{M^{y}}, \mathfrak{J}_{{}_x} \otimes \mathfrak{J}_{{}_y} \big)$). 
In this case for each idempotent $P_L$ acting in the representation space $(\mathcal{H}_L , \mathfrak{J}_L)$
and commuting with $L^x$ we could similarly define the corresponding operator $P^L$:
$(P^L f)_{{}_x} = P_L f_{{}_x}$ for $f \in \mathcal{H}^{L^{x}}$. (Similarly we can define $P^{{}^{N^{x,y}}}$
for each idempotent $P_{{}_{N^{x,y}}}$ commuting with $N^{x,y}$.) However in this case with non-standard fundamental
symmetry $\big( \mathfrak{J}_{{}_{x}} g \big)_{{}_{s}} \, = \, L_{h_{{}_1}(xs)} \,  \mathfrak{J}_L \,  L_{h_{{}_1}(xs)^{-1}} \big( g \big)_{{}_{s}}$, $g \in \mathcal{H}^{L^x}$ (resp. $\big( \mathfrak{J}_{{}_{x,y}} w \big)_{{}_{s}}
= \big( L_{h_{{}_1}(xs)} \,  \mathfrak{J}_L \,  L_{h_{{}_1}(xs)^{-1}} \big) \otimes 
\big( M_{h_{{}_2}(ys)} \, \mathfrak{J}_M \,  M_{h_{{}_2}(ys)^{-1}} \big) \big( w \big)_{{}_{s}}$, $w \in 
\mathcal{H}^{N^{x,y}}$) the operator $P^L$ (or $P^{{}^{N^{x,y}}}$) is in general
\emph{unbounded}. Moreover $P^L$ (resp. $P^{{}^{N^{x,y}}}$) is non 
self-adjoint in this case even if $P_L$ (resp. $P_{{}_{N^{x,y}}}$) is self-adjoint.
We hope the slightly misleading (unjustified) notation $U^{N^{x,y}}$ will cause no serious troubles. 

Thus in particular the Theorem \ref{twr:decomposable_L.1} (and its generalizations with Dunford-Gelfand-Mackey
spectral measure decompositions) cannot in general be immediately applied to the representations
$U^{N^{x,y}}$ standing in the Kronecker Product Theorem for the tensor product of {\L}opusza\'nski
representations of the double covering $\mathfrak{G} = T_4 \circledS SL(2,\mathbb{C})$
of the Poincar\'e group. But $U^{N^{x,y}}$ as a Krein-isometric representation
of the semi-direct product $\mathfrak{G} = T_4 \circledS SL(2,\mathbb{C})$ defines an imprimitivity
system in Krein space (in the sense of Sect \ref{lop_ind}) which is concentrated on a single orbit.
We then restore the form of the ordinary induced representation to $U^{N^{x,y}}$
by applying Theorem \ref{lop_ind:twr.1} of Sect. \ref{lop_ind}  but we need the generalized version 
of this theorem with the finite multiplicity condition 3) discarded. We hope to present in a subsequent 
paper the full analysis of the component representations $U^{N^{x,y}}$ in the decomposition of tensor 
product of {\L}opusza\'nski representations.

The necessity of restoring the standard form of induced Krein-isometric representation to the component 
Krein-isometric representations in the decomposition
of tensor product of standard induced Krein-isometric representations is the main difference in comparison
to the Mackey theory of unitary induced representations. In case of the double covering $\mathfrak{G}$ 
of the Poincar\'e group this ``restoring'' is quite elaborate, but effectively computable. 
The case of tensor products of ordinary unitary induced representations may be rather effectively reduced to the harmonic
analysis on ``small groups'' $G_{\chi_p} = SU(2,\mathbb{C})$ or $SL(2,\mathbb{R})$ (see Sect. \ref{lop_ind}; for the harmonic analysis on $SL(2,\mathbb{C})$ compare 
\cite{Harish-Chandra,Ehrenpreis_Mautner1,Ehrenpreis_Mautner2,Ehrenpreis_Mautner3}.) 
and to the tensor products (computed in \cite{nai1,nai2,nai3}) 
of Gelfand-Neumark representations of $SL(2,\mathbb{C})$, with the help of the original Mackey's
Subgroup and Kronecker Product Theorems for unitary induced representations. Indeed for tensor products
of integer spin representations (for both versions of the energy sign) these decompositions
have indeed been computed by Tatsuuma \cite{tat}. Unfortunately the paper \cite{tat} presents only 
the results without proofs, and some of the results presented there are not correct, namely those
under X).  

\begin{rem}
Because the representation of the translation subgroup $T_4 \subset T_4 \circledS SL(2, \mathbb{C})$
in  {\L}opusza\'nski-type representation is equivalent to the representation of $T_4$ in 
direct sum of several 
(four in case of the  {\L}opusza\'nski representation) representations of, say helicity zero, ordinary unitary
induced representations of $T_4 \circledS SL(2, \mathbb{C})$, corresponding to the 
``light-cone orbit'' in the momentum space, and the representation of $T_4 \circledS SL(2, \mathbb{C})$ in 
the Fock space is the direct sum of symmetrized/anitisymmetrized
tensor products of one-particle representations, then investigation of the multiplicity of the representation of
$T_4$ in the Fock space is reduced to the decomposition of tensor products of ordinary unitary induced 
representations of $T_4 \circledS SL(2, \mathbb{C})$. 
\label{decomposable_L:uniform_mult}
\end{rem}

\vspace*{0.5cm}

{\bf ACKNOWLEDGEMENTS}

\vspace*{0.3cm}

The author is indebted for helpful discussions to prof. A. Staruszkiewicz.  
The author would especially like to thank prof. A. K. Wr\'oblewski, prof. A. Staruszkiewicz, 
and prof. M. Je\.zabek for the warm encouragement. I would like to thank prof. M. Je\.zabek 
for the excellent conditions for work at INP PAS where the most part of this paper 
has come into being.

\end{document}